\documentclass[german,12pt]{book}
\usepackage{axodraw}
\usepackage{amsfonts}
\usepackage[all]{xy}
\usepackage{babel}
\usepackage{a4wide}  
\usepackage{epsfig}
\newcommand{\mt}{M_\tau}
\newcommand{\mts}{M_\tau^2}
\newcommand{\ntb}{\bar{\nu}_\tau}
\newcommand{\nt}{\nu_\tau}

\newcommand{\Lg}{\langle \tau(p) \bar{\nu}_\tau(k) | S|  \bar{\nu}_\tau(k^\prime) \tau(p^\prime) \rangle}
\newcommand{\nn}{\nonumber}
\newcommand{\be}{\begin{equation}}
\newcommand{\ee}{\end{equation}}
\newcommand{\ba}{\begin{eqnarray}}
\newcommand{\ea}{\end{eqnarray}}
\newcommand{\als}{\frac{\alpha_s}{\pi}}
\newcommand{\alsb}{\left( \frac{\alpha_s}{\pi} \right) }
\newcommand{\alsbup}{\left( \frac{\alpha_s(\mu^\prime)}{\pi} \right) }
\newcommand{\MSsch}{\overline{\rm{MS}}}
\newcommand{\shp}{p \hspace{-2mm}{/}}
\newcommand{\shk}{k \hspace{-2mm}{/}}
\newcommand{\shq}{q \hspace{-2mm}{/}}
\newcommand{\shd}{D \hspace{-2.5mm}{/}}
\newcommand{\ep}{\epsilon}
\newcommand{\vr}{\varrho}
\newcommand{\ind}{{(\mu_1 \dots \mu_n)}}
\newcommand{\al}{\alpha}
\newcommand{\pr}{\prime}
\newcommand{\no}{$\rm{N}^3\rm{LO}$}

\newcommand{\intd}{ \int \!\!\!\! \int}
\newcommand{\bbbone}{\hbox{\rm 1\kern-3pt l}}

\begin{document}
\frontmatter

\thispagestyle{empty}
\begin{center}
\begin{huge}
\begin{bf}

QCD-Analyse von hadronischen $\tau$-Lepton-Zerf"allen


\end{bf}
\end{huge}

\vspace{3cm}

\begin{Large}
Diplomarbeit am \\
Institut f\"ur Physik \\
Johannes Gutenberg-Universit\"at \\
Mainz \\
\vspace{2cm}
Florian Krajewski\\
\vspace{4cm}
Betreuer: Prof. Dr. J.G. K"orner\\

\vspace{3cm}
Mainz, Juni 2000
\end{Large}
\end{center}


\tableofcontents
\thispagestyle{empty}
\mainmatter

\chapter{Einleitung}
Das Studium 
der Zerf"alle von $\tau$-Leptonen
bietet eine 
Vielzahl von Informationen "uber die niedrigenergetische 
Hadronphysik, wobei sich die Genauigkeit der experimentellen Daten 
in diesem Gebiet st"andig verbessert \cite{PDG,exp1al,exp1ms,exp2}. 
Die zentrale Gr"o"se ist hierbei die hadronische Spektraldichte 
bzw. die mit ihr verbundene Korrelationsfunktion,
die mit einem gro"sen Grad an Genauigkeit im Rahmen
der St"orungstheorie berechnet wurde.
Die Observablen des $\tau$-Systems sind durch Momente der Spektraldichte 
\cite{cont,cont1,cont2} gegeben.
Sie wurden im Rahmen der Operator-Produkt-Entwicklung (OPE)
in den letzten Jahren intensiv untersucht 
\cite{SchTra84,Bra88,Bra89,NarPic88,BraNarPic92}.
Dies macht die 
Physik der
$\tau$-Leptonen zu einem wichtigen 
Gebiet der ph"anomenologischen Elementarteilchenphysik,
in dem die theoretischen Vorhersagen der 
Quantenchromodynamik (QCD) mit experimentellen Daten 
bei hoher Pr"azission konfrontiert werden k"onnen.

In dieser Arbeit wird die St"orungsreihe von 
$\tau$-Lepton-Observablen 
in endlicher Ordnung (FOPT\footnote{FOPT steht f"ur ``finite order 
perturbation theory''}) untersucht.
Es zeigt sich,
dass sowohl f"ur die Cabibbo-bevorzugten
Zerf"alle $(s= 0)$ als auch f"ur die 
Strange-Quarkmassen-Korrektur der Cabibbo-unterdr"uckten 
Zerf"alle $(s = 1)$ die ultimative Genauigkeit
der St"orungsreihe aufgrund ihrer asymptotischen 
Struktur bereits erreicht ist \cite{one,two}.
Diese Schlussfolgerung ist in einer
vom Renormierungsschema unabh"angigen Art und Weise m"oglich.
Will man theoretische Vorhersagen erzielen,
die in ihrer Genauigkeit mit den heute schon 
vorhandenen experimentellen Daten der Cabibbo-bevorzugten
Zerf"alle und den in naher Zukunft verf"ugbaren
Daten der Cabibbo-unterdr"uckten    
Zerf"alle vergleichbar sind, so ist eine 
Neuinterpretation der QCD-St"orungsreihen
unumg"anglich.
Zwei Probleme von besonderem Interesse sind hierbei 
eine genauere Bestimmung der starken Kopplungskonstanten 
$\al_s$ und der Strange-Quarkmasse $m_s$ \cite{three,four}.

Die Bestimmung des nummerischen Wertes
der starken Kopplungskonstante bei niedrigen Energien
wie der $\tau$-Lepton-Masse erm"oglicht einen 
beeindruckenden Test der QCD. 
Der bei niedrigen Energien bestimmte Wert der 
starken Kopplungskonstante kann mit Hilfe des
Formalismus der Renormierungsgruppe (RG) auf gro"se 
Energien "ubertragen werden.  
So kann dieser Wert mit aus Hochenergieexperimenten 
bestimmten Werten f"ur $\al_s$ verglichen werden.
Auf diese Weise l"asst sich die G"ultigkeit der
QCD in einen Energiebereich von einem bis zu einigen 
hundert GeV auf ihre Konsistenz testen.
Aus theoretischer Sicht erwartet man 
aufgrund der gro"sen Anzahl bereits berechneter Terme der 
St"orungsreihe einen hohen 
Grad an Genauigkeit bei der Bestimmung der starken 
Kopplungskonstante $\al_s$
durch $\tau$-Zerf"alle.
Da jedoch der nummerische Wert der starken 
Kopplungskonstante im Energiebereich der 
$\tau$-Lepton-Masse nicht klein ist, sind die Beitr"age 
von Termen h"oherer Ordnung in $\al_s$ entscheidend.
Hier zeigt sich die asymptotische Struktur der St"orungsreihe,
welche die Genauigkeit der Kopplungskonstante
$\al_s$ beschr"ankt. In dieser Arbeit wird ein neues Verfahren
zur Bestimmung der starken Kopplungskonstante aus der 
St"orungsreihe vorgeschlagen. Ein entscheidender Vorteil dieses 
Verfahrens ist die Unabh"angigkeit der Resultate von dem
verwendeten Renormierungsschema.

Auch die bei der Bestimmung der Strange-Quarkmasse $m_s$ 
auftretende St"orungsreihe zeigt bereits in den heute
bekannten Termen asymptotischen Charakter.
Dies macht auch bei diesem Problem eine Neuinterpretation 
der St"orungsreihe notwendig. Die Wahl eines Resummationsverfahrens,
mit dem aus einer divergenten Reihe ein nummerischer Wert extrahiert 
werden kann, ist nicht eindeutig
und es gibt viele M"oglichkeiten, eine asymptotische
Reihe zu resummieren oder ihr Konvergenzverhalten
zu verbessern \cite{renRS, pivrho}.
Das in dieser Arbeit f"ur die Bestimmung von $m_s$ 
verwendete Resummationsverfahren wurde nach 
folgenden Kriterien ausgew"ahlt:
Das Renormierungsschema soll dem physikalischen Problem angemessen 
und so einfach wie m"oglich sein.
Ausserdem muss die durch die Renormierungsgruppe gegebenen Struktur 
der St"orungsreihe ber"ucksichtigt werden. Aus diesem Grund wird 
f"ur die Bestimmung der Strange-Quarkmasse die sogenannte 
konturverbesserte St"orungstheorie (CIPT\footnote{CIPT steht f"ur ``contour 
improved perturbation theory''}) \cite{Pivtau,DP}
verwendet und ein effektives  
Renormierungsschema 
(nicht das ``modified minimal substraction''-($\MSsch$)-Schema)
mit effektiver Ladung und Massen verwendet, welches 
f"ur das $\tau$-System nat"urlich ist.

Die vorliegende Arbeit gliedert sich folgenderma"sen: 
Im ersten 
Teil werden die f"ur die Analyse der St"orungsreihen 
notwendigen Aspekte der 
QCD als renormierbare Quantenfeldtheorie wiederholt und die 
durch die Renormierungsgruppe gegebene Struktur der St"orungsreihe 
beschrieben. 
Au"serdem werden die f"ur die Beschreibung von 
$\tau$-Lepton-Zerf"allen notwendigen theoretischen 
Grundlagen hergeleitet, 
wie der Ausdruck f"ur die relative 
$\tau$-Lepton-Zerfallsrate
und die Eigenschafen der Korrelationsfunktion zweier Hadronstr"ome. 
Insbesondere sind die f"ur die pertubative Berechnung der 
Korrelationsfunktion zweier Hadronstr"ome notwendigen 
Techniken -- das Berechnen von masselosen Graphen mit 
Gegenbauer-Polynomen (GPXT\footnote{GPXT steht f"ur ``Gegenbauer polynomial 
X space technique''}) sowie die Methode der partiellen Integration --  
beschrieben.

Im zweiten Teil werden die im Rahmen meiner Arbeit neu erzielten 
Ergebnisse vorgestellt.
Die St"orungsreihen f"ur $\tau$-Lepton-Observable
aus Cabbibo-bevorzugten-Zerf"allen und die 
St"orungsreihen der bei
Cabbibo-unterdr"uckten Zerf"allen notwendigen 
$m_s^2$-Korrektur aufgrund der Strange-Quarkmasse
werden auf ihr asymptotisches Verhalten analysiert.
In den beiden letzten Kapiteln werden die starke Kopplungskonstante
$\al_s$ und die Strange-Quarkmasse $(m_s)$ mit den oben erw"ahnten Verfahren 
bestimmt.

\chapter{Quantenchromodynamik (QCD)} \label{QCDkap}
\section{QCD - Die Theorie der starken Wechselwirkung}
Die Quantenchromodynamik (QCD) ist eine renormierbare Quantenfeldtheorie, 
die die starke
Wechselwirkung beschreibt.
Ein guter "Ubersichtsartikel "uber QCD ist \cite{marc}. 
Ihre fundamentalen Felder sind die Spin-$1/2$-Felder der 
Quarks, welche gebrochene elektrische Ladungen tragen 
und ein nicht-abelsches Spin-$1$-Eichfeld,  
das die Gluonen beschreibt. Das Gluonfeld wechselwirkt mit den Quarkfeldern,
mit zur Erhaltung der Unitarit"at 
der $S$-Matrix im Quantisierungsprozess einzuf"uhrenden
Geistfeldern  und mit sich selbst. 
Die Quarkfelder werden durch Dirac-Spinoren $q_{\alpha i}^A$ beschrieben. 
$\alpha$ bezeichnet den 
Spinorindex und wird mit der $4$-dimensionalen Spinordarstellung 
der Lorenzgruppe transformiert.
$i$ bezeichnet das Flavour des Quarkfeldes. 
In der QCD besitzt das Flavour keine dynamische Funktion,
sondern spielt nur die Rolle einer weiteren Quantenzahl.
Zur Zeit sind sechs verschiedene Flavours bekannt, die in drei
Generationen eingeteilt werden 
$((u ,d),\;(s,c),\;(b,t))$.
Nimmt man eine Symmetrie 
zwischen den Generationen der Leptonen und Quarks an, 
so l"asst sich 
die Existenz einer vierten Quarkgeneration mit gro"ser 
Wahrscheinlichkeit ausschlie\ss en. 
Experimentell kann kein Zerfallskanal des $Z_0$-Bosons 
in ein Neutrino-Antineutrino-Paar
der vierten Generation 
nachgewiesen werden. 
Dieses m"usste somit mehr als die halbe Masse des $Z_0$-Bosons 
besitzen\footnote{An dieser Stelle sei 
Prof. Dr. K. Jakobs f"ur ein hilfreiches Gespr"ach gedankt.}.
Der Index $A$ an den Quarkfeldern steht f"ur den Farbfreiheitsgrad. 
Es wir angenommen, dass es genau drei verschiedene 
Farbladungen gibt, die nach den Grundfarben mit  rot, 
gr"un und blau bezeichnet werden. 
F"ur genau drei Farbladungen sprechen verschiedene experimentelle Tatsachen, 
von denen hier 
zwei genannt werden sollen:
\begin{enumerate}
\item Baryonen sind gebundene Zust"ande aus drei Quarks. 
Da Quarks Fermionen sind, m"ussen sie 
dem Pauli-Prinzip Rechnung tragen und eine 
antisymmetrische Wellenfunktion bilden. 
Das $\Omega^-$
ist der gebundene Grundzustand aus drei Strange-Quarks mit Gesamtspin $3/2$. 
Das bedeutet, dass die
Spin-Wellenfunktion vollst"andig symmetrisch ist. 
Ausserdem wird angenommen, dass sich die 
Quarks in einem symmetrischen $s$-Zustand befinden. 
Will man an Quarks festhalten, welche 
das Pauliprinzip erf"ullen, so ist es notwendig, 
eine weitere Quantenzahl einzuf"uhren, 
bez"uglich derer die Wellenfunktion des $\Omega^-$ antisymmetrisch ist.
Diese Quantenzahl wird aus anschaulichen "Uberlegungen heraus als 
Farbquantenzahl bezeichnet. 
Dies und die experimentelle
Tatsache, dass Hadronen mit Farbladung in der 
Natur nicht beobachtet werden, f"uhrt zu der Annahme,
dass Quarks eine Farbladung besitzen und Hadronwellenfunktionen 
total antisymmetrisch in der 
Farbwellenfunktion sind. 
\item F"ur gro"se Energien und abseits der 
nichtperturbativen Resonanzen erwartet man bei einer 
asymptotisch freien Theorie wie der QCD, 
dass die Effekte der Wechselwirkung gering sind.
F"ur den $e^+e^-$-Vernichtungs-Wirkungsquerschnitt in 
Hadronen ist das Partonmodell unter diesen 
Bedingungen eine gute Approximation. 
F"ur das Verh"altnis des totalen Wirkungsquerschnittes der 
$e^+e^-$-Vernichtung in Hadronen und Myonen erh"alt man 
\be \label{eecross}
R = \frac{\sigma(e^+ e^- \rightarrow \rm{Hadronen})}
       {\sigma(e^+ e^- \rightarrow \mu^+ \mu^-)}
= N_c \sum_f^{n_f} Q_f^2\;,
\ee
wobei $N_c$ die Anzahl der Farbfreiheitsgrade und 
$Q_f$ die elektromagnetische Ladung der Quarks
in Einheiten der Elementarladung 
bezeichnet. Ein 
Vergleich mit experimentellen Befunden f"uhrt zu $N_c= 3$.       
\end{enumerate}
Wegen $N_c=3 $
sind die einzigen m"oglichen einfachen Lie-Gruppen, 
die als Eichgruppen f"ur die Farbladung in Frage kommen,
$SO(3)$ und $SU(3)$.
Es stellt sich heraus, 
dass Eichtheorien mit Eichgruppe $SO(3)$ und mehr als 2 Flavours keine 
asymptotische Freiheit besitzen 
$( \beta_0^{SO(3)}=11/3 - 4/3 n_f )$ \cite{beta4}, 
wogegen Eichtheorien mit Eichgruppe $SU(3)$ 
mit bis zu 16 Flavours 
noch asymptotisch frei sind $(\beta_0^{SU(3)} = 11 - 2/3 n_f)$. 
Ausserdem w"urde 
in einer Theorie mit Eichgruppe  
$SO(3)$ ein Zweiquarksystem als Farbsingulett existieren, welches
nicht beobachtet wird.

Die Dynanik der Wechselwirkung 
wird durch die Forderung nach lokaler Eichsymmetrie bestimmt. 
Um dies zu erreichen, muss ein 
Vektorfeld $(A_\mu^a , \; a=1,2,\dots 8)$ eingef"uhrt werden,
das sich mit der adjungierten Darstellung der $SU(3)$ transformiert. 
Als lokal eichinvariante Lagrangefunktion ergibt sich
\be\label{minimalL}
\mathcal{L}_{\rm{Klassisch}} 
= -\frac{1}{4} G^a_{\mu \nu} G_a^{\mu \nu} + \sum_f \bar{\Psi}_f (i \shd-m_f)  \Psi_f\;.
\ee
$\Psi_f$ bezeichnet das Farbtripel aus 
Quarkspinoren $(q_{\alpha f}^R,q_{\alpha f}^G,q_{\alpha f}^B)$ mit 
Flavour $f$, 
$ G^a_{\mu \nu}$ den Gluon-Feldst"arketensor,
\be
 G^a_{\mu \nu} =  
   \partial_\mu A_\nu^a - \partial_\nu A_\mu^a + g  f^{abc} A_\mu^b A_\nu^c 
\ee
 und $D_\mu$
die kovariante Ableitung,
\be
D_\mu =  \partial_\mu - i g  A_\mu^a \frac{\lambda^a}{2} \; .
\ee
$\lambda^a$ sind die infinitesimalen Generatoren der $SU(3)$. 
Sie erf"ullen die Kommutator-Relation
$[ \lambda^a,\lambda^b] =  2if^{abc} \lambda^c $, 
wobei $ f^{abc}$ die Strukturkonstanten der $SU(3)$
sind. 
Der erste Term in Gl.~(\ref{minimalL}) ist die Yang-Mills-Lagrangedichte, 
in der die Selbstwechselwirkung
der Eichfelder enthalten ist. 
Die St"arke der Wechselwirkungen in~(\ref{minimalL}) wird durch eine
einzige Kopplungskonstante bestimmt. 
Es ist nicht m"oglich, verschiedene Kopplungskonstanten 
einzuf"uhren, da die Eichgruppe eine einfache Liegruppe ist.
Die Lagrangefunktion~(\ref{minimalL}) 
ist invariant unter der lokalen Eichtransformation  
\ba
A_\mu(x) = A_\mu^a(x) \frac{\lambda^a}{ 2} &\rightarrow& U(x) A_\mu(x) U^{-1}(x) 
         + \frac{i}{g} U(x) \partial_\mu  U^{-1}(x) \nn \\
{\rm und \qquad \qquad \qquad}\Psi(x) &\rightarrow & U(x)\Psi(x) \nn \\
{\rm wobei \qquad \qquad \quad }U(x) &=& \exp \left( i \theta^a(x) 
  \frac{\lambda^a}{2} \right) \;. 
\ea
Die Parameter $ \theta^a(x)$ der lokalen $SU(3)$ 
Transformation sind Funktionen der Raumzeit.
Bei der Quantisierung der Theorie stellt sich heraus, 
dass die Lagrangedichte~(\ref{minimalL}) noch nicht
vollst"andig ist. 
Es m"ussen zwei weitere Terme $\mathcal{L}^{GF} $ (Eichfixierung)
und  $\mathcal{L}^{FP} $ (Fadeev-Popov-Counterterm)  hinzugef"ugt
werden, um die Unitarit"at der $S$-Matrix zu gew"ahrleisten.
\be
\mathcal{L}^{GF} = - \frac{1}{2 \zeta} (\partial_\mu A^\mu )^2 
\ee
\be
\mathcal{L}^{FP} = (\partial^\mu \bar{C}^a) \nabla_\mu^{ab} C^b
\ee
Hier bezeichnet $\nabla^{ab}_\mu$ die adjungierte kovariante Ableitung
\ba
\nabla^{ab}_\mu &=& \delta^{ab} \partial_\mu - g  f^{abc} A_\mu^c \;.
\ea 
Die Terme $\mathcal{L}^{GF}$ und $\mathcal{L}^{FP}$ 
zerst"oren die lokale Eichinvarianz der 
Theorie. 
An die Stelle der lokalen Eichinvarianz tritt eine neue Symmetrie, 
die BRS-Symmetrie. 
Die mit dieser 
Symmetrie verbundenen Relationen zwischen den Greenfunktionen 
sind die Slavnov-Taylor-Identit"aten, die f"ur den
Beweis der Renormierbarkeit der Theorie eine zentrale Rolle spielen.    
Die volle Lagrangedichte der QCD ist somit
\be \label{QCDlag}
\mathcal{L}_{QCD} 
= -\frac{1}{4} G^a_{\mu \nu} G_a^{\mu \nu} + \sum_f \bar{\Psi}_f (i \shd -m_f) \Psi_f
             + \mathcal{L}^{GF} + \mathcal{L}^{FP} \; .
\ee
\section{Renormierung}
An der Lagrangefunktion~(\ref{QCDlag}) lassen sich die Feynman-Regeln 
(vgl. Anhang~\ref{Frules})
der QCD
ablesen \cite{pessch}. 
Mit ihnen ist es m"oglich, 
Greenfunktionen f"ur Quark-Gluon-Prozesse perturbativ zu berechnen. 
Es zeigt sich, 
dass in der niedrigsten Ordnung die Parton-Modell-Resultate 
reproduziert werden.
Die Baumgraphenn"ahrung der QCD stellt 
im Fall der $\tau$-Lepton Zerf"alle gerade das Partonmodell dar.
Die eigentlichen Auswirkungen der 
Wechselwirkung werden erst durch Strahlungskorrekturen 
ber"ucksichtigt. 
Die Baumgraphenn"ahrung einer Quantenfeldtheorie ist eigentlich noch eine 
klassische Theorie, da sie in der Entwicklung des Erzeugendenfunktionals f"ur 
Einteilchen-irreduzible 
Greenfunktionen dem Term der Ordnung $(\hbar)^0$ entspricht. 
Die eigentlichen Quanteneffekte werden gerade durch Feynman-Graphen mit
Schleifen beschrieben, da die Anzahl der Schleifen der Ordnung  
im Planckschen Wirkungsquantum $\hbar$ entspricht,
von der die Korrektur zu dem 
Erzeugendenfunktional f"ur Einteilchen-irreduzible Greenfunktionen ist.
Bei der  Integration "uber interne Schleifenimpulse 
treten bei gro"sen Impulsen Divergenzen auf   
(UV-Divergenzen). 
Zus"atzlich entstehen in Theorien, die masselose Felder enthalten,
Divergenzen bei kleinen Impulsen 
(IR Divergenzen). 
Um diesen divergenten Ausdr"ucken eine Bedeutung zu geben,  
wurde das Verfahren der Renormierung entwickelt.

In dieser Arbeit soll die allgemeine Theorie der 
Renormierung nicht behandelt werden, 
sondern es sollen kurz die Grundideen und ihre 
Konsequenzen beschrieben werden, 
so dass die Basis f"ur
das folgende Kapitel "uber die  
Renormierungsgruppe (RG) gelegt wird.
Au"serdem sollen Methoden erkl"art werden, mit denen die 
f"ur die Beschreibung des $\tau$-Systems notwendigen perturbativen Ausdr"ucke
berechnet werden, insbesondere die Regularisierung von masselosen Graphen mit Hilfe der dimensionalen 
Regularisierung. 

Im Folgenden wird die Methode der multiplikativen Renormierung beschrieben.
Um mit der Lagrangefunktion~(\ref{QCDlag}) 
endliche Greenfunktionen zu berechnen, m"ussen 
zun"achst die in den Integrationen "uber die 
Schleifenimpulse auftretenden Divergenzen parametrisiert werden.
Das geschieht durch die sogenannte Regularisierung. 
Das Standardverfahren hierf"ur ist die  
dimensionale Regularisierung, 
in der die Raumzeitdimension des Schleifenimpulses als Regulator dient.
Der Vorteil dieser Regularisierung gegen"uber anderen Methoden ist, 
dass durch sie keine 
Symmetrien der QCD zerst"ort werden und sich 
UV- und IR-Divergenzen gleichzeitig regularisieren lassen.
Die Divergenzen der Integrale treten in den 
dimensional regularisierten Integralen als Pole in der 
Raumzeitdimension bei $d=4$ auf. 
Neben der dimensionalen Regularisierung 
gibt es noch eine Reihe weiterer Regularisierungsverfahren
(Impuls Cut-Off, Pauli-Villars-Regularisierung, 
analytische Regularisierung, Gitter-Regularisierung),
die aber Symmetrien der Theorie zerst"oren. 

Aus den regularisierten Greenfunktionen erh"alt man endliche Funktionen, 
indem die Felder und Parameter
der Theorie renormiert werden. 
Die nackten Felder und Parameter werden durch renormierte ersetzt, 
so dass die Greenfunktionen,
ausgedr"uckt durch die renormierten Gr"o\ss en, endlich sind. 
Die nackten Felder $A_{B\mu}^a$, $C_B^a$, 
und $\Psi_B$ werden durch die renormierten Felder ausgedr"uckt 
\be
A_{B \mu}^a = (Z_3)^{1/2} A_{R \mu}^a \; ,\quad  
C^a_B = ( \tilde{Z}_3)^{1/2} C^a_R \; ,\quad
\Psi_B = (Z_2)^{1/2} \Psi_R \; 
\ee
und die nackten Parameter $g_B$, 
$\alpha_B$ und $m_B$ durch die renormierten  Parameter
\be
g_B =  Z_{\al_s}^{1/2} g_R \; , \quad
\zeta = Z_\zeta  \zeta_R \; \quad
m = Z_m m_R \; .
\ee
Die Renormierungskonstanten 
k"onnen in der Kopplungskonstanten entwickelt werden.
Falls die Schleifenintegrale dimensional regularisiert werden, haben die 
Renormierungskonstanten die folgende Form:
\be
Z = 1 + \sum_{i,j}^{0<j \le i} Z_{ij} 
\left( \frac{\al_s}{\pi} \right)^i \frac{1}{\ep^j} \; 
\quad \al_s := \frac{g^2}{4 \pi} \; .
\ee
$\ep$ beschreibt die Abweichung der Dimension von 4 $(d = 4 - 2 \ep)$.
Ausgedr"uckt durch die renormierten 
Felder und Parameter, sind die Greenfunktionen endlich.
Durch diese Forderung sind die $Z_{ij}$ 
allerdings noch nicht eindeutig bestimmt, 
da a priori nicht klar ist, 
wie der divergente Anteil einer  Greenfunktion 
festgelegt wird, welcher durch Multiplikation  
mit den Renormierungsparametern subtrahiert werden muss.
Diese Freiheit 
bez"uglich der Hinzunahme endlicher Anteile  
wird am Ende dieses 
Kapitels und in dem Kapitel~\ref{rg} "uber die Renormierungsgruppe weiter 
erl"autert.
\section{Dimensionale Regularisierung}
Die Idee der dimensionalen Regularisierung besteht darin, die Integrale 
als analytische Funktion der 
Raumzeitdimension f"ur Dimensionen zu berechnen, in denen die Integrale 
konvergent sind, und diese Resultate analytisch fortzusetzen.
Die Dimension der 
Kopplungskonstante in der QCD-Lagrangedichte~(\ref{QCDlag}) ergibt sich zu 
\be
d_g = d - 2 
\left( \frac{d-1}{2}  \right)- \left(\frac{d-2}{2}\right) 
= \frac{1}{2} (4-d) =: \ep \;,
\ee
da die Wirkung dimensionslos ist und da 
Fermionfelder die kanonische Dimension $(d-1)/2$
und Bosonfelder die kanonische Dimension $(d-2)/2$ besitzen, 
was man an den kinetischen Termen der
Lagrangedichte ablesen kann. 
Um f"ur beliebige Raumzeitdimensionen die Dimensionslosigkeit 
der Kopplungskonstante zu erhalten, 
f"uhrt man eine Massenskala $\mu$ ein, die 't \nolinebreak Hooftsche 
Renormierungsskala, und macht in der Lagrangedichte die Ersetzung
\be
g \rightarrow  \mu^\ep g \;,
\ee
wobei $\ep = (4 - d)/2$ ist.
In der so modifizierten Lagrangedichte 
bleibt die Kopplungskonstante f"ur beliebige 
Raumzeitdimensionen dimensionslos. 
Durch die Verwendung der Methode der dimensionalen Regularisierung 
tritt also die Notwendigkeit auf, 
einen weiteren Parameter in die Theorie einzuf"uhren,
die Renormierungsskala $\mu$. 
Im n"achsten Kapitel "uber die Renormierungsgruppe wird sich 
zeigen, dass die Freiheit in der Wahl 
der Massenskala $\mu$ eine Untergruppe der 
Renormierungsgruppe parametrisiert, wobei die Renormierungsgruppe der Freiheit
in der Wahl des Renormierungsschemas entspricht.
\subsection{Einfaches Beispiel f"ur die dimensionale Regularisierung} 
Um 
Integrale in dem Rahmen der dimensionalen Regularisierung berechnen zu
k"onnen, ist es notwendig, 
zun"achst eine Wick-Rotation durchzuf"uhren, 
um im euklidischen Raum rechnen zu k"onnen.
Als Beispiel \cite{pessch} kann das Integral 
\be \label{dimregbsp}
\int \frac{d^d \!k}{(2 \pi)^d} \frac{k^{2 \alpha}}{(k^2 + \Delta )^\beta }= 
       \frac{ \Omega_d}{(2 \pi)^d} 
\int_0^\infty dk \frac{k^{d-1 + 2 \alpha}}{(k^2 +\Delta)^\beta} 
\ee
verwendet werden.
In Gl.~(\ref{dimregbsp}) wurde die Winkelintegration bereits ausgef"uhrt, 
was einfach ist, da
der Integrand nur vom Betrag der Integrationsvariablen abh"angt
\footnote{
\begin{samepage}
Die Fl"ache der d-dimensionalen Einheitssph"are l"asst sich folgenderma\ss en 
bestimmen \cite{pessch}:
\ba
(\pi)^{d/2} &=& 
\left( \int dx e^{-x^2} \right)^d =
      \int d^d\!x \exp \left(-\sum_{i=1}^d x_i^2 \right) \nn \\
&=& \Omega_d \int_0^\infty dx \, x^{d-1} e^{-x^2} =
  \Omega_d \frac{1}{2} 
    \int_0^\infty d(x^2) (x^2)^{\frac{d}{2}-1} e^{-(x^2)}\nn \\
&=& \Omega_d \frac{1}{2} \Gamma(d/2)\;.
\ea 
Daraus folgt
\be \label{sph}
\Omega_d = \frac{2 \pi^{d/2}}{\Gamma(d/2)} \;.
\ee
\end{samepage}}.
Um das radiale Integral aus Gl.~(\ref{dimregbsp}) zu berechnen, substituiert
man f"ur $k^2$
\be 
k^2 = \frac{\Delta}{x} - \Delta \;,
      \quad \quad 2 k \, d k = - \frac{\Delta}{x^2} d x
\ee
und erh"alt f"ur Gl.~(\ref{dimregbsp})
\be \label{dimregbsp2}
\frac{\Omega_d}{2 (2 \pi)^d } \Delta^{d/2 - \beta + \alpha} 
           \int_0^1 dx \, 
        x^{\beta - d/2 - \alpha -1} (1-x) ^{d/2 + \alpha -1}\;.
\ee
Das Integral "uber $x$ kann mit Hilfe der Definition der Beta-Funktion 
ausgewertet werden
\be \label{beta}
\int_0^1 x^{\alpha-1}(1-x)^{\beta-1} = 
B(\alpha, \beta) = \frac{\Gamma(\alpha) \, \Gamma(\beta) }{\Gamma(\alpha + \beta)} \;.
\ee
Setzt man Gl.~(\ref{beta}) und Gl.~(\ref{sph}) 
in Gl.~(\ref{dimregbsp2}) ein, so erh"alt man das Resultat
\be \label{resdim}
  \int \frac{d^d \! k}{(2 \pi)^d} \frac{k^{2 \alpha}}{(k^2 + \Delta )^\beta} 
   = \frac{1}{(4 \pi)^{d/2} \Gamma(d/2)} (\Delta)^{d/2 + \alpha - \beta} 
    \frac{\Gamma(\beta-\alpha - d/2) \Gamma(d/2 + \alpha)}{\Gamma(\beta)} \;.
\ee
Das urspr"unglich zu berechnende Integral 
Gl.~(\ref{dimregbsp}) war f"ur $\alpha = 0 $,
$\beta = 2$ und $d=4$ divergent und f"ur $d=1,2,3$ konvergent.
Um das Verhalten von Gl.~(\ref{dimregbsp}) 
in den N"ahe von $d = 4$ zu untersuchen, setzt man 
$d=  4 - 2 \ep$ und beachtet, 
dass $\Gamma(z)$ bei $z = 0,-1,-2,\dots $ Pole besitzt.
Die Laurent-Entwicklung von $\Gamma(z)$ bei $z=0$ ist
\be \label{gammalaur}
\Gamma(2 - d/2) = \Gamma(\ep) 
       = \frac{1}{\ep} - \gamma_E + \mathcal{O}(\ep) \; ,
\ee
wobei $\gamma_M \approx 0.58  $ die Euler-Mascheroni-Konstante ist. 
Damit ergibt sich f"ur das Integral aus Gl.~(\ref{dimregbsp}) 
\be
\int \frac{d^d \! k}{(2 \pi)^d} 
\frac{1}{(k^2 + \Delta )^2 } \stackrel{d \rightarrow 4}{=}\frac{1}{(4 \pi)^2}
                  \left(\frac{1}{\ep} - \ln (\Delta)
             -\gamma_M + \mathcal{O}(\ep) \right)  \;.
\ee
Die urspr"ungliche Divergenz des 
Integrals wird durch den $1/\ep$-Pol in der Laurent-Entwick"-lung 
parametrisiert.
An dem Ergebnis Gl.~(\ref{resdim}) l"asst sich eine 
Eigenschaft der dimensionalen 
Regularisierung ableiten. Der Grenzwert von Gl.~(\ref{resdim})
f"ur  $ \beta \rightarrow 0 $ 
ergibt 
\be
\int \frac{d^d\!k}{(2 \pi)^d} \;  k^{2 \alpha} = 0 \; .
\ee
\subsection{Definition des ``Integrals'' in komplexwertigen Dimensionen}
Der oben beschriebenen 
Idee der analytischen Fortsetzung des Wertes eines Integrals
in komplexwertige Dimensionen folgend,   
definiert man die $d$-dimensionale Integration als Funktional mit folgenden 
Eigenschaften \cite{coll} 
\begin{enumerate}
\item Linearit"at
\item Skalierungsverhalten anlog zu dem gew"ohnlicher Integrale
\item Translationsinvarianz
\item Um das so definierte Integral eindeutig zu machen, 
ben"otigt man noch eine 
Normierungsbedingung, die das Ma"s festlegt.
Hierf"ur verwendet man:
\be
\int d^d \! p \, e^{-p^2} = \pi^{d/2} \;.
\ee
\item Falls der Integrand nicht nur vom Betrag der 
Integrationsvariablen, sondern 
auch von weiteren Vektoren abh"angt,
wird das $d$-dimensionale Integral durch
die Zerlegung der Integration 
"uber den Parallel- und Orthogonalraum erkl"art.  
Der Parallelraum wird hierbei von 
den "au"seren Vektoren aufgespannt und besitzt
eine ganzzahlige positive Dimension. 
Die Integration "uber den Parallelraum "uber $p_\parallel$ kann
im herk"ommlichen Sinne ausgef"uhrt werden.
Im Orthogonalraum h"angt der Integrand nur vom Betrag der Integrationsvariablen
$p_T$ ab, so dass die Winkel- und Radialintegration 
faktorisieren und analog zum
obigen Beispiel vorgegangen werden kann. 
Man definiert
\be \label{parsenk}
\int d^d\!p \, f(p) 
      = \Omega_{d-j} \int d^j p_\parallel \int_0^\infty dp_T \, 
           p_T^{d-j-1} f(p) \;.
\ee
\item Konvergiert das Integral aus 
Gl.~(\ref{parsenk}) f"u $0 < Re(d) < d_{max}$, so l"asst sich f"ur 
$-2 l -2 < Re(d) < -2 l$ zeigen:
\be \label{fort}
\int d^d \!p \, f(p) = \Omega_d  \int_0^\infty dp \, d^{d-1} 
   \Bigg( f(p^2) - f(0) - 
   p^2 f'(0)  \dots - (p^2)^l \, \frac{f^{(l)}(0)}{l!} \Bigg) \;.
\ee
Verhalten sich Funktionen 
polynomial bei $p \rightarrow \infty$ und konvergiert 
Gl.~(\ref{parsenk}) f"ur  kein $d$,
ist es sinnvoll, 
Gl.~(\ref{fort}) als Definition f"ur 
das Integral zu verwenden und es analytisch 
zu gro"sen $ {\rm Re} (d)$ hin fortzusetzen.
\end{enumerate}
\subsection{Clifford-Algebra in $d$ Dimensionen}
Die $d$-dimensionale Regularisierung stellt eine 
Verallgemeinerung der Minkowski-Dimension des 
Schleifenimpulses dar, nicht aber der Dimension 
der Matrixdarstellung der Clifford-Algebra.
F"ur die Clifford-Algebra in $d$ Dimensionen  
gelten
folgende Relationen:
\be
\{\gamma^\mu,\gamma^\nu\} 
= 2 g^{\mu \nu} \bbbone,  \quad g_\mu^\mu = d, 
      \quad \gamma^\mu \gamma_\mu = d \bbbone\;.
\ee
\section{Berechnung von Schleifenintegralen}
Bei der Berechnung von QCD-Strahlungskorrekturen 
treten verschiedene Typen von Integralen auf.
Vernachl"asigt man die Massen der Quarks oder 
entwickelt die auftretenden Integranden in 
den Quarkmassen, 
so treten nur sog. 
masselose Schleifenintegrale auf, in denen die Masse nicht erscheint.
Diese Integrale h"angen nur noch von einer einzigen Skala ab, 
dem "au"serem Impuls.
Diese Abh"angigkeit ist durch die Dimension des Integrals gegeben.
Die homogene Abh"angigkeit der Integrale von den 
"au"seren Impulsen ist eine wichtige Eigenschaft, die
es erm"oglicht, viele Graphen h"oherer 
Ordnung auf verschachtelte Ein- und Zwei-Schleifen-Integrale zur"uckzuf"uhren.
Zur Berechnung der masselosen Integrale wird die  
Gegenbauer-Polynom-X-''Space''-Technik (GPXT) \cite{GPXT} verwendet,
in der die Integrale im 
Ortsraum berechnet werden. Sie wird im Anhang~\ref{AGPXT} erkl"art.
\subsection{Das Ein-Schleifen-Integral}
Das Ein-Schleifen-Integral ist durch 
\be \label{einschleifint}
\int \frac{d^{4 - 2 \ep  }}{(2 \pi)^{4-2\ep}} 
\frac{ p^{(\mu_1 \dots \mu_n)}}{p^{2 \alpha} (p-q)^{2 \beta}}
      =\frac{G^{(n)}(\alpha, \beta)}{(4 \pi)^{2-\ep}} 
\frac{ q^{(\mu_1 \dots \mu_n)}}{(q^2)^{\alpha + \beta -2 + \ep}} 
\ee
gegeben, 
wobei $ p^{(\mu_1 \dots \mu_n)} $ den symmetrischen, 
spurlosen Tensor n-ter Stufe bezeichnet, der aus dem 
Impuls und dem metrischen Tensor gebildet werden kann.
F"ur $n = 1,2,3 $ ist
\ba
q^{(\mu_1)} &=& q^{\mu_1} \;,\nn \\
q^{(\mu_1\mu_2)} &=& 
       q^{\mu_1}  q^{\mu_2} - \frac{1}{d} q^2 g^{\mu_1 \mu_2} \;,\\
q^{(\mu_1\mu_2 \mu_3)} &=&  q^{\mu_1}  q^{\mu_2} q^{\mu_3}
   - \frac{1}{d+2} q^2 
    \Bigg( g^{\mu_1 \mu_2} q^{\mu_3} + g^{\mu_2 \mu_3} q^{\mu_1} 
              +  g^{\mu_3 \mu_1} q^{\mu_2} \Bigg) \;. \nn
\ea 
F"ur die Funktion $G^{(n)}(\al , \beta)$ 
l"asst sich ein geschlossener Ausdruck finden, der in 
Anhang~(\ref{A1loop}) hergeleitet ist.
\be
G^{(n)}(\alpha, \beta) = B(n + 2 - \alpha - \ep  ,\; 2 - \beta - \ep) 
     \frac{\Gamma( \alpha + \beta - 2 + \ep)}{\Gamma(\alpha) \, \Gamma(\beta)}
\ee
\subsection{Das Zwei-Schleifen-Master-Integral}
Mit der L"osung f"ur das Ein-Schleifen-Integral lassen sich bis auf das
Zwei-Schleifen-Master-Integral\footnote{Als ``Master-Integral'' 
bezeichnet man ein Integral, das den Z"ahler 1 besitzt und sich nicht
als Verschachtelung oder Produkt von einfacheren Integalen schreiben l"asst.}
alle Zwei-Schleifen-Integrale berechnen.
Das Zwei-Schleifen-Master-Integral ist durch
\be
\int   \frac{d^{4 - 2 \ep }\!p \, d^{4 - 2 \ep }\!k }{(2 \pi)^{2(4-2\ep)}} 
     \frac{1}{p^{2\alpha} (p-q)^{2 \beta} k^{2 \gamma} (k-q)^{2 \delta}(p-k)^{ 2 \eta}} = 
 \frac{F(\alpha, \beta,\gamma,\delta, \eta) }{(4 \pi)^{4-2\ep}} 
         \frac{1}{(q^2)^{\alpha + \beta + \gamma + \delta + \eta - 4 + 2 \ep}}
\ee
gegeben.
F"ur die Funktion $F(\alpha,\beta,\gamma,\delta,\eta)$ 
existiert kein geschlossener Ausdruck f"ur beliebige 
Argumente. Allerdings lassen sich mit der 
GPTX-Technik geschlossene Ausdr"ucke f"ur einige Spezialf"alle
berechnen. Zum Beispiel kennt man
\be \label{falbt}
F(\alpha ,\beta,1,1,1) = \frac{G(1,1)}{2-2 \ep  - \alpha- \beta} \bigg( \alpha 
      \left( G(\alpha + 1 ,\beta) - G(\alpha + 1 , \beta+ \ep )\right) 
+ (\alpha \leftrightarrow \beta) \bigg)\;.
\ee
Ein analoges Resultat kann f"ur ganzzahlige Werte von 
$( \gamma, \delta, \eta)$ gefunden werden.
In Anhang~(\ref{A2loop}) ist die Berechnung von 
$F(1,1,1,1,1)$ mit der GPXT-Technik skizziert.
In der Praxis ben"otigt man oft nur eine 
begrenzte Anzahl von Termen der Laurententwicklung um $(d-4)/2 =\ep$.
Diese ist von $F(\alpha, \beta,\gamma,\delta, \eta)$ 
f"ur eine begrenzte, aber relativ hohe Ordnung in $\ep$ f"ur beliebige Werte von 
$ \alpha,\beta,\gamma,\delta,\eta$ bekannt \cite{GPXT,partint}.
\subsection{Drei-Schleifen-Diagramme}
Grunds"atzlich ist es m"oglich, die 
GPXT-Technik auch auf nichttriviale Drei-Schleifen-Integrale anzuwenden. 
In der Praxis ist dies aber sehr 
m"uhsam und aufgrund der gro\ss en Anzahl von Drei-Schleifen-Diagrammen,
die bei der Berechnung von Greenfunktionen auftreten, nicht praktikabel. 
Der Durchbruch bei der Berechnung von 
Drei-Schleifen-Korrektu"-ren gelang mit der Entwicklung der Methode
der partiellen Integration \cite{partint} von dimensional regularisierten 
Integralen. Die entscheidende Identit"at hierf"ur ist
\be \label{intparts}
\int d^d \!p \frac{d}{d p_\mu} f(p, \dots ) = 0 \;.
\ee
Gl. (\ref{intparts}) 
l"asst sich leicht beweisen, indem man mit einem beliebigen Vektor 
kontrahiert: 
\be
\int d^d \! p \,  k_\mu \frac{d}{d p_\mu} f(p, \dots ) = 0 \;.
\ee
Um dieses Integral zu definieren, muss $k$ 
im Parallelraum liegen. Somit ist das Problem auf ein gew"ohnliches 
Integral zur"uckgef"uhrt.
Die Identit"at~(\ref{intparts}) erlaubt es,
dimensional regularisierte Integrale
partiell zu integrieren, 
ohne dass dabei Randterme auftreten. 
Dies erm"oglicht es, Rekursionsbeziehungen zwischen 
Graphen herzuleiten.
Ein einfaches Beispiel 
f"ur eine Rekursionsbeziehung dieser Art erh"alt man, indem man von 
\be
0 = \int \frac{d^d \! p \, d^d \! k}{(2 \pi)^{2 d}} 
                 \frac{\partial}{\partial p^\mu} 
              \left( \frac{(p-k)^\mu}{p^2 (p-q)^2 k^2(k-q)^2(p-k)^2} \right)
\ee
ausgeht. 
Durch Ausf"uhren der Differentation, 
Erg"anzen der Terme im Z"ahler zu quadratischen Ausdr"ucken und
Nutzung der Translationsinvarianz des Integrals 
(explizite Rechnung siehe Anhang~\ref{ARb}) erh"alt
man die in Abbildung~(\ref{rekursion}) dargestellte 
Identit"at.
\begin{center}
\begin{figure}[h]
\begin{picture}(700,100)(0,0)
\Text(25,50)[]{$\ep$}
\Line(50,50)(60,50)
\Oval(100,50)(40,40)(0)
\Line(100,10)(100,90)
\Line(140,50)(150,50)
\Text(165,50)[]{$=$}
\Line(175,50)(185,50) 
\Oval(225,50)(40,40)(0)
\CArc(270,10)(40,90,180)
\Vertex(224,90){3}
\Line(265,50)(275,50)
\Text(295,50)[]{$-$}
\Line(315,50)(325,50)
\Oval(350,50)(25,25)(0)
\Oval(400,50)(25,25)(0)
\Vertex(400,75){3}
\Line(425,50)(435,50)
\end{picture}
\caption{\label{rekursion}
Darstellung des nichttrivialen Zwei-Schleifen-Diagramms durch zwei triviale 
Zwei-Schleifen-Diagramme. 
Einfache Linien symbolisieren Propagatoren der 
Form $1/p^2$ und Linien mit Punkt Propagatoren der Form $1/p^4$.}
\end{figure}
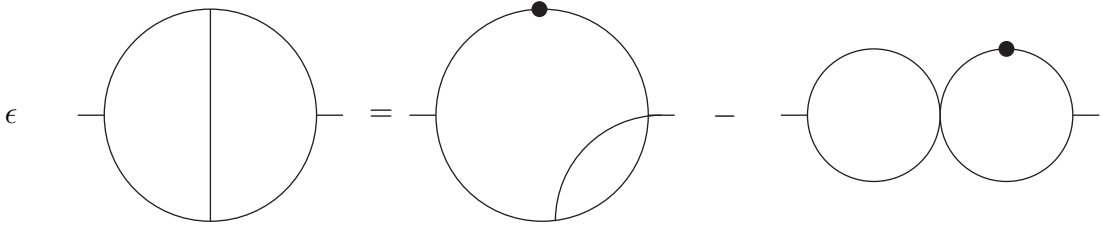
\end{center}
Mit Hilfe von Rekursionsbeziehungen dieser Art ist es m"oglich, 
einen Algorithmus zu entwickeln, um ein 
beliebiges Drei-Schleifen-Diagramm in eine 
Summe von Ein-Schleifen-Integralen und zwei fundamentalen 
Drei-Schleifen-Diagrammen 
(das nicht planare und das planare Drei-Schleifen-Master-Diagramm) 
zu transformieren.
\begin{center}
\begin{figure}[h]
\begin{picture}(400,120)(0,0)
\Oval(100,70)(50,50)(0)
\Text(100,10)[]{$N_0$}
\Line(30,70)(50,70)
\Line(150,70)(170,70)
\Line(70,110)(130,30)
\Line(70,30)(90,60)
\Line(130,110)(110,80)
\CArc(100,70)(15,45,230)
\Oval(300,70)(50,50)(0)
\Text(300,10)[]{$L_0$}
\Line(230,70)(250,70)
\Line(350,70)(370,70)
\Line(270,110)(270,30)
\Line(330,110)(330,30)
\end{picture}
\caption{Das nicht planare und das planare Drei-Schleifen-Master-Diagramm
         (Linien symbolisieren skalare Propagatoren $1/p^2$)}
\end{figure}
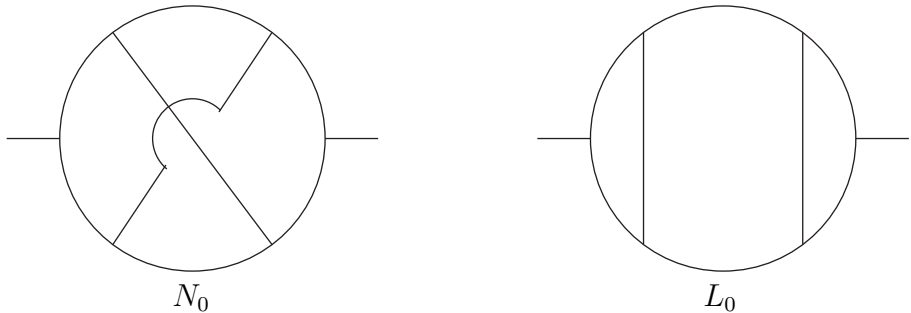
\end{center}
Die beiden Drei-Schleifen-Master-Diagramme 
sind mit GPXT bis einschliesslich der Ordnung $\ep^0$ berechenbar.
Es stellt sich heraus, 
dass sie sich nur in Termen der Ordnung $\ep$ unterscheiden.
Man findet
\be
L_0 = N_0 + \mathcal{O}(\ep) = 
             \frac{20 \zeta(5)}{(4 \pi)^6 k^4 } + \mathcal{O}(\ep) .
\ee
\subsection{Vier Schleifen-Diagramme}
Im Augenblick existiert noch kein 
Algorithmus, um ein beliebiges Vier-Schleifen-Integral zu berechnen. 
Allerdings gen"ugt die Kontrolle der Drei-Schleifen-Integrale bis 
zu Ordnung $\ep^0$,
um die Polstellen der Vier-Schleifen-Diagramme 
zu bestimmen und so die 
Vier-Schleifen-Koeffizienten der $\beta$- und $\gamma$-Funktion
zu berechen\footnote{Die $\beta$- und $\gamma$-Funktion werden 
in Kapitel \ref{rg} eingef"uhrt.}.

\section{Beispiele f"ur Renormierungsschemata}
Um die in der St"orungstheorie auftretenden Divergenzen zu beseitigen, 
werden die Felder und Parameter der Theorie renormiert.
Dies geschieht in der multiplikativen Renormierung durch die
Multiplikation der Felder und Parameter mit den Renormierungskonstanten
$Z_i$, die so gew"ahlt werden m"ussen, dass die Greenfunktionen,
ausgedr"uckt duch renormierte Gr"o\ss en, endlich werden. Diese Bedingung 
gen"ugt nicht, um die Renormierungskonstanten eindeutig 
festzulegen. Der endlichen Anteil der 
$Z_{i}$ mit $ i = ( 2,3,\tilde{3},g,\zeta,m)$ ist frei w"ahlbar,
was der Ambivalenz der Definition des 
divergenten Anteils einer Greenfunktion entspricht.
Um diesen eindeutig festzulegen, ben"otigt man Renormierungsbedingungen,
welche das Renormierungsschema festlegen. 
Die nummerischen Vorhersagen
der Theorie werden sich in verschiedenen Renormierungsschemata voneinander
unterscheiden, wobei die Unterschiede von der Ordnung h"oherer Terme in der
St"orungstheorie sind.
Am Beispiel der Quark-Zwei-Punkt-Funktion soll hier die Freiheit 
der Wahl eines angemessenen Renormierungsschemas illustriert werden \cite{mut}.
Der volle Quark-Propagator, welcher die QCD-Wechselwirkung ber"ucksichtigt,
ist durch
\be
S_{AB}(p) =
\int d^4 \! x \, \langle 0 | T( q_A(x) \bar{q}_B(0)) | 0 \rangle 
    e^{i p \cdot x} 
\ee
definiert.
Hierbei bezeichnet $|0 \rangle$ das QCD-Vakuum und $q_A$ den vollen 
Quark-Feld-Operator f"ur ein Quark mit Farbladung $A$.
Die Flavour- und Spinorindices werden in diesem Abschnitt unterdr"uckt. 
$S_{AB}(p)$ ist in  nullter Ordnung der St"orungstheorie durch 
\be
\begin{array}{ccc}
\begin{picture}(100,50)(0,0)
\ArrowLine(0,25)(100,25)
\end{picture}
&
\raisebox{3.7ex}[-3.7ex]{$\displaystyle{\;=\;}$}
&
\raisebox{3.7ex}[-3.7ex]{$\displaystyle{
\frac{i \delta_{AB}\, \shp  }{p^2  + i  \ep}  = G_{AB}(p) }$}
\end{array}
\ee
gegeben, wenn die Quarkmasse vernachl"assigt wird.
In der ersten Ordnung in $\al_s$ tr"agt zu $S_{AB}(p)$ nur 
ein einziges Diagramm bei:
\begin{equation}
\begin{array}{ccc}
\begin{picture}(100,70)(0,0)
\ArrowLine(0,20)(30,20)
\ArrowLine(70,20)(100,20)
\ArrowLine(30,20)(70,20)
\Vertex(30,20){2}
\Vertex(70,20){2}
\GlueArc(50,20)(20,180,0){3}{8}
\Text(15,10)[]{p}
\Text(85,10)[]{p}
\Text(50,10)[]{p+k}
\Text(50,30)[]{k}
\end{picture}
&
\raisebox{3.2ex}[-3.2ex]{$ \displaystyle {\; = \; }$}
&
\raisebox{3.2ex}[-3.2ex]{$\displaystyle{  G_{AB}(p) 
\delta_{BC} \; \left(- i \Sigma(\shp) \right) G_{CD}(p) \;.} $}
\end{array}
\end{equation}
Der sich aus den Feynman-Regeln in der Feynman-Eichung ($\zeta=1$) 
ergebende analytische Ausdruck
f"ur $\Sigma$ ist
\be \label{Quarkselbst}
\Sigma(\shp) =  - i C_F g^2 \int \frac{d^dk \mu^{2 \ep}}{(2 \pi)^d } 
      \frac{\gamma_\mu (\shp + \shk) \gamma^\mu }{ (p+k)^2 k^2} \;.
\ee
$\Sigma(\shp)$ l"asst sich mit den f"ur das Ein-Schleifen-Integral
angegebenen Formeln im Rahmen der dimensionalen Regularisierung berechnen, 
wenn man beachtet, dass f"ur die Gamma-Matrizen $\gamma_\mu \gamma^\nu
 \gamma^\mu = (2-d) \gamma^\nu$ gilt. F"ur die Laurententwicklung von $\Sigma$
um $d=4$ ergibt sich
\be
\Sigma(\shp) = - C_F \; \shp \;\frac{\al_s}{4 \pi} \left( \frac{1}{\ep} + 
   1 + \ln \left( 4 \pi \right) - \gamma_E 
               - \ln\left(\frac{-p^2}{\mu^2}\right)  
         + \mathcal{O}(\ep) \right) \;,  
\ee
wobei $\ep = (d-4)/2$ ist.
Die amputierte (mit ``trunc'' f"ur truncated bezeichnete), 
nicht renormierte Quark-Zwei-Punkt-Funktion 
bis zur Ordnung 
$\al_s$ erh"alt man, nachden die "au"seren Propagatoren herausdividiert 
wurden:
\ba
S_{AB}^{B\;\rm{trunc}}(p) &=& -i \delta_{AB} \, \shp \Bigg(
       1 - C_F\frac{\al_s}{4 \pi} \Bigg( \frac{1}\ep + 
   1 + \ln(4 \pi ) \nn \\
 &-& \gamma_E - \ln \left( \frac{-p^2}{\mu^2} \right)
       + \mathcal{O}(\ep) \Bigg) + \mathcal{O}(\al_s^2) \Bigg) \;. 
\ea
Die renormierte Quark-Zwei-Punkt-Funktion 
ergibt sich aus der nicht renormierten 
durch die Renormierung der Quark-Wellenfunktion der Kopplungskonstante.
Die Renormierung der Kopplungskonstante spielt erst ab der Ordnung
$\al_s^2$ eine Rolle, da sich renormierte und nackte Kopplung
nur in Termen der Ordnung $\al_s^2$ unterscheiden.
F"ur die renormierte Quark-Zwei-Punkt-Funktion ergibt sich so
\be
S_{AB}^{R \; \rm{trunc}}(p) = Z_2^{-1} \, S_{AB}^{B \;\rm{trunc}}(p) \;.
\ee
Setzt man f"ur 
\be
Z_2^{-1} = 1  + \frac{\al_s}{4 \pi} C_F   z + \mathcal{O}(\al_s^2) \;,
\ee
so erh"alt man
\ba
S_{AB}^{R \; \rm{trunc}}(p) &=& -i \delta_{AB} \shp \Bigg(
       1 -   C_F\frac{\al_s}{4 \pi} \Bigg( \frac{1}\ep -  z +
   1 \nn \\
  &+& \ln(4 \pi ) - \gamma_E - \ln \left( \frac{-p^2}{\mu^2} \right)
       + \mathcal{O}(\ep) \Bigg) + \mathcal{O}(\al_s^2) \Bigg) \;.
\ea
Um die Divergenz zu beseitigen, muss 
$z = \frac{1}\ep  -  z^\prime$
gew"ahlt werden.
F"ur die renormierte Quark-Zwei-Punkt-Funktion ergibt sich
\ba
S_{AB}^{R \; \rm{trunc}}(p) &=&  -i \delta_{AB} \shp \Bigg(
       1 - C_F\frac{\al_s}{4 \pi} \Bigg( z^\prime +
   1 \nn \\
  &+& \ln(4 \pi ) - \gamma_E - \ln \left( \frac{-p^2}{\mu^2} 
      \right) + \mathcal{O}(\ep) \Bigg) + \mathcal{O}(\al_s^2) \Bigg)\;.
\ea
Der Parameter $z^\prime$ ist v"ollig unbestimmt und
muss durch Renormierungsbedingungen festgelegt weden,
die das Renormierungsschema definieren.
\begin{enumerate}
\item Im ``minimal subtraction``-($\rm{MS}$)-Schema wird 
nur der $1/\ep$-Term subtrahiert, was zu 
$z^\prime = 0$ f"uhrt.
F"ur die Quark-Wellenfunktion-Renormierungskonstante und die 
Quark-Zweipunktfunktion ergibt sich
\be
Z_2^{\rm{MS}} = 1 - \frac{\al_s}{4 \pi} C_F  \frac{1}\ep \;,
\ee 
\ba
S_{AB}^{\rm{MS} \;\rm{trunc}}(p) &=& -i \delta_{AB} \shp \Bigg(
       1 +   C_F\frac{\al_s}{4 \pi} \Bigg( 
   1 + \ln(4 \pi ) \nn \\
    &-& \gamma_E - \ln \left( \frac{-p^2}{\mu^2} 
      \right) + \mathcal{O}(\ep) \Bigg) + \mathcal{O}(\al_s^2) \Bigg) \;.
\ea
\item Als Standard-Renormierungsschema f"ur die QCD hat sich das 
``modified  minimal subtraction''-($\MSsch$)-Schema  
etabliert, in dem au"ser dem $1/\ep$-Pol noch 
die durch die Entwicklung der Gammafunktion und des Faktors
$1/(4\pi)^\ep$ 
auftretenden Terme $(\ln(4 \pi) - \gamma_E)$
subtrahiert werden.
Hier erh"alt man
\be
Z_2^{\MSsch} = 1 - \frac{\al_s}{4 \pi} C_F  
  \left( \frac{1}\ep + \ln(4 \pi) - \gamma_E \right) \;,
\ee
\be
S_{AB}^{\MSsch \;\rm{trunc}}(p) = -i \delta_{AB} \shp \left(
       1 + C_F\frac{\al_s}{4 \pi} \left( 
   1  - \ln \left( \frac{-p^2}{\mu^2} 
      \right) + \mathcal{O}(\ep) \right) + \mathcal{O}(\al_s^2) \right) \; .
\ee 
\end{enumerate}

\chapter{Die Renormierungsgruppe (RG)} \label{rg}
In diesem Kapitel sollen die bereits bekannten, 
durch die Renormierungsgruppe gegebenen, 
Resultate "uber die 
Struktur der St"orungsreihen rekapituliert werden. 
Gute Ausf"uhrungen zu diesem 
Thema finden sich in \cite{marc,mut, coll}.
Durch die Remormierung werden alle Divergenzen der Greenfunktionen 
systematisch in jeder Ordnung der St"orungstheorie 
subtrahiert. In dieser Subtraktionsprozedur existiert die Unbestimmtheit 
der Definition des divergenten Anteils einer Greenfunktion oder, anders
ausgedr"uckt, wieviel des endlichen Anteils der Greenfunktion mit  
der Divergenz zusammen subtrahiert werden soll.
Diese Unbestimmtheit, die im ersten  Kapitel am Beispiel der 
Quark-Zwei-Punkt-Funktion
illustriert wurde,
ist "aquivalent zu der Freiheit, die nackte Lagrangedichte in die 
renormierte Lagrangedichte und die Counterterme aufzuteilen,
und resultiert in 
der Freiheit, ein Renormierungsschema zu w"ahlen. 

Durch die Subtraktion von Divergenzen wird unvermeidlich eine unbestimmte 
Massen"-skala $\mu$ in die Theorie eingef"uhrt, 
die sogenannte Renormierungsskala.
%
Die Renormierungsskala im MS- bzw. $\MSsch$-Schema 
besitzt eine implizite Struktur.
In diesem Schema wird nur der Pol in der 
Raumzeitdimension (MS) bzw. der Pol in der 
Raumzeitdimension in Kombination 
mit $(\ln(4 \pi) - \gamma_E)$ ($\MSsch$) subtrahiert, 
und auf den ersten Blick scheint es so,
als ob keine zus"atzliche Renormierungsskala 
ben"otigt wird. Allerdings zeigt sich, dass es notwendig ist, 
eine zus"atzliche Skala einzuf"uhren, um die 
Dimensionslosigkeit der Kopplungskonstante 
f"ur beliebige Raumzeitdimensionen zu erhalten. 
Diese Skala spielt die Rolle 
der Renormierungsskala und ist vollkommen beliebig.

Insgesamt entsteht durch die Subtraktion 
von Divergenzen im Rahmen der Renormierung
eine Unbestimmtheit in zweifacher Sicht:
\begin{enumerate}
\item
Die Freiheit, Renormierungsbedingungen zu w"ahlen, welche festlegen, wie die 
Divergenzen zu subtrahieren sind.
\item
Die Freiheit, die Renormierungsskala $\mu$ festzulegen.
\end{enumerate}
Es wird sich zeigen, dass die Unbestimmtheit der Renormierungsskala in der 
Freiheit der Wahl einer Renormierungsbedingung enthalten ist, und so die 
$\mu$-Abh"angigkeit von renormierten Gr"o\ss en eine Einparameter-Untergruppe 
der Renormierungsgruppe darstellt.

Aufgrund dieser Unbestimmtheit gibt es viele m"ogliche Ausdr"ucke f"ur eine 
physikalische Gr"o\ss e, die von der Wahl des Renormierungsschemas und der
Renormierungsskala abh"angen. Diese unterschiedlichen Resultate der Theorie 
sind durch endliche Renormierungen miteinander verkn"upft. 
Eine sich aufdr"angende Frage ist, ob die unterschiedlichen Ausdr"ucke 
f"ur eine physikalische Gr"o\ss e wirklich "aquivalent sind, da sie 
alle dieselbe Observable beschreiben sollen und mit derselben 
nackten Lagrangefunktion berechnet wurden.
\section{Die Gruppe der endlichen Renormierungen}
Die Freiheit in der Wahl eines Renormierungsschemas 
wird durch die Gruppe der endlichen Renormierungen beschrieben.
Greenfunktionen sind, ausgedr"uckt durch renormierte 
Gr"o\ss en, endlich. Die renormierten Gr"o\ss en h"angen mit den
nackten durch die Renormierungsparameter zusammen, 
welche durch das Renormierungsschema festgelegt werden.
Angenommen, es werden zwei verschiedene 
Renormierungsschemata verwendet, die zu den renormierten Gr"o\ss en
\ba
A_{B \mu}^a &=& (Z_3)^{1/2} A_{R \mu}^a \; ,\quad  
C^a_B = ( \tilde{Z}_3)^{1/2} C^a_R \; ,\quad
\Psi_B = (Z_2)^{1/2} \Psi_R \;,  \nn \\
\al_{sB} &=&  Z_{\al_s} \al_{sR} \; , \quad
\zeta = Z_\zeta  \zeta_R \; \quad
m_B = Z_m m_R \; ,
\ea 
und 
\ba
A_{B \mu}^{ a} &=& (Z_3^\pr )^{1/2} A_{R \mu}^{a \pr} \; ,\quad  
C^a_B = ( \tilde{ Z}_3^\pr)^{1/2} C^{ \pr a}_R \; ,\quad
\Psi_B = (Z_2^\pr)^{1/2}  \Psi_R^\pr \;,  \nn \\
\al_{sB} &=&  Z_{\al_s}^{\pr} \al_{sR}^\pr  \; , \quad
\zeta = Z_\zeta^\pr  \zeta_R^\pr \; \quad
m_B = Z_m^\pr  m_R^\pr
\ea
f"uhren.
Von dem Schema 
$( A_{R \mu}^a, C^a_R, \Psi_R ,\al_{sR}, \zeta_R, m_R$)
nach
$( A_{R \mu}^{\pr a} , C^{\pr a}_R, \Psi_R^\pr, 
\al_{sR}^\pr, \zeta_R^\pr, m_R^\pr)$
gelangt man durch eine endliche Renormierung, 
die durch 
\ba \label{endlren}
A_{R \mu}^{\pr a} &=& (z_3 )^{1/2} A_{R \mu}^{a} \; ,\quad  
C^{\pr a}_R = ( \tilde{ z}_3)^{1/2} C^{  a}_R \; ,\quad 
\Psi_R^\pr = (z_2)^{1/2}  \Psi_R \;,  \nn \\
\al_{sR}^\pr &=&  z_{\al_s} \al{s_R}  \; , \quad
\zeta^\pr = z_\zeta  \zeta_R \; \quad
m_R^\pr = z_m  m_R \; 
\ea
gegeben ist.
Die Parameter der endlichen Renormierung sind 
\ba
z_3 &=& Z_3 / Z_3^\pr \;, \quad 
\tilde{z}_3 = \tilde{Z}_3 / \tilde{Z}_3^\pr \;, \quad
z_2 = Z_2 / Z_2^\pr \;,  \nn \\
z_{\al_s} &=& Z_{\al_s} / Z_{\al_s}^\pr \;, \quad
z_\zeta = Z_\zeta / Z_\zeta^\pr \;, \quad
z_m = Z_m / Z_m^\pr\;.
\ea
Die Parameter $z_i$, $(i = 2,3,\tilde{3},\al_s,\zeta,m) $ 
sind endlich, da sich die divergenten Anteile 
der $Z_i^\pr $ gegen die der $Z_i $ 
wegheben. Wie die urspr"unglich eingef"uhrten 
divergenten Renormierungskonstanten $Z_i$
sind auch die endlichen Renormierungskonstanten 
Potenzreihen in der Kopplungskonstante.
Die Anzahl der Parameter der Gruppe h"angt somit von der 
Ordnung ab, bis zu der die St"orungstheorie betrieben wird.

Physikalische Resultate (im Folgenden mit $P$ bezeichnet)
d"urfen nicht von der Wahl des 
Renormierungsschemas abh"angen. Dies impliziert die 
Invarianz physikalischer Vorhersagen unter endlichen 
Renormierungen, was bedeutet
\be \label{reninv}
P_{\rm{exakt}}^\pr(p,\al_{sR}^\pr,m_{R}^\pr) = P_{\rm{exakt}}(p,\al_{sR},m_{R}) \;.
\ee
Die Funktionen $P$ und $P^\pr$ unterscheiden sich in 
ihrer funktionalen Form, so dass die experimentelle 
Bestimmung der Parameter $\al_{sR},m_{R}$ durch das Anpassen
von $P$ an experimentelle Daten andere Resulate liefern wird 
als die Bestimmung von $\al_{sR}^\pr,m_{R}^\pr$ mit $P^\pr$.
Die Resultate f"ur  $\al_{sR},m_{R}$ und $\al_{sR}^\pr,m_{R}^\pr$
m"ussen allerdings Gl.~(\ref{endlren}) erf"ullen.
In der Regel lassen sich Observable nur im Rahmen der 
St"orungstheorie als St"orungsreihe in der Kopplung $\al_s$ berechnen.
Durch das Abbrechen der der St"orungsreihe nach einer endlichen 
Anzahl von Termen wird die durch die Renormierungsgruppe beschriebene 
Invarianz physikalischer Resultate bez"uglich des Renormierungsschemas
gebrochen.
Anstelle von Gl.~(\ref{reninv}) erh"alt man in $n$-ter Ordnung der St"orungstheorie
\be
P^{\pr}_n (p,\al_{sR}^\pr,m_{R}^\pr) -  P_n(p,\al_{sR},m_{R})  
  = \mathcal{O}(\al_{sR}^{n+1}) \;.
\ee
Die nummerischen Vorhersagen der Theorie h"angen also in h"oheren Ordnungen 
von dem gew"ahlten Renormierungsschema ab.
\section{Die Abh"angigkeit von der Renormierungsskala}
Durch die Renormierung tritt eine weitere Skala auf, die 
Renormierungsskala $\mu$. Diese Skala ist zun"achst vollkommen beliebig, so 
dass man von Observablen fordern muss, dass
\be \label{renskalinv}
\mu^2 \frac{d}{d \mu^2} P_{\rm{exakt}}(p,\al_{sR},m_R,\mu) = 0
\ee
gilt,
wobei die Differentiation nach der Renormierungsskala $\mu$ so zu 
verstehen ist, dass die nackten Parameter als konstant angenommen werden
\be
\mu^2 \frac{d}{d \mu^2} \equiv \mu^2 \frac{d}{d \mu^2} \Bigg|_{\al_{sB} , m_B} \;.
\ee
Auch hier wird die Invarianz bez"uglich der 
Renormierungsskala durch das Abbrechen
der St"orungsreihe nach endlich vielen Termen gebrochen, 
so dass man anstelle von 
Gl.~(\ref{renskalinv}) nur noch
\be
\mu^2 \frac{d}{d \mu^2} P_n(p,\al_{sR},m_R,\mu) = \mathcal{O}(\al_{sR}^{n+1})
\ee 
hat. Die Renormierungsskalenabh"angigkeit der 
in der St"orungstheorie berechneten 
Resultate ist von der Ordnung h"oherer Terme der St"orungsreihe.
Die totale Differentation nach $\mu$ l"asst sich 
ausschreiben als
\be
\mu^2 \frac{d}{d \mu^2} = \mu^2 \frac{\partial}{\partial \mu^2}
            + \beta(\al_s)  \frac{ \pi \partial}{\partial \al_s} 
            + 2 m^2 \gamma (\al_s) \frac{\partial}{\partial m^2} \;,
\ee
wobei hier und im Folgenden Parameter ohne die Indizes $(R,B)$
immer die renormierten Gr"o\ss en bezeichnen. 
Die $\beta$- und $\gamma$-Funktionen beschreiben die Abh"angigkeit der 
renormierten Kopplungskonstante $\al_s$ und der renormierten Masse 
$m$ von der Renormierungsskala.
Sie sind definiert als
\be \label{RGGk}
\mu^2 \frac{d}{d \mu^2} \left( \als \right) = \beta(\al_s) = - \sum_{i \ge 0} \beta_i \alsb^{i+2} 
\ee
und
\be \label{RGGm}
\mu^2 \frac{d}{d \mu^2} m = m \, \gamma(\al_s)  = - m \sum_{i \ge 0} \gamma_i \alsb^{i+1}\;.
\ee
Die Gleichungen~(\ref{RGGk}) und~(\ref{RGGm})  
werden als Renormierungsgruppen-Gleichungen
(RG-Gleichungen) f"ur die Kopplungskonstante und die Quarkmassen bezeichnet.
Die Koeffizienten der $\beta$- und $\gamma$- Funktion wurden bis zu 
Vier-Schleifen-N"aherung berechnet.
Dies ist m"oglich, da es gen"ugt, die Polstellen der Graphen zu berechnen,
um die Koeffizienten der $\beta$- und $\gamma$-Funktion zu 
bestimmen\footnote{Die Koeffizienten der $\beta$-Funktion
lassen sich aus der Renormierungskonstanten f"ur die Kopplungskonstante
$Z_{\al_s}$ bestimmen, die sich widerum aus den Polstellen der 
Feynman-Graphen ergibt
\cite{beta4}:
\be
\mu^2 \frac{d}{d \mu^2} \al_s = 
 = - \ep \al_s - \al_s \frac{\partial}{\partial \al_s} 
   \left( \al_s Z^{(1)}_{\al_s} \right) =
 - \ep \al_s + \beta(\al_s)\; ,
\ee
wobei $Z^{(1)}_{\al_s}$ der erste Term der Laurententwicklung von $Z_{\al_S}$
\be
Z_{\al_s}(\al_s) = 1+ \sum_{n=1}^\infty \frac{Z_{\al_s}^{(n)} (\al_s)}{\ep^n}
\ee
ist.
F"ur die $\gamma$-Funktion hat man
\be
\gamma(\al_s) = \al_s \frac{\partial Z^{(1)}_m}{\partial \al_s}
\ee
mit dem in Analogie zu $Z^{(1)}_{\al_s}$ definierten Koeffizienten 
$Z^{(1)}_m$ \cite{gamma3v}.
}.
Die Resultate f"ur die Koeffizieten der $\beta$- und $\gamma$-Funktion sind
\cite{beta4,gamma3c,gamma3v}
\ba
\beta_0 &=&\frac{1}{4}\left(11-\frac{2}{3}n_f\right)\;, \quad \quad
\beta_1 = \frac{1}{16}\left(102-\frac{38}{3}n_f\right)\nn \\
\beta_2 &=& \frac{1}{64}\left(\frac{2857}{2}-\frac{5033}{18}n_f+\frac{325}{54}n_f^2\right) \;, \nn \\
\beta_3 &=& \frac{1}{256}
\left(\frac{149753}{6}+3564\;\zeta(3)-\left[\frac{1078361}{162}+\frac{6508}{27}\;\zeta(3)\right]n_f \right. \nn \\
&& + \left.\left[\frac{50065}{162}+\frac{6472}{81}\;\zeta(3)\right]n_f^2+\frac{1093}{729}n_f^3\right) \;; \\
&& \nn \\
\gamma_0 &=&  1 \;, \quad \quad
\gamma_1 =\frac{1}{6}\left[\frac{202}{3}-\frac{20}{9}n_f\right] \;, \nn \\
\gamma_2 &=&\frac{1}{64}\left[1249-\left(\frac{2216}{27}+
   \frac{135680}{27}\zeta(3)\right)-\frac{140}{81}n_f^2\right] \;, \nn \\
\gamma_3&=&\frac{1}{256}\left[\frac{4603055}{162}+\frac{135680}{27}\zeta(3)-8800\zeta(5)\right.\nonumber\\
&&\left. -\left(\frac{91723}{27}+
\frac{34192}{9}\zeta(3)-880\zeta(4)-\frac{18400}{9}\zeta(5)\right)n_f\right. \nn \\
&&\left.+\left(\frac{5254}{243}+
\frac{800}{9}\zeta(3)-\frac{160}{3}\zeta(4)\right)n_f^2
-\left(\frac{332}{243}-\frac{64}{27}\zeta(3)\right)n_f^3\right].
\end{eqnarray}
Hierbei bezeichnet $n_f$ die Anzahl der Flavours.
Die Tatsache, dass sowohl die Kopplungskonstante als auch die Quarkmassen 
von der Renormierungsskala $\mu$ abh"angen, zeigt deutlich, 
das es sich hierbei  
um Parameter handelt, die nur indirekt mit 
messbaren  Gr"o\ss en in Bezug stehen. 

In der St"orungstheorie berechnete Observable besitzen, wenn man von Quarkmassen 
absieht, die Form
\be
P= \sum_{n_1 < n_2} c_{n_1,n_2} \left( \ln \frac{\mu^2}{Q^2}\right)^{n_1} 
\left(\frac{\al_s(\mu)}{\pi} \right)^{n_2} \;.
\ee
Da $P$ von der Renormierungsskala $\mu$ nur durch h"ohere Ordnungen in 
der Kopplung $\al_s$ abh"angt, kann man $\mu^2 = Q^2 $ w"ahlen und erh"alt so
\be
P= \sum_{n_2} c_{0,n_2}\left(\frac{\al_s(Q)}{\pi} \right)^{n_2} \;.
\ee  
Die Energieabh"angigkeit von $P$ l"asst sich also durch die Energieabh"angigkeit 
der Kopplungskonstante beschreiben.
\section{L"osungen der RG-Gleichungen}
Um die RG-Gleichungen 
f"ur die Kopplungskonstante~(\ref{RGGk}) 
und die Quarkmasse~(\ref{RGGm}) zu l"osen, kann man auf verschiedene Arten 
vorgehen. Werden die $\beta$- und die $\gamma$-Funktion 
nur in der f"uhrenden Ordnung
ber"ucksichtigt, so lassen sich beide Gleichungen exakt l"osen. 
Im Rahmen der St"orungstheorie
bietet es sich an, die RG-Gleichungen iterativ als Reihe in $\al_s$ zu 
l"osen, was zu beliebig hohen Korrekturen 
der $\beta$- bzw. $\gamma$- Funktion m"oglich ist.
Im Rahmen der konturverbesserten St"orungstheorie (CIPT) ist es notwendig, 
die RG-Gleichungen~(\ref{RGGk})
und~(\ref{RGGm}) nummerisch zu l"osen, 
wobei die RG-Funktionen als exakte Funktionen behandelt werden.
\subsection{Exakte L"osung der RG-Gleichungen} 
Die RG-Gleichung der Kopplung hat, 
falls die $\beta$-Funktion nur in f"uhrender Ordnung 
verwendet wird, die Form
\be
\mu^2 \frac{d}{d \mu^2} \left(\frac{\al_s(\mu)}{\pi} \right)  
          = - \beta_0 \left( \frac{\al_s(\mu)}{\pi} \right)^2\;.
\ee
F"ur die Ableitung nach $\mu$ l"asst sich auch schreiben
\be
\mu^2 \frac{d}{d \mu^2} = - \frac{d}{d \ln(\mu^{\pr 2} /\mu^2 ) } 
          \equiv -\frac{d}{d l} 
\;\;, \rm{wobei} \quad l = \ln\left(\frac{\mu^{\pr 2}}{\mu^{ 2 }}\right).
\ee
So ergibt sich 
\be \label{DGL1}
\frac{d}{d l} \al_s(\mu) =  \frac{\beta_0}{\pi} 
\left(\al_s(\mu) \right)^2 \;.
\ee
Die Differentialgleichung~(\ref{DGL1}) 
l"asst sich separieren und integrieren, so dass man  
\be \label{integDglexakt}
\frac{1}{\al_s(\mu)} = - \frac{\beta_0}{\pi} l  + \frac{1}{\al_s(\mu^\pr)} 
\ee
erh"alt.
Als L"osung f"ur $\al_s(\mu)$ ergibt sich durch 
Aufl"osen von~(\ref{integDglexakt})
\be \label{kopplung}
\al_s(\mu) = \frac{\al_s(\mu^\pr)}{1- \beta_0/ \pi  \ln(\mu^{\pr 2} /\mu^2) 
           \al_s(\mu^\pr) }  \;.
\ee
Die RG-Gleichung f"ur die Quarkmasse (\ref{RGGm}) 
l"asst sich f"ur beliebige $\gamma$-Funktionen
integriegen, so dass man 
\be \label{mlosung}
m(\mu) = m(\mu^\pr) \;\exp \left(-\int_0^{\ln(\mu^{\pr 2} /\mu^{ 2})} 
\gamma\left(  \al_s(l) \right)
d l \right) 
\ee
erh"alt.
Werden die $\beta$- umd $\gamma$-Funktionen in f"uhrender Ordnung verwendet, so
ist das Integral in Gl.~(\ref{mlosung}) analytisch l"osbar. Es ergibt sich
\be
\int_0^{l} 
\gamma\left( \frac{ \al_s(l^\prime)}{\pi} \right)
dl^\pr =
 \frac{\gamma_0}{\beta_0} 
   \ln \left( 1 - \beta_0 l \frac{\al_s}{\pi} \right) \;,
\ee
wobei die L"osung f"ur die laufende Kopplungskonstante~(\ref{kopplung}) 
f"ur $\al_s(\mu)$ verwendet wurde.
\subsection{Iterative L"osung der RG-Gleichungen}
Die RG-Gleichungen f"ur die Kopplungskonstante 
und die Quarkmasse k"onnen als Anfangswertproblem der Form
\be \label{kopppic}
 - \frac{d}{d l} \frac{\al_s(l)}{\pi} 
      = \beta \left(\frac{\al_s(l)}{\pi} \right) \;, \quad
\al_s(l= 0) = \al_s(\mu = \mu^\pr) 
\ee
und
\be \label{massepic}
-\frac{d}{d l} m(l) =  m(l) \; \gamma  \left(\frac{\al_s(l)}{\pi} \right) \;, 
  \quad
m(l = 0) = m(\mu = \mu^\pr) 
\ee
geschrieben werden, wobei $l = \ln(\mu^{\pr 2}/\mu^{ 2})$ ist.
Anfangswertprobleme dieser Form lassen sich mit der 
Picard'schen Iterationsfolge
l"osen.
F"ur das Anfangswertproblem 
\be \label{AWP}
\dot{X}(t) = F(t,X(t))\;, \quad X(t_0) = x_0 
\ee
definiert man gem"a\ss 
\ba
X_0(t) &:=&  x_0 \nn \\
X_{n+1}(t) &:=& x_0 + \int_{t_0}^t F(t^\prime,X_n(t^\prime))dt^\prime 
\ea
die Picard'sche Iterationsfolge, 
die gegen die L"osung des Anfangswertproblems (\ref{AWP}) konvergiert.
Wird dieses L"osungsverfahren auf die 
RG-Gleichungen~(\ref{kopppic}, \ref{massepic}) 
angewendet, 
so erh"alt man 
als L"osungen f"ur die laufende Kopplungskonstante
\ba \label{runcoupl}
\frac{\alpha_s(\mu)}{\pi} &=& \alsbup + \beta_0 l \alsbup^2 
 +  \Bigg(  \beta_1 l +
\beta_0^2 l^2 \Bigg) \alsbup^3 \nn \\
&&+ \Bigg( \beta_2 l +\frac{5}{2}\beta_1\beta_0 l^2 + \beta_0^3 l^3 \Bigg)  
 \alsbup^4
\nonumber \\
 &&+ \Bigg( \beta_3 l + 3 \beta_2 \beta_0 l^2 + \frac{3}{2} \beta_1^2 l^2 + 
\frac{13}{3} \beta_1 \beta_0^2 l^3 + \beta_0^4 l^4 \Bigg)  \alsb^5 + \ldots \;. 
\ea
F"ur die Masse ergibt sich analog
\ba 
\label{runmass}
\frac{m(\mu)}{m(\mu^{\prime })} &=& 
         1 +   \gamma_0 l \alsbup 
     +  \Bigg(
          \gamma_1 l  +  \frac{1}{2} \beta_0 \gamma_0 l^2 
       + \frac{1}{2}  \gamma_0^2 l^2   \Bigg) \alsbup^2  \nn \\
  &&+  \Bigg( 
        \gamma_2 l + 
        \gamma_0 \gamma_1 l^2 +
          \frac{1}{2} \beta_1 \gamma_0 l^2 +
              \beta_0 \gamma_1 l^2 + \nn \\
     && \quad       \frac{1}{6}  \gamma_0^3 l^3+
                \frac{1}{3} \beta_0^2 \gamma_0 l^3 + 
  \frac{1}{2}  \beta_0 \gamma_0^2 l^3  
              \Bigg ) \alsbup^3 + \ldots\;. 
\ea  
\section{Die Gruppe der Skalentransformationen}
An den perturbativen L"osungen f"ur die laufende 
Kopplungskonstante  und  Masse
sieht man, dass Kopplungskonstanten und Massen 
bei verschiedenen Renormierungsskalen 
durch eine endliche Renormierung miteinander verkn"upft sind.
Aus den Gl.~(\ref{runcoupl}, \ref{runmass}) l"asst sich ablesen, dass
\ba
\al_s(\mu) &=& z_{\al_s}(\mu, \mu^\prime) \al_s(\mu^\prime) , \\
m(\mu) &=& z_m(\mu, \mu^\prime) m(\mu^{\prime }) ,
\ea
wobei
\ba
 z_{\al_s}(\mu, \mu^\prime) &=& 1 +   \beta_0 l \alsbup 
    +  \Bigg(  \beta_1 l +
\beta_0^2 l^2 \Bigg) \alsbup^2 + \ldots \;, \\
  z_m(\mu, \mu^\prime) &=& 1+  l  \gamma_0 \alsbup 
 +  \Bigg( 
          \gamma_1 l  +  \frac{1}{2} \beta_0 \gamma_0 l^2 
       + \frac{1}{2}  \gamma_0^2 l^2   
       \Bigg) \alsbup^2 + \ldots \;. 
\ea
Dies illustriert, dass ein Wechsel der Renormierungsskala einer endlichen 
Renormierung entspricht. 
Die Renormierungsskala $\mu$ parametrisiert gerade eine 
Ein-Parameter-Unter"-gruppe der Renormierungsgruppe, die durch die Elemente
$\{z(\mu, \circ ) \}$ gegeben ist.
Damit die Menge der $\{z(\mu, \mu^\prime ) \}$ 
zu einer abgeschlossenen Gruppe wird, 
definiert man die 
Komposition zweier Elemente so, dass sie einem
doppelten 
Renormierungsskalenwechsel 
$\mu \rightarrow \mu^\prime \rightarrow \mu^{\prime\prime}$
entspricht. 
Das mit $\{\circ\}$ abgek"urzte Argument der $z(\mu,\circ)$ muss f"ur die 
Kompositon entsprechend angeglichen werden. Es gilt
\be
z(\mu^{\prime \prime }, \mu^\prime ) \, z(\mu^{\prime}, \mu ) =
z(\mu^{\prime\prime}, \mu)
\;.
\ee
Die Invertierung eines Elementes aus $\{z(\mu,\circ) \}$ geschieht durch Vertauschen der Argumente 
\be
z^{-1}(\mu, \circ ) =  z( \circ, \mu) \;,
\ee
so dass
\be  
z(\mu^{\prime}, \mu) \, z(\mu, \mu^{\prime}) = 1, 
\ee 
und die Identit"at entspricht
\be
z(\mu, \mu) = 1 \; .
\ee
Auf diese Weise werden die Skalentransformationen zu einer 
abelschen Ein-Parameter-Untergruppe
der Gruppe der endlichen Renormierungen.
\section{Das Transformationsverhalten der \\ $\beta$- und $\gamma$-Funktionen}
Unter endlichen Renormierungen ver"andern sich die Kopplungskonstante, 
die Quarkmassen und die 
Greenfunktionen, 
wohingegen Observable invariant bez"uglich den Transformationen 
der Renormierungsgruppe sind.
In diesem Abschnitt soll das Transformationsverhalten 
der $\beta$- und $\gamma$-Funktion untersucht werden.
Die Definition der $\beta$-Funktion ist (Gl.~\ref{RGGk})
\be
\mu^2 \frac{d}{d \mu^2} \alsb = \beta(\al_s) \;.
\ee
Wird eine neue Kopplungskonstante gem"a\ss 
\be \label{kopptrans}
\al_s^\prime = z_{\al_s} \al_s
\ee
eingef"uhrt, wobei der Reskalierungsfaktor $z_{\al_s}$ durch 
\be
z_{\al_s}= 1 + \kappa_1 \als 
         + \kappa_2 \alsb^2 + \kappa_3 \alsb^3 + \kappa_4 \alsb^4 + \ldots
\ee 
gegeben ist, so wird sich die $\beta$-Funktion der Kopplungskonstanten
$\al_s^\prime$ von der urspr"unglichen $\beta$-Funktion unterscheiden.
Man erh"alt
\ba \label{betaprime}
\beta^\prime(\al_s^\pr) &=& 
    \mu^2 \frac{d}{d \mu^2} \left(z_{\al_s} \als \right) \nn \\
&=& \alsb  \mu^2 \frac{d}{d \mu^2}(z_{\al_s}) +  
         z_{\al_s} \mu^2 \frac{d}{d \mu^2}  \alsb  \nn \\
&=& \als  \left( \frac{\pi \partial }{\partial \al_s} z_{\al_s}\right)  
                  \mu^2 \frac{d}{d \mu^2 }\alsb
      + z_{\al_s} \mu^2 \frac{d}{d \mu^2}\alsb \nn \\
&=& \left( \al_s \left( \frac{ \partial }{\partial \al_s} z_{\al_s} \right) 
                 + z_{\al_s} \right) \beta(\al_s) 
\Bigg|_{\al_s = \al_s(\al_s^\pr)} \;.
\ea
Um die transformierte $\beta$-Funktion $\beta^\pr(\al_s^\pr)$ 
in der entsprechenden Kopplungskonstanten auszudr"ucken, muss
Gl.~(\ref{kopptrans}) invertiert werden.
Dies geschieht mit dem Standard-Ansatz
\ba \label{stansatz}
\als &=& \left( \frac{\al_s^\prime}{\pi} \right)
       + a_1 \left( \frac{\al_s^\prime}{\pi} \right)^2 + 
       a_2 \left( \frac{\al_s^\prime}{\pi} \right)^3 + 
         a_3 \left( \frac{\al_s^\prime}{\pi} \right)^4 + 
           \mathcal{O}(\al_s^5) \;. 
\ea
Einsetzen von Gl.~(\ref{stansatz}) in 
Gl.~(\ref{kopptrans}) und Koeffizientenvergleich 
ergibt
\be \label{loes}
\als = \left( \frac{\al_s^\prime}{\pi} \right) 
       - \kappa_1 \left( \frac{\al_s^\prime}{\pi} \right)^2 + 
      ( 2 \kappa_1^2 - \kappa_2) \left( \frac{\al_s^\prime}{\pi} \right)^3 + 
      ( - 5 \kappa_1^3 + 5 \kappa_1 \kappa_2 - \kappa_3)
             \left( \frac{\al_s^\prime}{\pi} \right)^4
        + \mathcal{O}(\al_s\prime^5) \;.
\ee
Mit (\ref{loes}) erh"alt man aus (\ref{betaprime}) die Koeffizienten der  
transformierten $\beta$-Funktion $\beta^\prime(\al_s^\prime)$
\ba \label{betatransform}
\beta_0^{\pr} &=& \beta_0 \;,\nn \\
\beta_1^{\pr} &=& \beta_1 \;,\nn \\
\beta_2^{\pr} &=& \beta_2 - \kappa_1 \beta_1 
                       + (\kappa_2 - \kappa_1^2)\beta_0 \;,\nn \\
\beta_3^{\pr} &=& \beta_3 - 2 \kappa_1 \beta_2 + \kappa_1^2 \beta_1
         + (2 \kappa_3 -  6 \kappa_2 k_1 +  4 \kappa_1^3 ) \beta_0 \;. 
\ea
Die ersten beiden Koeffizienten der 
$\beta$-Funktion $\beta_{0}$ und $\beta_1$ sind Invarianten der 
Renormierungsgruppe. 
Koeffizienten h"oherer Ordnung transformieren sich gem"a{\ss} 
Gl.~(\ref{betatransform}).

Um das Transformationsverhalten der $\gamma$- Funktion zu bestimmen,
geht man
analog vor. Wird die Masse gem"a\ss 
\be \label{mtransf}
m^\prime = z_m m
\ee
transformiert, 
wobei die Reskalierung $z_m$ durch 
\be \label{}
z_{m}= 1 + \vr_1 \als + \vr_2 \alsb^2 + \vr_3 \alsb^3 + \vr_4 \alsb^4 + \ldots
\ee
gegeben ist, so ergibt sich f"ur die transformierte $\gamma$-Funktion
\ba \label{gammaprime}
 m^\prime  \, \gamma^\prime(\al_s^\pr) &=&\mu^2 \frac{d}{d \mu^2} m^\prime  \nn \\
&=& \mu^2 \frac{d}{d \mu^2} (z_m m) \nn \\
&=& m \left( \mu^2 \frac{d}{d \mu^2} z_m \right) 
     + \left( \mu^2 \frac{d}{d \mu^2} m \right)  z_m \nn \\
&=& m\left( \frac{\pi \partial}{\partial \al_s} z_m \right) 
           \left(  \mu^2 \frac{d}{d \mu^2} \als \right)
  + \left( \mu^2 \frac{d}{d \mu^2} m \right) z \nn \\
&=& m\left( \frac{\pi \partial}{\partial \al_s} z_m \right) \beta(\al_s) 
            + m^\prime \gamma(\al_s)
\Bigg|_{\stackrel{m = m(m^\pr,\al_s^\pr)}{\al_s = \al_s(\al_s^\pr)}} \;. 
\ea 
Um das Transformationsverhalten der Koeffizienten der 
$\gamma$-Funktion zu bestimmen, muss
Gl.~(\ref{mtransf}) invertiert werden:
\ba \label{mloes}
m  &=& m^\prime \Big( 1 
       - \vr_1 \als + 
      ( \vr_1^2 - \vr_2) \alsb^2   + 
      ( -\vr_1^3  +  2  \vr_1 \vr_2 -\vr_3 )  \alsb^3 \nn \\ 
&&+
      ( \vr_1^4 + 2 \vr_1 \vr_3 - 3 \vr_1^2 \vr_2 + \vr_2^2 -\vr_4) \alsb^4  
        +   \mathcal{O}(\al_s)^5 \Big) \;.
\ea
Aus Gl.~(\ref{gammaprime}) und (\ref{mloes}) 
ergibt sich das Transformationsverhalten der 
$\gamma$-Funktion bei einer Transformation der 
Kopplung und Masse gem"a{\ss} 
(\ref{kopptrans}, \ref{mtransf}) zu
\ba \label{gammatransform}
\gamma_0^\prime &=& \gamma_0 \;,  \nn \\
\gamma_1^\prime &=& \gamma_1 - \kappa_1 \gamma_0 + \vr_1 \beta_0  \;,\nn \\
\gamma_2^\prime &=& \gamma_2  - 2 \kappa_1 \gamma_1 +( - \kappa_2 + 2 \kappa_1^2  )\gamma_0  
 + \vr_1 \beta_1  + (- 2 \kappa_1 \vr_1 + 2 \vr_2 - \vr_1^2     )\beta_0 \;,
        \nn \\
\gamma_3^\pr &=& \gamma_3 - 3 \kappa_1 \gamma_2 
         + ( -2 \kappa_2 + 5 \kappa_1^2) \gamma_1  + ( - \kappa_3 + 5 \kappa_2
   \kappa_1 - 5 \kappa_1^3 ) \gamma_0 \nn \\
&&     + \vr_1 \beta_2 + ( - 3 \kappa_1 \vr_1 + 2 \vr_2 - \vr_1^2 )\beta_1 \nn \\
&& + ( - 2 \kappa_2 \vr_1 + 5 \kappa_1^2 \vr_1 - 6 \kappa_1  \vr_2 
       + 3 \kappa_1 \vr_1^2 - 3 \vr_2 \vr_1 + 3 \vr_3 + \vr_1^3)\beta_0 \;.
\ea 
Anders als im Fall der  der $\beta$-Funktion ist nur  $\gamma_0$  invariant
unter RG-Transformationen.
\subsection{Invarianz der RG-Funktionen  unter \\ Skalen-Transformationen}
Die RG-Gleichungen (\ref{RGGk}, \ref{RGGm}) 
sind autonome Differentialgleichungen. Das bedeutet, dass 
die $\beta$- und $\gamma$-Funktion nicht explizit von der 
Renormierungsskala $\mu$ abh"angen und 
deswegen invariant unter den Skalen-Transformationen sind, 
welche eine Ein-Parameter-Untergruppe
der Renormierungsgruppe bildet.
Dies l"asst sich anhand des im vorhergehenden 
Abschnittes bestimmten Transformationsverhaltens der
Koeffizienten der $\beta$- und $\gamma$-Funktion "uberpr"ufen.
Betrachtet man eine beliebige Transformation der Renormierungsskala 
$\mu^{ 2} \rightarrow \mu^{\pr 2}  = \mu^{2} e^\xi$,
so transformieren sich die Kopplungskonstante und die Masse gem"a{\ss}
Gl.~(\ref{runcoupl}, \ref{runmass})
zu
\ba 
\frac{\alpha_s(\mu)}{\pi} &=& \alsbup + \beta_0 \xi  \alsbup^2 +  \Bigg(  \beta_1 \xi +
\beta_0^2 \xi ^2 \Bigg) \alsbup^3 \nn \\
&&+ \Bigg( \beta_2 \xi +\frac{5}{2}\beta_1\beta_0 \xi^2 + \beta_0^3 \xi^3 \Bigg)  \alsbup^4
\nonumber \\
 &&+ \Bigg( \beta_3 \xi + 3 \beta_0 \beta_2 \xi^2 + \frac{3}{2} \beta_1^2 \xi^2 + 
\frac{13}{3} \beta_0^2 \beta_1 \xi^3 + \beta_0^4 \xi^4 \Bigg)  \alsb^5 + \ldots  
\ea
und 
\ba 
\frac{m(\mu)}{m(\mu^{\prime})} &=& 
      1 + \xi  \gamma_0 \alsbup +  \Bigg(  \gamma_1 \xi + \frac{1}{2} 
\beta_0 \gamma_0 \xi^2  + \frac{1}{2}  \gamma_0^2 \xi^2
           \Bigg) \alsbup^2  \nn \\
  &&+  \Bigg(\gamma_2 \xi + 
        \gamma_0 \gamma_1 \xi^2 +
          \frac{1}{2} \beta_1 \gamma_0 \xi^2 +
              \beta_0 \gamma_1 \xi^2 + 
           \frac{1}{6}  \gamma_0^3 \xi^3\nn\\
&&+
                \frac{1}{3} \beta_0^2 \gamma_0 \xi^3 
 + \frac{1}{2}  \beta_0 \gamma_0^2 \xi^3 
           \Bigg) \alsbup^3 + \ldots  \;.
\ea  
Dies ergibt die Koeffizienten der Transformationsformeln f"ur die
$\beta$- und $\gamma$-Funktionen  (\ref{betatransform}, \ref{gammatransform}):
\ba \label{kappatrans}
\kappa_1 &=& \beta_0 \xi \;,\nn \\
\kappa_2 &=& \beta_1 \xi + \beta_0^2 \xi ^2  \;,\nn \\
\kappa_3 &=& \beta_2 \xi +\frac{5}{2}\beta_1\beta_0 \xi^2 + \beta_0^3 \xi^3\;,
\ea
und 
\ba \label{vrtrans}
\vr_1 &=&   \gamma_0 \xi \;,\nn \\
\vr_2 &=&  \gamma_1 \xi  +  \frac{1}{2} \beta_0 \gamma_0 \xi^2 
       + \frac{1}{2}  \gamma_0^2 \xi^2  \;,\nn \\
\vr_3 &=& \gamma_2 \xi + 
        \gamma_0 \gamma_1 \xi^2 +
          \frac{1}{2} \beta_1 \gamma_0 \xi^2 +
              \beta_0 \gamma_1 \xi^2 + 
           \frac{1}{6}  \gamma_0^3 \xi^3+
                \frac{1}{3} \beta_0^2 \gamma_0 \xi^3  
 + \frac{1}{2}  \beta_0 \gamma_0^2 \xi^3  \;.
\ea
Einsetzen von Gl.~(\ref{kappatrans}) in Gl.~(\ref{betatransform})  und
Gl.~(\ref{kappatrans}, \ref{vrtrans}) in Gl.~(\ref{gammatransform}) ergibt 
explizit die Invarianz der $\beta$- und der $\gamma$-Funktion 
bez"uglich Skalen-Transformationen. 
F"ur $\gamma_1$ hat man z.B. 
\ba
\gamma_1^\pr &=& \gamma_1 - \kappa_1 \gamma_0 + \vr_1 \beta_0 \nn\\
&=& \gamma_1 - \beta_0 \xi \gamma_0 + \gamma_0 \xi \beta_0 = \gamma_1 \;,
\ea
was die Invarianz von $\gamma_1$ bez"uglich Skalen-Transformationen 
explizit zeigt.

\chapter{Theoretische Grundlagen des $\tau$-Zerfalls} \label{tautheo}
In diesem Kapitel soll der theoretische Rahmen f"ur die Theorie des 
$\tau$-Zerfalles dargestellt werden.
Obwohl die in diesem Kapitel hergeleiteten Ergebnisse bekannt sind,
erscheint es mir sinnvoll, sie mit einer ausf"uhrlichen Herleitung 
hier darzustellen, da sie zu speziell sind, um sie in 
einem Lehrbuch der Quantenfeldtheorie zu finden und da sie in den aktuellen 
Ve"offentlichungen nicht mehr erkl"art werden. 
Das  $\tau$-Lepton ist mit einer Masse von  1.777 GeV das schwerste der 3 Leptonen.
Die Masse des $\tau$-Leptons reicht aus, um in hadronische Endprodukte, 
bestehend aus einem $\tau$-Neutrino und Mesonen, zu zerfallen.
Zerf"alle des $\tau$-Leptons in hadronische Endprodukte
werden durch die schwache und die starke Wechselwirkung beschrieben.
Da die St"orungsreihe der schwachen Wechselwirkung schnell abnimmt, 
gen"ugt es, in f"uhrender Ordnung der schwachen Wechselwirkung zu 
rechnen. Korrekturen der schwachen Wechselwirkung
besitzen nur eine Gr"o"se von 1\% und k"onnen als multiplikative 
und additive Korrekturen sp"ater ber"ucksichtigt werden.
In der niedrigsten Ordnung der schwachen Kopplungskonstanten wird der Zerfall 
des $\tau$-Leptons
durch folgendes Feynman-Diagramm beschrieben: 
   \begin{center}
    \begin{picture}(300,125)(0,0)
       \Line(165,73)(195,73)
       \Line(165,67)(195,67)
       \Text(180,85)[]{$p_W^2 \ll m_W^2$}
       \ArrowLine(0,70)(40,70)
       \Photon(40,70)(90,35){5}{10}
       \Vertex(40,70){2}
       \Vertex(90,34){2}
       \ArrowLine(40,70)(100,100)
       \ArrowArcn(110,35)(20,180,90)
       \ArrowArc(110,35)(20,180,270)
       \Line(105,55)(135,55)
       \Line(105,15)(135,15)
       \Text(20,80)[]{$\tau^-$}
       \Text(64,35)[]{$W^-$}
       \Text(90,80)[]{$\nu_\tau$}
       \Text(125,48)[]{$q$}
       \Text(125,23)[]{$\bar{q}$} 
       \Text(128,36)[l]{Hadronen}
       \ArrowLine(220,70)(260,70)
       \Text(235,80)[]{$\tau^-$}
       \Vertex(260,70){2}
       \ArrowLine(260,70)(280,110)
       \ArrowArcn(270,53)(20,127,30)
       \ArrowArc(270,53)(20,127,230)
       \Line(286,65)(300,45)
       \Line(256,39)(266,25)
       \Text(292,45)[]{$q$}
       \Text(270,33)[]{$\bar{q}$}
       \Text(286,97)[]{$\nu_\tau$}
       \Text(300,20)[]{Hadronen}
    \end{picture}
   \end{center}
Zerf"alle des $\tau$-Leptons in leichtere Leptonen $(e^-, \mu^-)$ 
werden ausschlie\ss lich durch die schwache Wechselwirkung beschrieben.
In der Baumgraphenn"ahrung mu"s  man f"ur die Berechnung der 
Zerfallsrate des $\tau$-Leptons in leichte Leptonen nur den 
folgenden Graphen berechnen:
    \begin{center}
     \begin{picture}(310,125)(0,0)
       \Line(165,73)(195,73)
       \Line(165,67)(195,67)
       \Text(180,85)[]{$p_W^2 \ll m_W^2$}
       \ArrowLine(0,70)(40,70)
       \Photon(40,70)(90,35){5}{10}
       \Vertex(40,70){2}
       \Vertex(90,34){2}
       \ArrowLine(40,70)(100,100)
       \ArrowLine(90,34)(135,55)
       \ArrowLine(90,34)(135,15)
       \Text(125,28)[]{$l$}
       \Text(125,61)[]{$\bar{\nu}_l$}
       \Text(35,80)[]{$\tau^-$}
       \Text(64,35)[]{$W^-$}
       \Text(90,80)[]{$\nu_\tau$}  
       \ArrowLine(220,70)(260,70)
       \Text(245,80)[]{$\tau^-$}
       \Vertex(260,70){2}
       \ArrowLine(260,70)(320,110)
       \ArrowLine(260,70)(320,60)
       \ArrowLine(260,70)(320,15)
       \Text(315,96)[]{$\nu_\tau$}
       \Text(315,70)[]{$\bar{\nu}_l$}
       \Text(315,31)[]{$l$}
    \end{picture}
    \end{center} 
\section{Die effektive Lagrangedichte der schwachen Wechselwirkung}
Um den Anteil der $\tau$-Zerfallsrate zu berechnen, 
der durch die schwache Wechselwirkung beschrieben wird,
ben"otigt man die Wechselwirkungs-Lagrangedichte der schwachen Wechselwirkung.
Da die in $\tau$-Zerf"allen durch das Eichboson "ubertragene Energie 
im Vergleich zu der Masse der Eichbosons $(m_W)$ gering ist, gen"ugt es, eine 
ph"anomenologische Lagrangefunktion zu verwenden, 
die eine effektive Strom-Strom-Wechselwirkung 
beschreibt. Diese ergibt sich durch Entwicklung des $W$-Boson-Propagators in 
$q^2$ in nullter Ordnung, in der die Propagatoren durch Konstanten gem"a"s
   \begin{equation}
      \frac{- i(g_{\mu \nu} -q_\mu q_\nu / m_W^2 )}{q^2 -m_W^2}
            \begin{array}{c}
                   \\
                   \longrightarrow \\
                    q^2  \ll m_W^2
            \end{array} 
              \frac{i g_{\mu \nu}}{m_W^2}
   \end{equation}
ersetzt werden.
Ein konstanter Propagator im Impulsraum entspricht einer 
Delta-Funktion im Ortsraum.
Die resultierende Wechselwirkung ist eine punktf"ormige Kopplung 
von vier Fermionen,
die bereits aus der Fermi-Theorie des Betha-Zerfalls bekannt ist.
 \begin{figure}[h]
  \begin{center}
  \begin{picture}(200,80)(0,0)
      \ArrowLine(0,0)(30,40)
      \ArrowLine(0,80)(30,40)
      \Photon(30,40)(70,40){3}{6}
      \Text(50,50)[]{$W^{-}$}
      \ArrowLine(70,40)(100,80)
      \ArrowLine(70,40)(100,0)
      \Line(120,40)(180,40)
      \Line(180,40)(173,45)
      \Line(180,40)(173,35)
      \Text(150,20)[]{$p_W^2 \ll m_W^2$}
      \ArrowLine(200,0)(230,40)
      \ArrowLine(200,80)(230,40)
      \ArrowLine(230,40)(260,80)
      \ArrowLine(230,40)(260,0)
      \Vertex(30,40){2}
      \Vertex(70,40){2}
      \Vertex(230,40){2}
 \end{picture}
        \caption{4-Fermion-Vertex der effektiven Wirkung }
    \end{center}
 \end{figure}
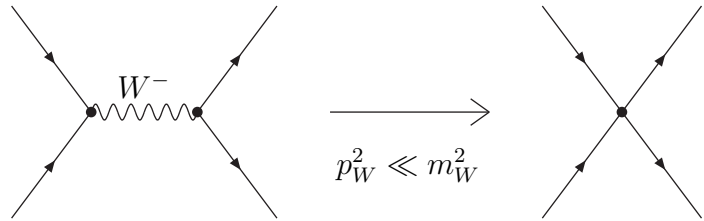
F"ur den Teil der effektiven Lagrangefunktion, 
die leptonische $\tau$-Zerf"alle 
beschreibt, ergibt sich so
 \begin{equation} 
     \mathcal{L}_{ \rm eff} = 
 - \frac{G_F}{\sqrt{2}} \left( (\bar{\tau} \gamma^\mu(1-\gamma_5) \nt )
      (\bar{l} \gamma_\mu(1-\gamma_5)\nu_l ) 
             + h.c.\right) \,, 
\ee
wobei $l$ f"ur $e$ und $\mu$ steht.
Hadronische Zerf"alle werden durch folgenden Teil der 
effektiven Lagrangefunktion 
beschrieben:
  \begin{equation} \label{leffektiv}
     \mathcal{L}_{ \rm eff} = 
 - \frac{G_F}{\sqrt{2}} \left( V_{ij} (\bar{\tau} \gamma^\mu(1-\gamma_5) \nt )
      (\bar{q}_i \gamma_\mu(1-\gamma_5)q_j ) + h.c.\right) \;.
   \end{equation} 
Hier steht $V_{ij}$ f"ur das Cabbibo-Matrixelement und 
 \begin{equation}\label{fermikonst}
     \frac{1}{\sqrt{2}} G_F := \frac{e^2}{8 m^2_W \sin^2 \vartheta_W}
   \end{equation}
f"ur die Fermikonstante.
\section{Berechnung der $\tau$-Zerfallsrate}
\footnote{Beim Aufschreiben der in diesem Abschnitt dargestellten Herleitung 
der relativen $\tau$-Lepton Zerfallsrate waren viele Hinweise von A.A.~Pivovarov
sehr hilfreich.}
Um die Zerfallsrate des $\tau$-Leptons in semihadronische Endprodukte 
zu berechnen, muss das "Ubergangsmatrixelement zwischen einem $\tau$-Lepton
und den Endprodukten des Zerfalls berechnet werden. 
Aus dem Absolutquadrat des Matrixelements 
ergibt sich dann die Zerfallsrate durch Integration "uber den Phasenraum der 
Zerfallsprodukte und Mittelung "uber alle Anfangszust"ande 
(Spinzust"ande des $\tau$-Leptons).
Au"serdem tritt noch ein Faktor $(2 \mts)^{-1}$ auf.
F"ur die Zerfallsrate eines $\tau$-Leptons in ein $\tau$-Neutrino 
und Hadronen ergibt sich so
\ba \label{zerfallr}
 && \Gamma(\tau \rightarrow \nu_\tau + H) \nn \\ 
   & = & \frac{1}{2 \mt} \:\frac{1}{2} \!\!
             \sum_{\tau \ntb Spin} \sum_H \int 
              \frac{d^3\!k}{(2\pi)^3 2 k_0} (2\pi)^4 
                  \delta(p_H+k-p)
               |\langle H \, \nu_\tau(k)|\mathcal{T}|\tau(p)\rangle|^2 \;. 
\ea
Die Summe "uber $H$ steht f"ur die Summation "uber alle m"oglichen Endzust"ande
der hadronischen Zerfallsprodukte. 
Sie entspricht der Summe "uber alle m"oglichen hadronischen Zerfallskan"ale 
und der Integration "uber den Phasenraum der hadronischen Zerfallsprodukte. 
Die Integration "uber $k$ 
entspricht mit der Summation "uber den $\bar{\nu}_\tau$-Spin der 
Integration "uber 
den Phasenraum des $\tau$-Neutrinos. Die Delta-Distribution beschreibt die 
Energie-Impulserhaltung des Zerfalls.
$p_H$ steht f"ur den Gesamtimpuls der hadronischen Zerfallsprodukte, 
$k$ f"ur den Impuls des Neutrinos
und $p$ f"ur den Impuls des zerfallenden $\tau$-Leptons.
\subsection{Das Optische Theorem} \label{opkap}
Das Optische Theorem erm"oglicht eine 
elegante Berechnung der $\tau$-Zerfallsrate.
Um das Optische Theorem herzuleiten, 
spaltet man zun"achst von der $S$-Matrix den
Wechselwirkungsanteil $T$ ab, 
aus dem sich zus"atzlich
die Delta-Distribution herausziehen l"asst, 
welche die Energie- und Impulserhaltung beschreibt.
So ergibt sich
\be
    S = I + i T = I + i (2\pi)^4 \delta(p_f-p_i)\mathcal{T}. 
\ee
Aus der Unitarit"at der $S$-Matrix $(S^\dagger S = I)$ 
ergibt sich f"ur den abgespaltenen 
Wechselwirkungsanteil $T$ die Relation
\ba
    I = S^\dagger S = (I - i T^\dagger)(I + i T)&=& 
            I + i T - i T^\dagger + T^\dagger T \\  
 \Longrightarrow \quad \quad  i(T - T^\dagger) 
               &=& - T^\dagger T \label{operatoroptheo} \;.
\ea
F"ur ein beliebiges Diagonalmatrixelement von Gl.~(\ref{operatoroptheo}) 
erh"alt man
\ba
   i(\langle A | T | A \rangle -\langle A | T | A \rangle^\star )&=&
                 -  \langle A |  T^\dagger T | A \rangle \\ 
    2\, i^2 \, {\rm Im} \langle A | T | A \rangle &=&
         - \sum_n  \langle A | T^\dagger | n \rangle \langle n | T | A \rangle \\
  {\rm Im}  \langle A | T | A \rangle  &=&  \frac{1}{2} \sum_n |\langle n | T | A \rangle|^2 
  \label{optheo}\;. 
\ea
Gl.~(\ref{optheo}) ist das Optische Theorem. 
Die Summe "uber alle Zwischenzust"ande $n$ entspricht gerade der 
in dem Ausdruck f"ur die Zerfallsrate (\ref{optheo}) auftretenden 
Phasenraumintegration und der Summation "uber die 
Endzust"ande.
Setzt man f"ur 
\be
|A \rangle = |  \bar{\nu}_\tau(k) \tau(p) \rangle
\ee
einen Zustand aus einem $\tau$-Lepton und einem $\tau$-Antineutrino, 
so ergibt sich
mit Gl.~(\ref{optheo})
\be \label{optau}
  {\rm Im} \, \langle \tau(p) \,  \bar{\nu}_\tau(k) |\mathcal{T}|  
    \bar{\nu}_\tau(k) \tau(p) \rangle =
          \frac{1}{2}\sum_{H} (2 \pi)^4 \delta(p_H - k-p) \,  
              |\langle H|\mathcal{T}|
                              \ntb(k) \, \tau(p)  \rangle |^2 \;.
\ee
Hierbei l"auft die Summation $H$ "uber alle 
m"oglichen hadronischen Zerfallsprodukte 
mit Gesamtimpuls $p_H$. 
Gleichung~(\ref{optau}) 
ist in Abb.~\ref{optheorembild} diagrammatisch dargestellt.
\begin{figure}
\begin{center}
\begin{picture}(340,40)(0,0)
 \Text(5,15)[l]{Im $\displaystyle \Bigg( $ }
 \GCirc(60,15){10}{.9}
 \ArrowLine(35,0)(53,9)
 \ArrowLine(35,30)(53,21)
 \Text(37,8)[]{$\tau$}
 \Text(37,21)[]{$\bar{\nu}_\tau$}
 \ArrowLine(67,9)(85,0)
 \ArrowLine(67,21)(85,30)
 \Text(87,8)[]{$\tau$}
 \Text(87,21)[]{$  \bar{\nu}_\tau$}
 \Text(90,15)[l]{$ \displaystyle \Bigg) = \sum_{H} \int d \Pi_H \Bigg($ }
 \GCirc(207,15){10}{.9}
 \ArrowLine(182,0)(200,9)
 \ArrowLine(182,30)(200,21)
 \Text(186,8)[]{$\tau$}
 \Text(186,21)[]{$\bar{\nu}_\tau$}
 \ArrowLine(217,15)(230,15)
 \ArrowLine(215,20)(230,22)
 \ArrowLine(215,10)(230,8)
 \ArrowLine(212,23)(230,29)
 \ArrowLine(212,7)(230,1)
 \Text(233,15)[l]{$\displaystyle H \Bigg)$}
 \Text(270,15)[r]{$\Bigg( H $}
 \GCirc(300,15){10}{.9}
 \ArrowLine(307,9)(325,0)
 \ArrowLine(307,21)(325,30)
 \Text(327,8)[]{$\tau$}
 \Text(327,21)[]{$\bar{\nu}_\tau$}
 \ArrowLine(290,15)(273,15)
 \ArrowLine(292,20)(273,22)
 \ArrowLine(292,10)(273,8)
 \ArrowLine(295,23)(273,29)
 \ArrowLine(295,7)(273,1)
 \Text(330,15)[l]{$ \displaystyle \Bigg )$}
\end{picture}
\end{center}
\caption{\label{optheorembild} Darstellung von Gleichung (\ref{optau}). 
   Die Summe "uber $H$ l"auft "uber alle hadronischen Zerfallskan"ale und
$\int d\Pi_f$ steht f"ur die Integration "uber den Phasenraum der Hadronen. }
\end{figure}
Mit der Crossingsymmetrie l"asst sich ein Teilchen im Anfangszustand 
eines Marixelementes 
durch das entsprechende Antiteilchen mit 
umgekehrtem Viererimpuls im Endzustand ersetzen. 
Bei Fermionen tritt hierbei
ein zus"atzliches Vorzeichen auf.
Aus Gl.~(\ref{optau}) ergibt sich, 
wenn das $\tau$-Antineutrino auf der rechten Seite
von Gl.~(\ref{optau}) durch ein $\tau$-Neutrino im Endzustand ersetzt wird,
somit
 \begin{equation} \label{crossing}
   {\rm Im} \, \langle \tau(p) \bar{\nu}_\tau(-k) 
 |\mathcal{T}|  \bar{\nu}_\tau(-k) \tau(p) \rangle = 
        - \frac{1}{2} \sum_H (2 \pi)^4 \delta(p_H + k  -p )
              | \langle H \nu_\tau(k)|\mathcal{T}|\tau(p) \rangle |^2 \;.
 \end{equation}
Setzt man Gl.~(\ref{crossing}) in den Ausdruck f"ur die 
Zerfallsrate Gl.~(\ref{zerfallr}) 
ein,
so ergibt sich f"ur die $\tau$-Zerfallsrate
 \begin{eqnarray} \label{gammamatrixelem}
   \Gamma(\tau \rightarrow \nu_\tau + H) 
   & = & \frac{1}{2 \mt} \frac{1}{2} 
             \sum_{\tau \ntb Spin} \sum_H \int \frac{d^3\!k}{(2\pi)^3 2 k_0} 
       (2\pi)^4 \delta(p_H+k-p)
               |\langle H \, \nu_\tau(k)|\mathcal{T}|\tau(p)\rangle|^2 \nn \\
   & = & - \frac{1}{ 2\mt}\sum_{\tau \ntb Spin} \int 
        \frac{d^3\!k}{(2\pi)^3 2 k_0} {\rm Im} \, 
        \langle \tau(p) \bar{\nu}_\tau(-k) |\mathcal{T}|  \bar{\nu}_\tau(-k) \tau(p) \rangle \;.
 \end{eqnarray}
Diese Zerfallsrate kann auch direkt mit den Resultaten f\"ur den drei Teilchen Phasenraum 
berechnet werden \cite{watermellon}.

\subsection{Berechnung des Matrixelementes}
Um die $\tau$-Zerfallsrate zu berechnen, 
mu"s das in Gl.~(\ref{gammamatrixelem}) 
auftretende Matrixelement berechnet werden.
In der f"uhrenden Ordnung  
ergibt sich f"ur das Matrixelement 
aus Gl.~(\ref{gammamatrixelem})  
 \ba
  &&\Lg = \nn \\
  &=& \frac{1}{2} \intd d^4\!x\,d^4\!y \;
        \langle \tau(p) \, \ntb(k) | -T  (\mathcal{L}_{ \rm eff}(x) 
           \mathcal{L}_{ \rm eff}(y))
        | \ntb(k) \tau(p) \rangle +  \mathcal{O}(G_F^2)
 \ea   
Hierbei wurde f"ur die $S$-Matrix der erste Term der Reihenentwicklung in $G_F$
eingesetzt, 
der einen nicht verschwindenden Beitrag liefert.
$\mathcal{L}_{ \rm eff}$ entspricht dem
f"ur den $\tau$-Zerfall relevanten Anteil der 
effektiven Lagrangedichte aus Gl.~(\ref{leffektiv}).
F"ur das gesuchte Matrixelement ergibt 
sich so in niedrigster Ordnung in $G_F$
 \begin{eqnarray} \label{kkkkkkk}
  \Lg  &=& - \frac{G_F^2 |V_{ij}|^2}{2}  
    \,  \intd d^4\!x\,d^4\!y \, \langle 0| b_\tau(p) d_{\nt}(k) \times \nn \\
        && \times \; \; T ( : \!(\bar{\tau}(x)\gamma^\mu(1-\gamma_5)\nt(x))\!: 
                    \times \nn \\
  && \times \quad \; : \!(\ntb(y)\gamma^\nu(1-\gamma_5)\tau(y)) \!: \times \nn\\
  && \times \quad \; : \!(\bar{q}_i(x)\gamma_\mu(1-\gamma_5)q_j(x))\!:  \times \nn \\
  && \times \quad \, \left. : \!(\bar{q}_i(y)\gamma_\nu(1-\gamma_5)q_j(y))^\dagger \! 
                  :\right) \,
                d^\dagger_{\nt}(k^\prime) b^\dagger_\tau(p^\prime)|0\rangle \;.
 \end{eqnarray}
In Gl.~(\ref{kkkkkkk}) wurden bereits die beiden beim Ausmultiplizieren des Produktes 
der Lagrangedichten auftretenden Summanden zusammengefasst, 
die f"ur diesen Prozess relevant sind.
Die Kontraktion der Erzeuger bzw. 
Vernichter der "au"seren Teilchen mit den entsprechenden Spinorfeldern 
in der Lagrangedichte ergibt
 \begin{eqnarray} \label{matrixelemxx}
   \Lg  &=&  - \frac{G_F^2 |V_{ij}|^2}{2}
             \intd  d^4 \!x \, d^4\!y \; 
        (\bar{u}(p)\gamma^\mu(1-\gamma_5)v(k))e^{ix(p+k)} \times \nn \\
        & &  \times \Pi_{ij\mu\nu}(x-y)
      (\bar{v}(k^\prime)\gamma^\nu(1-\gamma_5)u(p^\prime))
                     e^{-iy(p^\prime+k^\prime)}  \;,
 \end{eqnarray}            
wobei die Korrelationsfunktion $\Pi_{ij\mu\nu}(x)$ 
definiert ist als
 \begin{eqnarray}
    \Pi_{ij\mu\nu}(x)
           &=& \langle 0 |
:\!(\bar{q}_i(x)\gamma_\mu(1-\gamma_5)q_j(x)) \! : \, : \!
             (\bar{q}_i(0)\gamma_\nu(1-\gamma_5)q_j(0))^\dagger \!
      : |0 \rangle \nn \\
           &=&  \langle 0 |
        : \! (\bar{q}_i(x)\gamma_\mu q_j(x)) \! : \, : \! 
                (\bar{q}_i(0)\gamma_\nu q_j(0))^\dagger \! 
       : |0 \rangle \nonumber\\  
          & &+\langle 0 |
        :\! (\bar{q}_i(x)\gamma_\mu(1-\gamma_5)q_j(x)) \! :  \, : \! 
                 (\bar{q}_i(0)\gamma_\nu(1-\gamma_5)q_j(0))^\dagger \! 
             : |0 \rangle.  
 \end{eqnarray} 
Gl.~(\ref{matrixelemxx}) ist in Abb.~\ref{Quarkblase} als Diagramm
dargestellt.
\begin{figure}
\begin{center}
\begin{picture}(150,80)(0,0)
\GOval(75,40)(25,33)(0){.9}
\Vertex(42,40){2}
\Vertex(108,40){2}
\ArrowLine(0,0)(42,40)
\ArrowLine(0,80)(42,40)
\ArrowLine(108,40)(150,0)
\ArrowLine(108,40)(150,80)
\Text(10,18)[]{$\tau$}
\Text(10,60)[]{$\bar{\nu}_\tau$}
\Text(140,18)[]{$\tau$}
\Text(140,60)[]{$\bar{\nu}_\tau$}
\end{picture}
\end{center}
\caption{\label{Quarkblase} Diagramm f"ur Gl.~(\ref{matrixelemxx})}
\end{figure}
Alle QCD-Effekte werden 
durch die Korrelationsfunktion $\Pi_{ij\mu\nu}(x)$ beschrieben.  
Da in der QCD die Parit"at erhalten ist, verschwinden
die Vakuumerwartungswerte der Produkte aus Vektor- und Axialvektorstr"omen.

Die Korrelationsfunktion ist ein Lorenztensor zweiter Stufe. Sie l"asst sich in einen 
transversalen und einen longitudinalen Anteil zerlegen: 
 \begin{eqnarray} \label{korrelator}
   \Pi_{ij}^{\mu\nu}(x-y) &=& 
                - i \int \frac{d^4\!q}{(2\pi)^4} 
                            \Pi_{ij}^{\mu\nu}(q)e^{-iq(x-y)}  \\
                          &=& -i \frac{N_c}{6  \pi^2} \int 
                               \frac{d^4\!q}{(2\pi)^4}e^{-iq(x-y)}
                  \{(q^\mu q^\nu - g^{\mu \nu} q^2 ) \Pi_{Tij}(q^2) 
                     + q^\mu q^\nu \Pi_{Lij}(q^2)\} \nn \;.
 \end{eqnarray}
Der transversale Anteil beschreibt 
Spin-1-Hadronen und der longitudinale Anteil Spin-0-Hadronen.   
Setzt man die Zerlegung der Korrelationsfunktion aus Gl.~(\ref{korrelator}) 
in den Ausdruck
f"ur das Matrixelement aus Gl.~(\ref{matrixelemxx}) ein, so ergibt sich
 \begin{eqnarray}
   &&  - \frac{ N_c G_F^2 | V_{ij}|^2 }{12 \pi^2} 
        \intd  d^4\!x d^4\!y \; 
            (\bar{u}(p)\gamma^\mu(1-\gamma_5)v(k))e^{ix(p+k)} \times \nn \\
   &&\qquad \quad  \times ( -i) \int \frac{d^4q}{(2\pi)^4}
                 e^{-iq(x-y)}\left((q_\mu q_\nu - g_{\mu \nu} q^2 ) 
        \Pi_{Tij}(q^2) 
       + q_\mu q_\nu \Pi_{Lij}(q^2)\right) \times \nn \\  
      &&  \qquad\qquad \qquad\qquad \qquad \times
  (\bar{v}(k^\prime)\gamma^\nu(1-\gamma_5)u(p^\prime))e^{-iy(p^\prime+k^\prime)} \;.
 \end{eqnarray}
Um die beiden Integrationen im Ortsraum auszuf"uhren, transformiert 
man in Schwerpunkt- und Relativkoordinaten: 
 \begin{eqnarray}
  X            &:=& \frac{1}{2} (x+y) \\
  r            &:=& (x-y)\\
  d^4\!x \, d^4\!y & =& d^4\!X \, d^4\!r \;.
 \end{eqnarray}
Die Integrationen "uber $d^4\!x\,d^4\!y = d^4\!X\,d^4\!r$ ergeben 
zwei Delta-Distributionen,
welche die Energie-Impulserhaltung des gesamten Prozesses und an den Vertizes 
(vgl. Abb.~\ref{Quarkblase})
beschreiben. F"ur das gesuchte Matrixelement ergibt sich so
 \begin{eqnarray} 
     &&   i (2 \pi)^4 \delta(p+k-p^\prime -k^\prime)  
               \frac{N_c G_F^2 |V_{ij}|^2}{12 \pi^2}  
        \int \frac{d^4\!q}{(2\pi)^4} (2\pi)^4 \delta(p+k-q) \times \nonumber \\
       && \qquad \qquad \qquad\qquad \qquad \times
                (\bar{u}(p)\gamma^\mu(1-\gamma_5)v(k))
                 (\bar{v}(k^\prime)\gamma^\nu(1-\gamma_5)u(p^\prime))\times \nn \\
       &&  \qquad \qquad \qquad \qquad \qquad \times
         \{(q_\mu q_\nu - g_{\mu \nu} q^2 ) \Pi_{T ij}(q^2) 
            + q_\mu q_\nu \Pi_{L ij}(q^2)\} \;.
 \end{eqnarray} 
In dem Ausdruck f"ur die $\tau$-Zerfallsrate (Gl.~(\ref{gammamatrixelem}))
tritt eine Summation "uber die Spinzust"ande des
$\tau$-Leptons und des $\tau$-Antineutrinos auf.
Diese kann mit Casimirs Trick berechnet werden,
der im wesentlichen eine Konsequenz der 
Vollst"andigkeitsrelationen der Spinoren ist,
welche durch
 \begin{equation}\label{vs1}
   \sum_{s=1,2} u^{(s)}(p)\bar{u}^{(s)}(p) = ( \shp + m)
 \end{equation}
 \begin{equation}\label{vs2}
   \sum_{s=1,2} v^{(s)}(p)\bar{v}^{(s)}(p) = ( \shp -  m) 
 \end{equation}
gegeben sind.
Mit den Vollst"andigkeitsrelationen (Gl.~(\ref{vs1}, \ref{vs2})) 
lassen sich die Summationen 
"uber die Spins 
des $\tau$-Leptons und $\tau$-Antineutrinos berechnen. Es ergibt sich
 \begin{eqnarray} \label{leptens}
   & & \sum_{\tau,\ntb \;{\rm Spin}} 
            \left(\bar{u}(p)\gamma^\mu(1-\gamma_5)v(k) \right)
              \left( \bar{v}(k)\gamma^\nu(1-\gamma_5)u(p) \right) = \nn \\   
   &=& \sum_{\tau \;{\rm Spin}}
         \left( \bar{u}(p) 
        \gamma^\mu(1-\gamma_5) \shk  \gamma^\nu(1-\gamma_5)
         u(p) \right) \nn \\ 
   &=& {\rm Tr } \left(
         \gamma^\mu(1-\gamma_5) \shk \gamma^\nu(1-\gamma_5)
           \sum_{\tau \; {\rm Spin}} ( u(p) \bar{u}(p)  ) \right) \nn \\       
   &=&  {\rm Tr} \left( \gamma^\mu(1-\gamma_5)
         \shk  \gamma^\nu(1-\gamma_5)
                                       (\shp + \mt) \right) \;.
 \end{eqnarray}
Die Berechnung der Spuren und die Kontraktion des Leptontensors 
(Gl.~(\ref{leptens}))
mit der Tensorstruktur der Korrelationsfunktion $\Pi_{ij}^{\mu \nu}$ 
l"asst sich von Hand mit den Spurtheoremen f"ur $\gamma$-Matritzen 
\cite{pessch} 
oder bequemerweise mit
Hilfe eines Programmpaketes unter MATHEMATICA \cite{HipPaket}
ausf"uhren.
Man erh"alt f"ur die Spin-1-Komponente
 \begin{equation}
   (q_\mu q_\nu -g_{\mu \nu }q^2)
      {\rm Tr} \left(\gamma^\mu(1-\gamma_5) \shk  \gamma^\nu
        (1-\gamma_5)(\shp +\mt) \right)
         = 8 \left( 2 (k \cdot q)(  p \cdot q)+(k \cdot p)q^2 \right)
 \end{equation}
und f"ur die Spin-0-Komponente
 \begin{equation}
   q_\mu q_\nu 
  { \rm Tr} \left( \gamma^\mu(1-\gamma_5) \shk  \gamma^\nu(1-\gamma_5)
           (\shp + \mt) \right) 
        = 8 \left( 2 (k \cdot q)(  p \cdot q)- (k \cdot p)q^2 \right) \;.
 \end{equation}
F"ur das gesuchte Matixelement ergibt sich 
 \begin{eqnarray}
    && \frac{1}{2}\sum_{\tau,\ntb Spin}  \Lg = \\
    &=& i (2 \pi)^4  \delta(p+k-p^\prime -k^\prime)
      \frac{ N_c G_F^2 |V_{ij}|^2 }{3 \pi^2 } 
                  \int \frac{d^4 \! q}{(2 \pi)^4} \delta(p +k-q) \times \nn \\
    &&        \times  \left(  (2 (k \cdot q)(  p \cdot q)+
          (k \cdot p)\, q^2 ) \Pi_{Tij}(q^2) + 
                        (2 (k \cdot q)(  p \cdot q)- (k \cdot p)\,q^2)  
                      \Pi_{Lij}(q^2) \right) \;.\nn
 \end{eqnarray}
Nach Ausf"uhren der $q$-Integration mit der Delta-Distribution erh"alt man
\begin{samepage}
 \begin{eqnarray} \label{matrixfast}
   && \frac{1}{2}\sum_{\tau,\ntb Spin} \Lg \nn \\
   &=& i (2\pi)^4 \delta(p+k-p^\pr -k^\pr)
       \;\frac{1}{2}\!\!\!\! 
           \sum_{\tau,\ntb Spin}\langle \tau(p) \bar{\nu}_\tau(k) 
             | \mathcal{T}|  \bar{\nu}_\tau(k^\prime) \tau(p^\prime) \rangle \nn\\
       &=& i (2\pi)^4 \delta(p+k-p^\pr -k^\pr)
           \frac{ N_c G_F^2 |V_{ij}|^2 }{3 \pi^2}  \times\nn \\ 
     &&  \qquad  \qquad \times \left( (3(k^2+p^2)(k \cdot p) 
               + 4 (k \cdot p)^2 + 2 k^2 p^2) \Pi_{Tij}((p+k)^2) \right.\nn \\
     && \qquad \qquad \quad \left.
          + ( (k^2+p^2)(k \cdot p) + 2 k^2 p^2 ) \Pi_{Lij}((p+k)^2) \right) \;.
 \end{eqnarray}
\end{samepage}
Einsetzen von Gl.~(\ref{matrixfast}) in den Audruck f"ur die Zerfallsbreite 
(Gl.~(\ref{gammamatrixelem}))
ergibt 
 \begin{eqnarray} \label{ggg}
    \Gamma_{\tau ij}  
    &=& - \frac{N_c G_F^2 |V_{ij}|^2}{3 \pi^2 M_\tau}
       \int \frac{d^3\!k}{(2 \pi)^3 2 k_0} \times \nn \\ 
    & & \times {\rm Im} \{ (-3(k^2+p^2)(k \cdot p) 
            + 4 (k \cdot p)^2 + 2 k^2 p^2) \Pi_{T ij}((p-k)^2)  \nn \\
    & & \qquad \qquad \qquad \; \,+ ( -(k^2+p^2)(k \cdot p) 
            + 2 k^2 p^2 ) \Pi_{L ij}((p-k)^2) \} \;. 
 \end{eqnarray}
Die Integration "uber den $\tau$-Neutrino-Impuls $k$ 
l"asst sich am einfachsten im Ruhesystem 
des $\tau$-Leptons ausf"uhren. 
Dann gilt f"ur den Impuls des $\tau$-Leptons $p$ und den Impuls 
des $\tau$-Neutrinos $k$
 \begin{eqnarray}
   p &=& (\mt,0) \;,\quad \quad p^2 = \mts\label{mruhe1} \;,\\
   k &=& (|\vec{k}|,\vec{k}) \;, \: \quad \quad k^2 = 0 \label{mruhe2}\;.
 \end{eqnarray}
In Gl.~(\ref{matrixfast}) tritt der Impuls $k$ 
nur als Quadrat oder als Lorenzskalarprodukt mit
$p$ auf. Da $k^2=0$ und $(k \cdot p) = \mt |\vec{k}|$ sind, 
h"angt der Integrand nur von 
dem Betrag von $\vec{k}$ ab. 
Die Winkelintegration l"asst sich somit trivial ausf"uhren.
Das Integrationsma"s in Gl.~(\ref{ggg}) wird zu
 \begin{equation} \label{mruhe3}
  \frac{d^3\!k}{(2\pi)^3 2 k_0} = 
        \frac{4\pi \vec{k}^2 \: d|\vec{k}|}{(2 \pi)^3 2 |\vec{k}|} = 
           \frac{|\vec{k}| \: d|\vec{k}|}{4 \pi^2} \;.
 \end{equation}
Anstelle von $|\vec{k}|$ f"uhrt man als Integrationsvariable 
das Quadrat des Gesamtimpulses
der hadronischen Zerfallsprodukte $s$ ein: 
 \begin{eqnarray} 
   s   &:=&  (p-k)^2 = \mt^2 -2\mt |\vec{k}|\;, \\
   ds  &=& -2 \mt \: d|\vec{k}| \label{intmass}\;.
 \end{eqnarray}
Da es keine Hadronen mit negativer Energie gibt, sind die Imagin"arteile der 
Korrelationsfunktionen
$\Pi_{Tij}$ und $\Pi_{Lij}$ proportional zu der Stufenfunktion $\Theta(s)$.
F"ur die Integrationsgrenzen 
bez"uglich der neuen Integrationsvariablen $s$ ergibt sich 
deswegen
 \begin{equation} \label{intgrenz}
   \int^\infty_0 d|\vec{k}| \: \Theta(s) = 
          \int_{\mt^2}^{-\infty} \Theta(s) \frac{-d\!s}{2 \mt} = 
             \int^{\mt^2}_0\Theta(s) \frac{d\!s}{2 \mt} \;.
 \end{equation}
Einsetzen von Gl.~(\ref{mruhe1}, \ref{mruhe2}, \ref{mruhe3}) in 
Gl.~(\ref{ggg}) ergibt f"ur den Spin-1-Teil
 \be  \label{s1}
    |\vec{k}| (-3(k^2+p^2)(k \cdot p) + 
            4 (k \cdot p)^2 + 2 k^2 p^2) 
     = -\frac{\mt^5}{4} \left( 1-\frac{s}{\mt^2} \right)^2  
                        \left(1+\frac{2s}{\mt^2} \right)
  \ee
und f"ur den Spin-0-Teil
 \be \label{s0}
   |\vec{k}| (-(k^2+p^2)(k \cdot p)  + 2 k^2 p^2) 
    \Pi_{L ij}((p-k)^2) 
 =            -\frac{\mt^5}{4} 
  \left( 1-\frac{s}{\mt^2} \right)^2 
\ee
Mit Gl.~(\ref{s1}, \ref{s0}), 
der Transformation des Differentials Gl.~(\ref{intmass}) und den 
Integrationsgrenzen Gl.~(\ref{intgrenz}) erh"alt man schlie"slich 
aus Gl.~(\ref{ggg}) den 
Ausdruck f"ur die Zerfallsrate
 \begin{equation} \label{gammatau}
  \Gamma_{ij} = \frac{ N_c |V_{ij}|^2 G_F \mt^5}{192 \pi^4} 
            \int_0^{\mt^2} 2
         \left( 1-\frac{s}{\mt^2} \right)^2
               \left( \left(1+\frac{2s}{\mt^2} \right) {\rm Im} \: \Pi_{Tij}(s) 
              + {\rm Im} \: \Pi_{L ij}(s) \right)\frac{ds}{\mt^2} \;. 
 \end{equation}
\section{Leptonische Zerf"alle des $\tau$-Leptons}
Die Zerfallsrate des $\tau$-Leptons in leichtere Leptonen $(e,\mu)$ 
kann mit Gl. (\ref{gammatau})
bestimmt werden. Die Korrelationsfunktion f"ur zwei Leptonstr"ome ist
\be \label{lepkorrfk}
 \Pi^{{\rm Lepton}}_{\mu \nu} = i \int d^4 \! x \, e^{i q x} \: 
  \langle T j_\mu^L(x)  j_\nu^{\dagger L}(0) \rangle = \frac{1}{6 \pi^2} 
       \left( (q_\mu q_\nu - g_{\mu \nu} q^2 ) \Pi_{T}^{{\rm Lep}} + q_\mu q_\nu \Pi_L^{{ \rm Lep}} \right)\;,
\ee
wobei der Strom durch $j_\mu^L(x) = \bar{l}(x) \gamma_\mu (1-\gamma_5) \nu(x)$ 
gegeben ist.
In f"uhrender Ordnung ergibt sich f"ur die Korrelationsfunktionen 
aus Gl.~(\ref{lepkorrfk})
\footnote{Die Berechnung der f"uhrenden Ordnung der Korrelationsfunktion 
f"ur zwei Leptonstr"ome ist vollkommen analog zu der in Abschnitt 
\ref{berkorrf}
beschriebenen Berechnung der Korrelationsfunktion zweier Quarkstr"ome,
wenn man von dem Nichtauftreten des Farbfaktors $N_c$ absieht.}
 \begin{equation} 
   \Pi^{ {\rm Lep} }_T(q^2) = \ln \left( \frac{\mt}{-q^2} \right)
 \end{equation}
und 
 \begin{equation}
    \Pi^{ \rm Lep}_{L}(q^2) = 0\;.
 \end{equation}
Hierbei wurde die Masse des leichten Leptons vernachl"assigt.
F"ur die Imagin"arteile der Korrelationsfunktionen erh"alt man im 
Integrationsbereich (${\rm Re}(q^2) > 0 ;\; {\rm Im}(q^2)= 0$)
\footnote{Die 
analytische Fortsetzung des 
Logarithmus f"ur komplexe Argumente $z= r e^{i\phi}$
ist f"ur den Hauptzweig durch 
\be
 \ln z = \ln r + i \phi \;, \quad\quad  -\pi < \phi \leq \pi
\ee
gegeben.} 
 \begin{equation}
    {\rm Im} \:\Pi_{T ij}(q^2) = \pi \quad 
        {\rm und \quad Im} \:\Pi_{L ij}(q^2) = 0\;.
 \end{equation}
F"ur die Zerfallsrate des $\tau$-Leptons in ein 
als masselos angenommenes Lepton
ergibt sich somit 
in Baumgraphenn"aherung
 \begin{equation} \label{gammalepton}
   \Gamma(\tau \rightarrow \nt \bar{\nu}_l l)= 
   \frac{G_F^2 \mt^5 }{192  \pi^4} \int_0^{\mt^2} 2 
      \left( 1-\frac{s}{\mt^2} \right)^2  \left(1+\frac{2s}{\mt^2} \right)\pi
       \frac{ds}{\mt^2}  =
    \frac{G_F^2 M_\tau^5}{192 \pi^3} \;.
 \end{equation}
\section{Die relative Zerfallsrate $R_\tau$}
Die Zerfallsrate des $\tau$-Leptons in semihadronische Zerfallsprodukte 
l"asst sich relativ zu der leptonischen Zerfallsrate ausdr"ucken.
Man definiert in Analogie zu dem 
relativen $(e^+e^-)$-Wirkungsquerschnitt (Gl.~\ref{eecross}).
\be \label{Rtau}
R_\tau = \frac{\Gamma(\tau \rightarrow \nu_\tau {\rm Hadronen})}
             {\Gamma(\tau \rightarrow \nu_\tau l \bar{\nu}_l)}
\ee
Verwendet man die im letzen Abschnitt erzielten Resultate f"ur die 
Zerfallsrate des $\tau$-Leptons in hadronische (Gl.~(\ref{gammatau})) und 
leptonische (Gl.~(\ref{gammalepton})) Zerfallsprodukte, so ergibt sich 
aus (Gl.~(\ref{Rtau}))
 \begin{equation} \label{Rtauexpl}
   R_{\tau  ij}=
       {N_c} S_{EW}|V_{ij}|^2  \int_0^{\mt^2} 2  
           \left( 1-\frac{s}{\mt^2} \right)^2 
                     \left( \left(1+\frac{2s}{\mt^2}\right) 
    \frac{{\rm Im} \,  \Pi_{T ij}(s)}{\pi} 
           +  \frac{ {\rm Im} \, \Pi_{L ij}(s)}{\pi} 
        \right) \frac{ds}{\mt^2} \;.
 \end{equation}
$S_{EW}$ beschreibt elektroschwache Korrekturen, die im wesentlichen durch 
Quarkladungen zustandekommen, welche
von den ganzzahligen elektromagnetischen Ladungen der Leptonen abweichenden.
Die relative Zerfallsbreite l"asst sich als 
\be \label{rtaurr}
R_\tau =R_{\tau ud} +R_{\tau us}=
          N_c S_{EW} (( |V_{ud}|^2 (1+ \delta_{ud}) + |V_{us}|^2( 1+\delta_{us})) 
\ee
schreiben.
Die f"uhrenden Terme sind hierbei die Parton-Modell-Resultate. 
$\delta_{ud}$ und $\delta_{us}$ stehen f"ur Effekte der starken 
Wechselwirkung h"oherer Ordnung.
$V_{ud}$ und $V_{us}$  sind Elemente der 
Cabbibo-Kobayashi-Maskawa-Mischungsmatrix.
$R_{\tau ud }$ ist die Zerfallsrate in Hadronen mit Strangeness 0 
und $R_{\tau us}$
die Zerfallsrate in Hadronen mit Strangeness 1.
Nichtperturbative und additive elektroschwache Korrekturen zu $R_\tau$
wurden in Gl.~(\ref{rtaurr}) unterdr"uckt. 
\section{Die Zerlegung der Korrelationsfunktion in einen $q$- und einen $g$-Teil}
Alternativ zu der in der Herleitung der Wichtungsfunktion 
verwendeten Zerlegung der
Korrelationsfunktion in 
transversale und longitudinale Anteile l"asst sich die Korrealtionsfunktion
auch in einen $q$- und einen $g$-Teil zerlegen. 
Die Zerlegung der Korrelationsfunktion lautet dann
\begin{equation}
 \Pi^{\mu \nu}_{ij}(q^2) = \frac{N_c}{6  \pi^2} \left(
 q^\mu q^\nu \Pi_{qij}(q^2) + g^{\mu \nu} \Pi_{gij}(q^2) \right) \;.
\end{equation}
Der Zusammenhang zu der physikalischen Zerlegung ist 
\begin{eqnarray}
 \Pi_{Tij}(q^2) &=& \frac{\Pi_{gij}(q^2)}{-q^2} \;, \nn \\ 
 \Pi_{Lij}(q^2) &=& \Pi_{q ij}(q^2)- \frac{\Pi_{gij}(q^2)}{-q^2}  \;.
\end{eqnarray}
Ausgedr"uckt durch die Korrelationsfunktionen des $q$- und des $g$-Teils 
ergibt sich f"ur die relative $\tau$-Zerfallsrate (Gl.~(\ref{Rtauexpl}))
\begin{equation} \label{Rtauexplqg}
   R_{\tau ij}=
       {N_c} S_{EW} |V_{ij}|^2 \int_0^{\mt^2} 2
          \left( 1-\frac{s}{\mt^2} \right)^2 
                \left( \frac{{\rm Im} \, \Pi_{q ij}(s)}{\pi} 
     - \frac{2}{\mts} \frac{{\rm Im } \, \Pi_{g ij}(s)}{\pi} 
        \right)\frac{ds}{\mt^2}   \;.  
 \end{equation}
\section{Die Integration "uber die Kontur}
Der Ausdruck f"ur die relative $\tau$-Zerfallsrate l"asst sich umformulieren,
indem man die analytischen Eigenschaften der Korrelationsfunktionen 
$\Pi_{L/T}$ bzw. $\Pi_{q/g}$ ausnutzt. 
Diese sind analytische Funktionen
von $s$ in der komplexen Ebene. Auf der positiven reellen Achse besitzen sie 
einen Schnitt, an dem der Imagin"arteil der Korrelationsfunktionen 
eine Diskontinuit"at besitzt.
Dies l"asst sich folgendermassen einsehen:
Jedes Feynman-Diagramm, dass zu einem $S$-Matrixelement beitr"agt, 
ist reell, falls keiner der Nenner verschwindet, so dass der $i\ep$-Term
in den Propagatoren relevant wird. Ein Feynman-Diagramm tr"agt also nur dann  
einen imagin"aren Anteil zu einem $S$-Matrixelement bei, wenn das virtuelle
Teilchen auf der Massenschale liegt.
Betrachtet man ein $S$-Matrixelement $M(s)$ als Funktion der 
Gesamtenergie im Schwerpunksystem 
$s = E_{\rm cm}^2$, so k"onnen die virtuellen Zwischenzust"ande 
nicht auf der Massenschale liegen, falls $s<s_0$ ist, wobei 
$s_0$ der leichteste Vielteilchenzustand ist.
F"ur reelle $s<s_0$ ist $M(s)$ also reell und es gilt
\be \label{ssp}
M(s) = (M(s^*))^*\;.
\ee
Aus dem Schwarz'schen Spiegelungsprinzip folgt, dass die analytische 
Fortsetzung von $M(s)$ zu komplexen Argumenten $s$ 
die Gleichung~(\ref{ssp}) erf"ullt. 
In der N"ahe der reellen Achse impliziert dies: 
\ba
{\rm Re} \, M(s + i \ep) &=&{\rm Re} \, M(s - i \ep)\; , \\
{\rm Im} \, M(s + i \ep) &=&- {\rm Im} \, M(s - i \ep) \;. 
\ea
So l"asst sich der Imagin"arteil eines Matrixelementes mit 
der Diskontinuit"at in Verbindung bringen
\be
M(s+i\ep) - M(s-i \ep)  = {\rm Disc} \, M(s) = 2 i \, {\rm Im}\, M(s+i\ep) \;.
\ee
Aufgrund der analytischen Eigenschaften der Korrelationsfunktionen 
verschwindet 
das Konturintegral entlang der in Abbildung 
\ref{kontour} abgebildeten Kontur nach 
dem Cauchy-Theorem, da der Integrand innerhalb der Schleife analytisch ist und 
keine Singularit"aten besitzt. 
\begin{figure}
\begin{center}
\begin{picture}(200,200)(0,0)
\ArrowArc(100,100)(90,5,355)
\DashLine(0,100)(200,100){8}
\DashLine(100,0)(100,200){8}
\ArrowLine(94,107)(191,107)
\ArrowLine(191,93)(94,93)
\Line(94,107)(94,93)
\Text(100,210)[]{${\rm Im} (s)$}
\Text(210,90)[]{${\rm Re} (s)$}
\Line(100,100)(210,100)
\end{picture}
\caption{\label{kontour}Integrationskontur in der komplexen s-Ebene }
\end{center}
\end{figure}
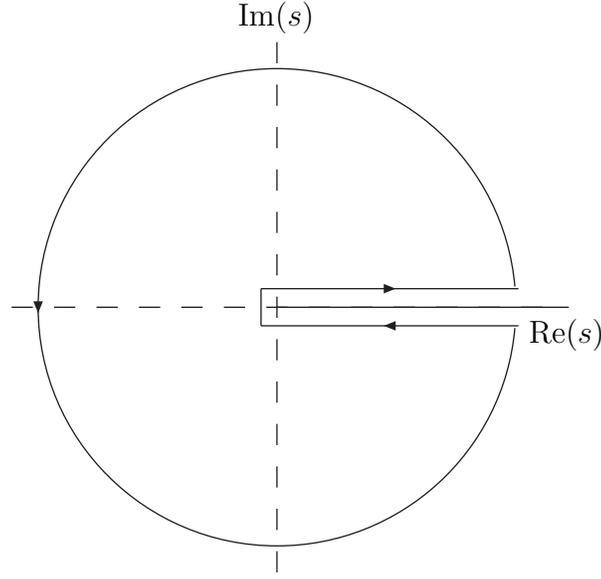
F"ur die Korrelationsfunktionen gilt die 
Dispersionsrelation
\be
  \frac{ {\rm Im} \: \Pi_{r}(q^2+ i \ep )}{\pi} = 
     \frac{-i}{2 \pi} (\Pi_r(q^2 + i \ep) -\Pi_r(q^2 - i \ep)) 
\ee
mit
$r = ( L,T,q,g)$.
F"ur die relative Zerfallsrate (\ref{Rtauexpl}) ergibt sich so
 \begin{equation} \label{Rtauexplcont}
   R_{\tau ij} =
       {N_c} S_{EW}|V_{ij}|^2  \frac{i}{2 \pi}  
     \oint_{|s|^2=\mt^2} 2 \left( 1-\frac{s}{\mt^2} \right)^2 
                    \left( \left(1+\frac{2s}{\mt^2}\right) 
       \Pi_{T ij}(s) +  \Pi_{L ij}(s) 
        \right)\frac{ds}{\mt^2} 
 \end{equation}
und f"ur (\ref{Rtauexplqg})
\begin{equation} \label{Rtauexplqgcont}
   R_{\tau ij} =
       {N_c} S_{EW}|V_{ij}|^2  \frac{i}{2 \pi} \oint_{|s|^2 = \mt^2} 
               2  \left( 1-\frac{s}{\mt^2} \right)^2 
                     \left( \Pi_{q ij}(s) - \frac{2}{\mts}   \Pi_{g ij}(s) 
        \right) \frac{ds}{\mt^2} \;.  
 \end{equation}
Die Integrationskontur der Integrationen in den Gleichungen
(\ref{Rtauexplcont}, \ref{Rtauexplqgcont}) 
verl"auft auf einem Kreis mit dem Radius $\mts$ gegen den Uhrzeigersinn
von $s= \mts + i \ep$ nach $s= \mts - i \ep$.

\chapter{Die Korrelationsfunktion} \label{kf}
Die in dem Ausdruck f"ur die $\tau$-Zerfallsrate 
auftretende Korrelationsfunktion $\Pi_{ij}^{\mu\nu}$
beschreibt alle QCD-Effekte dieses Prozesses.
In diesem Kapitel sollen die bereits im vorherigen 
Kapitel verwendeten analytischen Eigenschaften der 
Korrelationsfunktion weiter erl"autert werden und
die Berechnung der Korrelationsfunktion 
im Rahmen der perturbativen QCD durch die Operator-Produkt-Entwicklung (OPE)
und deren Resultate dargestellt werden.

\section{Herleitung der Dispersionsrelation}

Die Dispersionsrelationen sind im wesentlichen 
eine Konsequenz von Kausalit"atsbedingungen, aus denen Aussagen "uber 
die Analytizit"atsbereiche der Korrelationsfunktionen getroffen werden
k"onnen. Die Herleitung einer Dispersionsrelation soll in diesem Abschnitt
am Beispiel der Korrelationsfunktion zweier hermitischer, skalarer Felder
dargestellt werden.
Die Darstellung ist an \cite{bodrell} angelehnt.
An diesem Beispiel lassen sich alle wesentlichen Ideen darstellen.
\subsection{Die Spektraldichte f"ur skalare Felder}
Die Spektraldichte zweier Skalarfelder wird als 
\be \label{specdens}
\rho(q)= \frac{1}{2 \pi} 
           \int d^4 \! x \, e^{iqx} \langle 0| \phi(x) \phi(0) |0 \rangle 
\ee
definiert.
Durch das Einsetzen einer mit der Vollst"andigkeitsrelation 
ausgedr"uckten Eins $(I = \sum_H | H \rangle \langle H|)$ 
erh"alt man
\be
\rho(q)=\frac{1}{2 \pi} \int d^4\!x \, e^{iqx}
                \sum_H \langle 0| \phi(x) | H \rangle  \langle H| \phi (0) | 0 \rangle \;.
\ee
Die $x$-Abh"angigkeit des ersten Matrixelements l"asst sich durch die 
Transformationseigenschaft das Feldes $\phi$ 
bez"uglich der Translationen isolieren.
Es gilt
\be
\phi(x)  = e^{i \hat{p} x} \phi(0) e^{- i \hat{p} x} \;.
\ee
$\hat{p} $ bezeichnet den Impulsoperator, 
der der infinitesimale Generator einer Translation ist.
Das Vakuum und die Zust"ande $|H\rangle $ sind Impulseigenzust"ande mit den 
Eigenwerten
\be
\hat{p}|0\rangle = 0 \;, \quad \quad {\rm und} \quad \hat{p}|H\rangle = p_H|H\rangle \;.
\ee
F"ur die Spektraldichte ergibt sich so
\be
\rho(q)=\frac{1}{2 \pi} \sum_H \int d^4\!x \,e^{i(q- p_H)x}
                \langle 0| \phi(0) | H \rangle  \langle H| \phi(0) | 0 \rangle \;.
\ee
Die $x$-Integration 
l"asst sich ausf"uhren, so dass man f"ur die Spektraldichte 
\ba \label{spekdensabs}
\rho(q)&=&(2 \pi)^3 \sum_H \delta(q- p_H)
            \langle 0| \phi(0) | H \rangle  \langle H| \phi(0) | 0 \rangle \nn \\
  &&    (2 \pi)^3 \sum_H \delta(q- p_H)
           | \langle 0| \phi(0) | H \rangle |^2 
\ea
erh"alt.
Aus Gl.~(\ref{spekdensabs}) l"asst sich sofort ablesen, 
dass $\rho(q)$ eine reellwertige positiv semidefinite Funktion 
ist. Ausserdem l"asst sich zeigen, 
dass $\rho(q) = \rho(A q)$ ist, falls $A$ ein Element der 
orthochronen Lorentzgruppe ist, also das Vorzeichen der 0-Komponente invariant l"asst.
Um dies einzusehen, "uberlegt man sich, dass 
die Delta-Distribution invariant unter Transformationen ist, die
$|{\rm det} \,( A)| = 1 $ erf"ullen.
Es gilt
\be
\delta(A q - p_H ) =\delta(A (q - A^{-1} p_H) ) = \frac{ \delta(q - A^{-1} p_H )}{|{\rm det}(A)|}\;.
\ee
Mit der Eigenschaft der Delta-Distribution l"asst sich zeigen, dass die Spektraldichte 
nur vom lorentzinvarianten Quadrat des Impulses $q$ abh"angen kann.
Es gilt
\ba
\rho(A q) &=&(2 \pi)^3 \sum_H \delta( A q - p_H)
                |\langle 0| \phi(0) | H \rangle|^2  \nn \\
 &=&(2 \pi)^3 \sum_H \delta(  q -  A^{-1}p_H)
                |\langle 0| \phi(0) | H \rangle |^2 \nn \\
 &=&(2 \pi)^3 \sum_{H^\pr} \delta(  q -  p_H^\pr)
                |\langle 0| \phi(0) | U (A)\,  H^\pr \rangle |^2 \;, 
\ea
wobei $U(A)$ die Darstellung der Lorentzgruppe auf dem Zustandsraum bezeichnet.
Beachtet man das Transformationsverhalten von skalaren Feldern
\be
\phi(0) = U(A) \Phi(0) U^{-1}(A)
\ee
und die Invarianz des Vakuumzustandes unter Lorentztransformationen 
$( U(A) |0 \rangle = |0 \rangle)$, so ergibt sich $\rho(q) = \rho(Aq) $.
Da alle Energie-Impuls-Eigenwerte im Vorw"artslichtkegel liegen, ist
\be
\rho(q) =\rho(q^2) \Theta(q_0)
\ee
wobei $\rho(q^2) =0$ ist, falls $q^2<0$ gilt. 
\subsection{Die Dispersionsrelation f"ur skalare Felder}
Die Korrelationsfunktion zweier hermitischer Skalarfelder ist durch  
\be \label{shermkorr}
\Pi (q^2)= i \int d^4\!x \, e^{iqx} \langle 0| T( \phi(x) \phi(0) )|0 \rangle
\ee
gegeben.
Das zeitgeordnete Produkt l"asst sich mit der Stufenfunktion $\Theta(x_0)$
ausschreiben:
\be
 \label{tprod}
T(\phi(x) \phi(0)) =  \phi(x) \phi(0)  \Theta(x_0) 
                    +  \phi(0) \phi(x) \Theta(-x_0) \;.
\ee
Invertiert man den definierenden Ausdruck f"ur die Spektraldichte
 Gl.~(\ref{specdens}),
so ergibt sich
\ba
 \langle 0| \phi(x) \phi(0) |0 \rangle &=& (2 \pi)^3 
            \int d^4\! q \,e^{- i q x } \rho(q^2) \Theta(q_0) \label{z1}\\
 {\rm und} \quad
 \langle 0| \phi(0) \phi(x) |0 \rangle &=& (2 \pi)^3 
            \int d^4 \!q \,e^{ i q x } \rho(q^2) \Theta(q_0) \label{z2}\;,
\ea
wobei Gl.~(\ref{z2}) durch komplexe Konjugation aus Gl.~(\ref{z1}) hervorgeht,
wenn man beachtet, dass $\rho(q^2)$ reell ist und f"ur Operatoren 
$(AB)^\dagger = B^\dagger A^\dagger $ gilt.
Durch Einsetzen von Gl.~(\ref{tprod}) und Gl.~(\ref{z1}, \ref{z2}) in die 
Definition der Korrelationsfunktion Gl.~(\ref{shermkorr}) ergibt sich
\be 
\Pi(q^2) = \frac{i}{(2 \pi)^3} \int d^4\!x \, d^4\!k \, \rho(k^2) \Theta(k_0) \left(
                      \Theta(x_0) e^{-ikx} +\Theta(-x_0) e^{ikx} 
                     \right) e^{iqx} \;.
\ee
Die
Integration "uber $\vec{x}$ ergibt
\be \label{x0integral}
\Pi(q^2) = i \int dx_0 \, d^4\!k \, \rho (k^2) \Theta(k_0) \left(
                        \Theta(x_0) \delta(\vec{q}-\vec{k}) e^{i(q_0-k_0)x} 
                     +  \Theta(-x_0)\delta(\vec{q}+\vec{k}) e^{i(q_0- k_0)x} 
                     \right) \;.
\ee
Die Fourierdarstellung der Stufenfunktion lautet
\be
\Theta(x_0) = \frac{i}{2 \pi} \int dp_0 \,
  \frac{e^{-i p_0 x_0}}{p_0 + i \ep} \;,
\ee
woraus
\be \label{thetaxoint}
\int d x_0 \, \Theta(x_0) \, e^{i q_0 x_0} = \frac{i}{q_0 + i \ep} \;
\ee
folgt.
Mit Gl.~(\ref{thetaxoint}) l"asst sich die Integration "uber $x_0$ in
Gl.~(\ref{x0integral}) ausf"uhren.
\be
\Pi(q^2) = - \int  d^4\!k \, \rho(k^2)\Theta(k_0) \left(
                  \frac{ \delta(\vec{q}-\vec{k}) }{q_0 - k_0 + i  \ep} 
               +  \frac{\delta(\vec{q}+\vec{k})  }{-(q_0 + k_0) + i  \ep}  
                     \right)
\ee
Die Integration "uber $d^3\! \vec{k}$ ergibt
\ba
\Pi(q^2) &=& - \int  dk_0 \, 
                 \rho(k_0^2- \vec{q}^2) \Theta(k_0)\left(\frac{1}{q_0 - k_0 + i  \ep} 
                              +  \frac{1}{-(q_0 + k_0) + i  \ep}   
                     \right)\\
 &=& 
    \int  dk_0 \,
                 \rho(k_0^2- \vec{q}^2)\Theta(k_0)
               \left( \frac{2 k_0}{ k_0^2 - q_0^2 }  
                     \right) \;.
\ea
Durch eine Transformation der Integrationsvariablen von $k_0$ zu $s$
\ba
s  &:=& k_0^2 - \vec{q}^2 \\
ds &=& 2 k_0 \, d k_0 
\ea
ergibt sich die Dispersionsrelation
\be
\Pi(q^2) = 
    \int_0^\infty  \frac{\rho(s) ds}{ s - q^2} \;,  
\ee
da $\rho(s) \sim \Theta(s) $ ist.

\section{Dispersionsrelationen}
Die Korrelationsfunktion $\Pi^{\mu \nu}_{ij}$ l"asst sich 
durch zwei skalare Funktionen beschreiben, so dass
\be
\Pi^{\mu \nu}_{ij}(q^2) = \frac{N_c}{6 \pi^2}(T^{\mu \nu}_{r_1} \Pi_{r_1}(q^2)
                          + T^{\mu \nu}_{r_2} \Pi_{r_2}(q^2)) \;,
\ee
wobei $(r_1,r_2) = (T,L)\;  {\rm oder } \;(q,g) $ sind.
Die skalaren Funktionen $\Pi_r(q^2)$ sind analytisch in $q^2$
und besitzen einen Schnitt entlang der positiven reellen 
Achse. Sie lassen sich mit den entsprechende Spektraldichten 
$\rho_r$ durch die Dispersionsrelation 
\be \label{disprelation}
\Pi_r(q^2) = \int_0^\infty \frac{\rho_r(s) \, ds}{s- q^2 } 
\quad \quad{\rm modulo \;\;Subtraktionen}
\ee
darstellen,
wobei Subtraktionen notwendig sind, um das Integral in Gl.~(\ref{disprelation})
endlich zu machen, was im folgenden Absatz erkl"art wird.
Die Spektraldichten $\rho_r(s)$ sind reelle, positive Funktionen.
Die Dispersionsrelationen (\ref{disprelation}) lassen sich invertieren,
wenn man die sog. Sokhotsky-Plemelj-Gleichung
\be \label{so}
\frac{1}{x \pm i \ep} = P  \left( \frac{1}{x} \right) \mp i \pi \delta(x) 
\ee
beachtet.
$P$ steht f"ur den Hauptwert.
Mit Gleichung~(\ref{so}) ergibt sich aus der Dispersionsrelation 
Gl.~(\ref{disprelation})
\be \label{dispzwischen}
\Pi_r(q^2 \pm i \ep) = \int \frac{ \rho_r(s) ds}{s-q^2  \mp i \ep} 
                    = P \int \frac{ \rho_r(s) ds }{s-q^2} \pm i \pi \rho_r(q^2)
 \quad { \rm wobei \quad   Im} (q^2) =0 \;.
\ee
Aus Gl.~(\ref{dispzwischen}) erkennt man, dass  
der Imagin"arteil der Korrelationsfuntionen auf der reellen 
Achse unstetig ist, da er sein Vorzeichen wechselt.
Durch Subtraktion von Gl.~(\ref{dispzwischen}) mit unterschiedlichem
Vorzeichen des $ i \ep$-Terms ergibt sich die inverse 
Dispersionsrelation
\be \label{inversdisprel}
\frac{1}{2 \pi i} \left( \Pi_r(s + i \ep) -\Pi_r(s - i \ep) \right)=
  \frac{1}{2 \pi i} {\rm Disc }(\Pi_r(s)) 
      = \frac{{\rm Im} \Pi_r(s + i \ep)}{\pi} = \rho_r(s) \;.
\ee
\section{Die Adlerfunktion}
Die Korrelationsfunktion $\Pi_r(q^2)$
ist nicht multiplikativ renormierbar, da sie keinen
endlichen Term in der f"uhrenden Ordnung besitzt.
Bereits der erste bei der Berechnung der Korrelationsfunktion 
auftretende Graph enth"alt eine divergente Schleife
(vgl. Abb. \ref{korrfunkdiag}).
Um die Korrelationsfunktion endlich zu machen, ist es notwendig, den 
divergenten Anteil zu subtrahieren.
Es ergibt sich so
\ba \label{korrfsub}
\Pi_{\rm Sub}(q^2) &=& \Pi(q^2) -\Pi(0) \nn \\
       &=& \int_0^\infty \frac{\rho(s)ds}{s-q^2} 
      - \int_0^\infty \frac{\rho(s) ds}{s} \nn \\
   &=& \int_0^\infty \rho(s) ds \left( \frac{1}{s-q^2} - \frac{1}{s} \right)
       = q^2 \int_0^\infty \frac{\rho(s) ds}{s(s-q^2)} \;.
\ea
Da die Spektraldiche $\rho(s)$ f"ur gro"se Impulsquadrate $s$ konstant
ist (siehe Gl.~(\ref{densold}))   
und der 
Nenner des Integrals in Gl.~(\ref{korrfsub}) sich f"ur gro"se $s$
wie $s^{-2}$ verh"alt, ist die
subtrahierte Korrelationsfunktion endlich. 

Anstelle der Spektraldichte kann man auch die Adlerfunktion $D_r(Q^2)$
betrachten.
Diese ist durch 
\be
D_r(Q^2) = -Q^2 \frac{d}{d Q^2} \Pi_r(Q^2)
\ee
definiert, wobei $Q^2 = -q^2$ das euklidische Impulsquadrat bezeichnet.
F"ur die Adlerfunktion erh"alt man die Spektraldarstellung
\ba
D_r(Q^2) &=& - Q^2 \frac{d}{d Q^2} \Pi_r(Q^2)  \nn \\
   &=& -Q^2 \int_0^\infty \frac{d}{d Q^2} \frac{\rho(s) ds}{s+Q^2} = 
 Q^2 \int_0^\infty \frac{\rho(s) ds }{(s + Q^2)^2} \;.
\ea
Aus der Spektraldarstellung f"ur die Adlerfunkion sieht man wiederum, 
dass diese endlich ist, da die Spektraldichte f"ur gro"se 
Werte von $s$ konstant ist. 

\section{Die Operator-Produkt-Entwicklung (OPE)}
Das asymptotische Verhalten der
Korrelationsfunktion $\Pi_{ij}^{\mu \nu} $ l"asst sich 
im asymptotischen Grenzfall $-q^2 = Q^2 \rightarrow \infty$
im Rahmen der St"orungstheorie berechnen. Es ist durch die 
Operator-Produkt-Entwicklung (OPE) des Produktes der beiden 
Str"ome in der Korrelationsfunktion gegeben,
\be \label{OPExp}
\Pi_{ij}^{\mu \nu}(q) = i \int d^4x e^{i q x}
        \langle 0| T (j_{ij}^\mu(x) j_{ij}^{\nu \dagger }(0)) |0 \rangle  
     \begin{array}{c} 
\\
    
\longrightarrow \\
{\scriptstyle -q^2 \rightarrow \infty} 
\end{array}
                 \sum_d \sum_n^{n_d} 
          C_{n ij}^{(d)\mu \nu} \langle  \mathcal{O}^{(d)}_{n ij}  \rangle \;.
\ee
Die $ C_{nij}^{(d)\mu \nu}$ sind komplexe Funktionen, 
welche im Rahmen der St"orungstheorie 
berechnet werden k"onnen. 
Die $ \mathcal{O}_{nij}^{(d)}$ sind Operatoren, 
die aus Feldern der QCD-Lagrangefunktion zusammengesetzt sind. In der 
OPE (Gl.~(\ref{OPExp})) 
treten nur skalare Operatoren auf, 
da alle anderen Operatoren verschwindende Beitr"age
liefern. 
Die Operatoren werden nach ihrer Massendimension $(d)$ geordnet. 
Bis zur Dimension vier treten die
folgenden Operatoren auf
\ba
 \mathcal{O}^{(0)} &=& 1, \quad  \quad \, \mathcal{O}^{(2)} = m^2, \nn \\
 \mathcal{O}^{(4)}_1 &=& G_{\mu \nu}^2, \quad  
 \mathcal{O}^{(4)}_2 = m \bar{\psi}_i \psi_j,  \quad  
 \mathcal{O}^{(4)}_3 = m^4,
\ea
wobei $G_{\mu \nu}$ den Feldst"arketensor des Gluonfeldes und 
$\psi$ ein Quarkfeld bezeichnen.
Wenn die Operatoren als normalgeordnet angenommen werden,
verschwinden die Vakuumerwartungswerte der nicht-trivialen 
Operatoren in der St"orungstheorie. Im diesem Ansatz werden die 
Operatoren verwendet, um nicht perturbative-Effekte  
zu parametrisieren \cite{RHarl}.  
\section{Die Massenentwicklung der Korrelationsfunktion} \label{berkorrf}
F"ur die in den nachfolgenden Kapiteln beschriebene Analyse  
der asymptotischen Struktur der St"orungsreihe von $\tau$-Lepton 
Observablen wird insbesonders die Entwicklung der Korrelationsfunktion
in der Strange-Quark-Masse ben"otigt. 
Diese wurde in \cite{eek20,eek21,eek2c,eek2d,eek2e} Berechnet und
im Rahmen dieser Arbeit bis zu der Ordnung $\al_s$ auf Richtigkeit 
"uberpr"uft.
Bis zur Ordnung $m^2$ entspricht diese 
den Koeffizienten $C^{(0)\mu \nu}_{ij}$ und $C^{(2)\mu \nu}_{ij}$
der OPE (Gl.~(\ref{OPExp})). 
Die Massenentwicklung l"asst sich mit den im Anhang~\ref{Frules} angegebenen 
Feynman-Regeln berechnen, wenn man beachtet, dass der Operatoreinsetzung 
$ j^\mu_{V/A ij}(x) = \psi_{i}(x) \gamma^\mu (\gamma_5)\psi_j^\dagger(x) $  
ein zus"atzlicher Vertex entspricht. 
Die Feynman-Regeln 
f"ur diesen lauten\\
\begin{tabular}[t]{lcr}
&&\\
\raisebox{1.8ex}[-1.8ex]{
  \begin{tabular}{l}
    (Axial)Vektor-Strom \\ 
    Operatoreinsetzung
    \end{tabular}}
  &
\begin{picture}(110,30)(0,0)
\ArrowLine(0,15)(55,15)
\ArrowLine(55,15)(110,15)
\Vertex(55,15){3}
\Text(5,5)[l]{$A,i$}
\Text(5,25)[l]{$p,\al$}
\Text(105,5)[r]{$B,j$}
\Text(105,25)[r]{$p^\prime,\beta$}
\end{picture}
  &  
\raisebox{1.8ex}[-1.8ex]{ $
     (\gamma^\mu (\gamma_5))_{\al \beta}  \, \mu^\ep \, \delta_{AB} \;,$} 
\\ &&
\end{tabular}
\hfill
\parbox{8mm}{\be \label{opfeyman} \ee}\\
wobei bereits die bei der Verwendung der dimensionalen 
Regularisierung notwendige
Massenskala $\mu$ mit aufgef"uhrt wurde.
Die Erhaltung der Farbladung wird durch das Kronecker-Delta $\delta_{AB}$ 
beschrieben.
Das Quarkflavour  wird an diesem Vertex nicht erhalten.
In der relativen $\tau$-Zerfallsrate treten 
die Korrelationsfunktionen $\Pi_{ud}^{\mu \nu}$ und  $\Pi_{us}^{\mu \nu}$
auf, welche f"ur die Beschreibung des $\tau$-Leptons in Hadronen mit 
``strangeness'' 0 und 1 ben"otigt werden. Die Up- und Down-Quarks
besitzen im Vergleich zu dem $\tau$-Lepton eine sehr geringe Masse
$(m_u \approx 5 \,{\rm MeV}, \;  m_d \approx 10 \,{\rm MeV} )$ \cite{PDG}, 
so dass diese Massen vernachl"assigt werden k"onnen.
Die Masse des Strange-Quarks $(m_s)$ ist wesentlich gr"o"ser 
als die der Up- und Down-Quarks und betr"agt ungef"ahr 10\% der 
$\tau$-Lepton-Masse \cite{four}. Aus diesem Grund ist es sinnvoll, die 
Korrelationsfunktion in $m_s$ zu entwickeln.
Um die f"uhrende Ordnung und die erste QCD-Korrektur der Korrelationsfunktion 
zu berechnen, m"ussen 
die in Abb.~\ref{korrfunkdiag} dargestellten Feynman-Graphen berechnet werden.
\begin{figure}
\begin{center}
\begin{picture}(350,50)(0,0)
\Text(30,25)[]{$\displaystyle (-i) \Pi^{\mu \nu}_{u s} = $}
\Vertex(60,25){3}
\Vertex(90,25){3}
\ArrowArc(75,25)(15,180,0)
\ArrowArc(75,25)(15,360,180)
\Text(102,25)[l]{ $\displaystyle + \;  \Bigg[$} 
\Vertex(135,25){3}
\Vertex(165,25){3}
\Vertex(142,38){1.2}
\Vertex(158,38){1.2}
\ArrowArc(150,25)(15,180,0)
\ArrowArc(150,25)(15,360,180)
\GlueArc(150,40)(7,200,340){2}{4}
\Text(180,25)[]{$+$}
\Vertex(195,25){3}
\Vertex(225,25){3}
\Vertex(202,11){1.2}
\Vertex(218,11){1.2}
\ArrowArc(210,25)(15,180,0)
\ArrowArc(210,25)(15,360,180)
\GlueArc(210,10)(7,20,160){2}{4}
\Text(240,25)[]{$+ $}
\Vertex(255,25){3}
\Vertex(285,25){3}
\Vertex(270,10){1.2}
\Vertex(270,40){1.2}
\ArrowArc(270,25)(15,180,0)
\ArrowArc(270,25)(15,360,180)
\Gluon(270,10)(270,40){2}{6}
\Text(295,25)[l]{$ \displaystyle \Bigg] + \; \mathcal{O}(\al_s^2) $}
\end{picture} 
\begin{picture}(350,50)(0,0)
\Text(30,25)[]{$\displaystyle (-i) \Pi^{\mu \nu}_{u s} = $}
\Vertex(60,25){3}
\Vertex(90,25){3}
\ArrowArc(75,25)(15,180,0)
\ArrowArc(75,25.3)(14.3,180,0)
\ArrowArc(75,25)(15,360,180)
\Text(102,25)[l]{ $\displaystyle + \;  \Bigg[$} 
\Vertex(135,25){3}
\Vertex(165,25){3}
\Vertex(142,38){1.2}
\Vertex(158,38){1.2}
\ArrowArc(150,25)(15,180,0)
\ArrowArc(150,25.3)(14.3,180,0)
\ArrowArc(150,25.3)(14.15,180,0)
\ArrowArc(150,25)(15,360,180)
\GlueArc(150,40)(7,200,340){2}{4}
\Text(180,25)[]{$+$}
\Vertex(195,25){3}
\Vertex(225,25){3}
\Vertex(202,11){1.2}
\Vertex(218,11){1.2}
\ArrowArc(210,25)(15,180,0)
\ArrowArc(210,25.3)(14.3,180,0)
\ArrowArc(210,25.3)(14.15,180,0)
\ArrowArc(210,25)(15,360,180)
\GlueArc(210,10)(7,20,160){2}{4}
\Text(240,25)[]{$+$}
\Vertex(255,25){3}
\Vertex(285,25){3}
\Vertex(270,10){1.2}
\Vertex(270,40){1.2}
\ArrowArc(270,25)(15,180,0)
\ArrowArc(270,25.3)(14.3,180,0)
\ArrowArc(270,25.3)(14.15,180,0)
\ArrowArc(270,25)(15,360,180)
\Gluon(270,10)(270,40){2}{6}
\Text(295,25)[l]{$ \displaystyle \Bigg] + \; \mathcal{O}(\al_s^2) $}
\end{picture} 
\caption{ \label{korrfunkdiag} Bei der Berechnung der f"uhrenden Ordnung  
und der ersten QCD-Korrektur
der Korrelationsfunktion auftretende Feynman-Diagramme.
Fettgedruckte Quarkpropagatoren bezeichnen hierbei
das als massiv behandelte Strange-Quark mit Masse $m_s$ 
und d"unngedruckte Quarkpropagatoren die Up- und Down-Quarks, 
deren Massen vernachl"assigt werden k"onnen.
}
\end{center}
\end{figure}
Verwendet man die Feynman-Regeln der QCD aus Anhang \ref{Frules}
und die Regeln f"ur die Operatoreinsetzung des Stromes 
$j_{V/A ij }^{\mu \nu} (x) $
aus Gl.~(\ref{opfeyman}), 
so ergibt sich f"ur die Korrelationsfunktion
$\Pi_{us}^{\mu \nu}$ bis zur Ordnung $\al_s$ der Ausdruck
\ba \label{piusfe}
    \Pi_{us}^{\mu \nu} 
&=& 
    2 N_c \Bigg\{ i \int \frac{d^dp \mu^{2 \ep}}{(2 \pi)^d} \,
            \frac{ {\rm Tr} \left( \gamma^\mu (\shp - \shq) 
                     \gamma^\nu (\shp + m_s)  
          \right) }
       { ((p-q)^2 + i \ep )( p^2 -m_s^2 + i \ep ) } + \nn \\
&& + g^2 C_F \Bigg(
    \int \frac{d^dp \mu^{2 \ep}}{(2 \pi)^d}\int 
               \frac{d^dk \mu^{2 \ep}}{(2 \pi)^d} \, 
  \frac{ {\rm Tr } \left( 
            \gamma^\mu \shp \gamma^\sigma \shk \gamma_\sigma 
           \shp \gamma^\nu( \shp- \shq + m_s) \right) }
     {(p^2 + i \ep )^2(k^2 + i \ep ) ((p-k)^2+ i \ep)} \times \nn \\
 && \qquad \qquad\qquad \qquad \qquad \qquad \qquad \qquad \qquad \times 
\frac{1}{((p-q)^2 - m_s^2 + i \ep) } \nn \\
 && +
      \int \frac{d^dp \mu^{2 \ep} }{(2 \pi)^d}\int \frac{d^dk \mu^{2 \ep} }
            {(2 \pi)^d} \,
           \frac{ {\rm Tr } \left(  
       \gamma^\mu (\shp + m_s ) \gamma^\sigma (\shk + m_s ) 
         \gamma_\sigma (\shp + m_s) \gamma^\nu
                (\shp - q) \right) }
       { (p^2 - m_s^2 + i \ep)^2 (k^2 - m_s^2 + i \ep) } \times \nn \\
 && \qquad \qquad \qquad \qquad \qquad \qquad \quad \times
     \frac{1}{ ((p-q)^2  + i \ep)((p-k)^2 + i \ep) } \nn \\
 && + 
   \int \frac{d^dp \mu^{2 \ep} }{(2 \pi)^d}\int 
         \frac{d^dk \mu^{2 \ep} }{(2 \pi)^d} \, 
             \frac{ {\rm Tr } \left(
     \gamma^\mu (\shk + m_s) \gamma^\sigma (\shp + m_s ) \gamma^\nu (\shp - \shq) 
                  \gamma_\sigma (\shk - \shq) \right) }
     { (k^2- m_s^2 + i \ep) ( p^2 - m_s^2 +  i \ep) ((k-q)^2 + i \ep )} \times \nn\\
  && \qquad \qquad \qquad \qquad \qquad \qquad \qquad \quad  \times 
    \frac{1}{ (( p -q)^2 +  i \ep)((k-p)^2 + i \ep ) } \Bigg) \Bigg\} \;.
\ea
In Gl.~(\ref{piusfe}) wurde bereits ber"ucksichtigt, dass der eingesetzte 
Vektor-
und der Axialvektorstrom dieselben Resultate liefern, 
woraus sich der Faktor $2$ ergibt.
Der entsprechende Ausdruck f"ur $\Pi_{ud}^{\mu \nu}$  
ergibt sich, indem man in Gl.~(\ref{piusfe}) $m=0$ setzt.
F"ur die Berechnung von Gl.~(\ref{piusfe})
ist es sinnvoll, folgenderma"sen vorzugehen:
\begin{enumerate}
\item Anstelle des Tensors $\Pi_{us}^{\mu \nu}$ 
berechnet man besser die Skalare
$ \Pi_1 = g_{\mu \nu} \Pi_{us}^{\mu \nu}$ 
und $ \Pi_2 = q_\mu q_\nu \Pi_{us}^{\mu \nu}$.
\item Die Integranden werden in der Masse um $m_s=0$ entwickelt. 
\item Die im Z"ahler nach der 
Spurbildung auftretenden Skalarprodukte werden quadratisch
erg"anzt, so dass man anschlie"send mit dem Nenner k"urzen kann. \newline
Z.B.: $(p \cdot q) = -\frac{1}{2}((p-q)^2 -p^2 -q^2)$ 
\item Durch eine Wickrotation, 
bei der alle Skalarprodukte ihr Vorzeichen wechseln und sich 
das Integrationsma"s gem"a"s $d^dk_L = + i d^d k_E$ transformiert, 
lassen sich direkt die 
in Kapitel \ref{QCDkap} angegebenen Formeln f"ur die
Ein- und Zweischleifenintegrale anwenden.
\item Die Zerlegung der Korrelationsfunktion in den transversalen
und den longitudinalen Anteil bzw. in den $q$- und 
den $g$-Anteil ergibt sich aus den 
berechneten Kontraktionen $\Pi_{1/2}$ durch
\be
\Pi_L = \frac{6 \pi^2}{N_C}\frac{\Pi_2}{Q^4} \,,
  \quad \quad \Pi_T =  \frac{6 \pi^2}{N_C}\frac{\Pi_1 + \Pi_2 /Q^2}{Q^2 (3- 2 \ep)}
\ee
und
\be
 \Pi_q = \frac{6 \pi^2}{N_C} \Bigg[\frac{\Pi_2}{Q^4} 
+ \frac{\Pi_1 + \Pi_2 / Q^2}{Q^2(3- 2 \ep)} \Bigg] \;,\quad \quad
\Pi_g = \frac{6 \pi^2}{N_C} \frac{\Pi_1 +  \Pi_2 / Q^2}{3 - 2 \ep } \;.
\ee
Hier ist wieder $Q^2 = -q^2$. 
Der Regulator $\ep$ entsteht durch die Kontraktion des 
metrischen Tensors mit sich selbst $(g_{\mu \nu} g^{\mu \nu} = 4 - 2 \ep)$.
\end{enumerate}
Dieser Vorgang kann mit einem im Rahmen meiner Diplomarbeit geschriebenen 
\newline
MATHEMATICA-Programmes 
halbautomatisch ausgef"uhrt werden.
Man erh"alt auf diese Weise f"ur die Entwicklung der Korrelationsfunktionen in der Masse 
des schweren Quarks
\ba
\Pi_q  &=&
 \frac{1}{\ep} \left[ 1 + C_F \frac{3}{8}  \als \right]
 + \Bigg[ \frac{5}{3}
        + L + C_F \left( \frac{55}{16} - 3 \zeta(3)  + \frac{3}{4}L \right)\als \Bigg]\nn \\
&& + \ep \Bigg[ \frac{10}{9} + \frac{5}{3}L + \frac{1}{2}L^2 +
        C_F \Bigg( \frac{1567}{96} -\frac{19}{2}\zeta(3) - \frac{1}{20} \pi^4 
            + \Bigg(\frac{55}{8} - 6 \zeta(3) \Bigg) L 
       + \frac{3}{4} L^2  \Bigg) \als  \Bigg] \nn \\
&& + \frac{m^2_B}{Q^2} \Bigg\{ 
         - \frac{1}{\ep}  C_F\frac{9}{2} \als + \Bigg[ -3 + C_F \Bigg(
         - \frac{57}{4} - 9 L \Bigg)  \als  \Bigg] \nn \\
 && +  \ep \Bigg[ 
         - 6 - 3 L + C_F \Bigg( -\frac{295}{8} + 6 \zeta(3) - \frac{57}{2} L - 9 L^2 
        \Bigg) \als \Bigg] \Bigg\} 
\ea
und 
\ba
\Pi_g &=&  
  Q^2 \Bigg\{ \frac{1}{\ep} \left[ 1 + C_F \frac{3}{8}  \als \right]
 + \Bigg[ \frac{5}{3}
        + L + C_F \left( \frac{55}{16} - 3 \zeta(3)  + \frac{3}{4}L \right)\als \Bigg]\nn \\
&& + \ep \Bigg[ \frac{10}{9} + \frac{5}{3}L + \frac{1}{2}L^2 +
        C_F \Bigg( \frac{1567}{96} -\frac{19}{2}\zeta(3) - \frac{1}{20} \pi^4 
            + \Bigg(\frac{55}{8} - 6 \zeta(3) \Bigg) L 
       + \frac{3}{4} L^2  \Bigg) \als  \Bigg] \Bigg\} \nn \\
&& + 
     m_B^2 \Bigg\{  \frac{1}{\ep^2} C_F \frac{9}{8} \als
    + \frac{1}{\ep} \Bigg[ \frac{3}{2} +  C_F \Bigg( \frac{15}{16} + \frac{9}{4} L \Bigg) \als \Bigg] \nn \\
&& + 
         \Bigg[  \frac{3}{2} L + C_F  \Bigg( \frac{81}{32} - \frac{9}{2} \zeta(3)  +\frac{15}{8} L
      + \frac{9}{4} L^2 \Bigg)  \als  \Bigg] + \nn \\
&& + 
   \ep \Bigg[ -3 + \frac{3}{4} L^2 + C_F \Bigg( \frac{691}{64} - \frac{39}{4} \zeta(3) \nn \\
&&
      - \frac{3}{40} \pi^4 + \left( \frac{81}{16} - 9 \zeta(3) \right)L + 
     \frac{15}{8} L^2  + \frac{3}{2} L^3 \Bigg) \als \Bigg] \Bigg\} \;.           
\ea
Hierbei bezeichnet $L =  \ln(\frac{\mu^2}{Q^2})$ und $Q^2 =-q^2$.
Durch Renormierung der Quarkmasse gem"a"s
\be
m_B = m \left( 1 - C_F \frac{3}{4} \als \frac{1}{\ep} + \mathcal{O}(\al_s^2) \right)
\ee
ergibt sich f"ur die Korrelationsfunktion
\ba 
\Pi_q &=& \frac{1}{\ep} \Bigg[ 1 + C_F  \frac{3}{8} \als \Bigg] +  \frac{5}{3} + L 
       +C_F \Bigg( \frac{55}{16} - 3 \zeta(3) + \frac{3}{4} L \Bigg) \als \nn \\
&& + 
      \frac{m^2}{Q^2}  \Bigg\{
        -3 + C_F \Bigg( - \frac{21}{4} - \frac{9}{2}L \Bigg) \als \Bigg\}
\ea
und
\ba
\Pi_g &=& Q^2 \Bigg\{
\frac{1}{\ep} \Bigg[ 1 + C_F  \frac{3}{8} \als \Bigg] +  \frac{5}{3} + L 
      +  C_F \Bigg( \frac{55}{16} - 3 \zeta(3) + \frac{3}{4} L \Bigg) \als   \Bigg\}  \nn \\
&& +   m^2  \Bigg\{
             -\frac{1}{\ep^2}  C_F\frac{9}{8} \als + \frac{1}{\ep} \Bigg[  \frac{3}{2} + 
       C_F \frac{15}{16}  \als  \Bigg]  \nn \\
&& + 
        \frac{3}{2} L + C_F\Bigg( \frac{225}{32} - \frac{9}{2} \zeta(3) + \frac{15}{8}L + \frac{9}{8} L^2
         \Bigg) \als \Bigg\}  \,.
\ea
Die so erzielten Resultat stimmen mit den im folgenden Abschnitt zitierten
Ergebnissen aus \cite{eek20,eek21,eek2c,eek2d,eek2e} "uberein.
\section{Resultate f"ur die  Massenentwicklung der Korrelationsfunktion}
Die Massenentwickling der Korrelationsfunktion in $m^2/q^2$ l"asst sich als
\be
\label{expmq}
\Pi_q(q^2)=\Pi(q^2)+3\frac{m_s^2}{q^2}\Pi_{mq}(q^2)
\ee
\be
\label{expmg}
\Pi_{g}(q^2)=-q^2 \Pi(q^2) + \frac{3}{2} m_s^2   \Pi_{mg}(q^2)
\ee
schreiben.
Die Resultate der St"orungstheorie f"ur die Adlerfunktion und die Korrelationsfunktion sind 
\begin{eqnarray} 
\label{fullinf}
-Q^2\frac{d}{dQ^2}\Pi(Q^2)\Big|_{\mu^2= Q^2}&=& \nn\\
=D(Q^2)\Big|_{\mu^2= Q^2} 
&=&1 + \als + k_1 \left( \als \right)^2 + k_2 \left(\als \right)^3 
+k_3 \left( \als \right)^4 + \mathcal{O}(\alpha_s^5) \nonumber \\
-Q^2\frac{d}{dQ^2}\Pi_{mg}(Q^2)\Big|_{\mu^2= Q^2} &=& \nn \\ 
=D_{mg}(Q^2)\Big|_{\mu^2= Q^2}
&=&1+\frac{5}{3}\als + k_{g1}\left(\als \right)^2 
+ k_{g2}  \left( \als \right)^3 + k_{g3} 
\left(\als \right)^4 + \mathcal{O}(\alpha_s^5)\nonumber \\
\Pi_{mq}(Q^2)\Big|_{ \mu^2= Q^2}
&=& 1 + \frac{7}{3} \als + k_{q1}  \left( \als \right)^2
+ k_{q2}  \left( \als \right)^3 + \mathcal{O}(\alpha_s^4) \;,
\end{eqnarray}
wobei die nummerischen Werte der Koeffizienten im 
masselosen Grenzfall durch \cite{eek20,eek21,eek2c}
\ba \label{koeff1}
k_1 &=&  \frac{299}{24}- 9 \zeta(3), \nonumber \\
k_2 &=& \frac{58057}{288} - \frac{779}{4}\zeta(3) +\frac{75}{2}\zeta(5)
\ea
und f"ur die $m^2/q^2$-Korrektur durch
\cite{eek2d,eek2e}
\ba \label{koeff2}
k_{q1} &=& \frac{13981}{432} +
\frac{323}{54}\zeta(3)-\frac{520}{27}\zeta(5), 
                             \nonumber \\
k_{g1} &=& \frac{4591}{144} - \frac{35}{2} \zeta(3), \nonumber \\
k_{g2} &=& \frac{1967833}{5184} - \frac{\pi^4}{36}
-\frac{11795}{24}\zeta(3) +  \frac{33475}{108} \zeta(5) \, 
          \nonumber
\ea
gegeben sind.
Alle Resultate sind 
f"ur 3 Flavours $(N_f = 3)$ und Eichgruppe $SU(3) \;(C_F = 4/3,\; N_c = 3)$ 
im $\MSsch$-Schema angegeben. Die Koeffizienten
$k_3$, $k_{q2}$,und $k_{g3}$ sind noch nicht bekannt.
Da die Funktionen $D$, $(m_s^2 D_{mg})$ und $(m_s^2 \Pi_{mq})$ RG-invariant sind
\footnote{
Str"ome besitzen eine verschwindende anormale Dimension,
wenn sie Generatoren einer Symmetrie sind \cite{mut}. Dies sieht man 
ein, indem man die kanonischen Ladungsoperatoren
\be
Q^a(t) = \int d^3x J^a_0(x)
\ee
bildet, welche die Kommutatorrelation
\be \label{kommutator}
[Q^a(t),Q^b(t) ] = i f^{abc} Q^c(t)
\ee
erf"ullen. $Q^q(t)$ muss dimensionslos sein,
damit (Gl.~(\ref{kommutator})) gilt. Hieraus 
ergibt sich f"ur $J^a_0$ die Dimension $3$,
woraus mit Kovarianz folgt, dass $J_\mu^a$
keine anormale Dimension besitzen kann, was gleichbedeutend mit 
RG-Invarianz ist.}, 
kann die 
Renormierungsskala $\mu$ frei gew"ahlt werden. Wird wie 
in Gl.~(\ref{fullinf}) $\mu^2 = Q^2$ gesetzt, so 
verschwinden alle Logarithmen, und die $Q^2$-Abh"angigkeit 
der Funktionen $D$, $(m_s^2 D_{mg})$ und $(m_s^2 \Pi_{mq})$
wird durch das Laufen der Kopplungskonstante und der 
Quarkmasse beschrieben.
Die explizite $Q^2$-Abh"angigkeit von Gl.~(\ref{fullinf}) 
kann durch Einsetzen der Ausdr"ucke f"ur die laufende 
Kopplungskonstante $\al_s(Q^2)$ (Gl.~(\ref{runcoupl}))
und f"ur die laufende 
Masse $m(Q^2)$ (Gl.~(\ref{runmass})) in Gl.~(\ref{fullinf})
gefunden werden.

\chapter{Der masselose Teil} 
\label{Art1} 
\label{ART1}
In diesem Kapitel werden die St"orungsreihen von $\tau$-Lepton-Observablen 
im Limes verschwindender Quarkmassen analysiert.
In diesem Fall verschwindet der longitudinale Anteil der Spektraldichte 
$\Pi_L$ aufgrund der masselosen Diracgleichung, 
die die Quarkspinoren erf"ullen.
F"ur die Zerfallsbreite in Hadronen mit Strangeness Null ergibt sich 
im Limes masseloser Quarks
\ba
  \label{int}
R_{\tau S=0} &=&
     N_{c } S_{EW} |V_{ud}|^2 
     \int_0^{M_\tau^2}2 \left(1-{s\over M_\tau^2} \right)^2
         \left(1+2{s\over M_\tau^2}\right)
\rho(s){ds\over M_\tau^2} \nn \\ 
 &=&  N_c S_{EW} |V_{ud}|^2 (1 + \delta_P)\ ,
\ea  
wobei die additiven elektroschwachen und die nichtperturbativen 
Korrekturen zu $R_{\tau S=0}$ nicht aufgef"uhrt sind.
\section{Interne st"orungstheoretische Beschreibung}
Die zentrale Gr"o"se f"ur $\tau$-Lepton-Zerf"alle
ist die hadronische Spektraldichte, die in dem Energieintervall
 $(0,M_\tau=1.777~{\rm GeV})$ gemessen werden kann.
Im Rahmen der St"orungstheorie kann die hadronische Spektraldichte 
nicht punktweise mit den experimentellen Daten verglichen
werden. Anstelle dessen m"ussen Momente 
(oder allgemeiner Fourier-Komponenten "uber einen 
vollst"andigen Satz von Testfunktionen)  
mit dem Experiment verglichen werden.
Die Momente der Spektraldichte sind durch 
\begin{equation}
  \label{intmom}
M(n)=(n+1)\int_0^{\mts} \rho(s) 
       \left(\frac{s}{\mts}\right)^n \frac{ds}{\mts} \equiv 1+m_n  
\end{equation}
gegeben $(n = 0,1, \ldots , \infty)$.
Aufgrund der Vollst"andigkeit der Basis $\{s^n:n=0,1, \ldots,\infty\}$
enthalten die Momente $m_n$ alle Informationen der
Funktion $\rho(s)$.  
Das Spektrum l"asst sich RG-invariant, d.h. unabh"angig
von der Definition der Ladung, untersuchen, indem alle 
Momente gleichzeitig analysiert werden.
Es ist wichtig festzustellen, dass im Rahmen der St"orungstheorie 
in endlicher Ordnung die Momente aus Gl.~(\ref{intmom})
mit den Resultaten aus der Konturintegration (\ref{Rtauexplcont}) 
\cite{cont,cont1,cont2,Pivtau}
aufgrund der analytischen Eigenschaften des Logarithmus
"ubereinstimmen.

Um alle vom Renormierungsschema abh"angigen Konstanten in den 
Ausdr"ucken der St"orungstheorie f"ur die Momente  
zu beseitigen, wird durch die Relation 
\be
  \label{defofa}
\rho(s) = 1+a(s) \,; \quad 
a(\mts) =
       \als \left(1 + \dots \right)
\ee
eine effektive Ladung $a(s)$ definiert.
Die masselose Spektraldichte erh"alt man, indem man zun"achst
mit dem Ausdruck f"ur 
die laufende Kopplungskonstante
(Gl.~(\ref{runmass})) die explizite Impulsabh"angikeit der Adlerfunktion $D(Q^2)$
(Gl.~(\ref{fullinf})) bestimmt.
Die Korrelationsfunktion $\Pi(Q^2)$ ergibt sich aus der Adlerfunktion durch
logarithmische Integration gem"a"s
\be
\Pi(L) = \int_0^L dL^\pr D(L^\pr)\,,
\ee
wobei $L = \ln(\mts / Q^2) $ ist. 
Die Spektraldichte $\rho$ h"angt mit dem Imagin"arteil 
der Korrelationsfunktion "uber
\be
\rho = \frac{ {\rm Im} \,( \Pi(Q^2)) }{\pi}
\ee
zusammen,
wobei die Imagin"arteile der einzelnen Potenzen der Logarithmen  durch
\be
\frac{{\rm Im } \,( L^n )}{\pi} = \frac{{\rm Im } \, (l+ i \pi)^n }{\pi}
  = \left\{ \begin{array}{cc}
       1       &    n=1 \\
      2l       &    n=2 \\
 3l^2-\pi^2    &   n=3 \\
 4l^3 -4l\pi^2 &    n=4 
\end{array}
\right.
\ee
mit $l = \ln(\mts/q^2)$ gegeben sind.
F"ur die Spektraldichte ergibt sich so
\begin{eqnarray}
\label{densold}
\rho(s)&=&1+a+2.25 a^2 l + a^3(4l + 5.063l^2)+ a^4(-25.7 l + 22.5 l^2 +
11.4 l^3) \nonumber \\
&&+a^5((-409.5 + 4.5 k_3)l - 149.4 l^2 + 87.75 l^3 + 25.63 l^4)
+ \mathcal{O}(a^6)  \;.
\end{eqnarray}
Die effektive Kopplungskonstante $a$ ist an der Stelle $a = a(\mts) $
genommen.
Alle Konstanten, die durch die Wahl eines 
Renormierungsschemas auftreten, werden in 
der Definition der effektiven Ladung $a$ 
absorbiert \cite{effsch,ksch,kksch,effDh}. 
G"abe es kein Laufen der Kopplungskonstante 
(wie z.B. im ``conformal limit'' der QCD f"ur 
$(n_f \rightarrow 22/3)$ mit verschwindender
$\beta$-Funktion oder bei dem Infrarot-Fixpunkt),
so w"urde die gesamte Physik des $\tau$-Leptons
in der masselosen Approximation (ohne 
Teilchen mit Strangeness und bei Vernachl"assigung der 
nichtperturbativen Kondensate)
sich auf die Bestimmung einer einzigen
Zahl $a(M_\tau)\equiv a$ reduzieren, und es g"abe keine
Probleme mit der Konvergenz der St"orungsreihe.   
Aufgrund der Abh"angigkeit der Kopplungskonstante $a(s)$ von der Energieskala $s$
ergeben sich f"ur verschiedene Observable, welche durch Momente der 
Spektraldichte gegeben sind, verschiedene St"orungsreihen 
aus dem urspr"unglichen Objekt $\rho(s)$ in Gl.~(\ref{densold}).
Ohne das Laufen der Kopplungskonstante h"atte man   
\be
\label{norunmom}
M(n)=1+a(M_\tau)\equiv 1+a\quad {\rm oder} \quad   m_n\equiv a \;,
\ee
womit die Analyse der St"orungstheorie abgeschlossen w"are
(Kondensatkorrekturen werden in diesem Kapitel nicht ber"ucksichtigt).
Der ganze Satz von Momenten muss RG-invariant analysiert werden
\cite{effDh,brodsky,prl}. 
Das Einf"uhren einer nat"urlichen internen 
Kopplungskonstante $a(s)$ erlaubt es,
die St"orungsreihe
eine Ordnung weiter fortzuf"uhren,
als dies im $\MSsch$-Schema m"oglich ist \cite{prl,renRS}. 
Durch die Definition der effektiven Ladung mit $\rho_T(s)$
ergeben sich perturbative Korrekturen ausschlie"slich aufgrund 
des Laufens der Koppplungskonstante.
In jeder Ordnung der St"orungstheorie 
ist die Skalenabh"angigkeit der Kopplungskonstante $a(s)$
(Def.: Gl.~(\ref{defofa}))
durch die Koeffizienten der effektiven 
$\beta$-Funktion gegeben. Die Koeffizienten der effektiven $\beta$-Funktion
lassen sich 
aus Gl.~(\ref{betatransform}) mit der Relation zwischen der 
effektiven Kopplung $a$ und der Kopplungskonstante im
$\MSsch$-Schema $\al_s$ (\ref{defofa}) bestimmen:
\be
\label{run}
a(s) = a + \beta_0 l a^2 + (\beta_1 l + \beta_0^2 l^2) a^3
+ (\beta_2 l +\frac{5}{2}\beta_1\beta_0 l^2 + \beta_0^3 l^3) a^4 + \ldots\;.
\ee
Hier bezeichnet $a=a(M_\tau^2)$ und die $\beta_i$
die Koeffizienten der $\beta$-Funktion f"ur die effektiven Ladung $a$.
Die Beitr"age der verschiedenen Potenzen der Logarithmen 
ergeben sich zu
\footnote{Das Integral aus Gl.~(\ref{logs}) l"asst sich analytisch berechnen,
indem man die Substitution $s = \mts e^{-t}$ verwendet \cite{grootecomp}:
\begin{eqnarray}
 (n+1)\int_0^1 \left( \frac{s}{\mts} \right)^n 
     \ln^p \left( \frac{s}{\mts} \right) \frac{ds}{\mts}
  & =& (n+1)\int_0^\infty t^p e^{-(n+1)t}dt\ =\nonumber\\
  &=&\Big[-t^p e^{-(n+1)t}\Big]_{t=0}^\infty
  +p \int_0^\infty t^{p-1}e^{-(n+1)t}dt\ =\nonumber\\
  &=&\frac{p(p-1)}{n+1}\int_0^\infty t^{p-2}e^{-(n+1)t}dt\ =\ \ldots\\\ldots\
  &=&\frac{p!}{(n+1)^{p-1}}\int_0^\infty e^{-(n+1)t}dt \nn \\
  &=&\frac{p!}{(n+1)^p}\Big[-e^{-(n+1)t}\Big]_{t=0}^\infty \nn \\
  &=& \frac{p!}{(n+1)^p}\nonumber
\end{eqnarray} }
\begin{equation}
\label{logs}
(n+1)\int_0^{\mts} \left( \frac{s}{\mts} \right)^n 
  \ln^p \left( \frac{\mts}{s} \right) \frac{ds}{\mts} = \frac{p!}{(n+1)^p} \ .
\ee
Gleichung~(\ref{logs}) zeigt, dass die Effekte des Laufens 
der Kopplungskonstante mit gro"sen $n$ abnehmen und sich so 
die Konvergenz der St"orungsreihe verbessert.
Durch die in Gl.~(\ref{defofa}) gegebene Definition 
f"ur die Ladung verschwinden alle QCD-Korrekturen 
h"oherer Ordnung f"ur $n\rightarrow \infty$.
Wird das Laufen der Kopplungskonstante ber"ucksichtigt, 
so ergibt sich anstelle von Gl.~(\ref{norunmom}) 
\begin{eqnarray}
\label{momff}
m_0&=&a + 2.25 a^2  + 14.13 a^3  + 87.66 a^4 
+ (433.3 + 4.5 k_3) a^5 \;, \nn \\
m_1&=&a + 1.125 a^2  + 4.531 a^3  + 6.949 a^4 + (-175.2 + 2.25
k_3) a^5 \;,  \nonumber \\
m_2&=&a + 0.75 a^2  + 2.458 a^3  - 1.032 a^4  + 
(-142.6 + 1.5 k_3) a^5 \;, \nonumber \\
m_3&=&a + 0.563 a^2  + 1.633 a^3  - 2.542 a^4  + 
(-110.4 + 1.125 k_3) a^5 \;, \nonumber \\
&&\vdots \nonumber \\
m_{100}&=&a + 0.022 a^2  + 0.041 a^3  - 0.25 a^4  + 
(-4.08 + 0.045 k_3) a^5 \;.
\end{eqnarray}
F"ur gro"se $n$ verhalten sich die Momente besser, da der niedrigenergetische 
Bereich des Integrals (\ref{int}) 
unterdr"uckt ist. 
Die Koeffizienten der Reihen in Gl.~(\ref{momff})
werden f"ur gro"se $n$ durch den Beitrag des Logarithmus 
der niedrigsten Potenz 
dominiert.
Dies bedeutet, dass jede QCD-Korrektur f"ur gro"se $n$ durch den
h"ochsten auftretenden Koeffizienten der QCD-$\beta$-Funktion 
dominiert wird.

H"ohere Momente sind aus experimenteller Sicht unerw"unscht.
Sie werden durch den hochenergetischen Bereich des 
$\tau$-Lepton-Zerfallsspektrums dominiert.
Die experimentellen Daten f"ur diesen Bereich des Spektrums
sind sehr ungenau (vgl. Abb. \ref{s0rate}). 
Aus diesem Grund nimmt die experimentelle
Genauigkeit mit steigendem $n$ ab. 
Um experimentelle Fehler aus dem hochenergetischen 
Teil des Spektrums zu unterdr"ucken,
kann das modifizierte System aus gemischten Momenten
\begin{equation}
  \label{intmomkl}
\tilde M(k,l)=\frac{(k+l+1)!}{k!\,l!}\int_0^{\mts} \rho(s) 
      \left(1-\frac{s}{\mts} \right)^k \left(\frac{s}{\mts} \right)^l 
  \frac{ds}{\mts}
\equiv 1+\tilde m_{kl}
\end{equation}
benutzt werden \cite{DP}.
Die Wichtungsfunktion $(1-s/\mts )^k (s/\mts)^l$ 
hat ihr Maximum bei $\bar s = \mts \, l/(l+k)$.
Das Integral in Gl.~(\ref{intmomkl}) 
wird durch Beitr"age aus dem Bereich um das  
Maximum der Wichtungsfunktion  bestimmt. 
Der Nachteil von Momenten der Form 
Gl.~(\ref{intmomkl}) ist, dass der Faktor 
$(1-s/\mts )^k$ die Infrarot-Region stark betont und
die 
Konvergenz\footnote{Der Begriff Konvergenz wird in diesem und
den folgenden Kapiteln f"ur die Beschreibung des Verhaltens von
asymptotischen und somit im streng mathematischen Sinne divergenten
Reihen verwendet. ``Gute Konvergenz'' meint hier das schnelle Abnehmen der
Terme der Reihe wogegen ``schlechte'' oder ``gar keine Konvergenz'' ein 
langsames Abnehmen oder sogar ein Anwachsen der 
Terme beschreibt. Diese Begriffsbildung ist in der Fachliteratur dieses
Gebietes "ublich.}
der St"orungsreihe zerst"ort.
Es ergibt sich zum Beispiel 
\begin{eqnarray}
\label{altmom}
\tilde m_{00}&=&a + 2.25 a^2  + 14.13 a^3  + 87.66 a^4 + (433.3 + 4.5
k_3) a^5  \;, \nonumber \\
\tilde m_{10}&=&a + 3.375 a^2  + 23.72 a^3  + 168.4 a^4  + 
 (1042. + 6.75 k_3) a^5 \;,\nonumber \\ 
\tilde m_{20}&=&a + 4.125 a^2  + 31.24 a^3  + 241.1 a^4  + 
(1683. + 8.25 k_3) a^5 \;,\nonumber \\ 
\tilde m_{30}&=&a + 4.688 a^2  + 37.51 a^3  + 307.3 a^4  + 
 (2324. + 9.375 k_3) a^5 \; .
\end{eqnarray}
Die Koeffizienten der Reihen aus Gleichung~(\ref{altmom})
lassen sich f"ur beliebige Werte von $k$ in 
jeder Ordung der St"orungstheorie bestimmen.
Der Beitrag des logarithmischen Terms ergibt zum 
Beispiel
\be
\label{logalt}
(k+1)\int_0^{\mts} \left(1-\frac{s}{\mts} \right)^k 
        \ln \left( \frac{\mts}{s} \right) \frac{ds}{\mts} 
    = \sum_{j=1}^{k+1} \frac{1}{j} \ .
\end{equation}
Im Gegensatz zu Gl.~(\ref{logs}) verh"alt sich Gl.~(\ref{logalt}) f"ur gro"se
$k$ wie $\ln(k)$ und erzeugt so gro"se Koeffizienten in der 
St"orungsreihe.
Der Beitrag des $\ln^2$-Terms ist 
\be
\label{logalt2}
(k+1)\int_0^{\mts} \left(1-\frac{s}{\mts} \right)^k 
      \ln^2 \left( \frac{\mts}{s} \right) \frac{ds}{\mts} 
= \left(\sum_{j=1}^{k+1} \frac{1}{j} \right)^2
+ \sum_{j=1}^{k+1} \frac{1}{j^2} \;. 
\end{equation}
Dieser w"achst f"ur gro"se Werte von $k$ wie $\ln^2(k)$.

In der Anwendung ist 
das formale Kriterium f"ur die Genauigkeit einer asymptotischen Reihe
durch den nummerischen Wert des letzten Terms der Reihe 
gegeben. Dieses Kriterium muss allerdings mit gro"ser 
Vorsicht verwendet werden.
Aufgrund der Freiheit, die Kopplungskonstante neu zu 
definieren, kann der letzte Term einer St"orungsreihe 
f"ur eine gegebene Observable immer beliebig klein gemacht werden.
Um die Qualit"at der St"orungsreihe unabh"angig von der 
Definition der Kopplungskonstante zu beurteilen und so RG-invariante
Schlussfolgerungen zu ziehen, m"ussen immer
mehrere Observable gleichzeitig in Betracht gezogen werden.

Bevor die FOPT-Analyse weiter fortgesetzt wird, ist es notwendig,
die Beitr"age der Kondensatkorrekturen aus der 
Operatorproduktentwicklung (Gl.~(\ref{OPExp})) zu dem System von Momenten
Gl.~(\ref{intmom}, \ref{intmomkl})
zu kommentieren.
F"ur das System von Momenten aus Gl.~(\ref{intmom}) 
reduzieren sich die Beitr"age der Kondensatkorrekturen zu einem 
einzigen Term (falls man die schwache $\ln(Q^2)$-Abh"angigkeit 
der Koeffizientenfunktionen der OPE vernachl"assigt, 
was allgemein "ublich ist) von der Form $(\Lambda^2/M_\tau^2)^n$,
der mit wachsendem $n$ sehr schnell abnimmt
($\Lambda$ ist eine typische Skala der Kondensatkorrekturen, die mit der 
nicht-perturbativen QCD-Skala $\Lambda_{QCD} < \mts $ zusammenh"angt).
Dies macht den perturbativen Beitrag im Endergebnis dominant.
Der perturbative Term ist f"ur Momente mit gro"sem $n$ durch den 
hochenergetischen Bereich dominiert und konvergiert gut.
Die Konvergenz verbessert sich mit steigendem $n$, au"serdem sind die 
Momente durch den perturbativen Bereich bestimmt und aus diesem 
Grund pr"azise.
Im Gegensatz dazu werden die gemischten Momente aus Gl.~(\ref{intmomkl})
mit $l\sim 0$ und gro"sem $k$ haupts"achlich durch Beitr"age aus dem 
niedrigenergetischen Bereich dominiert und sind deswegen 
perturbativ nur schlecht berechenbar, 
was durch die schlechten St"orungsreihen reflektiert wird.
Die Momente aus Gl.~(\ref{intmomkl}) erhalten Kondensatkorrekturen der 
Dimension $d = 2l+2 $ bis $d = 2 (k+l +1)$.
"Uber eine solche Summe von Kondensatkorrekturen 
kann keine genaue nummerische Aussage gemacht werden, falls $k$ gro"s ist. 
Dies zeigt die Wichtigkeit der Kondensatkorrekturen f"ur gemischte Momente.
Die St"orungsreihen f"ur den Vektor- und Axialvektor-Beitrag sind 
identisch \footnote{In der Entwicklung der Korrelationsfunktion in der
Strange-Quarkmasse treten nur gerade Potenzen von $m_s$ auf, so dass
unterschiedliche Vorzeichen f"ur den Vektor- und Axialvektor-Teil der 
Korrelationsfunktion  aus der Spurbildung im Z"ahler nicht auftreten.
Unterschiede zwischen Vektor- und Axialvektor-Teil zeigen sich nur in den 
nichtperturbativen Korrekturen wie z.B. dem Strange-Quarkmassen Kondensatterm, 
welcher linear in $m_s$ ist.},
w"ahrend die Beitr"age der niedrigsten Resonanzen vollkommen 
verschieden sind (das $\pi$-Meson anstelle des $\rho$-Mesons ).
Deshalb kann keine Summation der St"orungsreihen diese 
Diskrepanz mit dem Experiment beheben. In diesem Fall ist die 
St"orungstheorie keine gute Approximation und Kondensatkorrekturen 
ergeben das richtige Resultat f"ur gro"se $k$.
Aus diesem Grund sind Momente mit gro"sem $k$ in der St"orungstheorie
nicht brauchbar, auch wenn sie aus experimenteller Sicht attraktiv sind. 

Hier zeigt sich in der Wahl der optimalen Observablen aus der Menge der 
Momente $(k,l)$
wieder der Konflikt zwischen experimenteller und theoretischer 
Genauigkeit.
Sind die expliziten st"orungstheoretischen Formeln verf"ugbar,
so l"asst sich die durch den asymptotischen Charakter
der St"orungsreihen gegebene ultimative Genauigkeit 
unabh"angig von den experimentellen Daten bestimmen.
Dies erlaubt es vorherzusagen, welcher Fehler -- experimenteller oder 
theoretischer -- die Unbestimmtheit der Observablen dominieren wird.

Im Folgenden wird f"ur die
effektive Kopplungskonstante $a$ der Wert $a=0.111$
verwendet, welcher aus dem entsprechenden Wert 
der Kopplungskonstante im $\MSsch$-Schema bestimmt wurde. 
Aus dem Satz von Momenten $\{m_n;n=0,\ldots,\}$
besitzt das Moment $m_0$ den gr"o"sten Beitrag aus dem 
niedrigenergetischen Bereich der Spektraldichte.  
Aus diesem Grund besitzt ein Satz von Observablen die 
schlechteste Genauigkeit, falls das Moment $m_0$ in dem 
Satz enthalten ist. F"ur $m_0$ ergibt sich
\be
\label{num0}
m_0=0.111 + 0.0277 + 0.0193 + 0.0133 + 
(0.0073 + 0.000076 k_3) \ .
\ee
Wie schon erw"ahnt, ist der Wert des Koeffizienten $k_3$
noch nicht bekannt.
In einigen der folgenden Ausdr"ucke wird f"ur $k_3$
ein nummerischer Wert eingesetzt, um einen Eindruck 
f"ur die Wichtigkeit des letzten Terms der St"orungsreihe 
zu erhalten.
Ein haufig benutzter Wert ist $k_3=25$, der auf der 
Pad\'e-Approximation beruht.
F"ur $k_3 = 25 $ ergibt sich f"ur das nullte Moment
\be
\label{num025}
m_0=0.111 + 0.0277  + 0.0193   + 0.0133   +  ``0.009'' \ .
\ee 
Das Konvergenzverhalten von Gl.~(\ref{num025}) ist sehr langsam.
Ausserdem betr"agt die Summe der vier QCD-Korrekturen
mehr als 40\% des f"uhrenden Terms der Reihe.
Das erste Moment zeigt ein deutlich besseres Konvergenzverhalten: 
\be
\label{num1} 
m_1= 0.111 + 0.014 + 0.006 + 0.001+(-0.003 + 0.00004 k_3) \;.
\ee
F"ur $k_3=25$ ergibt sich 
\be
\label{num125}
m_1= 0.111 + 0.014  + 0.006  + 0.001 - ``0.002'' \ .
\ee
Mit der Wahl $k_3 = 25 $ zeigt der $a^5$-Term 
bereits asymptotisches Wachstum.
Falls die Genauigkeit der Reihe durch ihren kleinsten 
Term abgesch"atzt wird, so ergibt sich ein formaler Fehler  
von ungef"ahr 1\%, und der gesamte Beitrag der Korrekturen h"oherer 
Ordnung betr"agt etwa 20\% des f"uhrenden Terms.
Der gro"se Unterschied in der Genauigkeit zwischen den 
Momenten $m_0$ und $m_1$ ist eine generelle Eigenschaft 
der Momente $m_n$ in der f"unften Ordnung der St"orungstheorie:
Es ist nicht m"oglich, durch die Wahl einer einzigen Zahl 
$k_3$ die $a^5$-QCD-Korrekturen aller Momente klein zu machen.
Aus diesem Grund muss man schlussfolgern, dass f"ur $a= 0.111$
in der f"unften Ordnung der St"orungstheorie bereits asymptotisches 
Wachstum zu beobachten ist.
Das Ansteigen der Korrekturen ist unabh"angig von jeder Definition
der Kopplungskonstante zu beobachten, indem man gleichzeitig mehrere 
Momente betrachtet. Diese Eigenschaft verschwindet 
durch keine Wahl des Koeffizienten $k_3$.
F"ur ein einzelnes Moment (z.B. $m_0$) kann 
die Kopplungskonstante immer
so gew"ahlt werden, dass die St"orungsreihe schnell bis zu einer 
gegebenen Ordnung abnimmt. Allerdings werden dann 
andere Momente, ausgedr"uckt durch diese Kopplungskonstante, 
ein schlechtes Konvergenzverhalten zeigen. 
Die RG-invariante Schlussfolgerung ist, dass  
das System von Momenten $m_n$,  eingeschlossen $m_0$,
st"orungstheoretisch nicht bis zur f"unften Ordnung  
behandelt werden kann, falls man Fehler von weniger als 
5\% - 10\% erreichen will. Hierbei wird f"ur die Kopplungskonstante 
ein Wert von ungef"ahr $a= 0.111$ vorausgesetzt. 
Diese Aussage "uber die ultimative Genauigkeit des Satzes von 
Momenten $m_n$ in f"unfter Ordnung der St"orungstheorie ist 
unabh"angig von dem Wert des unbekannten Koeffizienten $k_3$.
Falls das Moment $m_0$ ausgeschlossen wird, ist eine Pr"azision  
von mehr als 1\% im Rahmen der St"orungstheorie m"oglich. 
Durch Ausschlie"sen von $m_0$ und durch die Wahl $k_3 \sim 100$
erh"alt man ein System von Momenten,
in dem alle Korrekturen f"unfter Ordnung klein sind.
Um dies in RG-invarianter Weise zu demonstrieren,
kann man das zweite Moment als Definition einer
experimentellen Ladung verwenden. 
Dies ergibt
\begin{eqnarray}
\label{invmom}
m_{0}&=&m_2 + 1.5 m_2^2 +  9.417 m_2^3 +  59.28 m_2^4+(310.3 +
3 k_3) m_2^5 \;,
\nonumber \\
m_{1}&=&m_2 + 0.375 m_2^2 +  1.51 m_2^3 + 2.527 m_2^4
+(-54.45 + 0.75 k_3) m_2^5 \;,\nonumber \\ 
m_{2}&=&m_2 \;,\nonumber \\ 
m_{3}&=&m_2 - 0.19 m_2^2 - 0.544 m_2^3 + 0.742 m_2^4+(35.2 -
0.375 k_3) m_2^5\;,
\nonumber \\ 
m_{4}&=&m_2 - 0.3 m_2^2  - 0.803 m_2^3  + 1.69 m_2^4 +
(56.641 - 0.6 k_3) m_2^5  \ .
\end{eqnarray}
Die Konvergenz der Momente $m_1$-$m_4$
(und f"ur $n>4$) ist gut.
Der gesamte Beitrag der QCD-Korrekturen ist klein.
Die schlechteste Reihe ist die f"ur das nullte Moment.
Gl.~(\ref{invmom}) zeigt, dass keine Wahl von $k_3$ die 
Genauigkeit der Momente $m_0$ und $m_1$ in f"unfter Ordnung gegen"uber den 
Ausdr"ucken in vierter Ordnung wesentlich verbessern kann.  
Es gibt einen kleinen Bereich f"ur $k_3$ ($40<k_3<60$),
in dem das formale Kriterium der Konvergenz f"ur $m_0$ und $m_1$ 
erf"ullt ist. Es scheint aber unwahrscheinlich,
dass zuk"unftige Berechnungen des Koeffizienten $k_3$ 
einen Wert f"ur $k_3$
aus diesem Intervall ergeben. Au"serdem betr"agt bei dieser 
Wahl von $k_3$ die Genauigkeit des Moments $m_0$ nur ungef"ahr 10\%. 
Dies ist ein Anzeichen daf"ur, dass die ultimative Genauigkeit der 
St"orungstheorie f"ur das nullte Moment $m_0$ bereits erreicht ist.
Falls das Moment $m_0$ ausgeschlossen wird, erlaubt  die Wahl  
$k_3\sim 100$ eine schnelle Konvergenz bis einschlie"slich f"unfter Ordnung.
Dann sind keinerlei Schlussfolgerungen "uber asymptotisches Wachstum 
mehr m"oglich. 

Die St"orungsreihen f"ur ein System von Momenten mit Wichtungsfunktion
$(1-s/\mts)^k$ zeigen ein schlechteres Verhalten.
Mit dem "ublichen Kriterium f"ur die Genauigkeit liegt die 
Pr"azision der Reihen aus Gl.~(\ref{altmom}) in einem Bereich
von 10\% - 20\% f"ur den hier verwendeten nummerischen Wert f"ur $a$.
Dies ist f"ur den Vergleich mit den aktuellen experimentellen Daten 
nicht ausreichend. 
Die Entwicklung der h"oheren Momente aus Gl.~(\ref{altmom})
in dem ersten Moment (welches das perturbativste dieses Systems ist),
ergibt
\begin{eqnarray}
\label{altmomnum}
\tilde m_{00}&\equiv&m_0 = 0.17 \;,\nonumber \\
\tilde m_{10}&=&0.17  
+ 0.033 + 0.022 + 0.011 + (-0.005 + 0.00032 k_3) \;,\nonumber \\ 
\tilde m_{20}&=&0.17 
+ 0.054 + 0.043 + 0.027 + ( 0.0015+ 0.00053 k_3) \;,\nonumber \\
\tilde{m}_{30} &=& 0.17 + 0.070 + 0.061 + 0.046 + ( 0.014+ 0.0007 k_3 ) \;.
\end{eqnarray}
Diese Reihen besitzen eine formale Genauigkeit zwischen 6\%
und 25 \% und der Beitrag h"oherer Terme kann ebenso gro"s 
sein wie der des f"uhrenden Terms.
Aufgrund der langsamen Konvergenz scheinen sich die theoretischen 
Vorhersagen mit zunehmender Ordnung in der St"orungstheorie nicht 
zu verbessern. Die Reihenentwicklung erlaubt keine zuverl"assige 
Absch"atzung der Genauigkeit f"ur gro"se gemischte Momente. 
W"ahrend f"ur die Momente aus Gl.~(\ref{intmom})
der gesamte Beitrag der QCD-Korrekturen klein ist,
ist die Situation hier anders.
Die gesamte "Anderung des Resultates in f"uhrender Ordnung
aufgrund von Korrekturen h"oherer Ordnung ist wesentlich und 
und unterscheidet sich f"ur verschiedene Momente stark von einander.
Dies ist ein weiteres Anzeichen daf"ur, dass dieser Satz von Momenten 
nicht perturbativ behandelt werden kann. 
\section{$\tau$-Lepton-Zerfallsrate}
Die $\tau$-Lepton-Zerfallsrate ist durch eine spezielle Linearkombination von 
Momenten gegeben.
Aufgrund des Faktors $(1-s/\mts)^2$ in Gleichung~(\ref{int})
sind die Konvergenzeigenschaften der totalen Zerfallsrate
nicht optimal.
Der Faktor $(1-s/\mts)^2$ verst"arkt den Beitrag des Infrarot-Bereiches der 
Integration.
Die konkrete Form der Wichtungsfunktion mit dem Faktor $(1-s/\mts)^2$
ist der Hauptgrund f"ur die schlechte Konvergenz.
Es ergibt sich 
\be
\label{rtau}
\delta_P =a + 3.563 a^2  + 24.97 a^3  + 
    174.8 a^4  + (1041. + 7.125 k_3) a^5 \ .
\ee
Mit $a=0.111$ und $k_3=25$ erh"alt man
\be
\label{rtau0111}
 \delta_P= 0.111 + 0.044 + 0.034 + 0.027 + ``0.021'' \;. 
\ee
Die aufeinanderfolgenden Terme der Reihe nehmen zwar ab, diese 
Abnahme ist aber sehr langsam. Das Konvergenzverhalten der totalen 
Zerfallsrate $\delta_p$ folgt 
aufgrund des Faktors $(1-s/\mts)^2$ in Gl.~(\ref{int})
im wesentlichen dem des Momentes 
$\tilde m_{20}$ aus Gl.~(\ref{altmom}).

Gleichung (\ref{theoac})
zeigt das Wesentliche des sich stellenden Problems.
Wird St"orungstheorie in endlicher Ordnung (FOPT) verwendet,
so muss der experimentelle Wert $\delta^{exp}_P$ (vgl. Gl~
(\ref{expdec}, \ref{expdec0}))
mit dem 
theoretischen Ausdruck $\delta^{th}_P$ verglichen werden: 
\be 
\label{theoac}
0.203\pm 0.007 = \delta^{exp}_p  \quad {\rm vs.} \quad
\delta^{th}_p = 0.111 + 0.044 + 0.034 + 0.027 + ``0.021''\;.
\ee
Der Fehler des theoretischen Ausdrucks ``0.021'' 
(oder sogar 0.027) ist weitaus gr"o"ser als der experimentelle 
Fehler 0.007.
Die theoretische Ungenauigkeit aufgrund des Abbrechens der 
St"orungsreihe ist viel gr"o"ser als der experimentelle 
Fehler der entsprechenden Observablen.  
Die "ubliche praktische Erwartung an eine sinnvolle St"orungsreihe
ist die Kleinheit aller Korrekturen h"oherer Ordnung, falls 
nichts "uber die Konvergenz des Ausdruckes bekannt ist. 
Die Korrekturen der Zerfallsrate vergr"o"sern das Resultat der 
f"uhrenden Ordnung um den Faktor 2.
Es ist m"oglich, das explizite Konvergenzverhalten 
der zu der Zerfallsrate geh"orenden St"orungsreihe durch 
Redefinition des Entwicklungsparameters zu verbessern, 
was aufgrund der Freiheit, ein Renormierungsschema zu w"ahlen, 
m"oglich ist. Allerdings wird sich durch eine solche 
Definition des Entwicklungsparameters das Konvergenzverhalten 
des ersten Momentes der differentiellen Zerfallsrate wieder verschlechten.
Dieses Verhalten erm"oglicht Schlussfolgerungen "uber das asymptotische 
Wachstum der St"orungsreihen unabh"angig von der Wahl des 
Renormierungsschemas. 
Zwei verschiedenen S"atze von Observablen, wobei der eine das Moment $m_0$
enth"alt und der andere dieses Moment nicht enth"alt, k"onnen 
perturbativ nicht zu einander in Bezug gebracht werden, falls eine mit 
den experimentellen Daten vergleichbare Genauigkeit verlangt wird.
F"ur das erste Moment der differentiellen $\tau$-Zerfallsrate 
$d R_\tau/ds$ ergibt sich ein St"orungsreihe mit einem besseren
Konvergenzverhalten als in Gl.~(\ref{rtau}):
\be
\label{rtau1}
\delta_P^{(1)} =
a + 2.138 a^2  + 10.15 a^3  +  28.43 a^4  +
 (-268.3 + 4.275 k_3) a^5 \;.
\ee
Mit $k_3=25$, ergibt sich so
\be
\label{rtau10111}
 \delta_P^{(1)} =
0.111 + 0.026 + 0.014 + 0.004 - ``0.003'' \ .  
\ee       
Das zweite $s$-Moment besitzt eine noch bessere St"orungsreihe:
\be
\label{rtau2an} 
 \delta_P^{(2)}=a + 1.575 a^2  + 6.186 a^3  + 6.386 a^4 +
 (- 283.3  + 3.15 k_3)a^5
\ee
und mit $k_3=25$ hat man 
\be
\label{rtau2}
 \delta_P^{(2)}
=
0.111 + 0.0194 + 0.0085 + 0.001 - ``0.003'' \ .
\ee
Der f"unfte Term ist f"ur $k_3 = 25 $ 
gr"o"ser als der Term vierter Ordnung.  
Keine Wahl f"ur $k_3$ kann gleichzeitig 
alle drei Observable in der f"unften Ordnung konvergent machen.
Falls man $k_3 \sim 100$ w"ahlt, um eine bessere Konvergenz der
h"oheren Momente zu erzielen, zerst"ort man die St"orungsreihe 
f"ur die Zerfallsrate in Gl.~(\ref{rtau}).

Die Momente mit Wichtungsfaktor $(1-s/\mts)^n$ der differentiellen 
Zerfallsrate unterdr"ucken schlecht bekannte Daten aus dem 
hochenergetischen Bereich. Theoretisch ergibt sich f"ur diese 
\be
\label{alttauan} 
 \delta_P^{(1-s)}= a + 4.173 a^2  + 31.31a^3  + 237.6 a^4  + 
(1603.  + 8.35 k_3) a^5 \;.
\ee
Setzt man  $k_3=25$, so ergibt sich
\be
\label{alttau}
 \delta_P^{(1-s)}= 0.111 + 0.051 + 0.043   + 0.036 +  ``0.031'' \ .
\ee 
F"ur $k_3=100$ ergibt sich die Reihe
\be
\label{alttaunum}
 \delta_P^{(1-s)} = 0.111+ 0.051 + 0.043 + 0.036 + ``0.041'' \;,
\ee 
die nur eine Genauigkeit von 30\% besitzt, wobei ihr Wert
sich mehr als um den Faktor 2 von dem Resultat in f"uhrender Ordnung 
unterscheidet. 
In diesem Fall kann die theoretische Genauigkeit nicht mit der der 
experimetellen Daten verglichen werden.
 
Bei der Analyse der $\tau$-Zerf"alle treten zwei verschiedene Probleme auf. 
\begin{enumerate}
\item Die Beschreibung eines Satzes von Momenten mit Kopplungskonstante
und Massen, welche speziell f"ur dieses System definiert wurden und die 
gr"o"stm"ogliche Genauigkeit zulassen ($\tau$-Lepton-interner QCD-Test). 
\item Die Bestimmung der Standard-$\MSsch$-Paramter, welche mit Resultaten 
aus anderen Experimenten verglichen werden k"onnen.
\end{enumerate}
Es kann vorkommen, dass ein Satz von Observablen perturbativ mit einer 
gegebenen Genauigkeit miteinander 
verkn"upft ist, die $\MSsch$-Kopplungskonstante $\al_s$ aber nicht der 
bestm"ogliche Parameter f"ur die Entwicklung ist. 
Dies ist hier der Fall.
Intern l"asst sich das $\tau$-System mit einer gr"o"seren Genauigkeit und
mit einem Term mehr in der St"orungsreihe, 
als dies im $\MSsch$-Schema m"oglich ist,
durch die effektive Kopplung $a$
beschreiben. 
Trotzdem zeigt sich bereits das asymptotische Wachstum der St"orungsreihe
f"ur den experimentell gegebenen nummerischen Wert des Entwicklungsparameters. 

Der Ausdruck f"ur die Zerfallsrate im $\MSsch$-Schema besitzt nur eine 
Genauigkeit von der Ordnung-$\al_s^3$  
\be
\label{taumssch}
\delta_P = \left(\als\right)+5.20 \left(\als\right)^2
+26.4\left(\als\right)^3
+(78.0+k_3)\left(\als\right)^4\;,
\ee
mit einer nummerischen Pr"azision von nur 30\%.

Das $\tau$-System 
kann  in der Ordnung $a^4$ 
mit der internen Ladung $a$ ohne freie Parameter untersucht werden
und in der Ordnung $a^5$ mit nur einem freien Parameter $k_3$,
welcher die Schlussfolgerungen "uber die asymptotische Struktur der 
St"orungsreihen nicht beeinflusst.
Allerdings l"asst sich die $\MSsch$-Kopplungskonstante $\al_s$ 
aufgrund des unbekannten Parameters $k_3$ durch $a$
nur bis zur Ordnung $a^3$ ausdr"ucken.
Der Wert der $\MSsch$-Kopplungskonstante l"asst sich aus 
dem Wert von $a$ durch 
die Relation   
\be
\label{mstoarel}
 {\al_s^{\overline{\rm MS}}(M_\tau)\over \pi}=a - 1.64 a^2 + 15.7 a^3
+ (49.6 - k_3) a^4 +\ldots
\ee
bestimmen, welche f"ur $k_3 = 25$ oder $k_3 = 100$ 
vern"unftige Konvergenz zeigt. 
\section{Infrarot-Fixpunktmodell}
Die Genauigkeit, mit der eine Funktion durch eine endliche Summe von 
Termen ihrer asymptotischen Reihe approximiert werden kann, h"angt stark 
von der analytischen Struktur dieser Funktion ab.
Grunds"atzlich gibt es unendlich viele M"oglichkeiten, eine asymptotische 
Reihe zu summieren, welche relativ unterschiedliche Resultate liefern k"onnen.
Aus diesem Grund k"onnen Absch"atzungen der Genauigkeit, welche nur 
auf einer endlichen Anzahl der Terme beruhen, sehr irref"uhrend sein.
Um dies detaillierter zu untersuchen,
 wird hier ein Modell 
(zur Illustration) f"ur die
exakte Funktion verwendet, welche den Ursprung der asymptotischen Reihe
bildet (oder eine M"oglichkeit der Resummation darstellt). 
Dieses Modell verwendet die Existenz eines Infrarot-Fixpunktes
der laufenden Kopplungskonstante in dritter Ordnung der St"orungstheorie,
welcher erlaubt, 
das Laufen der Kopplungskonstante bis zum Ursprung fortzusetzen.
In diesem speziellen Fall ist es m"oglich, die St"orungsreihe mit dem exakten 
Resultat zu vergleichen.
Dieses Beispiel erlaubt es, die allgemeinen Schlussfolgerungen "uber die 
asymptotische Struktur und die Divergenz der Reihen 
an einem konkreten Beispiel zu "uberpr"ufen.
Die effektive $\beta$-Funktion f"ur die in Gl.~(\ref{defofa}) 
definierte Kopplungkonstante
$a$ ist durch  
\be
\label{effbeta}
\beta_{\rm eff}(a)= -\frac{9}{4} a^2  - 4 a^3  + 25.7 a^4  
+ (409.5 - \frac{9}{2} k_3) a^5 + \mathcal{O}(a^6) 
\ee
gegeben, 
wobei der einzige freie Parameter $k_3$ ist, da der Vierschleifen-Koeffizient 
$\beta_3$ im $\MSsch$-Schema bekannt ist \cite{beta4}. 
Die Approximation der $\beta$-Funktion (Gl.~(\ref{effbeta}))
bis zu der dritten Ordnung $\mathcal{O}(a^3)$
besitzt einen Infrarot-Fixpunkt bei dem Wert $a_f=a(0)=0.384$,
gegen den der Wert den Kopplungskonstante f"ur $s \rightarrow 0$
konvergiert, falls f"ur den Anfangswert $a(\mts) < a_f $ gilt.

Die exakten Momente des Fixpunktmodells erh"alt man, indem die 
Anfangswertaufgabe 
\be \label{fixmod}
q^2 \frac{d}{d q^2} a(q^2) = - a^2 ( \frac{9}{4} + 4 a - 25.7 a^2) 
  \quad {\rm mit} \quad a(\mts) = 0.111
\ee
nummerisch gel"ost wird, wobei die rechte Seite von Gl.~(\ref{fixmod})
als exakte Funktion behandelt wird.
Mit der so gewonnenen L"osung kann die Integration in Gl.~(\ref{intmom})
nummerisch ausgef"uhrt werden.
Es ergibt sich
$m_0^f=0.1605$ und $m_1^f=0.130954$, was mit den Resultaten der 
Gleichungen~(\ref{momff}, \ref{num0}, \ref{num1}) verglichen 
werden kann.
Die naive Absch"atzung der Genauigkeit ergibt nicht f"ur alle Momente
Resultate, die mit diesem Modell vertr"aglich sind.
Mit Hilfe dieses expliziten Modells ist es m"oglich, eine beliebige Anzahl 
von Termen der St"orungsreihe zu erzeugen.
F"ur das nullte Moment $m_0$ divergiert die Reihe folgenderma"sen:
\be
\label{pat}
m_0^{fix} = 0.111 + 0.028 + 0.019 + 0.013+0.014
+ 0.018 + 0.029 + 0.053 + 0.114+\dots\,,
\ee
was eine ultimative Genauigkeit von nur 10\% ergibt.
Diese Absch"atzung der Genauigkeit ergibt sich, indem wie 
"ublich, die Reihe aufsummiert wird, bis ihre Terme wieder zu 
wachsen beginnen. Hierbei dient der letzte Term zu Absch"atzung der
Genauigkeit. 
Die Summe der ersten vier Terme ergibt die beste Genauigkeit
\be
m_0^{fix,best} = 0.111 + 0.028 + 0.019 + 0.013 = 0.171\pm 0.013
\ee
und kann mit dem exakten Resultat $m_0^f=0.1605$ verglichen werden.
Der zentrale Wert ist etwas zu hoch, befindet sich aber noch in dem durch 
den letzten Term gegebenen Fehlerintervall.

F"ur $m_1^{fix}$ findet man eine divergente Reihe, deren erste Terme sehr viel 
schneller fallen. Das Konvergenzverhalten ist durch den folgenden gro"sen
Ausdruck gegeben, der illustriert, wie kompliziert das Verhalten solcher 
Reihen werden kann. Er ergibt sich zu 
\begin{eqnarray}
\label{m1f}
m_1^{fix}&=& 0.111 + 0.013861  + 0.006197  + 0.001054 
+0.000480 + 0.000088   \nonumber \\
&+&0.000053 + 0.000016+0.000015 + 0.000014 + 0.000019 + 0.000026\nonumber \\
&+&0.000042 + 0.000072
+0.000135 + 0.000268 + 0.000568 + 0.001277 \ \nonumber \\
&+&0.003 + 0.0076 + 0.01997 + 0.055 + \ldots
\end{eqnarray}
Die beste Approximation ist formal durch die Summe der ersten zehn Terme 
gegeben
\be
\label{best1}
m_1^{fix,best}=0.132795\pm 0.000014 \;,
\ee
wobei der formale Fehler wieder durch den
letzten Term gegeben ist.
Das exakte Resultat $m_1^f=0.130954$
f"allt allerdings nicht in dieses winzige, durch den Fehler aus Gl.~(\ref{best1}), 
gegebene Intervall.
Aus diesem Grund ist das formale Kriterium f"ur die Genauigkeit in diesem
Fall verletzt.
Die Diskrepanz
$m_1^{fix,best}-m_1^f=0.00184$ zwischen Approximation und exaktem Resutat 
wird nicht durch den kleinsten Term der asymptotischen Reihe 
aus Gleichung~(\ref{m1f}) kontrolliert. 
Die reale 
Genauigkeit des ersten Mometes ist 1.3\%, was  in der Praxis ausreichend ist.
Der Unterschied dieser beiden Observablen ($m_0$ und $m_1$) 
liegt in der unterschiedlich starken Betonung des
Infrarotbereichs, wecher sich in den verschiedenen ultimativen Genauigkeiten 
und dem schnelleren Abfallen der Reihe f"ur $m_1$ zeigt.
Die Gr"o"se des kleinsten Terms in dem Ausdruck in Gl.~(\ref{m1f})
ist sehr sensitiv auf den dritten Koeffizienten der $\beta$-Funktion. 
F"ur die Zerfallsrate findet man $\delta_P^f=0.1946$ und 
$\delta_P^{f(1)}=0.1527$,
was mit den entsprechenden Ausdr"ucken in den  
Gleichungen~(\ref{rtau}, \ref{rtau0111}, \ref{theoac}) und 
(\ref{rtau1}, \ref{rtau10111}) verglichen werden kann.

Aus Gleichung~(\ref{effbeta}) erkennt man, dass ein Infrarot-Fixpunkt
auch noch bei Ber"ucksichtigung der vierten Ordnung der $\beta$-Funktion
existiert, falls $k_3<95.9$ gilt.    
\section{Zusammenfassung von Kapitel \ref{Art1}}
Unter Verwendung der Standardabsch"atzung der Genauigkeit einer 
asymptotischen Reihe ergab sich, dass die theoretische Pr"azision 
der st"orungstheoretischen Beschreibung von $\tau$-Zerf"allen
durch das asymptotische Anwachsen der Koeffizienten in der f"unften 
Ordnung der St"orungstheorie beschr"ankt wird.
Diese Aussage ist unabh"angig von dem Renormierungsschema g"ultig.
Die Genauigkeit der St"orungsreihen f"ur einen allgemeinen Satz von 
Observablen kann nicht besser als 5\% - 10\% sein. 
Genauer bedeutet dies, dass das nullte Moment in Rahmen der 
St"orungstheorie nicht mit einer mit 
den aktuellen Experimenten vergleichbaren Genauigkeit berechenbar ist.
Jede konsistente Beschreibung der $\tau$-Zerfallsdaten in f"unfter Ordnung 
der St"orungstheorie verlangt den Ausschluss des nullten Momentes 
aus der Liste der Observablen oder dessen Unterdr"uckung in den theoretischen 
Ausdr"ucken.
In f"unfter Ordnung der St"orungstheorie und mit dem aktuellen Wert 
der Kopplungskonstante sind die ersten beiden Momente der Spektraldichte
zu verschieden, um gleichzeitig durch die St"orungstheorie mit einer 
Genauigkeit besser als 5\% - 10\% behandelt zu werden.  
Aus diesem Grund ist es notwendig, ein Verfahren jenseits der endlichen 
St"orungstheorie zu verwenden, falls man diese beiden Observablen mit 
einer Genauigkeit, die der der aktuellen experimentellen Daten entspricht,
vergleichen will.
Dies impliziert die Verwendung eines Resummationsverfahrens.
Die Resummation ist nicht eindeutig bestimmt und die 
Resultate h"angen von dem verwendeten Verfahren ab
\cite{Pivtau,renRS,groote}.

\chapter{Strange-Quarkmassen-Korrektur} \label{Art2} \label{ART2}
In Kaptel \ref{Art1} wurden 
$\tau$-Lepton-Observable im Limes verschwindender 
Quarkmassen untersucht. 
F"ur Up- und Down-Quarks ist diese N"aherung sehr gut, 
da ihre Masse gering im Vergleich zu der Masse des $\tau$-Leptons ist.
Entstehen Zerfallsprodukte, welche Strangenes tragen, so ist die 
Korrektur zur Zerfallsbreite
aufgrund der Strange-Quarkmasse von der Ordnung
$m_s^2/\mts$, was ungef"ahr 1\% entspricht.   
In diesem Kapitel soll diese Korrektur untersucht werden.
Aufgrund der endlichen Quarkmasse ist die Korrelationsfunktion nicht 
mehr transversal, und es ist notwendig, die Momente von zwei 
Spektraldichten zu untersuchen, die den beiden Komponenten der 
Korrelationsfunktion entsprechen.
F"ur die Zerfallsrate hat man
\begin{equation}
  \label{int-qg}
 R_{\tau us} =
       {N_c} S_{EW}|V_{us}|^2  \frac{i}{2 \pi} \oint_{|s|^2 = \mt^2} 
               2  \left( 1-\frac{s}{\mt^2} \right)^2 
                     \left( \Pi_{q us}(s) - \frac{2}{\mts}   \Pi_{g us}(s) 
        \right) \frac{ds}{\mt^2} \;,
\end{equation} 
wobei im Folgenden die Faktoren $N_c$ und $S_{EW}$ weggelassen werden.
\section{Der $g$-Teil der Korrelationsfunktion}  
Zun"achst wird der $g$-Teil der Korrelationsfunktion untersucht, welcher 
nur Zerf"alle in hadronische Zerfallsprodukte mit Spin-1 beschreibt.
Die Analyse des $g$-Teils verl"auft analog zu dem massenlosen Teil in 
Kapitel \ref{Art1}. 
Es wird ein neuer Massenparameter $m_g$ f"ur die Strange-Quarkmasse  
definiert, um alle Konstanten der Massenkorrektur der Spektraldichte
in der neuen Masse zu absorbieren.
Dies geschieht durch die Definition 
\ba \label{defofmg}
m_s^2(s)\rho_g^{\MSsch}(s;\al_s)\equiv
m_g^2(s;a)= m_g^2(\mts) \rho_g(s) .
\ea
Die Spektraldichte $\rho_g^{\MSsch}$ ist durch 
\be
\rho_g^{\MSsch} = \frac{ {\rm Im} \,( \Pi_{mg}(Q^2)) }{\pi}
\ee
definiert, wobei $\Pi_{mg}$ sich aus Gl.~(\ref{fullinf}) ergibt..
Der neue Massenparameter $m_g=m_g(\mts)$ 
h"angt mit der Masse im $\MSsch$-Schema $m_s=m_s(\mts)$
durch 
\ba 
\label{mg}
m_g^2&=&m_s^2 \left( 1+1.67 a - 5.87 a^2 - 51.0 a^3 
          +(-1342.5 -1.67 k_3 + k_{g3})
a^4+ \mathcal{O}(a^5) \right)  \nn \\
&=&m_s^2 \left( 1+1.67 \alsb  
      - 3.14 \alsb^2 - 87.4 \alsb^3 + \right. \nonumber \\ 
   &&  \quad \quad +\left. ( -1750 + k_{g3} ) \alsb^4 
        + \mathcal{O} \left(\alsb^5 \right) \right) 
\ea
zusammen.
Mit Gl.~(\ref{mg}) l"asst sich der nummerische Wert 
von $m_g^2$ aus dem Wert der $\MSsch$-Masse mit einer 
Genauigkeit von ungef"ahr 7\% bestimmen, falls Gl. (\ref{mg})
bis zu der dritten Ordnung ausgewertet wird.
Invertiert man die Reihe perturbativ, so kann die Strange-Quarkmasse 
im $\MSsch$-Schema mit einer Genauigkeit von 4\% 
aus der effektiven Masse mit der Relation 
\be
m_s^2 = m_g^2 (1 - 0.185 + 0.107 + 0.037 + (0.19 + 0.00025 k_3 - 0.00015
k_{g3})) 
\ee
bestimmt werden.
Diese Genauigkeit ist f"ur die 
derzeitigen
ph"anomenologischen 
Anwendungen ausreichend.
Die Spektraldichte $\rho_g(s)$
f"ur den $g$-Teil der Massenkorrektur enth"alt 
aufgrund ihrer Konstruktion nur Logarithmen.
Die St"orungsreihe der Spektraldichte ist
\begin{eqnarray}
\label{densg}
\rho_g(s)&=&1+2 a l + a^2 (8.05 l + 4.25 l^2) + a^3(5.3 l + 38.23
l^2 + 9.21 l^3 ) \nonumber \\
&&+a^4((-45.3  -2.0 k_3)l + 67.4  l^2 + 134.2 l^3 + 20.14l^4)+ 
 \mathcal{O}(a^5) .
\end{eqnarray}
Ist der Ausdruck f"ur die Spektraldichte gegeben (Gl.~(\ref{densg})),
so l"asst sich die in Kapitel \ref{Art1} ausgef"uhrte Analyse wiederholen.
Die Momente der Spektraldichte $\rho_g(s)$ verhalten sich grunds"atzlich
schlechter als die des masselosen Falls.
Dies ist verst"andlich, wenn man die Koeffizienten der Logarithmen 
in den Spektraldichten des masselosen Teils Gl.~(\ref{densold}) und 
des $g$-Teils Gl.~(\ref{densg}) vergleicht. Im $g$-Teil sind die 
Koeffizienten grunds"atzlich gr"o"ser und aufgrund der anormalen Dimension
der Masse tritt in jedem Term eine weitere Potenz des Logarithmus auf.
Die grundlegenden Objekte f"ur die Konstruktion von Observablen sind 
Momente der Spektraldichte $\rho_g(s)$, die in Analogie zum 
masselosen Fall durch    
\begin{equation}
  \label{intmomg}
M_g(n)=(n+1)\int_0^{\mts}  \rho_g(s) \left(\frac{s}{\mts}\right)^n \frac{ds}{\mts} 
\end{equation}
definiert sind.
Es ergibt sich
\begin{eqnarray}
\label{momfff}
M_{g}(0)&=&1+ 2a + 16.6 a^2  +  137.0  a^3  
+ (1378.5-2.0 k_3)a^4  \;, \nonumber \\
M_{g}(1)&=&1+ a + 6.15 a^2 + 28.67  a^3  
+ (141.97-1.0 k_3) a^4 \;, \nonumber \\
M_{g}(2)&=&1+\frac{2}{3}a+3.63 a^2 + 12.31 a^3   
+(35.69 -0.67 k_3) a^4 \;, \nonumber \\
M_{g}(3)&=&1+ \frac{1}{2}a +  2.54  a^2  + 6.97  a^3   
+ (11.58 -0.5  k_3 ) a^4  \;, \nonumber \\
M_{g}(4)&=&1+ \frac{2}{5}a +  1.95  a^2  + 4.56  a^3   
+ (3.55 -0.4  k_3 ) a^4  \;, \nonumber \\
&&\vdots \nonumber \\
M_{g}(100)&=&1+ \frac{2}{101} a + 0.081 a^2  + 0.060 a^3  
+ (-0.434 -0.0198 k_3) a^4 \;. 
\end{eqnarray}
Bemerkenswert ist, dass der unbekannte Koeffizient $k_{g3}$, welcher 
in der QCD-Korrektur vierter Ordnung der Momente im $\MSsch$-Schema
auftritt, in die Definition der Masse $m_g$ absorbiert wird.
Trotzden ist die Korrektur vierter Ordnung aufgrund ihrer 
Abh"angigkeit von $k_3$ nicht bekannt. 
Der Ursprung dieser Abh"angigkeit besteht in der Definition der Masse $m_g$
(\ref{defofmg}).
Der dritte Koeffizient der effektiven $\gamma$-Funktion $\gamma_{g3}$
h"angt von $k_3$ ab, wenn diese in der effektiven Kopplungskonstante
$a$ entwickelt wird.
Der Koeffizient $\gamma_{g3}$ geht in die Korrektur vierter Ordnung
von $\rho_{g}(s)$ durch das Laufen der Masse $m_g$ (\ref{runmass}) ein.
\be
\gamma_0 = 1, 
\quad \gamma_{g1} = 4.027,  
\quad \gamma_{g2} = 2.65,
\quad \gamma_{g3} = -22.65 - k_3 \, .
\ee
F"ur gro"se $n$ verhalten sich die Momente besser, da die
Infrarot-Region der Integration unterdr"uckt ist.
Wie schon im masselosen Fall werden die Koeffizienten von Gl.~(\ref{momfff})
f"ur gro"se $n$
durch den Beitrag des Logarithmus mit der niedrigsten Potenz, also
durch den h"ochsten auftretenden Koeffizienten der effektiven
$\beta$- und $\gamma$-Funktion dominiert.

Es ist instruktiv, die Resultate aus Gl.~(\ref{momfff})
mit den entsprechenden Ausdr"ucken im $\MSsch$-Schema zu vergleichen,
die durch 
\begin{eqnarray}
M_{g}^{\MSsch}(0) &=&1+ 3.67 \als + 20.0 \alsb^2 + 110.1 \alsb^3
+(-256.3 + k_{g3})\alsb^4 \;,\nn\\
M_{g}^{\MSsch}(1) &=&1+ 2.67 \als + 6.32  \alsb^2  -38.98 \alsb^3
+(-1779 + k_{g3})\alsb^4 \;,\nn\\
M_{g}^{\MSsch}(2) &=&1+ 2.33 \als + 2.70  \alsb^2  -64.26  \alsb^3
+(-1865 + k_{g3})\alsb^4 \;,\nn\\
M_{g}^{\MSsch}(3) &=&1+ 2.17 \als + 1.06  \alsb^2  -73.18  \alsb^3
+(-1863 + k_{g3})\alsb^4 \;,\nn\\
M_{g}^{\MSsch}(4) &=&1+ 2.07 \als + 0.14   \alsb^2  -77.46  \alsb^3
+(-1851 + k_{g3})\alsb^4 \;,\nn\\
&&\vdots \nn\\
M_{g}^{\MSsch}(100) &=&1+ 1.69 \als + 2.99 \alsb^2  -87.15  \alsb^3
+(-1755 + k_{g3})\alsb^4 
\end{eqnarray}
gegeben sind.
Der Vorteil des effektiven Schemas im Vergleich mit dem $\MSsch$-Schema
zeigt sich in dem Koeffizienten vor  $\al_s^3$.
F"ur Momente gr"o"ser als eins ($n>1$)
zeigen die Reihen im $\MSsch$-Schema bereits asymptotisches Wachstum
in der dritten Ordnung, wogegen im effektiven Renormierungsschema
die Korrektur dritter Ordnung noch kleiner als die vorherige ist.
Aufgrund der Konstruktion ist es klar, dass die Konvergenzeigenschaften
der Momente im effektiven Schema sich mit h"oheren Potenzen $s^n$ in der 
Wichtungsfunktion verbessern, wogegen im $\MSsch$-Schema hohe Momente 
ein schlechteres Konvergenzverhalten als niedrige Momente zeigen.
Der ungew"ohnlich kleine Koeffizient der dritten Ordnung des ersten Momentes
im $\MSsch$-Schema ist das Resultat von sich zuf"allig aufhebenden
Beitr"agen von logarithmischen und konstanten Termen aufgrund dieser 
speziellen Wahl des Renormierungsschemas.
Nun l"asst sich die Diskussion der Momente im masselosen Fall aus
Kapitel \ref{Art1} anwenden.
F"ur die nummerischen Absch"atzungen wird wieder $a=0.111$ verwendet.  
Im effektiven Schema ergibt sich so f"ur die St"orungsreihen der Momente
\begin{eqnarray}
M_{g}(0) &=&1+ 0.222 + 0.204 + 0.187 + (0.21-0.0003 k_3) 
\;,\nonumber \\
M_{g}(1) &=&1+ 0.111 + 0.076 + 0.039 + (0.022-0.00015 k_3)
\;,\nonumber \\
M_{g}(2) &=&1+ 0.074 + 0.045 + 0.017 + (0.0054-0.00010 k_3)
\;,\nonumber \\
M_{g}(3) &=&1+ 0.056 + 0.031 + 0.010 + (0.0018-0.000076 k_3)
\;,\nonumber \\
M_{g}(4) &=&1+ 0.044 + 0.024 + 0.006 + (0.00054-0.000061 k_3) \, .
\end{eqnarray}
Mit der Wahl $k_3 = 100$ 
\cite{one}
ist die Korrektur vierter Ordnung f"ur alle Momente mit $n<5$
kleiner als der dritte Term.
Falls das nullte Moment ausgeschlossen wird, kann bei $k_3 \sim 100$
eine formale Genauigkeit 
von ungef"ahr 0.7\% \footnote{
W"ahrend die relative Angabe der erzielbaren Genauigkeiten in Kapitel 
\ref{Art1}
sich auf den $a^1$-Term bezog, 
werden hier die m"oglichen Genauigkeiten relativ
zu dem Parton-Modell-Resultat der Ordnung $a^0$ angegeben. 
Dies ist sinnvoll, da diese relativen
Genauigkeiten denen f"ur die sp"ater zu 
bestimmenden Parameter $\al_s$ und $m_s$ 
entsprechen.} 
erzielt werden, wobei der Fehler jeweils durch den 
Beitrag des kleinsten Terms abgesch"atzt wurde.
Mit der Standard-Pad\'e-Absch"atzung f"ur $k_3 = 25 $ erh"alt man f"ur die 
Momente 
\begin{eqnarray}
M_{g}(0) &=&1+ 0.222 + 0.204 + 0.187 + 0.20  \;,\nonumber \\
M_{g}(1) &=&1+ 0.111 + 0.076 + 0.039 + 0.018 \;,\nonumber \\
M_{g}(2) &=&1+ 0.074 + 0.045 + 0.017 + 0.003 \;, \nonumber \\
M_{g}(3) &=&1+ 0.056 + 0.031 + 0.010 - 0.0001\;, \nonumber \\
M_{g}(4) &=&1+ 0.044 + 0.024 + 0.006 - 0.001 \, .
\end{eqnarray}
Falls das Moment $M_{g}(0)$ ausgeschlossen wird, ist ein Genauigkeit von 
2\% m"oglich.
Es gibt keinen Wert f"ur $k_3$, der die Korrekturen vierter Ordung 
aller Momente kleiner macht als die der vorherigen Korrektur.
Falls $k_3$ einen Wert zwischen $-46.5$ und $1.2$ hat, zeigt nur das 
nullte Moment asymptotisches Wachstum\footnote{Die genannten
Grenzen f"ur $k_3$ werden durch hohe Momente bestimmt und lassen sich
auch an den Koeffizienten der $l^1a^3$ und $l^1 a^4$ Terme der Spektraldichte 
$\rho_g$ aus Gl.~(\ref{densg}) ablesen, da f"ur hohe Momente die Logarithmen 
mit den niedrigsten Potenzen dominant sind (vgl. Gl.~(\ref{logs})).}. 
Es ist anzunehmen, dass der zuk"unftig berechnete Wert f"ur $k_3$ 
nicht in diesem Intervall liegen wird. 
Aus diesem Grund muss man schlu"sfolgern, dass asymptotisches
Wachstum in der vierten Ordnung im $g$-Teil wahrscheinlich unvermeidbar ist.
Die ultimative Genauigkeit h"angt von dem Wert von $k_3$ ab.
Werden nur Terme bis zur Ordnung $a^3$ verwendet, so ist die erzielbare 
Genauigkeit besser als 4\%, falls das nullte Moment ausgeschlossen wird.
Die vom Renormierungsschema unabh"angige Schlussfolgerung ist, dass    
asymptotisches Wachstum f"ur das System von Momenten 
$M_g(n)$, eingeschlosen $n=0$, in der vierten Ordnung unvermeidbar ist und
dass das System in dieser Ordnung nicht perturbativ behandelt werden kann,
falls der Entwicklungsparameter den Wert $a=0.111$ besitzt.
Diese Aussage ist unabh"angig von dem nummerischen Wert f"ur $k_3$.

Die St"orungsreihe f"ur ein System von Momenten mit der Wichtungsfunktion 
$(1-s/\mts)^n$
\be
\label{alsu}
\tilde M_g(n,0)= (n+1)\int_0^{\mts} \rho_g(s) 
         \left(1-\frac{s}{\mts} \right)^n \frac{ds}{\mts} =
(n+1)!\sum_{k=0}^n\frac{(-1)^k}{(k+1)!(n-k)!} M_g(k) \, 
\ee
zeigt ein deutlich schlechteres Verhalten.
Es ergibt sich
\begin{eqnarray}
\label{altmomnumm}
\tilde M_g(1,0)  &=& 1+0.333 + 0.332 + 0.336 
+ ( 0.397 - 0.00046 k_3) \;,\nonumber \\ 
\tilde M_g(2,0)  &=& 1+0.407 + 0.429 + 0.461 
+ ( 0.569 - 0.00056 k_3) \;,\nonumber \\
\tilde M_g(3,0)  &=& 1+0.463 + 0.509 + 0.572 
+ ( 0.728 - 0.00063 k_3) \, .
\end{eqnarray}
Das Konvergenzverhalten der Reihe ist offensichtlich relativ schlecht.
Alle Momente $\tilde M_g(n,0)$ enthalten Beitr"age von 
$\tilde M_{g}(0,0)\equiv M_{g}(0)$ in der Summe Gl.~(\ref{alsu}),
welche selbst ein schlechtes Konvergenzverhalten zeigt. Die anderen 
Summanden verschlechtern das Konvergenzverhalten weiter. 

Zusammenfassend l"asst sich sagen, dass die Struktur der St"orungsreihen 
der Momente des $g$-Teils im effektiven Renormierungsschema
mit dem neuen effektiven Massenparameter $m_g^2$ 
sehr "ahnlich der des masselosen Anteils ist.
Grunds"atzlich ist f"ur die Massenkorrektur  
aus den bereites genannten Gr"unden 
das Konvergenzverhalten etwas schlechter, 
es treten aber gegen"uber dem masselosen Fall
keine qualitativ neuen Eigenschaften auf.

\section{Der $q$-Teil der Korrelationsfunktion}
Die $q$-Amplitude $\Pi_q(q^2)$ enth"alt Beitr"age von Spin-1- und 
Spin-0-Zerf"allen.
Die Korrekturen des $q$-Teils unterscheiden sich von denen des $g$-Teils bez"uglich
ihrer analytischen Eigenschaften in der komplexen $q^2$-Ebene,
sie enthalten eine
$1/q^2$-Singularit"at im Ursprung,
welche eine besondere Behandlung erfordert.
Die Massenentwicklung des $q$-Teils der 
Korrelationsfunktion ist (Gl.~(\ref{expmq}))
\be \label{qqqexp}
\Pi_q(Q^2) = \Pi(Q^2) - 3 \frac{m_s^2}{Q^2} \Pi_{mq}(Q^2) \;,
\ee
wobei die explizite Impulsabh"angigkeit der Funktion $\Pi_{mq}(Q^2)$
durch
\be
\Pi_{mq}(Q^2) =1 +\left(\frac{7}{3} +2 L \right)\als 
  + \left( \frac{13981}{432} +\frac{323}{54} \zeta(3) - \frac{520}{27} \zeta(5) +
                \frac{35}{2}L + \frac{17}{4} L^2 \right) \alsb^2 + ...
\ee
gegeben ist $(L = \ln(\mts/Q^2) )$.
Die Singularit"at $1/Q^2$ der Funktion $\Pi_{mq}(Q^2)/Q^2$
macht die Formulierung der Momente der Spektraldichte des $q$-Teils direkt 
auf dem physikalischen Schnitt etwas schwieriger.
Aufgrund des Faktors $1/Q^2$ besitzt die Amplitude f"ur die $m_s^2$-Korrektur keine 
Standard-Dispersionsrelation.
Es l"asst sich aber eine Dispersionsrelation als  
\be
\frac{\Pi_{mq}(Q^2)}{Q^2} = \int\frac{d\sigma(s)}{s+Q^2}
\ee
mit einem nicht differenzierbaren Ma"s $d\sigma(s)$ schreiben,
was bedeutet, dass  es keine stetige Funktion $\sigma'(s)$ gibt, 
die $d\sigma(s)\ne \sigma'(s)ds$ erf"ullt.
Dies l"asst sich in eine bekanntere Form bringen, falls man ein anderes 
Gewicht benutzt:
\be \label{pirhof}
\frac{\Pi_{mq}(Q^2)}{Q^2}=\int_0^\infty\frac{\rho_F(s)ds}{(s+Q^2)^2}=
-\frac{d}{d Q^2}F(Q^2) \,,
\ee
wobei 
\be  \label{dispA2}
F(Q^2) = \int_0^\infty \frac{\rho_F(s) ds}{s+Q^2} 
\ee
und $\rho_F(s)$ eine stetige Spektraldichte ist.
Aus diesem Grund ist $F(Q^2)$ die Stammfunktion von
\be
-\Pi_{mq}(Q^2)/Q^2=\Pi_{mq}(q^2)/q^2\, .
\ee
Sie l"asst sich als 
\be
F(Q^2) = -\int^{Q^2} \frac{dQ^{\pr 2}}{Q^{\pr 2}}\Pi_{mq}(Q^{\pr 2})
 = \int^L d\!L^\pr \Pi_{mq}(L^\pr )  
\ee
schreiben.
Dies ergibt f"ur die Stammfunktion $F$
\be 
F(Q^2) = L + (\frac{7}{3} L + L^2)a +(15.757 L + 7.11 L^2 +1.417 L^3) a^2 + ...\;.
\ee
Mit der Diskontinuit"at am Schnitt der Funktion, welche durch 
\be
\rho_F(s)= \frac{i}{2\pi } \left(F(-s+i \ep)-F(-s-i \ep )\right)
\ee
gegeben ist, kann die Dispersionsrelation aus Gl.~(\ref{dispA2}) 
invertiert werden (vgl. Gl.~(\ref{inversdisprel})). 
Man erh"alt f"ur die Spektraldichte 
\ba
\label{roF}
\rho_F(s) &=& 1 + (\frac{7}{3}+2 l ) a + (1.77 +14.22 l +4.25 l^2)a^2 + \\
      && (-207.04 + k_{q2} + 62.21 l + 54.52 l^2 + 9.21 l^3 ) a^3 
    + ...\nonumber\,,
\ea
wobei die $\MSsch$-Kopplungskonstante $\al_s$
bereits durch die effektive Kopplungskonstante $a$  
ersetzt wurde und $l = \ln(\mts/q^2)$.
Gleichung~(\ref{roF}) besitzt die Standardform einer Spek"-traldichte
und erm"oglicht so einen Vergleich mit dem masselosen und dem $g$-Teil.
Um alle Konstanten zu absorbieren, wird f"ur den $q$-Teil 
ein neuer Massenparameter
$m_q^2$ in Analogie zum $g$-Teil eingef"uhrt,
so dass  
\be
m_q^2 =  m_s^2  \rho_F(M^2_\tau)
\ee
gilt,
wobei $m_q^2 =m_q^2(M^2_\tau)$ und  
$m_s^2= m_s^2(M^2_\tau)$ ist.
Die Relation zwischen der effektiven Masse f"ur den $q$-Teil und der 
$\MSsch$-Masse ist
\ba
\label{mfms}
m_q^2 &=& m_s^2 (1 +  {\textstyle \frac{7}{3}} a + 1.77 a^2 + (-207.044 +k_{q2})a^3 \nn \\ 
&&+ (-1335.5-2.33 k_3 - 4.92 k_{q2} + k_{q3}) a^4+...)\;.  
\ea
Der Term der Ordnung $a^4$ enth"alt nicht nur den unbekannten Koeffizienten
$k_{q2}$, sondern auch die Koeffizienten $k_3$ und $k_{q3}$, welche diese
Korrektur vollkommen unbekannt machen.
Die Definition dieser neuen Masse durch die $\MSsch$-Masse
(Gl.~(\ref{mfms})) besitzt eine Genauigkeit von ungef"ahr 2\%, wenn nur
Korrekturen bis zur zweiten Ordnung verwendet werden.
Wird Gl.~(\ref{mfms}) 
mit der $\MSsch$-Kopplungskonstante ausgedr"uckt,
so ergibt sich   
\be
\label{mfmsMS}
m_q^2=m_s^2\left(1+\frac{7}{3} \alsb + 5.60 \alsb^2 + 
        ( - 225.22 + k_{q2}) \alsb^3 + ...\right)\;.
\ee
Mit dieser neuen Masse $m_q^2$ kann eine neue Spektraldichte 
$\rho_q(s)$ in Analogie zu Gleichung~(\ref{defofmg}) definiert 
werden:
\be
m_s^2 \rho_F(s) = m_q^2 \rho_q(s) \;.
\ee
Das f"uhrt zu der Entwicklung
\ba 
\label{densF}
\rho_q(s) &=& 1 + 2 l a  + (9.55 l +4.25 l^2 )a^2 + (36.36 l +44.6 l^2
+9.21 l^3)a^3 + \nonumber \\
&& ((-1141  -2 k_3 +6.75 k_{q2})l \nn \\
&& \quad \! +253.6 l^2 + 154.96 l^3 +20.14
l^4)a^4 + \mathcal{O}(a^5)\, .
\ea
Die Reihe (\ref{densF}) enth"alt aufgrund der Definition der 
effektiven Masse nur Logarithmen und keine Konstanten.
Die Koeffizienten der Spektraldichte $\rho_q(s)$ aus Gl.~(\ref{densF})  
sind bis zur zweiten Ordnung denen des $g$-Teils der 
Massenkorrektur $\rho_g(s)$ (Gl.~(\ref{densg}))
"ahnlich.
Die Koeffizienten der Logarithmen mit den h"ochsten Potenzen 
sind in $q$- und $g$-Teil 
gleich,
wogegen sich die der Logarithmen mit niedrigeren Potenzen 
aufgrund der unterschiedlichen Koeffizienten der entsprechenden Adler-Funktion 
Gl.~(\ref{fullinf}) unterscheiden.
Der Koeffizient des Terms $l^1 a^3$ von $\rho_q(s)$ (Gl.~(\ref{densF}))
ist siebenmal gr"o"ser als der entsprechende Koeffizient aus Gl.~(\ref{densg})
von $\rho_g(s)$. Dieser Koeffizient dominiert die Korrektur dritter Ordnung
f"ur gro"se $n$, da Beitr"age von Logarithmen mit h"oheren Potenzen 
unterdr"uckt werden.   
Die Koeffizienten der Spektraldiche $\rho_q(s)$ sind generell gr"o"ser 
als die Koeffizienten der Spek"-traldichte $\rho(s)$ im masselosen Fall
aus Gl.~(\ref{densold}).

Auf diese Weise 
erlaubt das Verhalten  der Spektraldichten
($\rho,\rho_g,\rho_q$),
sofort Schl"usse auf das Konvergenzverhalten der 
entsprechenden Momente f"ur alle drei unabh"angigen Beitr"age zu ziehen.

Die Standardmomente von $\rho_q(s)$, 
wie sie in den Gleichungen~(\ref{intmom}, \ref{intmomg})
f"ur den masselosen und $g$-Teil definiert sind, 
stimmen nicht mit den physikalischen Momenten der 
$q$-Amplitude "uberein, weil der Faktor vor $\Pi_{mq}$ in der 
Entwicklung des $q$-Teils der Korrelationsfunktion (vgl. Gl.~\ref{qqqexp})
den Pol $m_s^2 / Q^2$ enth"alt.
Aufgrund dieses Poles m"ussen die physikalischen $q$-Momente $M_q^{ph}(n)$ 
durch ein Konturintegral folgenderma"sen 
definiert werden:
\be 
\label{defofqmom}
\frac{i m_s^2}{2 \pi}\oint \frac{\Pi_{mq}(q^2)}{q^2}
\left(\frac{q^2}{\mts}\right)^n dq^2 = m_q^2 M_q^{ph}(n)\;.
\ee 
Die Momente Gl.~(\ref{defofqmom})
k"onnen durch die Funktion $\rho_q(s)$
direkt ausgewertet werden.
F"ur das nullte Moment stellt sich heraus, dass
\be
\label{qmom0}
M_q^{ph}(0) = -\rho_q(M_{\tau}^2) =  -1 \, .
\ee
Dies ergibt sich folgenderma"sen:
\ba
m_q^2
M_q^{ph}(0) &=& \frac{i m_s^2}{2 \pi} \oint \frac{\Pi_{mq}(q^2)}{q^2} dq^2 \\
       &=& -\frac{i m_s^2}{2 \pi} \oint \frac{d}{dQ^2} (F(Q^2)) dQ^2 \nonumber \\ 
       &=& -\frac{i m_s^2}{2 \pi } (F(-M_\tau^2+i \epsilon) - F(-M_\tau^2 - i \epsilon)) \nonumber \\
       &=& - m_s^2  \rho_q(M^2_\tau) =: - m_q^2 \;.
\ea
Der Grund hierf"ur ist, dass die Integration mit $n=0$
in Gl.~(\ref{defofqmom}) 
genau die Beitr"age enth"alt, die durch die 
Neudefinition der Strange-Quarkmasse absorbiert werden.
H"ohere physikalische Momente h"angen mit den Standardmomenten 
von $\rho_q(s)$ "uber 
\be
\label{qmom1}
M_q^{ph}(n)|_{n >0} = n\int_0^{\mts} \rho_q(s) 
     \left( \frac{s}{\mts} \right)^{n-1} \frac{ds}{\mts} - 1 \equiv 
M_q(n-1)-1 
\ee
zusammen.
Die explizite Rechnung hierzu ist:
\ba
m_q^2 M_q^{ph}(n, n>0) &=&  \frac{i m_s^2}{2 \pi}  
                \oint \frac{\Pi_{mq}(Q^2)}{Q^2}\left(\frac{-Q^2}{\mts}\right)^n dQ^2  \\
        &=&  -\frac{i m_s^2}{2 \pi}  
                 \oint \left( \frac{d}{dQ^2}F(Q^2) \right) \left(\frac{-Q^2}{\mts}\right)^n
                                    d Q^2 \nonumber\\
        &=&  -\frac{i m_s^2}{2 \pi} 
                 \oint \left[ \frac{d}{d Q^2} \left\{F(Q^2) 
           \left(\frac{-Q^2}{\mts}\right)^n\right\} +
     n \frac{F(Q^2)}{\mts}  \left(\frac{-Q^2}{\mts}\right)^{n-1} \right] dQ^2 \nonumber\\
        &=& - m_q^2 -  \frac{i m_s^2}{2 \pi} n \
                  \oint  F(Q^2) \left( \frac{-Q^2}{\mts} \right)^{n-1} 
       \frac{dQ^2}{\mts}  \nonumber \\
        &=& - m_q^2  + n 
            \int_0^{\mts}  m_s^2 \rho_F(s) 
             \left( \frac{s}{\mts} \right)^{n-1} \frac{ds}{\mts}\nonumber \\
        &=& m_q^2 
          \left\{  n \int_0^{\mts} 
   \rho_q (s) \left( \frac{s}{\mts} \right)^{n-1}\frac{ds}{\mts}  
                 -1 \right\} \nn \\
  &=& m_q^2(M_q(n-1)-1) \;.
\ea 
Im drittletzten Schritt wurde verwendet, dass
\ba
 \oint  F(Q^2) \left(\frac{-Q^2}{\mts} \right)^{n-1}  dQ^2
              &=& \oint  \left( \frac{-Q^2}{\mts} \right)^{n-1} dQ^2
                      \int_0^\infty  \frac{\rho_F (s) ds}{s+ Q^2} \\
  &=&  \int_0^\infty \rho_F (s) ds  
     \oint \left(\frac{-Q^2}{\mts} \right)^{n-1}
         \frac{dQ^2}{s+Q^2} \nonumber\\
  &=&   2\pi i \int_0^{\mts}  \rho_F (s) 
      \left(\frac{s}{\mts} \right)^{n-1} ds \nonumber
\ea
ist,
wobei sich mit dem Residuensatz 
\ba 
  \oint \left(\frac{-Q^2}{\mts} \right)^n \frac{dQ^2}{s+Q^2} 
    &\begin{array}{c} 
          \\
           =\\
       { \scriptstyle  Q^{\pr^2} = s + Q^2 }
     \end{array} & 
          \Theta(\mts-s) \oint \left(\frac{s-Q^{\prime^2}}{\mts} \right)^n 
    \frac{ dQ^{\prime2}}{Q^{\prime^2}} \\
   &=&   2\pi i \left(\frac{s}{\mts} \right)^n \nonumber \Theta(\mts-s) 
\ea 
ergibt.
H"ohere Momente enthalten keine Parton-Modell-Beitr"age.
Die Momente $M_q(n)$ sind die in Gl.~(\ref{intmom})
definierten Standardobjekte.
Die gesamte Analyse aus Kaptel~\ref{Art1} ist auf die Momente
$M_q(n)$ mit $\rho_q(s)$
anstelle von $\rho_g(s)$ bzw. $\rho(s)$ anwendbar.
Die nummerischen Resultate f"ur die Momente sind
\ba
\label{qnmom}
M_q(0) &=& 1 + 2 a + 18.1 a^2 + 180.8 a^3 
+ (779.4 - 2 k_3 + 6.75 k_{q2}) a^4 \;,\nonumber\\ 
M_q(1) &=& 1 + 1a + 6.90 a^2 + 47.39  a^3 
+ (-297.3 -k_3 + 3.375 k_{q2}) a^4 \;,\nonumber \\
M_q(2) &=& 1 + \frac{2}{3}a + 4.13 a^2 + 24.08 a^3 
+ (-283.6 - 0.67 k_3 + 2.25 k_{q2})a^4 \;,\nonumber\\
M_q(3) &=& 1 + \frac{1}{2} a + 2.92 a^2 + 15.53 a^3 
+ ( -237.13 - 0.5 k_3 + 1.69 k_{q2}) a^4 \;,\nonumber \\
M_q(4)&=& 1 + \frac{2}{5}a + 2.25 a^2 + 11.28 a^3 
+ ( -199.699 - 0.4 k_3 + 1.35 k_{q2})a^4 \;,\nonumber \\
&& \vdots  \nonumber\\
M_q(100)&=&1+\frac{1}{50} a+0.095a^2+0.37a^3+(-11.25 - 0.020 k_3 
+ 0.067 k_{q2})a^4 \, .
\ea 
Die physikalischen $q$-Momente $M_q^{ph}(n)$
h"angen mit den Momenten $M_q(n)$ durch Gl.~(\ref{qmom0})
und (\ref{qmom1}) zusammen.
Sie "andern das Konvergenzverhalten der St"orungsreihe nicht.
Das sich aus $M_q(0)$ ergebende Moment $M_q^{ph}(1)$ 
zeigt bereits in der dritten Ordnung asymptotisches Wachstum.
Wie erwartet, zeigen h"ohere Momente ein besseres Konvergenzverhalten, 
da bei ihnen der niedrigenergetische Bereich unterdr"uckt ist.
\section{Massenkorrektur zu $\tau$-Lepton-Observablen} 
Nachdem die Technik f"ur die Analyse von Momenten der Ordnung $m_s^2$
eingef"uhrt worden ist, l"asst sich diese jetzt anwenden, um 
physikalische Observable zu untersuchen.
\subsection{Totale $\tau$-Lepton-Zerfallsrate}
Die $\tau$-Lepton-Zerfallsrate ist durch eine 
spezielle Linearkombination von Momenten gegeben.
Die Wichtungsfunktion enth"alt den Faktor $(1-s/\mts)^2$,
welcher die Konvergenz der $\tau$-Lepton-Zerfallsrate  
verschlechtert.
Die Form der Wichtungsfunktion mit dem Faktor $(1-s/\mts)^2$
ist wie vorher der Hauptgrund f"ur die schlechte Konvergenz der $m_s^2$-Korrektur 
der Zerfallsrate.
\be \label{decayrate}
R_{m\tau} = \frac{i}{2 \pi} \oint 2 \left( 1- \frac{q^2}{\mts} \right)^2 
3\left(\frac{m_s^2 \Pi_{mq}(q^2)}{q^2} - \frac{m_s^2}{\mts}
\Pi_{mg}(q^2)  \right)\frac{dq^2}{\mts} \, .
\ee
Die zwei verschiedenen Massen werden in der Zerfallsrate explizit
aufgef"uhrt, um die Struktur der beiden St"orungsreihen einzeln zu
untersuchen.  
Das Resultat f"ur die Massenkorrektur der totalen Zerfallsrate ist
\ba \label{decrthroumom}
R_{m\tau}&=&6\frac{m_q^2}{\mts}(M_q^{ph}(0)-2M_q^{ph}(1)+M_q^{ph}(2)) 
-6\frac{m_g^2}{\mts}(M_g(0)-M_g(1)+\frac{1}{3}M_g(2)) \nonumber \\
&=&-6 \frac{m_q^2}{\mts}(2 M_q(0)-M_q(1)) 
-6\frac{m_g^2}{\mts}( M_g(0)-M_g(1)+\frac{1}{3}M_g(2))   \, .
\ea
Dies ist durch die effektiven Massenparameter und die oben eingef"uhrten
Momente ausgedr"uckt.
In der Parton-Modell-Approximation sind alle Momente auf Eins normiert.
Nummerisch ergibt sich 
\ba 
\label{numdecayrate}
R_{m\tau} &=&-6 \frac{m_q^2}{\mts} \left( 1 + 3  a +  29.21 a^2 + 314.3 a^3 
+ (1856.1 - 3.0k_3 + 10.13 k_{q2}) a^4 \right) \nonumber \\
&&-2\frac{m_g^2}{\mts}
\left( 1 + 3.67 a + 34.84 a^2 + 337.3 a^3 + (3745.2-3.67 k_3) a^4 \right) \;.
\ea
Der Grund f"ur die schlechte Konvergenz von (\ref{numdecayrate})
sind die Beitr"age der niedrigen Momente 
$M_q(0)$, $M_q(1)$ und $M_g(0)$, $M_g(1)$
zu der Massenkorrektur der totalen Zerfallsrate.
Beide Reihen in Gl.~(\ref{numdecayrate}) konvergieren
nur schlecht. Der gesamte Ausdruck wird durch den Beitrag des 
$q$-Teils dominiert, der einen dreimal gr"o"seren Koeffizienten in
der f"uhrenden Ordnung besitzt.

Wenn man $m_q^2$, $m_g^2$ und  $a$ in Gl.~(\ref{numdecayrate})
durch die entsprechenden Gr"o"sen im
$\MSsch$-Schema $m_s^2$ und $\alpha_s$ ersetzt, so erh"alt man 
das Standardresultat
\be 
R_{m\tau} = - 8 \frac{m_s^2}{\mts} \left( 1 + 5.33  \als + 46.0  \alsb^2 +
(283.55 + 0.75 k_{q2}) \alsb^3 + ...\right) \, .
\ee

\subsection{Die '1+0'-Methode}
In \cite{exp1ms} wird der nummerische Wert der Strange-Quarkmasse
im $\MSsch$-Schema mit der '1+0'-Methode abgesch"atzt, welche 
die Darstellung der totalen Zerfallsrate als eine Summe des
$(L+T)$- und des $L$-Beitrages verwendet.
Der Ausdruck f"ur die totale Zerfallsrate Gl.~(\ref{decayrate}) l"asst sich 
folgenderma"sen umschreiben:
\be \label{LT}
R_\tau = \frac{i}{2\pi}\oint 2 \left(1-\frac{q^2}{\mts} \right)^2 
\left\{ \left( 1 + 2\frac{q^2}{\mts} \right) \Pi_{(L+T)}(q^2) -2
\frac{q^2}{\mts} \Pi_L(q^2) \right\} \frac{dq^2}{\mts}\, ,
\ee
wobei
\be
\Pi_{(L+T)}(q^2) = \Pi_q(q^2) 
\ee
ist.
Es wird angenommen, dass im $\MSsch$-Schema die Reihe f"ur den 
$(L+T)$-Teil der $m_s^2$-Korrektur in Gl.~(\ref{LT})
gut konvergiert \cite{exp1ms,pichprades}.
Die Konvergenz ist in der FOPT nicht sehr eindrucksvoll, aber 
die f"ur die Kontur-verbesserte St"orungsheorie (CIPT)   
gegebenen Resultate in Ref.~\cite{exp1ms,pichprades} 
zeigen schnelle Konvergenz. 
Die interessante Gr"o"se ist jetzt 
\ba 
\label{Rtlt}
R_{m \tau}^{L+T}&=&\frac{i}{2\pi} 
\oint 2\left( 1- \frac{q^2}{\mts}\right)^2 
\left(1 + 2\frac{q^2}{\mts} \right)3\frac{\Pi_{mq}(q^2)}{q^2}
\frac{dq^2}{\mts}\nonumber \\
&=&6\frac{m_q^2}{\mts}(M_q^{ph}(0)-3M_q^{ph}(2)+2M_q^{ph}(3))\nonumber \\
&=&-6\frac{m_q^2}{\mts}(3M_q(1)-2M_q(2)).
\ea
Aufgrund der in $R_{m\tau}^{L+T}$ (\ref{Rtlt}) enthaltenen Momente
ist zu vermuten, dass das Konvergenzverhalten von 
$R_{m\tau}^{L+T}$ besser ist als das der totalen Zerfallsrate,
da in $R_{m\tau}^{L+T}$ das nullte Moment nicht eingeht.
Dies ist ein invarianter (vom Renormierungsschema unabh"angiger)
Grund f"ur bessere Konvergenz: Die spezielle Kombination 
$R_{m\tau}^{L+T}$ erh"alt kleinere Infrarot-Beitr"age aus dem 
Integrationsbereich und ist in der St"orungstheorie besser berechenbar.
Trotzden ist die Konvergenz nicht besonders gut.
Die nummerischen Resultate f"ur den $(L+T)$-Teil sind:
\ba
R_{m \tau}^{L+T}|_{q-scheme} &=&
- 6 \frac{m^2_q}{\mts} \Bigg( 1 + 1.67 a + 12.448 a^2 + 94.01 a^3 \nn \\ 
&& + (-324.629 - 1.67 k_3 + 5.625 k_{q2})a^4 + \ldots\Bigg)\, .
\ea
F"ur Standardwerte der unbekannten Parameter $25<k_3<100$
und $0<k_{q2}<160$ fallen die nummerischen Werte der QCD-Korrekturen 
ab, so dass kein asymptotisches Wachstum auftritt.
Der gesamte Beitrag der ersten drei QCD-Korrekturen betr"agt 
0.45, was eine vern"unftige "Anderung des Resultates im Parton-Modell ist.
Im $\MSsch$-Schema erh"alt man den Ausdruck 
\ba
R_{m \tau}^{L+T}|_{\MSsch-scheme} &=&
-6 \frac{m_s^2}{\mts} \Bigg( 1+ 4.0 \alsb + 24.67 \alsb^2 + (-62.77 +
k_{q2})\alsb^3  + \nonumber  \\ 
 &&  \quad \quad + (-3110 + 7.29 k_{q2} + k_{q3})\alsb^4 +... \Bigg) \;. 
\ea
Die Summe der ersten beiden QCD-Korrekturen ist 0.65. Die "Anderung der 
Parton-Modell-Vorhersage durch die bekannten Korrekturen ist gr"o"ser 
als im effektiven $q$-Schema, obwohl in diesem eine QCD-Korrektur mehr 
bekannt ist. 
Der einzige Vorteil der $(L+T)$-Amplitude ist die Abwesenheit des 
nullten Moments $M_q(0)$, das das divergenteste aus dem Satz der 
$q$-Momente ist.
Allerdings ben"otigt man das Moment $M_q(1)$ aus Gl.~(\ref{qnmom})
zur Berechnung von Gl.~(\ref{Rtlt}), was ein gutes Konvergenzverhalten 
verhindert. 
F"ur den longitudinalen Teil erh"alt man die folgende Linearkombination 
aus $q$- und $g$-Momenten:
\ba 
\label{lll}
R_{m \tau}^{L}&=&\frac{i}{2\pi}\oint 2\left(1-\frac{q^2}{\mts}\right)^2 
(-2)\frac{q^2}{\mts} \Pi_L(q^2) \frac{dq^2}{\mts}  \\
&=&-12\frac{m_q^2}{\mts} ( M_q(0) - 2 M_q(1) + M_q(2))
-6\frac{m_g^2}{\mts} (M_g(0) - M_g(1) + \frac{1}{3}M_g(2))\;. \nn
\ea
Deren Konvergenzverhalten ist relativ schlecht,
da auch die nullten Momente in den Ausdruck eingehen.
Der $q$-Teil enth"alt hier keinen Beitrag des Parton-Modells.
Die nummerischen Resultate sind
\ba
\label{ll}
R_{m \tau}^{L}=&-&2\frac{m_g^2}{\mts}\left(1+3.67 a + 34.8 a^2 + 337.3 a^3 
+ (3745.2 - 3.67 k_3)a^4 \right) \nonumber \\  
&-&8\frac{m_q^2}{\mts} \left( 0+ a + 12.57 a^2 + 165.2 a^3 
+ (1635.6 - k_3 + 3.375 k_{q2})a^4 \right).
\ea
Aus Gl.~(\ref{ll}) kann man den Grund f"ur die schlechte St"orungsreihe
erkennen.
W"ahrend die Konvergenz des $g$-Teils normal ist, verh"alt sich 
der $q$-Teil ohne Parton-Modell-Beitrag sehr schlecht und wird gegen"uber 
dem $g$-Teil um den Faktor vier verst"arkt, was die Summe in 
Gl.~(\ref{ll}) vollkommen uninterpretabel macht.

Bisher werden die Massenparameter $m_q$ und $m_g$ 
f"ur den $g$- und den $q$-Teil
der Korrelationsfunktion als interne Massenskalen f"ur die 
$m_s^2$-Korrektur verwendet.
Es ist aber auch m"oglich einen anderen Satz von Parametern $m_{T,L}$
zu verwenden, die zu der Zerlegung der Korrelationsfunktion in den 
transversalen 
und den longitudinalen Anteil geh"oren.
In diesem Fall sind der transversale und der $g$-Teil identisch $(m_T=m_g)$,
die effektive Masse $m_L^2$ ergibt sich aus dem 
longitudinalen Teil $\Pi_L(q^2)$.
Der Ausdruck f"ur $m_T^2$ durch die $\MSsch$-Schema-Masse ist vern"unftig  
(Gl.~(\ref{mg})).
Hingegen ist die entsprechende Relation f"ur $m_L$ viel schlechter:
\ba
m_L^2 &=& m_s^2 \left( 1+ 5.67 \als + 31.9 \alsb^2 + 89.2 \alsb^3 + \right. \nonumber \\
      &&   \left. (- 5180 + k_{g3} + 17.5 k_{q2}) \alsb^4 + ... \right) \;.
\ea
Die unterschiedlichen Eigenschaften der St"orungsreihen f"ur die Parameter 
$m_T$ und $m_L$ k"onnen als Resultat des Unterschiedes der vollen QCD-Wechselwirkung 
im Spin-1- und Spin-0-Kanal verstanden werden. 
Der st"arkere und fr"uhere Zusammenbruch
der St"orungstheorie im Spin-0-Kanal kann mit den Beitr"agen von nichtperturbativen Effekten 
(Instantonen) zusammenh"angen, die im Spin-1-Kanal nicht vorhanden sind.

Insgesamt ist die Konvergenz der St"orungsreihe f"ur die $m_s^2$-Korrekturen
der nat"urlichsten und am pr"azisesten messbaren physikalischen Observablen 
immer schlecht.
Dies liegt daran, dass der Infrarot-Bereich der Integration nummerisch wichtig
f"ur den gegebenen experimentellen Wert der Kopplungskonstante ist. Eine
gute Konvergenz ist immer das Resultat einer speziellen Linearkombination 
der Momente oder einer speziellen Wahl des Renormierungsschemas.
Der erste Fall l"asst sich allerdings f"ur eine interessante, da gut 
messbare, Observable nicht realisieren.
Dies zeigt die Notwendigkeit der Resummation der Reihe 
f"ur eine solide Interpretation
der theoretischen Formeln.  
\section{Zusammenfassung von Kapitel \ref{Art2}}
In diesem Kapitel wurde die Asymptotik der St"orungsreihe
f"ur die $m_s^2$-Korrektur analysiert.
Unter Benutzung der Standardabsch"atzung der Genauigkeit einer 
asymptotischen Reihe ergibt sich, dass die theoretische Genauigkeit 
in der st"orungstheoretischen Beschreibung der Cabbibo-unterdr"uckten 
$\tau$-Lepton-Zerf"alle durch das asymptotische Wachstum der Koeffizienten in
der vierten Ordnung der St"orungsreihe limitiert ist.
Diese Schlussfolgerung ist unabh"angig von der Wahl des Renormierungsschemas.
Die Genauigkeit der St"orungsreihe der $m_s^2$-Korrektur in 
Cabbibo-unterdr"uckten Zerfallskan"alen ist bestenfalls 15\%-20\%.
Aus diesem Grund ist die Bestimmung des nummerischen Wertes der 
Strange-Quarkmasse
aus den $m_s^2$-Korrekturen der 
$\tau$-Zerfallsrate in Hadronen mit Strangeness 
durch die Pr"azision dieser St"orungsreihen limitiert.  
Bessere theoretische Genauigkeiten k"onnten 
durch die Verwendung von Observablen,
die nur h"ohere Momente enthalten, erreicht werden. 
Allerdings sind f"ur diese Observablen die experimentellen Daten im Augenblick 
noch zu schlecht.
Vom ph"anomenologischen Standpunkt aus unterscheidet sich die Analyse der 
$m_s^2$-Korrektur von dem masselosen Fall.
W"ahrend im letzterem die niedrigen Momente durch die Verwendung von 
experimentellen Daten vermieden werden k"onnen 
(z. B. aus der $e^+e^-$-Annihilation),
besitzen die $q$- bzw. 
$g$-Momente keine direkte physikalische Bedeutung und k"onnen nicht 
in dieser Weise behandelt werden.

Das Einf"uhren von zwei nat"urlichen Massen erlaubt es, den masselosen,
den $q$- und 
den $g$-Teil der Korrelationsfunktion bis zu der Ordnung 
$a^4$ mit nur zwei unbekannten 
Parametern $k_3$ und $k_{q2}$ zu beschreiben, 
anstelle von vier Parametern im $\MSsch$-Schema.
Die Existenz von zwei unterschiedlichen Massenskalen ist 
durch die Unterschiede der Wechselwirkung im Spin-1- und Spin-0-Teil
physikalisch motiviert.
F"ur die beiden unabh"angigen $m_s^2$-Korrekturen 
($g$-und $q$-Teil) ist die Konvergenz der niedrigen Momente langsam.
Die Beitr"age der Infrarotbereiche sind gro"s.
Nur diejenigen Momente konvergieren gut, bei denen die Infrarotregion 
des Integrationsbereiches unterdr"uckt ist.
Die QCD-Parameter $a(s)$ und $m_{q,g}(s)$
laufen zu schnell, um pr"azise St"orungsreihen f"ur einen Satz von
$\tau$-Lepton-Observablen zu erzeugen.
Wenn die Paramter langsamer laufen w"urden, w"aren die Koeffizienten der 
$\beta$- und der $\gamma$-Funktion kleiner, und es w"are m"oglich, die 
gro"se Genauigkeit der  experimentellen Daten f"ur niedrige Momente 
voll auszunutzen.
Observable, die nur hohe Monente enthalten, k"onnen mit der FOPT 
gut beschrieben werden, allerdings sind f"ur diese noch keine 
ausreichend genauen experimentellen Daten verf"ugbar.
Um einen pr"azisen Vergleich von Theorie und Experiment zu 
erm"oglichen, ist deswegen eine Resummationsprozedur notwendig
\cite{Pivtau,renRS,groote}.


\chapter{Bestimmung der starken Kopplungskonstante $\al_s$} 
\label{Art3} \label{ART3}
In diesem Kapitel wird eine Methode vorgeschlagen, mit der
der nummerische Wert der Kopplungskonstante $\al_s$ aus den $\tau$-Daten
und dem theoretischen Ausdruck f"ur diese in 
der St"orungstheorie in endlicher Ordnung (FOPT) bestimmt werden 
kann \cite{three}.
Die pr"azise Bestimmung der starken Kopplungskonstante 
$\al_s$ bei niedrigen Energien 
ist von besonderem Interesse.
Dieser Wert l"asst sich
mit der Renormierungsgruppe zu hohen Energien "ubertragen,
so dass ein QCD-Test m"oglich ist, der Hadronphysik in einem Energiebereich 
von 100 GeV vergleicht \cite{wilczek}.

Aufgrund der Existenz sehr genauer Formeln in der St"orungstheorie
und der Einfachheit der Renormierungsgruppe im masselosen Fall 
erwartet man eine sehr hohe Pr"azision des Wertes f"ur $\al_s$.
Da allerdings der nummerische Wert des Entwicklungsparameters $\al_s$
bei der $\tau$-Lepton-Masse $\mts$ nicht klein ist, sind die 
Beitr"age von Termen h"oherer Ordnung wichtig.
Die Analyse der $\tau$-Lepton-Observablen im Limes masseloser 
Quarks in Kapitel~\ref{Art1} \cite{one} hat ergeben, dass die 
St"orungsreihen sich bereits nahe der asymptotischen Grenze befinden,
und diese teilweise sogar schon "uberschritten worden ist.
Dies macht eine Interpretation der St"orungsreihe notwendig, welche die Terme
h"oherer Ordnungen einschlie"st.

\begin{figure}[h]
\begin{center}
       \epsfig{file=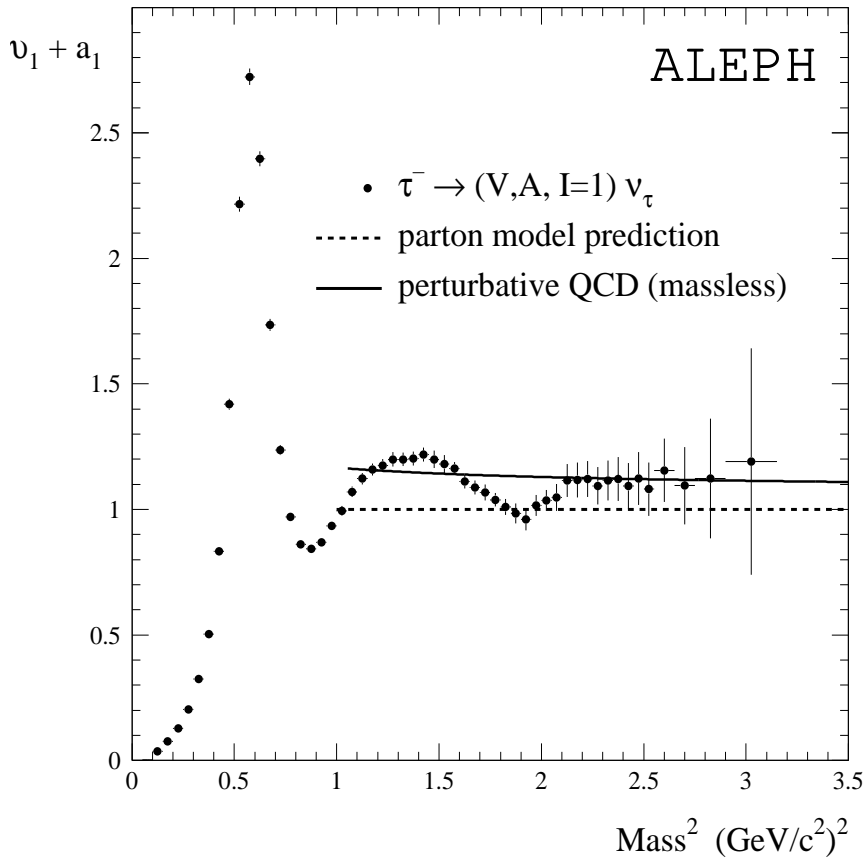,angle=0,width=.7
                  \textwidth,height=.7\textwidth}
\end{center}
       \caption{\label{s0rate}
Messung der Vektor-$(v_1)$-plus-Axialvektor-$(a_1)$-Spektraldichte aus $\tau$-Zerf"allen 
in strangelose Hadronen und Vorhersage des 
Parton-Modells f"ur masselose QCD.
(Quelle: ALEPH-Kollaboration \protect\cite{exp1al})}
    \end{figure}

Die entscheidende neue Eigenschaft der in dieser Analyse vorgeschlagenen 
Methode ist die explizite Unabh"angigkeit vom Renormierungsschema.
Wie schon in dem Kapitel "uber die Renormierungsgruppe (RG) 
(Kapitel~\ref{rg})
erl"autert wurde, ist RG-Invarianz eine fundamentale 
Eigenschaft der St"orungstheorie in der Quantenfeldtheorie,
die mit der Freiheit in der Wahl eines Subtraktionsschemas zusammenh"angt
\cite{mut}. 
Diese Eigenschaft sollte bei der Extraktion eines nummerischen Wertes
f"ur die Kopplungskonstante ber"ucksichtigt werden.
Die normierte $\tau$-Lepton-Zerfallsrate in Hadronen ohne Strangenes
$H_{S = 0}$ ist durch 
\ba 
\label{rate}
R_{\tau S=0}&=&\frac{\Gamma(\tau \rightarrow H_{S=0} \nu)}
{\Gamma(\tau \rightarrow l \bar{\nu} \nu )} \nn \\
&=& N_c |V_{ud}|^2 S_{EW}(1+\delta_P + \delta_{EW} + \delta_{NP})
\ea
gegeben,
wobei $N_c=3$ die Anzahl der Farbladungen ist.
F"ur das Element der Cabbibo-Matrix wird   
 $|V_{ud}|^2 = 0.9511 \pm 0.0014$ \cite{PDG}
verwendet.
Der Faktor $S_{EW} = 1.0194 $ ist eine multiplikative 
elektroschwache Korrektur 
\cite{ewcorr1}.
Der erste Term in Gleichung~(\ref{rate}) ist das Parton-Modell-Resultat 
und der zweite Term $\delta_P$ steht f"ur perturbative QCD-Effekte.
$\delta_{EW} = 0.001$ ist eine
additive elektroschwache Korrektur \cite{ewcorr2}.
Die nichtperturbativen Korrekturen sind relativ klein und mit Null 
konsistent; hier wird 
$\delta_{NP} = -0.003 \pm 0.003$ (siehe z.B. \cite{NarPic88})
verwendet.
Der Wert der Zerfallsrate $R_{\tau S=0}$ wurde von den 
ALEPH-\cite{exp1al}- und OPAL-\cite{exp2}-Kollaborationen 
mit sehr "ahnliche Resultaten gemessen.
Abbildung~\ref{s0rate} zeigt die gemessene Spektraldichte, aus der sich
durch Integration (Gl.~(\ref{int})) mit der Wichtungsfunktion 
die Zerfallsrate $R_{\tau S=0}$ 
ergibt.
Die Spektraldichte ist ausserdem eine der zentralen Gr\"ossen f\"ur 
Pr\"azisionstests des Standardmodells, insbesondere f\"ur die Berechnung 
des nummerischen Wertes der elektromagnetischen Kopplungskonstante bei $M_{\rm Z}$
und f\"ur das annormale magnetische Moment des M\"uons \cite{HighPres}.
In diesem Kapitel werden die ALEPH-Daten verwendet. 
Die entsprechenden
Resultate f"ur die OPAL-Daten 
werden am Ende genannt.
Mit dem experimentellen Resultat 
\be
\label{expdec}
R_{\tau S=0 }^{exp}=3.492 \pm 0.016
\ee
ergibt sich
\be
\label{expdec0}
\delta_{P}^{exp}=0.203\pm0.007 \ .
\ee
Der theoretische Ausdruck f"ur die $\tau$-Zerfallsrate wurde bereits
im Kapitel~\ref{Art1} untersucht. 
F"ur die perturbativen Korrekturen ergab sich im $\MSsch$-Schema 
(siehe Gl.~\ref{taumssch}) 
\ba
\label{taumsschaa}
\delta_P^{th} &=& \als + 5.20232 \left(\als\right)^2
+26.3659\left(\als\right)^3 \nn \\
&& +(78.0029+k_3)\left(\als\right)^4 + \mathcal{O}(\al_s^5) \;,
\ea
wobei die Kopplungskonstante $\al_S$ bei der Skala  
$M_\tau = 1.777~{\rm GeV}$
genommen ist.
Wie vorher schon erw"ahnt wurde, ist  
der Koeffizient $k_3$ ist nicht bekannt, was die Verwendung des 
letzten Terms in Gl.~(\ref{taumsschaa}) verhindert.
Er wird in dieser Analyse trotzdem mit aufgef"uhrt, um einen Eindruck
f"ur die m"ogliche Gr"o"se der $(\al_s^4)$-Korrektur
zu erhalten, wenn die nummerischen Werte
$0<k_3<50$ als konservative Absch"atzung verwendet werden.
\section{Bestimmung der Kopplungskonstante mit FOPT}
Normalerweise extrahiert man den nummerischen Wert f"ur
$\al_s(M_\tau)$, indem man die ersten drei Terme der Reihe 
Gl.~(\ref{taumsschaa}) als exakte Funktion behandelt.
Die Wurzel dieses Polynoms ist 
\be
\label{dirres}
\al_s^{st}(M_\tau)= 0.3404\pm 0.0073_{exp}.
\ee
Der genannte Fehler entsteht durch den Fehler des experimentellen 
Wertes  $\delta_P^{exp}$.
Dies ist die 
Standardmethode.
Falls eine andere Methode verwendet wird
(z.B. ein anderes Renormierungsschema),
kann sich der extrahierte nummerische Wert
deutlich unterscheiden.
In der Praxis k"onnen Korrekturen h"oherer Ordnung, 
die in der Standardmethode vernachl"assigt werden, 
aufgrund des relativ gro"sen nummerischen Wertes der Kopplungskonstante 
relevant sein \cite{renRS}. 

Im folgenden wird eine neue Methode vorgeschlagen,
um $\al_s$ aus Gl.~(\ref{taumsschaa}) zu extrahieren. 
Die RG-Gleichung 
\be \label{RGE}
\mu^2 \frac{d}{d \mu^2} a(\mu^2) = \beta(a(\mu^2))
\ee
wird durch das Integral 
\be 
\label{LQCD}
\ln \left(\frac{\mu^2}{\Lambda^2}\right) =
   \Phi(a(\mu^2)) + \int_0^{a(\mu^2)} 
\left( \frac{1}{\beta(\xi)} - \frac{1}{\beta_2(\xi)} \right)d\xi
\ee
implizit gel"ost,
wobei die Konstante des Integrals $\Phi(a)$ folgenderma"sen 
festgelegt wird:
\be 
\label{intb2}
\Phi(a) = \int^{a}  \frac{1}{\beta_2(\xi)} d \xi 
= \frac{1}{a \beta_0} 
       + \frac{\beta_1}{\beta_0^2} 
\ln \left( \frac{a \beta_0^2}{\beta_0 + a \beta_1} \right).
\ee
Hier bezeichnen $\beta_2(a)$ und $\beta(a)$
die $\beta$-Funktion in der zweiten Ordnung und die volle $\beta$-Funktion,
in der man so viele Terme verwendet, wie bekannt sind, d.h. es ist 
\ba
\beta_2(a) &=& -a^2(\beta_0 + \beta_1 a ) \nn \\
{\rm und} \qquad \qquad 
\beta(a) &=& -a^2(\beta_0 +  \beta_1 a +  \beta_2 a^2 + \beta_3
a^3) 
+ \mathcal{O}(a^6).
\ea
Der Vierschleifen-Koeffizient $\beta_3$ ist im $\MSsch$-Schema 
bekannt \cite{beta4}.
Die implizite L"osung (\ref{LQCD})
der RG-Gruppengleichung (\ref{RGE})
beschreibt die Entwicklung einer Trajektorie
der Kopplungskonstanten.
Diese Trajektorie wird durch den Skalenparameter $\Lambda$
und die Koeffizienten der $\beta$-Funktion $\beta_i$ mit
$i>2$ parametrisiert.
Gl.~(\ref{LQCD}) ist als Reihe RG-invariant unter der RG-Transformation
\be  \label{aatrans}
a  \rightarrow a(1 + \kappa_1 a + \kappa_2 a^2 + \kappa_3 a^3 + \dots )\;,
\ee
wenn gleichzeitig der Skalenparameter $\Lambda$ gem"a"s 
\be
\label{transform}
\Lambda^2 \rightarrow  \Lambda^2 e^{-\kappa_1/\beta_0}
\ee
und die $\beta$-Funktion transformiert werden.
Das Transformationsverhalten der $\beta$-Funktion unter RG-Transformationen
wurde in Gl.~(\ref{betatransform}) bestimmt. 
$\beta_{0}$ und $\beta_1$ bleiben unter der Transformation 
(\ref{aatrans}) invariant, 
$\beta_{2}$ und $\beta_3$ transformieren sich gem"a"s 
(siehe auch Gl.~(~\ref{betatransform}))
\ba \label{bbtrans}
\beta_2 &\rightarrow&\beta_2-\kappa_1^2\beta_0+
\kappa_2 \beta_0 - \kappa_1 \beta_1 \;,\nn \\
\beta_3 &\rightarrow&\beta_3+4 \kappa_1^3 \beta_0 
+ 2 \kappa_3 \beta_0 + \kappa_1^2  
\beta_1 - 2 \kappa_1 (3 \kappa_2 \beta_0 + \beta_2 )  .
\ea
Die RG-Invarianz von Gl.~(\ref{LQCD}) unter der 
in Gl.~(\ref{aatrans}, \ref{transform}, \ref{bbtrans}) 
beschriebenen Transformation bedeutet, 
dass der Ausdruck (\ref{LQCD})
als Reihe seine G"ultigkeit nach der Transformation beh"alt.
Die RG-Invarianz von Gl.~(\ref{LQCD}) wird nur durch Terme h"oherer 
Ordnung in der Kopplungskonstante gebrochen.
Die Integrationskonstante in Gl.~(\ref{LQCD}) ist so festgelegt, dass 
die asymptotische Entwicklung von $\al_s/\pi = a_{\MSsch}$ 
f"ur grosse Impulse 
\be \label{Lser}
a_{\MSsch}(Q^2) =  \frac{1}{\beta_0 L}\left(1 - \frac{\beta_1}{\beta_0^2} 
\frac{\ln(L)}{L} \right)+ \mathcal{O} \left(\frac{1}{L^3} \right)
  \qquad {\rm mit} \quad 
L=\ln\left(\frac{Q^2}{\Lambda^2_{\MSsch} }\right)  
\ee
ergibt.
Gl.~(\ref{Lser}) erh"alt man aus Gl.~(\ref{LQCD}) durch iteratives Aufl"osen
nach $a_{\MSsch}$.
Insbesondere tritt in Gl.~(\ref{Lser}) kein Term der Art $( {\rm const}/ L^2)$
auf.
Dies definiert die Wahl der Integrationskonstante in Gl.~(\ref{LQCD})
und damit den Parameter $\Lambda_{\MSsch}$.

Die hier vorgeschlagene Prozedur zu Extraktion von $\al_s$
respektiert die RG-Invarianz der Entwicklung der Trajektorien.
Im 1. Schritt wird eine neue effektive Ladung $a_\tau
$ durch die Relation 
\cite{prl,brodsky}
\be
\delta_P^{th} = a_\tau
\ee
eingef"uhrt.
Im 2. Schritt wird der Parameter $\Lambda_\tau$, der 
mit der Kopplungskonstante $a_\tau$
assoziiert ist, mit der effektiven $\beta$-Funktion
$\beta_\tau = -a_\tau^2(\beta_{\tau 0} + \beta_{\tau 1} a_\tau +
\beta_{\tau 2} a_\tau^2 + \beta_{\tau 3} a_\tau^3 + \dots) $ 
und dem experimentellen Wert f"ur $a_\tau(M_\tau) = \delta_P^{exp}$
bestimmt.
F"ur die Koeffizienten der effektiven $\beta$-Funktion 
ergibt sich 
$(\beta_{\tau 0} = \beta_0$ und $ \beta_{\tau 1 } = \beta_1 )$
\be \label{effbetaa}
\beta_{\tau 2 }= -12.3204\, ,
\quad
\beta_{\tau 3 }= -182.719 + \frac{9}{2} k_3 \, .
\ee
Der Wert f"ur $\Lambda_\tau$ wurde in der NNLO-N"aherung bestimmt,
so dass der Koeffizient $\beta_{\tau 3}$
nicht in die Analyse eingeht.

Im 3. Schritt bestimmt man den Parameter $\Lambda_{\MSsch}$  
mit Gleichung~(\ref{transform}).
Aus diesem ergibt sich die $\MSsch$-Kopplungskonstante $\al_s(\mts)$,
indem Gl.~(\ref{LQCD})
nach $a(M_\tau)$ mit $\ln(M_\tau^2 / \Lambda^2_{\MSsch})$ 
und der $\beta$-Funktion im $\MSsch$-Schema aufgel"ost wird.
Aus Konsistenzgr"unden kann die $\MSsch$-$\beta$-Funktion 
nur bis zur Drei-Schleifen-Korrektur verwendet werden,
da die $\beta$-Funktion $\beta_\tau$ nur bis zu dieser Ordnung bekannt 
ist (vgl. Gl.~(\ref{effbetaa})).

Dieses Verfahren ist RG-invariant.
Es ist m"oglich, den Ausdruck f"ur die Rate in einem beliebigen 
Renormierungsschema zu verwenden.
Die Qualit"at der Konvergenz der Reihen wird nicht 
durch unsichere Kriterien wie ``die Gr"o"se der Korrekturen 
nimmt ab'' abgesch"atzt.
Die einzigen perturbativen Objekte sind die $\beta$-Funktionen, welche 
beide relativ gut konvergieren.
Die Qualit"at der st"orungstheoretischen-Ausdr"ucke f"ur die $\beta$-Funktionen
zeigt auch die Beschr"ankung der erreichbaren Genauigkeit auf.
Die Entwicklung der Funktion $\beta_\tau$ 
ist asymptotisch wie alle Reihen der St"orungstheorie.
Betrachtet man $\beta_\tau$, so erkennt man, dass die asymptotische 
Grenze fast erreicht ist, wenn man $a_\tau(\mts)\sim 0.1$
annimmt.
\be
\label{betatau}
\beta_\tau(a_\tau)
=- a_\tau^2 \Bigg( \frac{9}{4} + 4 a_\tau - 12.3204 a_\tau^2 
  + a_\tau^3 \left(-182.719+\frac{9}{2} k_3 \right) \Bigg) 
+ \mathcal{O}(a_\tau^6) 
\ee
Die Konvergenz der Reihe h"angt stark von dem Wert f"ur $k_3$
ab.
Falls $k_3$ einen Wert besitzt, bei dem das asymptotische Wachstum 
bereits in der dritten Ordnung beginnt,
so ist eine weitere Verbesserung der Genauigkeit 
mit der St"orungstheorie in endlicher Ordnung nicht mehr m"oglich.

Mit der RG-invarianten Extraktionsmethode
(RSI), wie sie oben beschrieben ist, ergibt sich f"ur die Kopplungskonstante
im $\MSsch$-Schema
\ba
\label{finFO}
\al_{s}^{RSI}(M_\tau) &=& 0.3184 \pm 0.0060_{exp} \;.  
\ea
Der Wert ist kleiner als der entsprechende Wert mit der 
Standardmethode in Gl.~(\ref{dirres}).
Der Referenzwert f"ur die Kopplungskonstante wird gew"ohnlich 
bei der Energieskala des $Z$-Bosons $M_Z = 91.187~{\rm GeV}$
angegeben.
Die Entwicklung zu dieser Referenzskala wird mit der 
Vier-Schleifen-$\beta$-Funktion in $\MSsch$ Schema ausgef"uhrt
\cite{beta4}.
An den Schwellenenergien der schweren Quarks (Charm und Bottom)
werden Drei-Schleifen-''matching conditions'' \cite{matching}
verwendet, um
die Kopplungskonstante der effektiven Theorie ohne schweres 
Quark in Beziehung zu der Kopplungskonstanten der Theorie mit 
schwerem Quark zu bringen.
F"ur die Schwellenenergien wird  
$\mu_b=\bar{m}_b(\mu_b)=(4.21\pm 0.11)~{\rm GeV}$ \cite{bbmass}
und 
$\mu_c=\bar{m}_c(\mu_c)=(1.35\pm 0.15)~{\rm GeV}$ verwendet,
wobei 
$\bar{m}_q(\mu)$ die laufende Masse im 
$\MSsch$-Schema ist.

\begin{figure}[h]
\begin{center}
       \epsfig{file=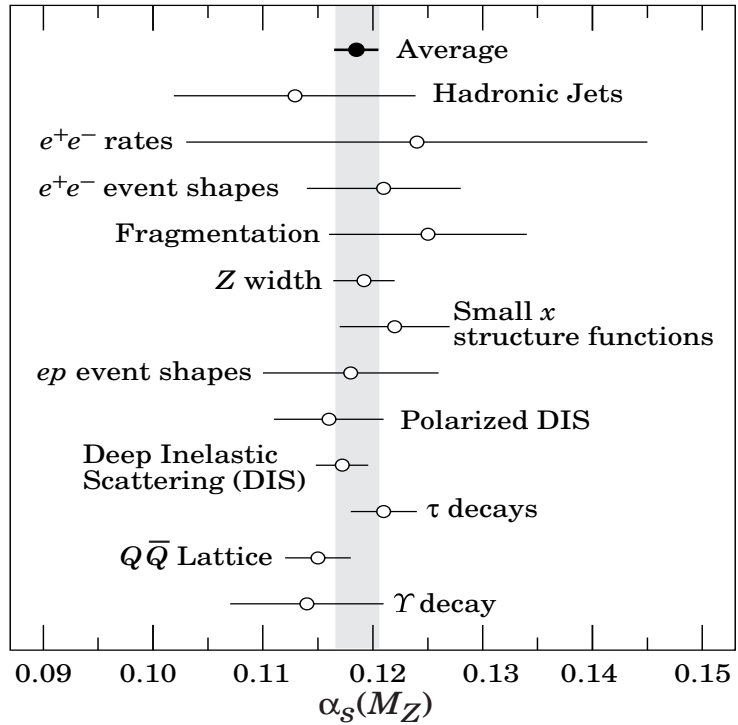,angle=0,width=.6
                  \textwidth,height=.6\textwidth}
\end{center}
       \caption{\label{asvergl}
Resultate f"ur die starke Kopplungskonstante $\al_s$ 
aus verschiedenen Experimenten. (Quelle:Paricle Data Book \protect\cite{PDG}) }
    \end{figure}

Das Laufen ergibt f"ur den mit der Standardmethode bestimmten
Wert das Resultat
\be
\label{dir0}
\al_s^{st}(M_Z)= 0.1202 \pm 0.0008_{exp} \pm 0.0006_c \pm 0.0001_b \;.
\ee
Hierbei bezeichnet der Index $exp$ den von 
$\delta_P^{exp}$ stammenden Fehler.
Die Fehler mit den Indizes $c,b$ entstehen durch die 
Unsicherheit in der Charm- und Bottom-Quarkmasse,
welche beim Laufen des Parameters zu der $Z$-Masse
in diesen eingehen. 
Der Mittelwert ist etwas gr"o"ser als der 
bei gro"sen Energien  
bestimmte Wert (vgl. Abb.~\ref{asvergl}) \cite{PDG}.
Die st"orungstheoretischen Ausdr"ucke f"ur Observable in 
Hochenergie-Experimenten konvergieren besser als Entwicklungen bei 
niedrigen Energien, da die Kopplungskonstante,
die der Parameter der asymptotischen Entwicklung ist,
wegen der asymptotischen Freiheit bei gro"sen Energien kleiner ist.
Diese Eigenschaft macht die Behandlung der Terme h"oherer Ordnung 
in Hochenergie-Experimenten weniger wichtig als bei den relativ 
niedrigen Energieskalen der $\tau$-Zerf"alle.
Allerdings sind die experimentellen Daten von Experimenten bei 
gro"sen Energien ungenauer als bei niedrigen Energien.
Die Tatsache, dass das Resultat in Gl.~(\ref{dir0})
gr"o"ser ist als der bei hohen Energien bestimmte Wert,
verursachte eine Diskussion "uber die Verl"asslichkeit der
Vorhersagen aus $\tau$-Lepton-Daten.
Diese Analyse l"ost dieses Problem. Das Laufen von 
$\al_s^{RSI}(M_\tau)$ zu $M_Z$ mit der Vier-Schleifen-$\beta$-Funktion 
und mit Drei-Schleifen-``heavy quark matching conditions''
ergibt
\be
\label{rgi0}
\al_{s}^{RSI}(M_Z) = 0.11768 \pm 0.00074_{exp} 
        \pm 0.00053_c \pm 0.00005_b \;, 
\ee
wobei f"unf Dezimalstellen aufgef"uhrt werden, um die verschiedenen
Gr"o"senordnungen der einzelnen Fehlerquellen zu unterscheiden.
Gleichung~(\ref{rgi0}) ist das Hauptresultat dieses Kapitels "uber die 
Bestimmung der starken Kopplungskonstanten $\al_{s}(M_Z)$ 
aus $\tau$-Lepton-Daten.

Die OPAL-Kollaboration berichtet von einem experimentellen Wert 
von $R_{\tau S=0}^{exp} = 3.484 \pm 0.024$ \cite{exp2}.
Dies f"uhrt zu 
$\delta_P^{exp} = 0.200 \pm 0.009_{exp}$
und
\be
\al_s^{RSI}(M_\tau) =0.3158 \pm 0.0078_{exp} \;.
\ee
Die Entwicklung zu $M_Z$ ergibt
\be
\al_s^{RSI}(M_Z) 
= 0.11737 \pm 0.00098_{exp} \pm 0.00052_c \pm 0.00005_b \, .
\ee
Dieser Wert liegt nahe bei dem aus Gl.~(\ref{rgi0}) aus den
ALEPH-Daten.
\section{Bestimmung der Kopplungskonstante mit CIPT}
Wie bereits festgestellt wurde, ist die Interpretation der Terme 
h"oherer Ordnung der St"orungstheorie f"ur die nummerischen Resultate
der Analyse von $\tau$-Lepton-Daten wichtig.
Die Standardmethode zu Resummation von Korrekturen h"oherer Ordnungen der 
St"orungstheorie besteht darin, die RG-verbesserten Korrelationsfunktionen 
auf der Kontur in der komplexen $Q^2$-Ebene zu integrieren \cite{Pivtau}.
Diese Methode resummiert Korrekturen, die durch das Laufen der 
Kopplungskonstante auf der Integrationskontur entstehen, und 
wird normalerweise bei der Analyse der $\tau$-Lepton-Daten verwendet.
Um die Kopplungskonstante mit resummierter St"orungstheorie 
zu bestimmen, wird der theoretische Ausdruck f"ur die Zerfallsrate 
in der konturverbesserten St"orungstheorie (CIPT)
an das experimentelle Resultat $\delta_{P}^{exp}$
Gl.~(\ref{expdec0}) angepasst \cite{groote}.
So ergibt sich 
f"ur die in Kapitel~\ref{Art4} definierte Kopplungskonstante 
\be \label{numbcopE}
a_E(\mts) = 0.1445  \;,
\ee
was sich mit der oben beschriebenen Methode
in die $\MSsch$-Kopplungskonstante f"ur die CIPT 
\be 
\label{CI}
\al_s^{CI}(M_\tau) = 0.343 \pm 0.009_{exp}
\ee
"ubersetzt.
Dieser Wert unterscheidet sich von dem Resultat der St"orungstheorie in
endlicher Ordnung in Gl.~(\ref{finFO}).
Die Fehlerbalken f"ur die starke Kopplungskonstante in der FOPT 
Gl.~(\ref{finFO}) und in der CIPT Gl.~(\ref{CI}) 
"uberlappen sich nicht. 
Diese Situation wurde in \cite{Pivtau}
bereits vorausgesagt,
wo eine Analyse mit resummierter St"orungstheorie erstmals
ausgef"uhrt wurde.
Resummation stellt eine besondere Absch"atzung der Terme 
h"oherer Ordnung dar.
In der FOPT 
wird ein Modell verwendet, in dem alle
Terme h"oherer Ordnung vernachl"assigt werden. In dem anderen 
Fall (CIPT) verwendet man ein Modell, in dem Terme h"oherer Ordnung
durch das Laufen der Kopplungskonstanten entlang der Integrationskontur
erzeugt werden.
Mit der augenblicklich zu Verf"ugung stehenden experimentellen 
Genauigkeit kann bereits zwischen den beiden Modellen unterschieden werden.
Es ist wichtig, sich dar"uber im Klaren zu sein, dass die zwei Resultate 
f"ur die Kopplungskonstante Gl.~(\ref{finFO}) (mit FOPT)
und Gl.~(\ref{CI}) (mit CIPT) mit verschiedenen Modellen erzielt wurden 
und nicht vermischt werden d"urfen. 
Um beispielsweise den nummerischen Wert der Kopplungskonstante mit 
Resultaten aus Experimenten bei hohen Energien zu vergleichen,
sollte der mit FOPT  
bestimmte Wert aus Gl.~(\ref{finFO}) verwendet werden, falls 
die zum Vergleich herangezogenen Resultate auch mit FOPT 
bestimmt wurden.   
\section{Zusammenfassung von Kapitel \ref{Art3} }
In diesem Kapitel wurde der Wert der starken Kopplungskonstante 
aus $\tau$-Lepton-Zerf"allen in strangelose Hadronen bestimmt.
Der mit der in diesem Kapitel vorgeschlagenen 
RG-Gruppen-invarianten Methode 
extrahierte Wert f"ur $\al_s$
ist systematisch kleiner als das mit dem 
Standardverfahren erzielte Resultat.
F"ur den nummerischen Wert der Kopplungskonstante bei $M_Z$ ergibt 
sich so 
\be
\al_s(M_Z) = 0.1177 \pm 0.0007_{exp} \pm 0.0006_{hq\; mass} \, .
\label{finresMZ}
\ee
Dieser Wert liegt n"aher bei den Werten f"ur $\al_s$ 
aus Experimenten bei hohen Energien als die vorherigen 
Bestimmungen von $\al_s$ mit $\tau$-Lepton-Zerfallsdaten.

\chapter{Bestimmung der Strange-Quarkmasse $m_s$} \label{Art4} \label{ART4}
Aufgrund der Analyse der Momente der hadronischen Spektraldichte 
in dem endlichen Energieintervall \mbox{$(0,M_\tau^2$ )}
aus den Kapiteln~\ref{Art1} und \ref{Art2}
mit St"orungstheorie in endlicher Ordnung (FOPT)
gibt es deutliche Hinweise daf"ur, dass
die ultimative theoretische Pr"azision der $\tau$-Lepton-Observablen
bereits erreicht ist \cite{one,two}.
Die St"orungsreihen sind asymptotisch und ihre ultimative Genauigkeit
h"angt von dem konkreten nummerischen Wert 
des Entwicklungsparameters ab.
Im Fall des $\tau$-Systems ist die starke Kopplungskonstante nicht klein, 
was bereits bei einer relativ niedrigen Ordnung zu asymptotischem 
Wachstum f"uhren kann.
Dies begrenzt die theoretische Genauigkeit der FOPT.
Die Analyse der Kapitel~\ref{Art1} und \ref{Art2} zeigt, 
dass asymptotisches Wachstum im masselosen Fall in der f"unften 
und f"ur die Massenkorrektur $m_s^2$ in der vierten 
Ordnung beginnt.
F"ur einige Observablen sind die derzeitigen experimentellen Daten 
bereits mit der ultimativen theoretischen Genauigkeit in der FOPT vergleichbar 
\cite{exp1al,exp1ms,exp2}.
Dies stellt das Problem, pr"azisere theoretische Formeln zu erlangen.
Weitere Terme der St"orungsreihe in der FOPT werden
die Genauigkeit der Resultate nicht verbessern,
sind aber vom theoretischen Standpunkt sehr interessant, 
da sie zus"atzliche
Informationen liefern. 
Mit der sich st"andig verbessernden experimentellen Genauigkeit wird es 
notwendig, eine neue Interpretation der St"orungsreihe zu finden,
falls man eine theoretische Genauigkeit erreichen will, 
die mit den experimentellen 
Daten vergleichbar ist.
Eine M"oglichkeit, 
nummerische Resultate aus den St"orungsreihen zu extrahieren, 
besteht in der Anwendung eines Resummationsverfahrens
(siehe z.B. \cite{Zakh}), das 
weiter entwickelt ist als das einfache Aufaddieren von Termen der 
St"orungsreihe, bis diese wieder zu wachsen beginnen.
Die Wahl eines solchen Resummationsverfahrens ist nicht eindeutig,
so dass es viele M"oglichkeiten gibt, eine 
asymptotische Reihe zu resummieren oder 
ihr Konvergenzverhalten zu verbessern \cite{renRS,pivrho}.
Es gibt zwei wichtige Kriterien f"ur die Wahl eines
angemessenen Resummationsverfahrens:
Die durch die Renormierungsgruppe gegebene Struktur der St"orungsreihe
muss respektiert werden, und die Definition von effektiven Parametern, 
die bei der Beschreibung der Observablen verwendet werden, muss physikalisch
motiviert sein.
Aufgrund dieser beiden Kriterien wird f"ur die Resummation der 
St"orungsreihen die konturverbesserte St"orungstheorie (CIPT) \cite{Pivtau}
(siehe auch \cite{DP})
in Verbindung mit 
einem effektiven Renormierungsschema (nicht $\MSsch$) verwendet.
Durch die Wahl der CIPT ist das Resummationsverfahren nicht 
eindeutig festgelegt.
Die Freiheit der Wahl eines Renormierungschemas bleibt bestehen und
beeinflusst die nummerischen Resultate.
F"ur diese Analyse wird ein effektives Schema verwendet,
das f"ur das $\tau$-System nat"urlich und einfach 
erscheint.
Die Wahl eines angemessenen Renormierungsschemas wird auch durch 
die Methode bestimmt, mit der das System beschrieben werden soll.
In der FOPT ist die Spektraldichte das fundamentale 
Objekt und dient somit als Definition 
f"ur die effektive Kopplungskonstante und
die effektiven Massen.
Wird die CIPT verwendet, 
so ist die Korrelationsfunktion selbst das fundamentale 
st"orungstheoretische Objekt. 
Aus diesem Grund werden in diesem Kapitel die noch einzuf"uhrenden 
effektiven Parameter "uber die Korrelationsfunktion definiert.
Ihre Definitionen unterscheiden sich von der in den Kapiteln~\ref{Art1}
und \ref{Art2} verwendeten Definition der effektiven Parameter auf dem 
physikalischen Schnitt. Die hier verwendeten Parameter erhalten deshalb zur 
Unterscheidung den zus"atzlichen Index $E$ f"ur euklidisch, der andeuten 
soll, dass diese Parameter f"ur euklidische Impulse $Q^2$ nat"urlich sind.
In dem effektiven Schema werden alle Korrekturen der Korrelationsfunktion
in der Definition der Kopplungskonstante (masseloser Fall
\cite{effsch,ksch,kksch,effDh,brodsky,prl})
und in zwei effektiven Parametern 
$m^2_{Eq}$ 
und $m^2_{Eg}$ absorbiert, 
die die zwei Massenskalen
in dem  $q$- und $g$-Teil der Korrelationsfunktion repr"asentieren.
Diese Massenparameter h"angen perturbativ mit der Strange-Quarkmasse zusammen.
F"ur verschwindende Quarkmassen wurde diese Analyse in 
in \cite{groote} bereits ausgef"uhrt.
In diesem Kapitel wird sie auf die $m_s^2$-Korrektur 
ausgedehnt \cite{four}.
Eine Analyse der $m_s^2$-Korrektur 
mit der CIPT im $\MSsch$-Schema wurde in  
\cite{msNPB} ausgef"uhrt (siehe auch \cite{pichprades}).
Die mit dem effektiven Schema erzielten Resultate 
best"atigen Ref. \cite{msNPB} im wesentlichen. 
Um ein Verst"andnis f"ur die Zuverl"assigkeit der hier verwendeten 
Prozedur zu erhalten, werden die Resultate auf ihre Stabilit"at
bez"uglich Korrekturen h"oherer Ordnung der RG-Funktionen
untersucht.
\section{Resummation in einem effektiven Renormierungsschema}
Im ersten Schritt der Analyse des $\tau$-Systems wird ein 
effektives Renormierungsschema definiert, das alle QCD-Korrekturen 
in der Kopplungskonstante und den beiden Massen absorbiert.
Wenn ein solches Renormierungsschema verwendet wird, 
sind die einzigen perturbativen Ausdr"ucke, die ben"otigt werden, 
um das System zu beschreiben, die $\beta$-Funktion und die beiden 
$\gamma$-Funktionen.
Mit diesen drei Funktionen kann die Entwicklung der
effektiven Kopplungskonstante und der zwei Massenkoeffizienten-Funktionen
auf der Kontur in der komplexen Ebene bestimmt werden.
Hierbei werden die $\beta$- und $\gamma$-Funktionen als exakte 
Funktionen behandelt.
Dies ist das Standard-Verst"andnis der 
RG-Resummation.
Hat man die expliziten L"osungen der laufenden Kopplungskonstante und der 
laufenden Massenkoeffizienten-Funktionen, so lassen sich Observable, 
wie die Momente der Zerfallsrate, durch Integration der Koeffizientenfunktionen
mit den f"ur die Observablen spezifischen Wichtungsfunktionen bestimmen.   
\subsection{Definition des effektiven Schemas}
Die grundlegende theoretische Gr"o"se f"ur 
$\tau$-Lepton-Zerf"alle ist die Korrelationsfunktion 
von zwei hadronischen Str"omen
\begin{equation}
\label{corrF}
\Pi_{\mu\nu}(q) = i \int dx e^{iqx}
\langle T j_{\mu}(x) j_{\nu}^{\dagger} (0) \rangle
= \frac{N_c}{6\pi^2}( q_{\mu}q_{\nu} \Pi_q(q^2)+g_{\mu\nu}  \Pi_g(q^2) )  
\label{correlator}
\end{equation}
wobei $j_{\mu}(x) = \bar{u}\gamma_{\mu}(1-\gamma_5) s$
ist.
$\Pi_q(q^2)$ und $\Pi_g(q^2)$
sind skalare RG-invariante Funktionen.
Es wird die QCD mit drei leichten Quarks verwendet.
Die Entwicklung von 
 $\Pi_q(q^2)$ und $\Pi_g(q^2)$
in Potenzen von $m_s^2/q^2$ bis zur ersten Massenkorrektur 
ergibt 
\be
\label{expmqq}
\Pi_q(q^2)=\Pi(q^2)+3\frac{m_s^2}{q^2}\Pi_{mq}(q^2)
\ee
und
\be
\label{expmgg}
\Pi_{g}(q^2)=-q^2 \Pi(q^2) + 
\frac{3}{2} m_s^2   \Pi_{mg}(q^2) \;,
\ee
wobei 
$\Pi(q^2)$ 
und $\Pi_{mg/q}$ die bereits aus Kapitel~\ref{kf}
bekannten Funktionen sind.

Die neuen effektiven Parameter 
$a_E, m_{Eq}^2, m_{Eg}^2$
werden so definiert, dass alle Informationen der 
St"orungstheorie in die efffektiven 
$\beta$- und $\gamma$-Funktionen absorbiert werden. 
Da die Korrelationsfunktion der Massenkorrektur nicht 
transversal ist, m"ussen zwei verschiedene Massenparameter 
eingef"uhrt werden.
Die Definitionen sind 
\begin{eqnarray} 
\label{effdef}
-Q^2\frac{d}{dQ^2}\Pi(Q^2)
&=&1 + a_E(Q^2)\;,   \nn \\
- m_s^2(\mts) Q^2\frac{d}{dQ^2}\Pi_{mg}(Q^2)
&=& m_{Eg}^2(\mts)C_g(Q^2)\;, \\
m_s^2(\mts) \Pi_{mq}(Q^2)
&=& m_{Eq}^2(\mts)C_q(Q^2) \nn \; .
\end{eqnarray}
Ausgedr"uckt durch die $\MSsch$-Parameter ($\al_s$ und $m_s$)
ergibt sich f"ur die effektiven Parameter aus Gl.~(\ref{effdef}).
\ba\label{effparamm}
a_E(\mts)     
&=& 
\als + k_1 \alsb^2  + k_2 \alsb^3 + 
      k_3 \alsb^4 + ...  \;, \\  
\label{effmgg}
m_{Eg}^2(\mts)   
&=& 
m_s^2(\mts) ( 1+ \frac{5}{3} \als  + 
   k_{g1} \alsb^2 + k_{g2} \alsb^3 +...) \;,\\
\label{effmqq}
m_{Eq}^2(\mts) 
&=& m_s^2(\mts)  ( 1+ \frac{7}{3} \als  + 
         k_{q1} \alsb^2 + ...) \; .
\ea
Wie vorher ausgef"uhrt ist der nummerische Wert f"ur $k_3$ unbekannt.
Die Werte der Koeffizienten $k_{1,2}$, $k_{g1,g2}$,  
$k_{q1}$ sind in Gl.~(\ref{koeff1}, \ref{koeff2}) 
\cite{eek20,eek21,eek2c,eek2d,eek2e}
angegeben.
\subsection{Das Laufen der Kopplungs- und der Koeffizientenfunktionen}
Das Verhalten der effektiven Kopplung $a_E(Q)$ und 
der Koeffizientenfunktionen der Massenparameter $C_{q,g}(Q)$
ist durch die effektive $\beta$- und die $\gamma$-Funktion 
bestimmt.
Die RG-Gleichungen sind
\ba
\label{betaf}
Q^2\frac{d}{dQ^2} a_E(Q^2)&=&\beta(a_E), \\
\label{gammagf}
Q^2\frac{d}{dQ^2}C_g(Q^2)&=&2 \gamma_g(a_E)C_g(Q^2), \\
\label{gammaqf}
Q^2\frac{d}{dQ^2}C_q(Q^2)&=&2 \gamma_q(a_E)C_q(Q^2) \;.
\ea
Die effektiven RG-Funktionen aus Gl.~(\ref{betaf}, \ref{gammagf}) und
(\ref{gammaqf}) lassen sich aus den RG-Funktionen im $\MSsch$-Schema
mit Gl.~(\ref{betatransform}, \ref{gammatransform}) bestimmen.
Die Koeffizienten der effektiven $\beta$- und der $\gamma$-Funktion 
$\beta_i^{\rm eff}$ und $\gamma_i^{\rm eff}$ h"angen mit den entsprechenden
Koeffizienten im $\MSsch$-Schema $\beta_i$ und $\gamma_i$ folgenderma"sen 
zusammen: 
F"ur die effektive $\beta$-Funktion erh"alt man
\ba
\beta_0^{\rm{eff}} &=& \beta_0 \;,\nn \\
\beta_1^{\rm{eff}} &=& \beta_1 \;,\nn \\
\beta_2^{\rm{eff}} &=& \beta_2 - k_1 \beta_1 
                       + (k_2 - k_1^2)\beta_0 \;,\nn \\
\beta_3^{\rm{eff}} &=& \beta_3 - 2 k_1 \beta_2 + k_1^2 \beta_1
         + (2 k_3 -  6 k_2 k_1 +  4 k_1^3 ) \beta_0  
\ea
und f"ur die $\gamma$-Funktion ergibt sich
\ba
\gamma_{n0}^{\rm{eff}} &=& \gamma_0 \;,\nn\\
\gamma_{n1}^{\rm{eff}} &=& \gamma_1 - k_1 \gamma_0 
             + \frac{1}{2} k_{n0}\beta_0 \;,\nn \\
\gamma_{n2}^{\rm{eff}} &=& \gamma_2 - 2 k_1 \gamma_1 
               +( - k_2  + 2 k_1^2 ) \gamma_0
               + \frac{1}{2} k_{n0} \beta_1
               + \left(
                  -  k_1 k_{n0}+
                  k_{n1}- \frac{1}{2} k_{n0}^2 
                 \right)\beta_0 \;,\nn \\
\gamma_{n3}^{\rm{eff}} &=&  \gamma_3  - 3 k_1 \gamma_2
               +( - 2 k_2 + 5 k_1^2) \gamma_1
               +( - k_3 +  5 k_2 k_1 - 5 k_1^3) \gamma_0 \nn \\
            && +\frac{1}{2} k_{n0} \beta_2
               + \left(
                  -\frac{3}{2}k_1 k_{n0} +  k_{n1}
                      -\frac{1}{2} k_{n0}^2 
                 \right) \beta_1  \\
             &&  + \left(
                   - k_2 k_{n0}+ \frac{5}{2} k_1^2 k_{n0}
                   - 3 k_1 k_{n1} + \frac{3}{2} k_1 k_{n0}^2
                   + \frac{3}{2} k_{n2} 
                   - \frac{3}{2}k_{n1} k_{n0} + \frac{1}{2} k_{n0}^3
                 \right) \beta_0 \;.\nn
\ea
Hier stehen die Koeffizienten $k_{nj}$ mit $n=q,g$ f"ur der Koeffizienten 
in Gl.~(\ref{effmgg}) f"ur $n=g$ $(k_{g0} = 5/3)$ 
und f"ur die Koeffizienten 
in Gl.~(\ref{effmqq}) f"ur $n=q$ $(k_{q0} = 7/3)$. 
Nummerisch ergibt sich f"ur die effektiven RG-Funktionen
\ba
\label{betafNum}
\beta(a_E) &=& - a_E^2 
     \left( 2.25 + 4 a_E + 11.79 a_E^2 + 
    a_E^3 \left( -76.36 + 4.5 k_3 \right) 
     \right) \, ,\\
\label{gammagfNum}
\gamma_g(a_E) &=& 
       - a_E \left( 1 + 4.027 a_E + 17.45 a_E^2  + 
    a_E^3 \left( 249.59 - k_3  \right)  \right) \, ,\\
\label{gammaqfNum}
\gamma_q(a_E) &=& 
        - a_E \left(1 + 4.78 a_E + 32.99 a_E^2  +   a_E^3  
\left( -252.47  - k_3  + 3.38  k_{q2} 
                        \right) \right) \;.
\ea
Im Folgenden wird f"ur die effektive Kopplungskonstante der
Wert   
$a_E(\mts)=0.1445$ \cite{three} aus Gl.~(\ref{numbcopE})
verwendet, der aus der $\tau$-Zerfallsrate in Hadronen ohne Strangeness
mit der CIPT und dem effektiven Renormierungsschema (\ref{effparamm})
bestimmt wurde (zu dem Verfahren siehe \cite{groote}).
Die effektive $\beta$-Funktion konvergiert f"ur den nummerischen Wert  
der effektiven Kopplungskonstanten $a_E(\mts)=0.1445$ gut.
Falls der Koeffizient $k_3$ in dem Bereich $0<k_3<50$ liegt, 
wird der $a_E^3$-Koeffizient in Gl.~(\ref{betafNum}) nicht 
sehr gro"s. Trotzdem zeigt die $\beta$-Funktion 
(\ref{betafNum}) f"ur $k_3>35$ in der \no-N"aherung asymptotisches Wachstum.
Die $\gamma_g$-Funktion verh"alt sich schlechter als die 
$\beta$-Funktion, es besteht aber noch Konvergenz bis zur 
NNLO:
Die \no-Korrektur wird asymptotisch anwachsen, falls $k_3$ kleiner 
als 129 ist.
Die $\gamma_q$-Funktion in Gl.~(\ref{gammaqfNum})
zeigt bereits in der NNLO-N"aherung asymptotisches Wachstum, 
was die Genauigkeit der Resultate beschr"anken wird.
Grunds"atzlich muss man die \no-bereits 
als asymptotisch betrachten, 
da keine Wahl f"ur $k_3$ dazu
f"uhrt, dass
die $\beta$- und $\gamma$-Funktionen beide gut konvergieren.
Das best"atigt die Schlussfolgerungen aus den Kapiteln~\ref{Art1} und
\ref{Art2} \cite{one,two},
wo asymptotisches Wachstum in den FOPT-Ausdr"ucken 
der Momente der Spektraldichten in der \no-N"aherung unabh"angig von der Wahl
f"ur  $k_3$ gefunden wurde.

Das Laufen der Kopplungskonstante und der Massenkoeffizienten-Funktionen 
$C_{q,g}(Q^2)$ auf der Kontur wird durch die 
RG-Gleichungen bestimmt.
Die Integrationskontur wird durch 
\mbox{$Q^2= \mts e^{i \phi}$}, $-\pi< \phi<\pi$
parametrisiert, was zu den Differentialgleichungen
\be \label{difa}
- i \frac{d}{d  \phi} a_E(\phi) = \beta(a_E(\phi))\; ; 
\quad a_E(\phi = 0) = a_{E0}
\ee
und 
\be \label{difm}
- i \frac{d}{d  \phi} C_n(\phi) 
= 2 \gamma_n(a_E(\phi)) C_n(\phi) \; ; 
\quad C_n(\phi = 0) = 1
\ee
f"uhrt, wobei $a_{E0}= a_E(\mts)$ und
$n=q,g$ ist.
Die L"osung der 
Gleichung f"ur das Laufen der Masse, das durch die Koeffizienten-Funktion 
$C_n(\phi)$ in Gl.~(\ref{difm}) beschrieben wird, 
l"asst sich durch das Integal  
\be
C_n(\phi) = \exp 
 \left(2i\int_0^\phi \gamma_n(a_E(\phi)) d \phi  \right) 
\ee
ausdr"ucken.
Die Anfangswerte f"ur 
$a_E(\phi)$ und $C_n(\phi)$ 
werden bei $Q^2 = \mts$
oder $\phi=0$ festgelegt.
Alle st"orungstheoretischen Korrekturen werden in den Koeffizienten der
$\beta$- and $\gamma$-Funktionen absorbiert,
falls das in Gl.~(\ref{effdef}) definierte Renormierungsschema verwendet wird.
Die L"osungen f"ur die Kopplungskonstante $a_E(\phi)$ 
und die Koeffizienten-Funktionen $C_q(\phi)$
werden in den Abbildungen~\ref{runc} und~\ref{runmCq}) dargestellt.
Das Laufen der effektiven Kopplungskonstante ver"andert sich nicht stark, 
falls Korrekturen h"oherer Ordnung der $\beta$-Funktion verwendet werden.
$C_g(\phi)$ ver"andert sich etwas st"arker von der LO- zu der 
NNLO-N"aherung wogegen 
$C_q(\phi)$ 
(Abb.~\ref{runmCq}) in der N"ahe von $\phi=\pm \pi$
(nahe dem physikalischen Schnitt) aufgrund der gro"sen "Anderung in der
$\gamma_q$-Funktion (Gl.~(\ref{gammaqfNum})) von der NLO zu der NNLO nicht
konvergiert.  
  \begin{center}
    \begin{figure}[!ht]
\label{fig1b}
       \epsfig{file=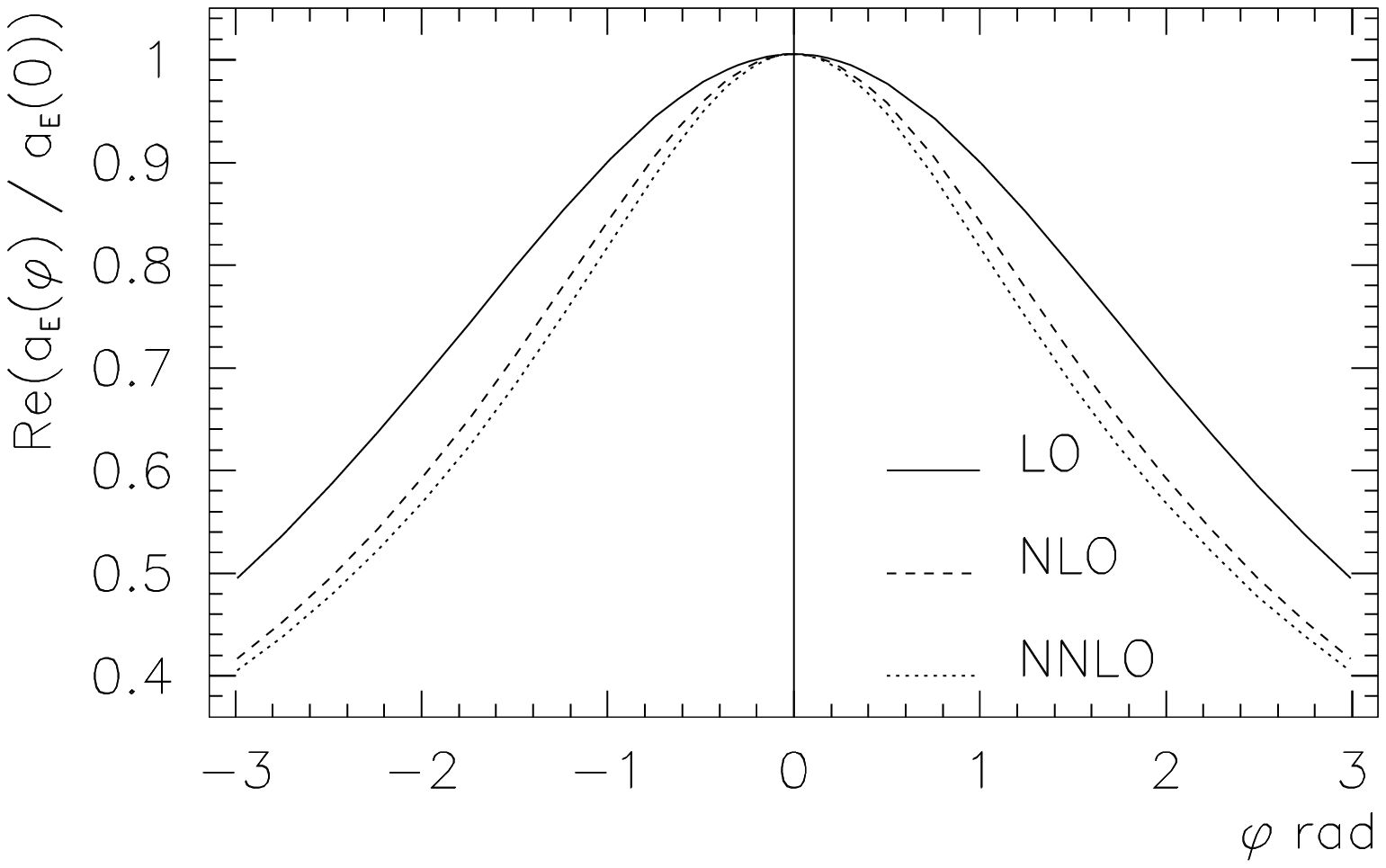,angle=0,width=.45
                 \textwidth,height=.33\textwidth}
       \quad \quad
       \epsfig{file=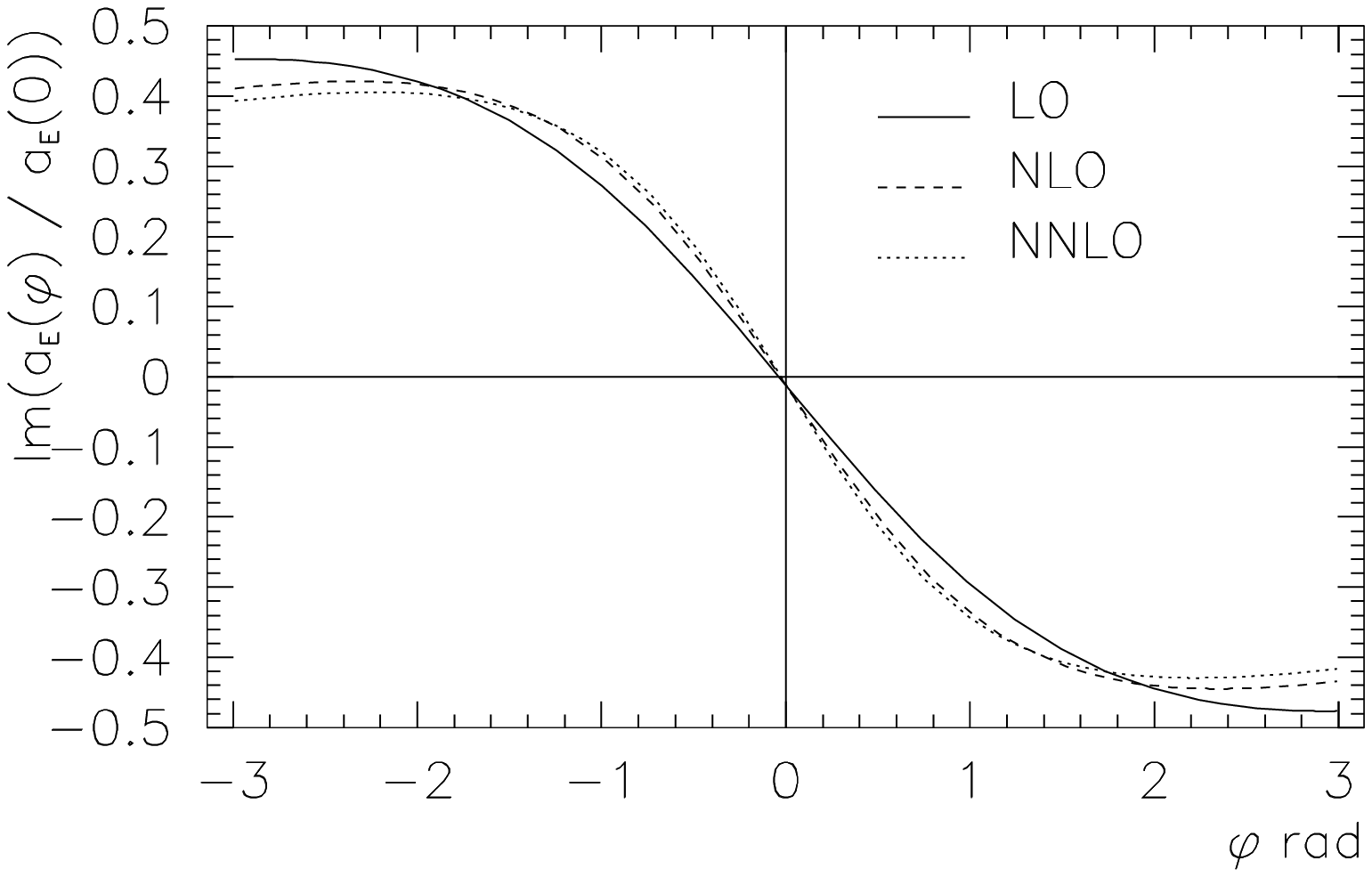,angle=0,width=.45
                  \textwidth,height=.33\textwidth}
       \caption{\label{runc}
Laufen der effektiven Kopplungskonstante $a_E(\phi)$
auf einer kreisf"ormigen Kontur in der komplexen Ebene 
($ Q^2 = \mts e^{i\phi }$), berechnet mit der $\beta$-Funktion 
in der LO-, NLO- und NNLO-N"aherung
(links: Realteil, rechts: Imagin"arteil).}
    \end{figure}
    \begin{figure}[!ht]
       \epsfig{file=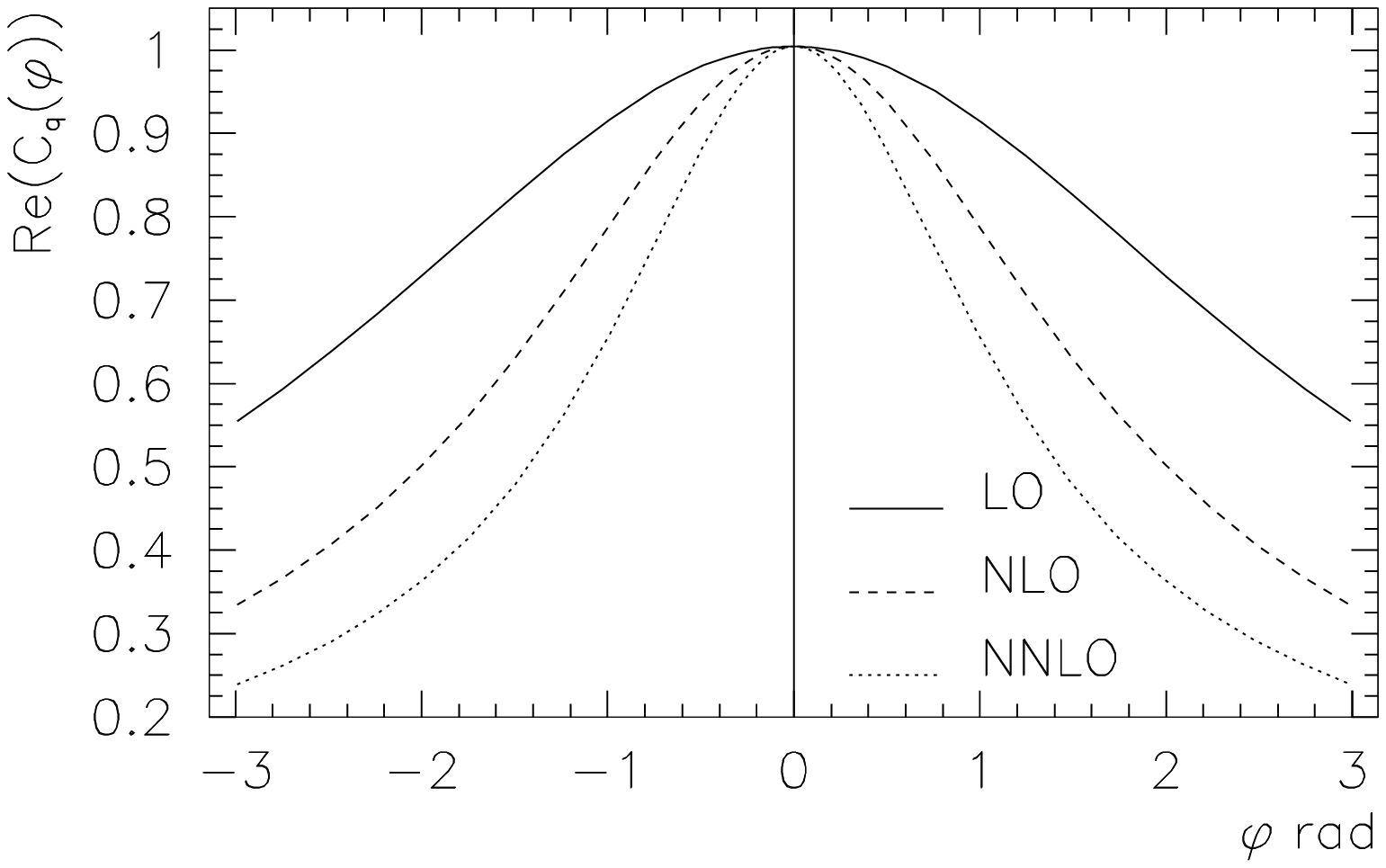,angle=0,width=.45
       \textwidth,height=.33\textwidth}
       \quad \quad
       \epsfig{file=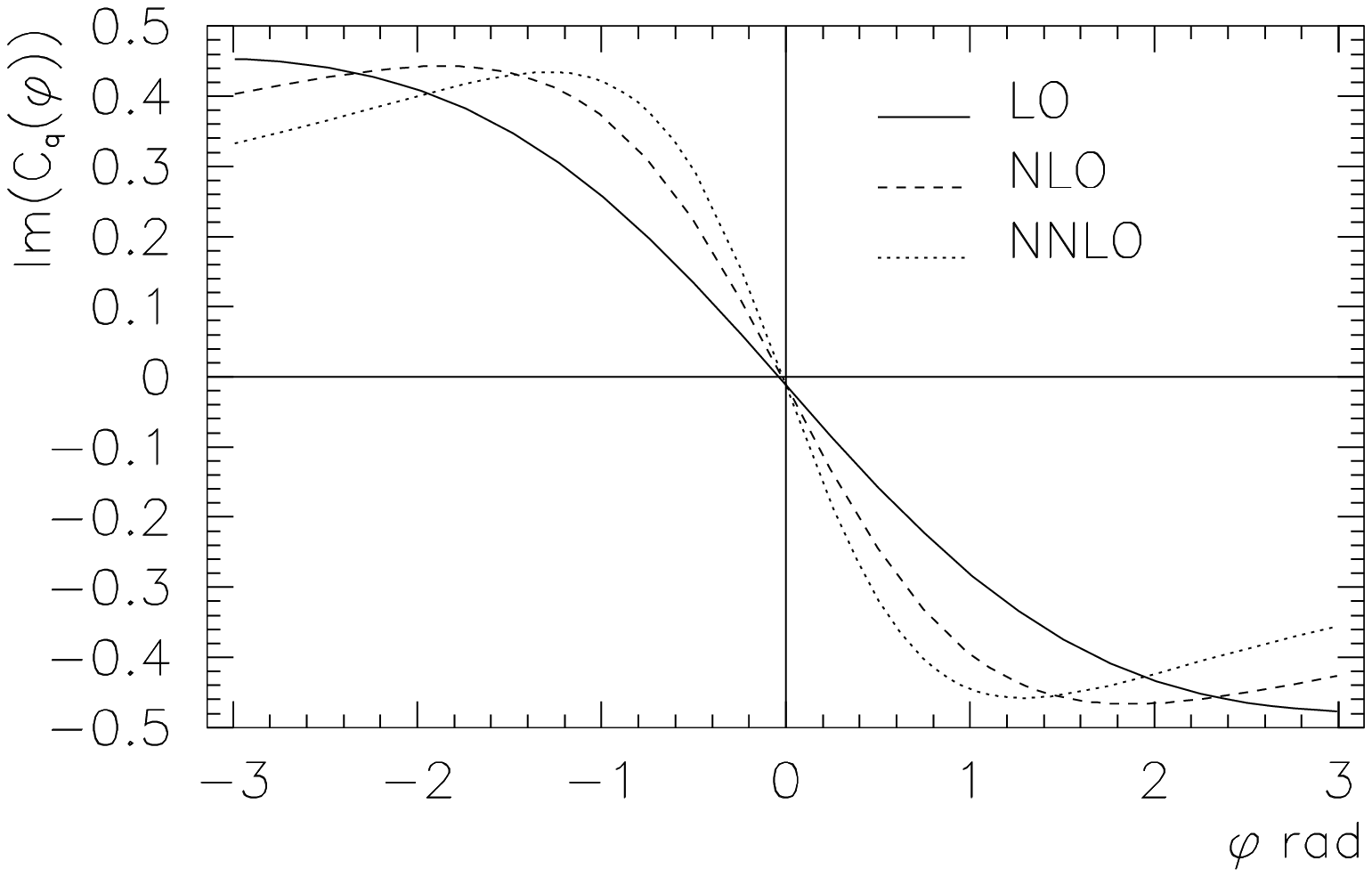,angle=0,width=.45
       \textwidth,height=.33\textwidth}
       \caption{\label{runmCq}
Laufen der effektiven Koeffizienten-Funktion  $C_q$
auf einer kreisf"ormigen Kontur in der komplexen Ebene 
($ Q^2 = \mts e^{i\phi }$), berechnet mit der $\gamma_q$-Funktion 
in der LO-, NLO- und NNLO-N"aherung
(links: Realteil, rechts: Imagin"arteil).}
    \end{figure}
   \end{center}     

\subsection{Resummation} 
\label{Resummation}
Es wird die direkte Verallgemeinerung des Resummationsverfahrens f"ur den 
masselosen Fall aus \cite{groote} verwendet.
Die Differentialgleichungen f"ur die laufenden Parameter~(\ref{difa}, \ref{difm})
werden gel"ost, wobei die st"orungstheoretischen Ausdr"ucke f"ur die 
RG-Funktionen wie exakte Funktionen behandelt werden.
F"ur den masselosen Teil der Korrelationsfunktion werden Momente der 
der Spektraldichte durch
\ba \label{masslessmom}
M(n) &=& (n+1) \frac{i}{2 \pi} \oint \Pi(q^2) 
    \left(\frac{q^2}{\mts} \right)^n \frac{dq^2}{\mts} \nn \\
&=&  (n+1) (-1)^{n+1} \frac{i}{2 \pi} \oint \Pi(Q^2) 
 \left( \frac{Q^2}{\mts} \right)^n \frac{dQ^2}{\mts} \nn \\
&=&  (n+1) (-1)^{n+1} \frac{i}{2 \pi} \oint \Pi(x \mts ) 
  (x)^n \frac{d x}{\mts} 
        \quad \quad \quad \quad \quad{\rm wobei } \; Q^2 = \mts x \nn \\
&=&  (-1)^{n+1} \frac{i}{2 \pi} \oint \Pi(x \mts ) 
     \frac{d}{dx}(x W(x)) dx \nn \\
&=& (-1)^{n+1} \frac{i}{2 \pi} \oint D(x \mts ) W(x) dx \nn\\
&=&  (-1)^{n+1} \frac{i}{2 \pi} \Bigg( \oint x^n D(x \mts ) dx 
+ \oint \frac{(-1)^n}{x} D(x \mts ) dx \Bigg) \\
&=&  1 + (-1)^n \frac{1}{2 \pi} \int_{- \pi}^{\pi}  
e^{i (n+1)\phi} a_E(\phi) d \phi + \frac{1}{2 \pi} 
      \int_{- \pi}^{\pi} a_E(\phi) d \phi  
                    \quad \quad{\rm wobei } \; x= e^{i \phi} \nn
\ea
definiert, wobei im 5. Schritt partiell integriert wurde.
Die Wichtungsfunktion $W(x)$ ergibt sich aus der Bedingung
\be
\frac{d}{dx} (x W(x)) = (n+1) x^n 
\ee
und der Forderung, dass diese Funktion an den Enden der Integrationskontur
$ x = -1 \pm i \ep$ verschwinden soll, so dass bei der partiellen 
Integration kein Randterm auftritt.
\be
W(x) = \frac{1}{x} \int_{-1}^x (n+1) x^n dy = x^n +\frac{(-1)^n}{x}
\ee
Die Ausdr"ucke f"ur den $g$-Teil ergeben sich in Analogie zum 
masselosen Fall.
Hier ist
\be\label{gmoments}
M_g(n)=\frac{(-1)^n}{2 \pi} 
\int_{-\pi}^{\pi} e^{i (n+1) \phi} 
C_g (\phi)d \phi \; +  
\frac{1}{2 \pi} \int_{-\pi}^{\pi} 
       C_g(\phi) d \phi \;.
\ee
Im $q$-Teil wird die bereits in Kapitel~\ref{Art2} Gl.~(\ref{defofqmom})
gegebene Definition der physikalischen Momente verwendet.
Der entsprechende Ausdruck auf der Kontur ist
\ba
\label{qmompart}
 M_q^{ph}(n) &=&  \frac{m_s^2(\mts)}{m_{Eq}^2(\mts)}
        \frac{i}{2 \pi} \oint \frac{\Pi_{mq}(q^2)}{q^2} 
                  \left( \frac{q^2}{\mts} \right)^n
               d q^2 \\
&=&  (-1)^n \frac{i }{2 \pi}  \oint \frac{C_q(Q^2)}{Q^2} 
       \left( \frac{Q^2}{\mts} \right)^n
               d Q^2 \nn \\
&=&  (-1)^{n+1} \frac{1}{2 \pi} \int_{- \pi}^{\pi} 
         C_q( \phi ) e^{i n \phi} d \phi \nn \;.
\ea
Die Momente des masselosen Teils $M(n)$
und die der Massenkorrektur $M_{q,g}(n)$ sind die 
fundamentalen berechenbaren Objekte.
Die qualitativen Schlussfolgerungen "uber die Stabilit"at 
der laufenden Parameter bez"uglich Korrekturen h"oherer Ordnung 
der $\beta$- and $\gamma$-Funktionen
werden durch das Verhalten der Momente best"atigt.
Im masselosen und $g$-Teil "andern sich die Resultate nicht stark,
wenn man von der LO- zu der NNLO-N"aherung geht, w"ahrend die NNLO-Korrektur 
f"ur Momente des $q$-Teils gr"o"ser ist als die vorherige Korrektur.
Im masselosen und $g$-Teil werden die Werte der Momente durch das
zweite Integral in den Gleichungen~(\ref{masslessmom}, \ref{gmoments})
dominiert, da die Werte der oszillierenden Integrale 
in den Gleichungen~(\ref{masslessmom}, \ref{gmoments})
f"ur h"ohere Momente abnehmen.
Dies stabilisiert das Verhalten der Momente bez"uglich ihrer Abh"angigkeit
von Termen h"oherer Ordnung in der Entwicklung der RG-Funktionen.
Der $q$-Teil (Gl.~(\ref{qmompart})) enth"alt keinen solchen Term und 
die nummerischen Werte f"ur die Korrekturen der Momente aufgrund von 
h"oheren Termen der RG-Funktion $\gamma_q$ sind relativ unregelm"a"sig.
\section{Bestimmung der $\MSsch$-Strange-Quarkmasse}
Eines der wichtigsten Ziele der Analyse von 
Cabbibo-unterdr"uckten Zerf"allen ist die Bestimmung des
Massenparameters $m_s$.
Zu diesem Zweck k"onnen verschiedene Observable verwendet werden.
In dieser Analyse wird die totale Zerfallsrate verwendet.
Die experimentellen Daten f"ur Cabbibo-unterdr"uckte
$\tau$-Lepton-Zerf"alle sind nicht sehr genau.
Im Gegensatz dazu sind die theoretischen Ausdr"ucke sehr pr"azise.
Der theoretische Ausdruck f"ur die $m_s^2$-Korrektur zu den Momenten $(k,l)$
der differentiellen Zerfallsrate ist durch das Konturintegral 
in der komplexen $q^2$-Ebene
\ba
\label{fincor}
R_{m\tau}^{kl} &=& \frac{i}{2 \pi} 
\oint 2 \left( 1- \frac{q^2}{\mts} \right)^{2+k} 
\left( \frac{q^2}{\mts} \right)^l 
3\left(\frac{m_s^2 \Pi_{mq}(q^2)}{q^2} 
- \frac{m_s^2}{\mts}
\Pi_{mg}(q^2)  \right)\frac{dq^2}{\mts} \nn \\
&=& -6\left(\frac{m_q^2}{M_\tau^2} A_{kl} 
     + \frac{m_g^2}{M_\tau^2}
B_{kl}\right) 
=
-6\frac{m_s^2}{M_\tau^2} 
\left(\omega_q A_{kl} + \omega_g B_{kl} \right) 
= -6\frac{m_s^2}{\mts} F_{kl} 
\ea  
gegeben, wobei die Bezeichnungen $(k,l)$
f"ur zus"atzliche Wichtungsfaktoren in der Integration
stehen, die den Bereich der hohen $(k>0,l=0)$
und der niedrigen $(k=0,l>0)$ Energien unterdr"ucken.
Hier wird das $(0,0)$-Moment als beste Absch"atzung f"ur $m_s$
verwendet.
Andere Momente werden kurz diskutiert, und es werden Gr"unde genannt,
warum diese nicht verwendet werden.
Die Wahl einer speziellen Linearkombination von Momenten ist 
gleichbedeutend
mit der Integration der laufenden Parameter mit einer speziellen 
Wichtungsfunktion.
Ob die QCD-Korrekturen zu einer gegebenen Linearkombination gro"s oder 
klein sind, kann stark von dem Integrationsbereich abh"angen, auf dem 
das Integral den gr"o"sten Beitrag erh"alt.
Die "Anderung der laufenden Parameter ist in dem Bereich nahe dem 
physikalischen Schnitt (bei $\phi= \pm\pi$) am gr"o"sten 
und im tief euklidischen Bereich klein, wo die Funktionen durch ihre
Anfangswerte fixiert sind.
Wenn bei einer Observablen diese Bereiche ($\phi= \pm\pi$)
stark unterdr"uckt sind, besitzt sie keine Beitr"age
des perturbativen Laufens der Kopplungskonstante 
und ist aus diesem Grund sehr stabil bez"uglich Korrekturen 
der RG-Funktionen.
Allerdings erhalten diese Observablen gro"se Beitr"age aus dem 
niedrigenergetischen Bereich und von Kondensat-Termen mit Operatoren hoher
Dimension, welche v"ollig unbekannt sind.
Aus diesem Grund sind die Momente $(k,l)$ mit gro"sen Werten
f"ur $k$ nicht unter zuverl"assiger Kontrolle der St"orungstheorie,
wogegen sie vom experimentellen Standpunkt aus mit gr"o"serer 
Pr"azision messbar sind.
\subsection{Relation zwischen der 
$\MSsch$-Masse und den effektiven Massenparametern}
Das $\tau$-System wird in dem effektiven Schema durch zwei 
Massenparameter $m_{q,g}$ beschrieben.
Um einen Wert f"ur $m_s$ zu erhalten, der mit 
Resultaten aus anderen Bestimmungen verglichen werden 
kann, ist es sinnvoll, die nat"urlichen Massenparameter des $\tau$-Systems
$m_{q,g}$ durch den $\MSsch$-Massenparameter  $m_s$ auszudr"ucken.
Dies ist nur notwendig, um die mit $\tau$-Daten bestimmten Parameter
mit Ergebnissen aus anderen Experimenten zu vergleichen.
Grunds"atzlich werden Observable des $\tau$-Systems am besten durch 
ihre internen Parameter $a_E,\; m_{Eq}$ und $m_{Eg}$ beschrieben.
Relationen zwischen Observablen des $\tau$-Systems k"onnen ohne jede
Referenz zu Parametern des $\MSsch$-Schemas aufgestellt werden. 

Die St"orungstheorie wird durch eine Massenskala und die Koeffizienten der
RG-Funk"-ti"-onen parametrisiert, die die Entwicklung 
der Massenskala beschreiben.
Im masselosen Fall sind dies der Skalenparameter $\Lambda$ und
die Koeffizienten der $\beta$-Funktion \cite{stevenson,stevenson1}.
Betrachtet man die Massenkorrektur,
so gibt es zus"atzlich die invariante Masse $M$, die in  
(\ref{invmass}) definiert ist,
und die Koeffizienten der $\gamma$-Funktion.
Wie der Skalenparameter $\Lambda$, so kann auch die 
invariante Masse auf verschiedene Arten definiert werden.
Die konkrete Definition wird durch das Verhalten bei gro"sen Momenten 
bestimmt.
Auf diese Weise wird $\Lambda_{\MSsch}$ festgelegt.
Hier wird die invariante Masse $\mu$-unabh"angig durch 
\be 
\label{invmass}
M = \frac{m(\mu^2)}{a_E(\mu^2)^{\gamma_0/\beta_0}} 
\exp \Bigg\{- \int_0^{a_E(\mu^2)} 
\left( \frac{\gamma(\xi)}{\beta(\xi)} -
       \frac{\gamma_0}{\beta_0 \xi} 
\right) d \xi \Bigg\} 
\ee
definiert.
$M$ ist RG-invariant.
Das bedeutet, dass wenn $m(\mu^2)$ gem"a"s 
Gl.~(\ref{effmgg},\ref{effmqq})
transformiert wird, die "Anderung in Gl.~(\ref{invmass})
durch die entsprechende 
"Anderung der $\gamma$-Funktion (Gl.~(\ref{gammatransform}))
\footnote{Gl.~(\ref{gammatransform}) 
gibt das Transformationsverhalten der $\gamma$-Funktion 
f"ur den allgemeinen Fall an, 
in dem sowohl die Masse als auch die Kopplungkonstante
transformiert werden. 
Auch unter solchen allgemeineren 
Transformationen ist 
$M$ aus Gl.~(\ref{invmass}) invariant, wobei nat"urlich auch die 
$\beta$-Funktion gem"a"s Gl.~(\ref{gammatransform}) transformiert werden muss.}
kompensiert wird, so dass $M$ bis zu der Ordnung in der Kopplungskonstanten,
in der gerechnet wurde, unver"andert bleibt.
Die RG-Invarianz von M kann dazu verwendet werden, Relationen zwischen
Massen in verschiedenen 
Renormierungsschemata aufzustellen.
Nach dem Quadrieren von Gl.~(\ref{invmass})
ergibt sich eine Relation zwischen Massen in unterschiedlichen Schemata
mit den $\gamma$-Funktionen $\gamma(a)$ und $\gamma^\prime(a)$,
die beide in derselben Kopplungskonstante $a_E$ entwickelt werden.
In dieser Analyse verwenden wir die in Gl.~(\ref{effdef})
definierte effektive Kopplungskonstante 
und stellen eine Relation zwischen  
$m_n^2$ und ${m'}_s^2$ durch 
\be 
\label{relmass}
m_{En}^{2}(\mu^2) = {m'}_s^2(\mu^2) 
\exp \Bigg\{ -2 \int_0^{a_E(\mu^2)} 
\frac{\gamma'(\xi)-\gamma_n(\xi)}{\beta(\xi)} d\xi  
\Bigg\} 
\ee
auf.
Gl.~(\ref{relmass}) setzt zwei Massen mit einer Kopplungskonstante in Relation.
Aus diesem Grund muss die $\gamma$-Funktion der $\MSsch$-Masse 
mit dem effektiven Parameter $a_E$ ausgedr"uckt werden.

Es ist auch m"oglich, Gl.~(\ref{invmass}) 
im effektiven Schema und im $\MSsch$-Schema zu verwenden.
Dann l"asst sich die Relation zwischen den Massen 
$m_{E \,q,g}$ und $m_s$ direkt bestimmen, indem man 
die invariante Masse $M$ eliminiert. 
\ba
\label{relmass2}
m_n^2(\mu^2) &=& m_s^2(\mu^2) 
\left(\frac{a_{En}(\mu^2)}{a_{\MSsch}(\mu^2)}
\right)^{2 \gamma_0/\beta_0} \times \nn \\
&& \times \exp \Bigg\{-2\int_0^{a_{\MSsch }(\mu^2)} 
\left( \frac{\gamma_{\MSsch }(\xi)}{\beta_{\MSsch }(\xi)} 
- \frac{\gamma_0}{\beta_0 \xi } 
\right) d \xi \Bigg\}  \times \nn \\ 
&& \times \exp \Bigg\{ 2 \int_0^{a_{ n}(\mu^2)} 
\left( \frac{\gamma_{ n}(\xi)}{\beta_{ \rm{eff}}(\xi)} 
- \frac{\gamma_0}{\beta_0 \xi } \right) d \xi \Bigg\} 
\ea         
Hier ist $a_{En}$ f"ur die beiden Massen $m_{Eq}$ und $m_{Eg}$ gleich.
Au"serdem wird die abk"urzende  Schreibweise 
$a_{\MSsch} = \al_s/\pi$
verwendet. 
Beide Verfahren f"uhren zu 
"ahnlichen nummerischen Werten 
f"ur die Koeffizienten $\omega_{q}$ und $\omega_g$, welche die effektiven 
Massen $m_{E \,q,g}$ mit $m_s$ in Relation setzen.
Der nummerische Unterschied der sich aus 
Gl.~(\ref{relmass}) und Gl.~(\ref{relmass2}) ergebenden
Werte f"ur  $\omega_{q}$ und $\omega_g$
betr"agt weniger als 3\%, was der Abh"angigkeit der Resultate vom
Renormierungsschema entspricht.
Die Gleichungen~(\ref{relmass}, \ref{relmass2}) erlauben es, 
die internen Parameter $m_{Eq}$ und $m_{Eg}$ durch den 
Standard-Parameter im $\MSsch$-Schema auszudr"ucken
(siehe Gl.~(\ref{effmgg}, \ref{effmqq}) und (\ref{gammagf}, \ref{gammaqf})).
F"ur die Koeffizienten, die die effektiven Massen $m_{Eq}$ und $m_{Eg}$  
mit der Referenzmasse im $\MSsch$-Schema $m_s$ in Verbindung setzen,
ergibt sich
\ba
m_q^2 &=&\omega_q m_s^2\, , 
\quad \quad\quad m_g^2=\omega_g m_s^2 \,, \nn \\
\omega_q &=&  1.73 \pm 0.04 , \quad \omega_g = 1.42 \pm 0.03\;.
\ea
Die Koeffizienten $\omega_{q}$ und $\omega_g$ sind nicht nahe bei 1, 
was zeigt, dass
das $\MSsch$-Schema relativ unnat"urlich f"ur die Beschreibung des 
$\tau$-Systems ist.
Der FOPT-Ausdruck f"ur die Koeffizienten $\omega_{q}$ und $\omega_g$ 
aus Gl.~(\ref{effmgg}, \ref{effmqq}) konvergiert nicht gut.
Dies macht 
die RG-Umrechnung mit Gl.~(\ref{relmass}, \ref{relmass2}) notwendig.
\subsection{Kondensatkorrekturen der Dimension $D=4$}
\begin{figure}[h]
\begin{center}
       \epsfig{file=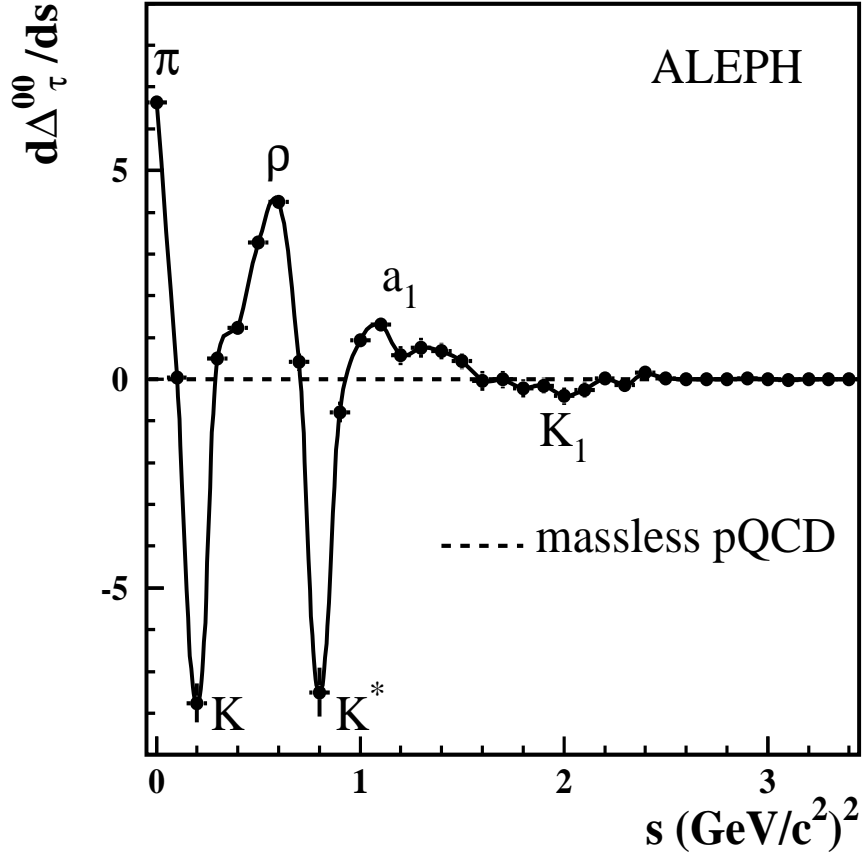,angle=0,width=.7
                  \textwidth,height=.7\textwidth}
\end{center}
       \caption{\label{diffrate} Experimenteller Integrand 
$d(\delta \! R_\tau)/ds$ f"ur Gl.~(\ref{diffmoments}) \protect\cite{exp1ms}} 
\end{figure}

Um den Massenparameter $m_s$ zu bestimmen, ben"otigt man nicht alle
($D=4$)-Kondensat"-kor"-rek"-turen zu der totalen Zerfallsrate, 
sondern nur diejenigen, die in die Differenz 
der Zerfallsraten in Hadronen ohne und mit Strangeness 
eingehen:
\be \label{diffmoments}
\delta \! R_\tau^{kl} 
= \frac{R_{\tau s=0}^{kl}}{|V_{ud}|^2} 
- \frac{R^{kl}_{\tau s=1}}{|V_{us}|^2}\, .
\ee
Hierbei  ist $R_{\tau s=0,1}^{kl}$ als 
\be
R_{\tau s=0,1}^{kl} 
= \int_0^{\mts} ds \left(1- \frac{s}{\mts} \right)^k
\left( \frac{s}{\mts} \right)^l 
\frac{d R_{\tau s = 0,1}}{ds} 
\ee
definiert und 
$(dR_{\tau s = 0,1}/ds)$ ist die differentielle $\tau$-Zerfallsrate 
in Hadronen mit Strangeness
$0,1$ und Energie $\sqrt{s}$. 
    \begin{figure}[h]
\begin{center}
       \epsfig{file=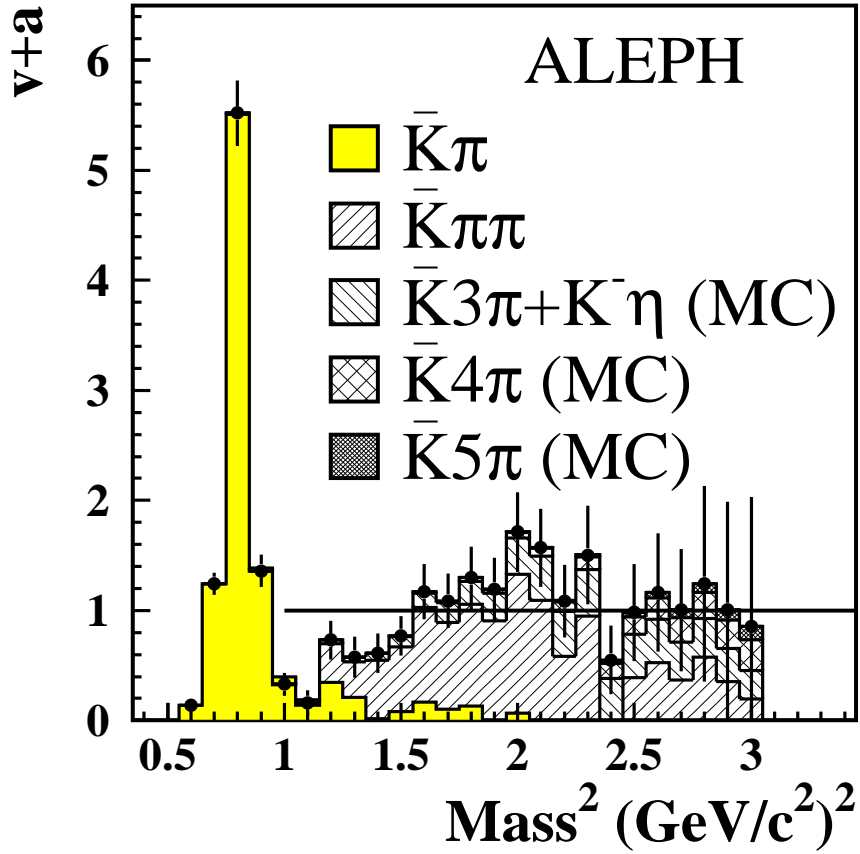,angle=0,width=.7
                  \textwidth,height=.7\textwidth}
\end{center}
       \caption{\label{s1rate} Aus $\tau$-Zerf"allen in Hadronen mit 
Strangeness 
bestimmte Vektor-$(v)$-plus-Axialvektor-$(a)$-Spektraldichte {\protect\cite{exp1ms}}. Die mit (MC)
bezeichneten Daten  stammen aus Monte-Carlo-Simulationen}
\end{figure} 
Der experimentelle Integrand f"ur die Berechnung der Momente 
in Gl.~(\ref{diffmoments}) ist in Abbildung~\ref{diffrate} dargestellt.
Vernachl"assigt man Terme der Ordnung 
$m_s^3/M_\tau^3$ und setzt die Up- und Down-Quarkmassen 
gleich null,
so treten  nur ein in $m_S$ linearer und ein in $m_s$ quadratischer 
Term in dem theoretischen Ausdruck
f"ur die 
Differenz $\delta \! R_\tau^{kl}$ 
auf, n"amlich
\be \label{finelres}
\delta \! R_\tau^{kl} 
= N_c S_{EW} \left( \frac{m_s^2}{M_\tau^2} 6 F_{kl}
 - \frac{m_s}{M_\tau} 
\frac{\langle \bar s s \rangle}{M_\tau^3} 
           4 \pi^2 T_{kl} \right) \;,
\ee
wobei $N_c=3$ und $S_{EW} = 1.0194$ 
die elektroschwachen Korrekturen \cite{ewcorr1,ewcorr2} beschreiben.
Die ($D=4$)-Korrektur wird in der f"uhrenden Ordnung 
im $\MSsch$-Schema verwendet.
Korrekturen h"oherer Ordnung werden nicht beachtet.
Verwendet man die Zerlegung der Korrelationsfunktion in
den (L+T)- und den (L)-Teil, so ergibt sich f"ur die in 
$m_s$ lineare Quark-Kondensat-Korrektur zu der Korrelationsfunktion
\cite{BraNarPic92}
\ba
 \Pi_{us(L+T)}^{V/A\, d=4} (Q^2) 
        &=& \frac{4 \pi^2}{Q^4} m_s \langle \bar{s} s \rangle \;, \nn \\
  \Pi_{us(L)}^{V/A \, d=4} (Q^2) 
        &=& \frac{4 \pi^2}{Q^4} m_s \left( \langle \bar{s} s \rangle     
    \mp \langle \bar{u} u \rangle \right) \;,
\ea
wobei $m_{u,d}$ und die QCD-Korrekturen der Koeffizientenfunktionen 
vernachl"assigt wurden.
Die entsprechende Korrektur f"ur die Momente der Zerfallsrate 
ergibt sich mit Gl.~(\ref{LT}) und $x=Q^2/\mts$ zu
\ba
\frac{1}{|V_{us}|^2 N_c S_{EW} } \, R_{\tau s=1}^{kl (D=4)} 
&=&  4 \pi^2 \frac{m_s}{M_\tau^4}
 (-1)^l \frac{-i }{2 \pi} \sum_{V/A} \oint
             \left( 1 + x \right)^{k+2} x^{l-2} \times \nn \\
    &&  \times \left\{ \left(1 -2 x \right) \langle \bar{s} s \rangle
       + 2 x \left\{ 
                  \begin{array}{cc}
          \langle \bar{s} s \rangle + \langle \bar{u} u \rangle & (V) \\
          \langle \bar{s} s \rangle - \langle \bar{u} u \rangle & (A) 
                  \end{array}
             \right\} \right\} dx \nn \\
 &=& 4 \pi^2 \frac{m_s}{M_\tau} \frac{\langle \bar{s} s \rangle}{M_\tau^3}
     2 (-1)^l \frac{-i}{2 \pi} \oint \left( 1+x \right)^{k+2} x^{l-2} dx \nn \\
 &=& 4 \pi^2 \frac{m_s}{M_\tau} \frac{\langle \bar{s} s \rangle}{M_\tau^3}
     2 (-1)^l  \frac{-i}{2 \pi} \oint \left( x^{l-2} + (k+2) x^{l-1} 
                                   + \mathcal{O}(x^l) \right) dx \nn \\
 &=& 4 \pi^2 \frac{m_s}{M_\tau} \frac{\langle \bar{s} s \rangle}{M_\tau^3}
       2 \left( \delta_{l,0}(k+2) - \delta_{l,1} \right)\;,
\ea
wobei im letzten Schritt der Residuensatz angewendet wurde, um die Integration
auszuf"uhren.
Aus diesem Resultat ergibt sich f"ur den 
Koeffizienten $T_{kl}$ der Kondensatkorrektur in der LO-Approximation 
\be
T_{kl} = 2\left( \delta_{l,0} (k+2) - \delta_{l,1} \right)\;.
\ee
F"ur die ersten Momente sind die nummerischen Werte 
der Koeffizienten vor dem Kondensat-Term
\be
T_{00} = 4, \quad T_{10}  = 6, \quad
T_{20} = 8,  \quad  T_{01} = -2, \quad T_{11} = -2\;.  
\ee
\subsection{Nummerische Resultate}
Alle Gr"o"sen f"ur die Evaluierung von Gl.~(\ref{finelres})
sind jetzt bekannt.
In Tabelle~\ref{T1} sind die Koeffizienten 
der $q$- und $g$-Masssenparameter aus Gl.~(\ref{fincor})
angegeben. Das Moment $(0,0)$ ist von dem Standpunkt
der St"orungstheorie das zuverl"assigste. Es verh"alt sich
sehr stabil bez"uglich der QCD-Korrekturen.
%
%
%

Die Koeffizienten $A_{kl}$ enthalten die QCD-Korrekturen des
q-Teils der Korrelationsfunktion. Dieser Teil enth"alt 
Beitr"age von Spin-0- und Spin-1-Teilchen.
Der Spin-0-Teil enth"alt m"ogliche nichtperturbative Beitr"age
von Instantonen und es ist zu erwarten, dass die St"orungstheorie
in diesem Teil fr"uher zusammenbricht.
Der g-Teil enth"alt nur Beitr"age von Spin-1-Teilchen. Er verh"alt sich 
regelm"a"siger, was man an den Koeffizienten $B_{kl}$ ablesen kann.
\begin{table}
\[ 
\begin{array}{|c|ccc|ccc|}
\hline 
(k,l) & A_{kl}^{\rm{LO}} & A_{kl}^{\rm{NLO}}  
& A_{kl}^{\rm{NNLO}} & B_{kl}^{\rm{LO}} & B_{kl}^{\rm{NLO}}  
& B_{kl}^{\rm{NNLO}}   \nn \\ \hline
(0,0)    & 1.361   &   1.445     &1.434   & 0.523 & 0.601  & 0.625  \\
(1,0)    & 1.568   &   1.843     & 1.976  & 0.441 & 0.552  & 0.601  \\
(2,0)    & 1.762   &   2.282     & 2.646  & 0.390 & 0.530  & 0.607 \\ \hline  
(0,1)    &- 0.207  & - 0.398     & -0.542 & 0.082 & 0.050  & 0.025 \\  
\hline  
\end{array}
\]
 \caption{\label{T1} Koeffizienten aus Gl.~(\ref{fincor})}
\end{table}
Die Koeffizienten $F_{kl}$ des $m_s^2$-Terms in Gl.~(\ref{finelres})
ergeben sich aus Gl.~(\ref{fincor}).
Der relevante Teil der Gleichung ist
\be
\label{abfdef}
F_{kl} =\omega_q A_{kl} + \omega_g B_{kl} .
\ee
F"ur den in $m_s$ linearen Term in Gl.~(\ref{finelres})
ergibt sich der Koeffizient aus einem ph"anomenologischen Wert
f"ur das Strange-Quark-Kondensat. F"ur die nummerische Analyse wird  
\be
\langle \bar s s \rangle 
= (0. 8 \pm 0.2 ) \langle \bar u u \rangle 
\ee 
verwendet \cite{narga,gammass,newgam}, wobei 
$\langle \bar u u \rangle = - (0.23~\rm{GeV} )^3$ ist.
Setzt man die Resultate f"ur $F_{kl}$ und den Kondensat-Term in 
Gl.~(\ref{finelres}) ein, so erh"alt man f"ur $X=m_s/(130~{\rm MeV})$
die Gleichung
\be
\label{finaleqms}
\frac{1}{N_c S_{EW}}\left(\frac{M_\tau}{130~{\rm MeV}}\right)^2
   \delta \! R_\tau^{kl} 
= X \left(6 F_{kl} \cdot X + 0.936 \cdot T_{kl}\right)\ .
\ee
Der Dimension-4-Kondensat-Term tr"agt zwischen 10\%
und 15\% zu dem gesamten theoretischen Resultat bei.
Die St"orungsreihe f"ur den Koeffizienten des Dimension-4-Terms
konvergiert gut. Die erste Korrektur in der FOPT ist von der Ordnung  
$\al_s^2$ mit einem kleinen Koeffizienten.
Diese Korrektur wird aufgrund des gro"sen Fehlers in dem nummerischen 
Wert des Strange-Quark-Kondensats nicht verwendet.

Es werden die experimentellen Daten der ALEPH-Kollaboration verwendet
\cite{exp1ms}. Die Resultate f"ur $m_s(\mts)$ aus unterschiedlichen
Momenten $\delta \! R_\tau^{kl}$ sind in Tabelle~\ref{T2} gegeben.
F"ur die Bestimmung von $m_s$ wird nur das Moment $(0,0)$
verwendet. 
\begin{table}
\[
\begin{array}{|c|c|c|}
\hline
(k,l) & \delta \! R_\tau^{kl} & m_s(\mts) \;\; \rm{[MeV]} \\ \hline
(0,0) & 0.394 \pm 0.137 & 130 \\
(1,0) & 0.383 \pm 0.078 & 111 \\
(2,0) & 0.373 \pm 0.054 &  95 \\ \hline
\end{array}
\]
 \caption{\label{T2} Resultate f"ur $m_s^2(\mts)$ 
aus verschiedenen Momenten von $\delta \! R_\tau$}
\end{table}
H"ohere Momente mit der Wichtungsfunktion $(1-s/M_\tau^2)^k$
sind weniger sensitiv auf das Laufen der Parameter  
(siehe Abb.~\ref{runc} und \ref{runmCq})
und haben unkontrollierbare Beimischungen 
von Kondensat-Termen h"oherer Dimension, die diese Momente 
stark nicht-perturbativ machen.
Das Endresultat f"ur den Zentralwert der Strange-Quarkmasse ist
\be
\label{finaleqms00}
\frac{1}{N_c S_{EW} }\left(\frac{M_\tau}{130~{\rm MeV}}\right)^2
\delta \! R_\tau^{00} 
=24.1 =  X \left(20.2 X + 3.74 \right) \;.
\ee
Die Gleichung wird von $X=m_s/(130~{\rm MeV})=1.00$
gel"ost (mit zwei Dezimalstellen Genauigkeit).
Dies f"uhrt zu der Vorhersage f"ur die Strange-Quarkmasse
bei $\mts$. Au"serdem wird der Wert f"ur $m_s(1 \rm{~GeV})$
angegeben. Dieser ergibt sich aus $m_s(\mts)$ mit dem durch 
die Vier-Schleifen-$\gamma$-Funktion bei drei Flavours ($n_f=3$)
gegebenen Laufen der Masse:
\ba 
m_s(M_\tau) &=& (130 \pm 27_{\rm{exp}} \pm 3_{\langle\bar{s}s \rangle} 
\pm 6_{\rm{th}})\;\rm{MeV}  \nn\\  
m_s(1\rm{~GeV}) 
 &=& (176 \pm 37_{\rm{exp}} \pm 4_{\langle\bar{s}s \rangle} 
\pm 9_{\rm{th }})\; \rm{MeV} \;.   
\ea
Das Endresultat f"ur den Koeffizienten vor der $m_s^2$-Korrektur
stimmt mit dem in \cite{msNPB} im $\MSsch$-Schema bestimmten 
Wert innerhalb einer Standardabweichung "uberein.
In dieser Analyse ergab sich $20.2\pm 2$ f"ur den   
Koeffizienten vor der $m_s^2$-Korrektur
w"ahrend die fr"uhere Absch"atzung $18.1\pm 2.6$ \cite{msNPB} betrug.
Der hier gegebene Wert ist das Resultat einer 
bez"uglich des Konvergenzverhaltens der St"orungsreihen 
sehr viel genaueren Analyse.
Die Interpretation der St"orungstheorie in einer geschlossenen Form
erlaubt es, die theoretische Ungenauigkeit des nummerischen 
Wertes der Strange-Quarkmasse zu reduzieren.

Der resultierende Wert der Strange-Quarkmasse liegt nahe bei den
"alteren Absch"atzungen aus 
\cite{narma,gassLeut,oldmass,oldmass1},
die weniger genaue theoretische Formeln und weniger pr"azise 
experimentelle Daten verwenden.
Die neuen Analysen auf der Basis von (pseudo)skalaren Korrelationsfunktionen
ergeben Werte, die nahe an dem hier erzielten Wert liegen
\cite{recent1,recent2}.
In Gitterberechnungen der Strange-Quarkmasse \cite{lattice1,lattice2} 
ist der theoretische Input nicht-perturbativer Natur.
Die j"ungsten Resultate auf der Basis von Gitterberechnungen sind kleiner 
als der hier bestimmte Wert, befinden sich allerdings noch innerhalb des 
durch die angegebenen Fehler gegebenen Intervalls.
F"ur die Strange-Quarkmasse wird ein Wert von 
$m_s^{\rm Lat }(2\, {\rm GeV}) = (97 \pm 4) {\rm MeV}$ \cite{lattice1}
vorhergesagt, was sich in 
$m_s^{\rm Lat }( \mt ) = (101 \pm 4)  {\rm MeV}$ bei der Masse des
$\tau$-Leptons "ubersetzt.
F"ur die in Gl.~(\ref{invmass}) definierte invariante Masse
ergibt sich der folgende Wert:
\be
M=(312 \pm 65_{\rm{exp}} \pm 7_{\langle\bar{s}s \rangle} 
\pm 14_{\rm{th}})\;\rm{MeV} \ .
\ee
\section{Zusammenfassung von Kapitel \ref{Art4}}
Es wurden die $m_s^2$-Korrekturen in einer auf der QCD basierenden 
Beschreibung des $\tau$-Systems untersucht.
Hierf"ur wird ein nat"urliches effektives Schema verwendet, das f"ur 
die Analyse des $\tau$-Systems mit der konturverbesserten St"orungstheorie
(CIPT) geeignet ist.
Die Qualit"at der Resultate wird durch die $\beta$- und $\gamma$-Funktionen
bestimmt, die die einzigen perturbativen Objekte der Analyse sind.
Die $\gamma_q$-Funktion zeigt bereits in der NNLO-N"aherung asymptotisches 
Wachstum, wogegen die $\beta$- und $\gamma_g$-Funktion in dieser Ordnung noch 
``konvergieren''. In der Diskussion der \no-Terme dieser Funktionen 
(welche von dem unbekannten Parameter $k_3$ abh"angen), ergaben sich 
starke Hinweise f"ur asymptotisches Wachstum in dieser Ordnung.
Dies zeigt, dass die ultimative Grenze der st"orungstheoretischen 
Genauigkeit f"ur diesen Satz von $\tau$-Lepton-Observablen 
bereits erreicht ist.  
Gleichwohl er"offnet diese Prozedur die M"oglichkeit,  
$\tau$-System-interne 
QCD-Tests mit einem gro"sen Grad an Genauigkeit durchzuf"uhren.
Dies ist aufgrund der noch nicht ausreichend guten Genauigkeit der 
experimentelln Daten f"ur Cabbibo-unterdr"uckte $\tau$-Zerf"alle 
noch kein aktuelles Problem, insbesondere wenn man dem hochenergetischen Teil
der differentiellen Zerfallsbreite betrachtet, f"ur den die theoretischen
Formeln besser konvergieren.
Die experimentelle Situation kann sich allerdings bald "andern.
Auf diesem Gebiet kann der Ansatz des effektiven Renormierungsschemas 
seine wirkliche Leistungsf"ahigkeit zeigen, da die Hauptquelle
der theoretischen Ungenauigkeit, das Umrechnen der effektiven internen
Parameter  $(m_{Eq},m_{Eg})$ in den $\MSsch$-Schema-Parameter 
$(m_s)$ entf"allt.
F"ur einen QCD-Test dieser Art m"ussen Relationen zwischen mindestens vier
$\tau$-Lepton-Observablen aufgestellt werden, um die drei Parameter 
$a_E, \; m_{Eq}$ und $m_{Eg}$ festzulegen.

F"ur die Strange-Quarkmasse ergab sich  
$m_s(\mts)  = (130 \pm 27_{\rm{exp.}} \pm 9_{\rm{th.}})\; \rm{MeV}$,
was 
$m_s(1~\rm{GeV})  = (176 \pm 37_{\rm{exp.}} \pm 13_{\rm{th.}})\; \rm{MeV}$
entspricht.
Dies ist konsistent mit vorherigen Resultaten, 
bei denen Resummation im $\MSsch$-Schema
verwendet wurde \cite{msNPB}.
Die Hauptursache der theoretischen Ungenauigkeit des Ergebnisses ist 
das Umrechnen von effektiven in $\MSsch$-Parameter.
Der Vorteil der hier pr"asentierten Methode liegt darin, dass 
die Absch"atzung der Genauigkeit nicht auf der Zerlegung 
des Resultats in Beitr"age aus QCD-Korrekturen der Korrelationsfunktion
im $\MSsch$-Schema beruht. Diese Zerlegung scheint unnat"urlich  
f"ur die resummierte St"orungstheorie zu sein, da sie eine zus"atzliche
Ungenauigkeit durch die Konvergenz der Korrelationsfunktion 
im $\MSsch$-Schema einf"uhrt.
Durch das Verfahren im effektiven Renormierungsschema werden alle 
Fehlerquellen
aufgrund der Konvergenz der St"orungstheorie 
in den $\beta$- und $\gamma$-Funktionen gesammelt, was eine 
solide Grundlage f"ur die Absch"atzung der erzielten Genauigkeit bietet.

\chapter{Zusammenfassung}
In der ersten H"alfte der Arbeit (Kapitel \ref{QCDkap} - \ref{kf}) 
wurden die notwendigen theoretischen 
Grundlagen dargestellt: die Grundz"uge der Renormierung (Kapitel \ref{QCDkap}), 
die Renormierungsgruppe (Kapitel \ref{rg}),
die Herleitung der $\tau$-Zerfallsrate (Kapitel \ref{tautheo}) 
und die Berechnung der Korrelationsfunktion 
im Rahmen der Operator-Produkt-Entwicklung (Kapitel \ref{kf}). 
Die QCD-Analyse der hadronischen Zerf"alle des $\tau$-Leptons wurde in 
der zweiten H"alfte der Arbeit (Kapitel \ref{Art1} - \ref{Art4}) in vier
Schritten durchgef"uhrt.

Im ersten Schritt (Kapitel \ref{Art1}) wurde die St"orungsreihe f"ur 
$\tau$-Lepton-Zerf"alle
in der N"aherung masseloser Quarks 
untersucht.
Hierf"ur wurde ein effektives Renormierungsschema verwendet, in dem 
die effektive Kopplungskonstante 
durch die 
Spektralfunktion des masselosen
Teils der Korrelationsfunktion definiert wurde.
Es zeigte sich, dass im masselosen Teil die Momente der Spektraldichte in der 
f"unften Ordnung der St"orungstheorie asymptotisch anzuwachsen beginnen, 
falls das nullte Moment der Spektraldichte nicht aus dem Satz der Observablen 
ausgeschlossen wird. 

Im zweiten Schritt (Kapitel \ref{Art2})wurde die quadratische Massenkorrektur untersucht.
Es wurden zwei effektive Massen durch den q- und den g-Teil der Korrelationsfunktion
so definiert, dass alle QCD-Korrekturen sich aus dem Laufen dieser Massen ergeben.
Die St"orungsreihen der Massenkorrektur zeigen bereits in der vierten Ordnung 
asymptotisches Wachstum.
Diese Resultate bez"uglich des asymptotischen Anwachsens 
sind unabh"angig von der Wahl des Renormierungsschemas 
und dem nummerischen Wert des unbekannten Vier-Schleifen-Parameters der
Korrelationsfunktion $k_3$.
Aufgrund des gefundenen asymptotischen Wachstums der St"orungsreihen
und der damit verbundenen Limitierung der theoretischen Genauigkeit
ist die Anwendung eines Resummationsverfahrens unausweichlich,
falls man eine theoretische Pr"azision erreichen will, 
die mit den experimentellen 
Daten vergleichbar ist.

Die experimentellen Daten f"ur Zerf"alle von $\tau$-Leptonen bieten die 
M"oglichkeit, die starke Kopplungskonstante und die Strange-Quarkmasse zu 
bestimmen. 
Im dritten Schritt (Kapitel \ref{Art3}) wurde die starke Kopplungskonstante $\al_s$ aus Cabbibo-bevorzugten
$\tau$-Lepton-Zerf"allen bestimmt.
Die bisher mit 
dem Standardverfahren und
St"orungstheorie in endlicher Ordnung 
gefundenen nummerischen Werte
f"ur die Kopplungskonstante $\al_s$ ergaben nach dem Laufen zu der 
Referenzskala $M_Z$
einen signifikant gr"o"seren 
Wert als Resultate aus Experimenten bei h"oheren Energien.
In dieser Arbeit wurde zur Extraktion der 
Kopplungskonstante aus der St"orungsreihe
ein neues renor"-mierungs"-gruppen-invariantes Verfahren verwendet, 
dass die Existenz der renormierungs"-gruppen-invarianten 
QCD-Skala nutzt.    
Die so erzielten nummerischen Resultate f"ur $\al_s(M_z)$ 
sind signifikant kleiner 
als die vorherigen und somit mit den Werten f"ur $\al_s$ aus 
anderen Experimenten besser vertr"aglich.
Das Resultat f"ur die starke Kopplungskonstante ist:
\[
\al_s(M_Z) = 0.1177 \pm 0.0007_{exp} \pm 0.0006_{hq\; mass} \, .
\]

Im vierten Schritt (Kapitel \ref{Art4}) der Analyse wurde die Strange-Quarkmasse bestimmt.
Aufgrund des bereits im zweiten Schritt gefundenen schlechten 
Konvergenzverhaltens der St"orungsreihen f"ur die quadratische Massenkorektur 
wird f"ur die Bestimmung 
der Strange-Quarkmasse $m_s$ konturverbesserte St"orungstheorie (CIPT)
in Verbindung mit einem effektiven Renormierungsschema 
verwendet.
Bei dieser Vorgehensweise sind die einzigen st"orungstheoretischen  
Objekte die effektive $\beta$-Fuktion und zwei effektive $\gamma$-Funktionen.
F"ur die Bestimmung der  Referenzmasse des Strange-Quarks im $\MSsch$-Schema
aus den effektiven Massenparametern wird 
in Analogie zu der Extraktion
von $\al_s$
im dritten Schritt die RG-invariante Masse verwendet.
Es zeigt sich, dass das Umrechnen von effektiven Parametern
in $\MSsch$-Parameter die 
Hauptquelle f"ur den theoretischen Fehler von $m_s$ ist. 
Aus diesem Grund sind genauere $\tau$-Lepton-interne QCD-Tests mit der 
hier verwendeten Methode zu erwarten, sobald die experimentelle 
Genauigkeit diese zul"asst.
F"ur die Strange-Quarkmasse ergab sich in "Ubereinstimmung mit den 
vorhergegangenen Resultaten aus \cite{narma,gassLeut,oldmass,oldmass1},
neuen Analysen auf der Basis von (pseudo)skalaren Korrelationsfunktionen
\cite{recent1,recent2},
und Gitterberechnungen \cite{lattice1,lattice2}:
\ba  
m_s(\mts)  &=& (130 \pm 27_{\rm{exp.}} \pm 9_{\rm{th.}})\; \rm{MeV} \;,\nn \\
m_s(1~\rm{GeV})  &=& (176 \pm 37_{\rm{exp.}} \pm 13_{\rm{th.}})\; \rm{MeV} 
\;.\nn
\ea

\begin{appendix}
\chapter{Feynman-Regeln der Quantenchromodynamik} \label{Frules}
Die Lagrangefunktion der QCD ist
\be
\mathcal{L} = -\frac{1}{4} G^a_{\mu \nu} G_a^{\mu \nu}
      + \sum_f \bar{\Psi}_f (i \shd -m_f) \Psi_f
             + \mathcal{L}^{GF} + \mathcal{L}^{FP} \;.
\ee
Hierbei bezeichnen $ G^a_{\mu \nu}$ , $\shd$ und $ \nabla_\mu^{ab}$ den 
Gluon-Feldst"arketensor,
die kovariante Ableitung und die adjungierte kovariante Ableitung. 
$\Psi_f$ steht f"ur ein Farbtripel
aus Quarkspinoren der Flavours $f$.
\ba
 G^a_{\mu \nu} &=&  
\partial_\mu A_\nu^a - \partial_\nu A_\mu^a 
+ g \mu^\ep f^{abc} A_\mu^b A_\nu^c \label{a2}\\
D_\mu &=&  \partial_\mu - i g \mu^\ep A_mu^a \frac{\lambda^a}{2} \label{a3}\\
\nabla^{ab}_\mu &=& \delta^{ab} \partial_\mu - g \mu^\ep f^{abc} A^c \label{a4}
\ea
Um die Kopplungskonstante bei der Benutzung der dimensionalen Regularisierung 
dimensionslos zu erhalten, 
ist in Gl.~(\ref{a2}, \ref{a3}, \ref{a4}) bereits die Massenskala $\mu$ 
eingef"uhrt worden und tritt auch in den Feynman-Regeln 
bei der Kopplungskonstante auf. 
$f^{abc} $ sind die Strukturkonstanten der Eichgruppe $SU(3)$.
$\mathcal{L}^{GF}$ bezeichnet den Term, der die Eichung fixiert,
\be
\mathcal{L}^{GF} = - \frac{1}{2 \zeta} (\partial_\mu A^\mu )^2 \;
\ee
und $\mathcal{L}^{FP}$ die Lagrangefunktion der Fadeev-Popov-Geister-Felder
\be
\mathcal{L}^{FP} = (\partial^\mu \bar{C}^a) \nabla_\mu^{ab} C^b \;.
\ee
Aus der Lagrangefunktion ergeben sich die folgenden 
Feynman-Regeln 
f"ur die pertubative Berechnung von Greenfunktionen \cite{pessch}:

\begin{tabular}{lcl}
\raisebox{1.8ex}[-1.8ex]{Quarkpropagator} &
\begin{picture}(70,25)
\ArrowLine(10,12)(60,12)
\Vertex(10,12){2}
\Vertex(60,12){2}
\Text(35,20)[]{p}
\Text(12,5)[]{A,i}
\Text(58,5)[]{B,j}
\Text(12,20)[]{$\alpha$}
\Text(58,20)[]{$\beta$}
\end{picture}
 &
\raisebox{1.8ex}[-1.8ex]{$ \displaystyle{\delta_{AB} \delta_{ij} i \left( \frac{\shp +m}{p^2 -m^2 + i \ep} 
   \right)_{\alpha \beta}  }  $}
\\ 
\\
\raisebox{1.8ex}[-1.8ex]{Gluonpropagator} 
 &
\begin{picture}(70,25)
\Gluon(10,12)(60,12){2}{7}
\Vertex(10,12){2}
\Vertex(60,12){2}
\Text(35,20)[]{k}
\Text(12,5)[]{a}
\Text(58,5)[]{b}
\end{picture}
&
\raisebox{1.8ex}[-1.8ex]{$\displaystyle{ -i \delta_{ab} 
      \left( \frac{g_{\mu \nu}}{k^2 + i \ep} - (1- \zeta)
           \frac{k_\mu k_\nu}{(k^2 + i \ep)^2 }\right)} $} 
\\
\\   
\raisebox{1.8ex}[-1.8ex]{Geistfeldpropagator}
&
\begin{picture}(70,25)
\DashArrowLine(10,12)(60,12){4}
\Vertex(10,12){2}
\Vertex(60,12){2}
\Text(35,20)[]{k}
\Text(12,5)[]{a}
\Text(58,5)[]{b}
\end{picture}
&
\raisebox{1.8ex}[-1.8ex]{ $\displaystyle{\frac{i \delta_{ab}}{k^2 + i \ep}} $ }
\\
\\ 
\raisebox{5ex}[-5ex]{Dreigluonvertex}  
&
\begin{picture}(70,70)(0,0)
\Vertex(35,35){2}
\Gluon(35,35)(35,70){2}{7}
\Gluon(0,0)(35,35){2}{7})
\Gluon(35,35)(70,0){2}{7}
\Text(20,5)[]{$\nu b $}
\Text(50,5)[]{$\rho c $}
\Text(5,20)[]{$p$}
\Text(65,20)[]{$q$}
\Text(22,63)[]{$k$}
\Text(45,63)[]{$a$}
\end{picture}
& 
\raisebox{5ex}[-5ex]{$\displaystyle{ g\mu^\ep  V_{abc}^{\mu \nu \rho}(p,q,r)} $ }
\\
\\
\raisebox{5ex}[-5ex]{Viergluonvertex}
&
\begin{picture}(70,70)(0,0)
\Vertex(35,35){2}
\Gluon(0,0)(35,35){2}{7}
\Gluon(35,35)(70,70){2}{7}
\Gluon(0,70)(35,35){2}{7}
\Gluon(35,35)(70,0){2}{7}
\Text(20,5)[]{$\sigma d $}
\Text(50,5)[]{$\rho c $}
\Text(20,65)[]{$\mu a $}
\Text(50,65)[]{$\nu b $}
\end{picture}
& 
\raisebox{5ex}[-5ex]{$\displaystyle{ - i g^2\mu^{2 \ep} W_{abcd}^{\mu \nu \rho \sigma}} $}
\\
\\
\raisebox{5ex}[-5ex]{Quark-Gluonvertex} 
&
\begin{picture}(70,70)(0,0)
\Vertex(35,35){2}
\Gluon(35,35)(35,70){2}{7}
\ArrowLine(0,0)(35,35)
\ArrowLine(35,35)(70,0)
\Text(8,25)[]{$p, \alpha$}
\Text(64,25)[]{$ p^\prime, \beta$}
\Text(15,5)[]{$A,i$}
\Text(53,5)[]{$B,j$}
\Text(25,63)[]{$k$}
\Text(45,63)[]{$a$}
\end{picture}
& 
\raisebox{5ex}[-5ex]{$\displaystyle{ i  \delta_{ij}g \mu^\ep (\gamma^\mu)_{\alpha \beta}  T_{AB}^a} $} 
\\
\\
\raisebox{5ex}[-5ex]{Geist-Gluonvertex } 
&
\begin{picture}(70,70)(0,0)
\Vertex(35,35){2}
\Gluon(35,35)(35,70){2}{7}
\DashArrowLine(0,0)(35,35){4}
\DashArrowLine(35,35)(70,0){4}
\Text(8,25)[]{$p$}
\Text(62,25)[]{$ q$}
\Text(15,5)[]{$b$}
\Text(55,5)[]{$c$}
\Text(25,63)[]{$k$}
\Text(45,63)[]{$a$}
\end{picture}
& 
\raisebox{5ex}[-5ex]{$\displaystyle{  g \mu^\ep  f_{abc} p_\mu}$}
\end{tabular}

Die Funktionen $V_{abc}^{\mu \nu \rho}(p,q,r)$ und $ W_{abcd}^{\mu \nu \rho \sigma}$
sind folgenderma"sen definiert: 
\be
V_{abc}^{\mu \nu \rho}(k,p,q) = f_{abc} \bigg( g^{\mu \nu} (k-p)^\rho 
         + g^{\nu \rho}(p-q)^\mu + g^{\rho \mu}(q-k)^\nu \bigg) \nn
\ee
\ba
W_{abcd}^{\mu \nu \rho \sigma} &=& 
           f_{eab} f_{ecd} (g^{\mu \rho} g^{\nu \sigma} - g^{\mu \sigma} g^{\nu \rho} ) \nn \\
      &&  +  f_{eac} f_{edb} (g^{\mu \sigma} g^{\rho \nu} - g^{\mu \nu} g^{\rho \sigma})\nn \\\ 
    && + f_{ead} f_{ebc} (g^{\mu \nu} g^{ \sigma \rho} - g^{\mu \rho} g^{\sigma \nu}). \nn
\ea

Die Indizes $(a,b,...)$ laufen von $1$ bis $8$ und transformieren sich mit der 
achtdimensionalen adjungierten Darstellung der Eichgruppe SU(3).
Die Indizes $(A,B,...)$ laufen von $1$ bis $3$ und bezeichnen die 
Farbladung. Sie transformieren mit der dreidimensionalen 
Standarddarstellung der Gruppe 
SU(3).
$i,j$ bezeichnen das Flavour der Quarks.
Um Greenfunktionen zu berechnen, muss folgenderma"sen vorgegangen werden:
\begin{enumerate}
\item Zeichnen aller 
m"oglichen topologisch unterschiedlichen Diagramme.
\item An jedem Vertex wird durch einen Faktor $(2 \pi)^d \delta(p_{\rm in} - p_{\rm out})$
die Energieerhaltung gew"ahrleistet.
\item "Uber interne Schleifenmomente muss mit dem Gewicht $ \int \frac{d^dp}{(2 \pi)^d} $
integriert werden, wobei die Divergenzen der Integrale 
angemessen regularisiert werden m"ussen, z.B. durch dimensionale Regularisierung.
\item F"ur jede Fermion-  und Geistschleife muss das Diagramm mit einem zus"atzlichen 
Faktor $(-1)$ multipliziert werden.
\item Das Diagramm erh"alt einen Symmetriefaktor \cite{pessch}.
\end{enumerate}

\chapter{Die Gegenbauer-Polynom-Technik} \label{AGPXT}
Die Gegenbauer-Polynom-Technik (GPXT) \cite{GPXT} stellt eine M"oglichkeit dar,
masselose Graphen zu berechnen.
Um ein Integral mit der Gegenbauer-Polynom-Technik zu berechnen, werden 
die Propagatoren durch 
ihre Fourier-Transformation ausgedr"uckt.
 Die Fourier-Transformation eines Propagators ist durch
\be \label{fourtrans}
\frac{1}{k^{2 \alpha}} 
      = \frac{\Gamma( \lambda + 1 - \alpha)}{\pi^{\lambda+1} \Gamma(\alpha)}
                \int \frac{e^{2 i k \cdot x} d^dx}{(x^2)^{\lambda + 1- \alpha}}
\ee
gegeben,
wobei  $\lambda = \frac{1}{2} (d-2) = 1  - \ep$ ist.
Anschlie"send lassen sich die Impulsintegrationen mit
\be \label{delt}
\int e^{2 i p \cdot x } d^dp = \pi^d \delta(x)
\ee
ausf"uhren.
Einige der durch die Fourier-Transformation der
Propagatoren eingef"uhrten Integrationen im 
Ortsraum lassen sich trivial mit Hilfe der Deltadistributionen ausf"uhren.
Der Integrand des Feynman-Integrals im $X$-Raum wird in 
Gegenbauer-Polynome entwickelt.
Um die Propagatoren zu entwickeln, verwendet man
\be \label{entw1}
\frac{1}{(x_1 - x_2)^{2 \lambda}} = \frac{1}{({\rm max}(r_1,r_2))^{2 \lambda}} \sum_{n=0}^\infty  
                        C_x^ \lambda (\hat{x}_1 \cdot \hat{x}_2) \left\langle \frac{r_1}{r_2} \right\rangle^n
 \ee
wobei $r = |x| $, $\hat{x} = x/r$ und
\begin{displaymath}
 \left\langle \frac{r_1}{r_2} \right\rangle = \left\{ \begin{array}{cc} 
                   \frac{r_1}{r_2} & r_1 \leq r_2 \\
                   \frac{r_2}{r_1} & r_2 \leq r_1 
                                                    \end{array}
\right .
\end{displaymath}
Die Entwicklung der Exponentialfunktion in Gegenbauer-Polynomen ist
\be \label{entw2}
e^{2 i p \cdot x } = 
\Gamma(\lambda) 
   \sum_{n=0}^\infty i^n (n+\lambda) 
          C_n^\lambda ( \hat{p} \cdot \hat{x} )(p^2 r^2)^{n/2}
             j_{\lambda +n}(p^2 r^2) \;,
\ee
wobei $j_\alpha$ die Potenzreihenentwicklung
\be
j_\alpha(z) = \sum_{n=0}^\infty \frac{(-z)^n}{n! \, \Gamma(n+ \alpha +1) }
\ee
besitzt und durch die Bessel-Funktionen $J_\alpha$ ausgedr"uckt werden kann
\be
J_\alpha(z) = \left(\frac{1}{2} z \right)^\alpha 
     j_\alpha \left(\frac{1}{4} z^2 \right) \;.
\ee
Gegenbauer-Polynome sind durch eine erzeugende-Funktion definiert:
\be
(1- 2 r^2 t  + r^4)^{-\lambda} = \sum _{n=0}^\infty C_n^\lambda(t) r^{2 n}\;.
\ee
Sie stellen eine Verallgemeinerung der Legendre-Polynome ($C_n^{1/2} = P_n$) und der Chebyshev-Polynome
($C^1_n = U_n$) dar.
Einige Spezialf"alle sind:
\be
\begin{array}{rclrcl} \label{spezgeg}
C_0^\lambda(x) &=& 1 \;, & C_1^\lambda(x) &=& 2 \lambda x \; , \\
C_2^\lambda(x) &=& - \lambda + 2 \lambda(\lambda +1)x^2 \; ,&
C_3^\lambda(x) &=& - 2 \lambda(\lambda +1)x 
                + \frac{4}3 \lambda (\lambda +1)(\lambda +2)x^3 \;.  
\end{array}
\ee  
Au"serdem gilt:
\ba
\label{BX}
C_n^\lambda(1) &=& \frac{\Gamma(n+2\lambda)}{n! \, \Gamma(2 \lambda)} \;,
\nn \\
C_n^\lambda(0) &=& \left\{ \begin{array}{cl}
 0 & n \quad {\rm \qquad ungerade} \\
\frac{(-1)^m \Gamma(\lambda + m)}{m! \, \Gamma(\lambda)} & n=2m  \quad
{\rm gerade}\;.
\end{array} \right.
\ea
Nach der Entwicklung des Integranden in Gegenbauer-Polynome 
faktorisiert sich der Integrand in einen
winkelabh"angigen und einen radialabh"angigen Anteil. 
Das Integrationsma"s l"asst sich als
\be
d^dx =  \Omega_d r^{2\lambda+1} d r d \hat{x}
\ee
schreiben, wobei $\Omega_d$ die Fl"ache einer Sph"are mit Radius $1$ im 
$d$-dimensionalen Raum bezeichnet.
Die Winkelintegrationen lassen sich mit der 
Orthogonalit"atsrelation der Gegenbauer-Polynome ausf"uhren
\be \label{orthogrel}
\int C_n^\lambda(\hat{x}_1 \cdot \hat{x}_2) d \hat{x}_2 
        C_m^\lambda(\hat{x}_2 \cdot \hat{x}_3) = \frac{\lambda}{n + \lambda} \delta_{nm} 
                  C_n^\lambda (\hat{x}_1 \cdot \hat{x}_3)   \;.
\ee
Um die Radialintegrationen auszuf"uhren,
setzt man $d$ in die komplexe Ebene fort und wertet die 
einzelnen Integrale in ihrem jeweiligen Konvergenzbereich aus.
Das wichtigste radiale Integral ist hierbei 
\be \label{mainradint}
\int_0^\infty z^b j_a(z) dz= \frac{\Gamma(b+1)}{\Gamma(a-b)}
\ee
f"ur ${\rm Re} (b) > -1$ und $ {\rm Re }(a) >2 { \rm Re} (b) + \frac{1}{2}$.

\chapter{Das Ein-Schleifen-Integral}\label{A1loop}
Um das Ein-Schleifen-Integral $I^{(\mu_1 \dots \mu_n)}(q^2)$ 
aus Gl.~(\ref{einschleifint})
zu berechnen, werden
die Propagatoren mit Gl.~(\ref{fourtrans}) Fourier-transformiert.
Es ergibt sich 
\ba
I^{\ind}(q^2) &=&\int \frac{d^d\!p}{(2 \pi)^d}\frac{p^\ind}{p^{2 \alpha}(p-q)^{2 \beta}}  \nn \\
&=& c(\alpha,\beta) \int \frac{d^d\!p}{(2 \pi)^d} p^{\ind} \int d^d\!x \, d^d \!y 
         \frac{e^{2 i p \cdot x} e^{2 i (p-q) \cdot y}}{x^{2(\lambda +1 - \alpha)}
          y^{2(\lambda +1 - \beta)}} \;,
\ea
wobei 
\be
c(\alpha,\beta) = \frac{ \Gamma(\lambda +1 - \alpha)\Gamma(\lambda +1 - \beta)}
                  {\pi^d \Gamma(\alpha) \Gamma(\beta)}
\ee
ist.
Der Tensor  $p^{(\mu_1 \dots \mu_n)}$  l"asst sich
als Ableitung der Exponentialfunktion schreiben, 
die mit der $p$-Integration vertauscht werden kann. 
\be
I^{\ind}(q^2) =c(\alpha,\beta) \int d^d \!x \,d^d \!y \,
       \frac{e^{- 2 i q \cdot y}}{y^{2( \lambda + 1 - \beta)}} 
  \frac{1}{x^{2(\lambda + 1 - \alpha )}} \frac{\partial^\ind_x}{(2 i)^n}
   \int \frac{d^d\!p}{(2 \pi)^d} e^{2 i (x+y) \cdot p }
\ee
Die $p$-Integration l"asst sich mit Hilfe von Gl.~(\ref{delt}) ausf"uhren. 
Mit einer $n$-fachen partiellen Integration ergibt sich 
\be
I^{\ind}(q^2) = 
\frac{ c(\alpha, \beta) }{i^n 2^d} \int d^d x d^d y 
        \frac{e^{- 2 i q \cdot y }     }{y^{2(\lambda +1 - \beta)}} 
         \left( \frac{ \partial^{\ind}_x}{(-2)^n}  \frac{1}{x^{2(\lambda+1-\alpha)}} \right) 
           \delta(x+y) \;.
\ee
Man rechnet nach, dass
\be \label{tensabl}
\frac{(-1)^n}{2^n} \left( \partial_x^{\ind} \frac{1}{x^{2\alpha}} \right) =
    \frac{\Gamma(n+\alpha)}{\Gamma(\alpha)} 
               \frac{x^{\ind}}{x^{2(n+\alpha)}} \;.
\ee
So ergibt sich:
\be 
I^{\ind}(q^2) = \frac{ (-i)^n c(\alpha,\beta)}{2^d} 
         \frac{\Gamma(\lambda+1- \alpha +n)}
                 {\Gamma(\lambda + 1 - \alpha )}
           \int d^d\!x \frac{e^{2 i q \cdot x } x^\ind }{x^{2( 2(\lambda + 1) 
                       -\alpha - \beta + n)}} \;.
\ee
Der symmetrische Tensor $ x^\ind$ l"asst sich als Ableitung nach dem "ausseren Impuls
$q$ schreiben, die mit der $d^d\!x$-Integration vertauscht weden kann.
\be
\frac{ (-i)^n c(\alpha,\beta)}{2^d} \frac{\Gamma(\lambda+1- \alpha +n)}
                 {\Gamma(\lambda + 1 - \alpha )}
                   \frac{\partial_q^\ind}{(2 i)^n}  
                 \int d^d\! x \frac{e^{2 i x \cdot q }}{x^{2( 2(\lambda + 1) 
                       -\alpha - \beta + n)}} 
\ee
Die Fourier-R"ucktransformation in den 
Impulsraum l"asst sich wieder mit Gl.~(\ref{fourtrans})
ausf"uhren.
\be
\frac{ c(\alpha,\beta)}{2^d} \frac{\Gamma(\lambda+1- \alpha +n)}
                 {\Gamma(\lambda + 1 - \alpha )}
                 \frac{\pi^{(\lambda +1)} 
             \Gamma(- \lambda -1 -n + \alpha + \beta)}
                         {\Gamma( 2 (\lambda+1) - \alpha - \beta + n )}
              \frac{\partial_q^\ind}{(-2)^n} 
                \frac{1}{q^{2(-\lambda-1-n + \alpha + \beta)}}     
\ee
Mit Gl.~(\ref{tensabl}) ergibt sich schlie"slich f"ur das 
Ein-Schleifen-Integral
\ba
I^\ind(q^2) &=& \frac{\pi^{\lambda +1}}{2^d}  c(\alpha,\beta)
            \frac{\Gamma(\lambda+1- \alpha +n)}
                 {\Gamma(\lambda + 1 - \alpha )} \times \nn  \\
     &&  \times   \frac{ \Gamma(- \lambda -1 -n + \alpha + \beta)}
                 {\Gamma( 2 (\lambda+1) - \alpha - \beta + n )}  
            \frac{ \Gamma(- \lambda - 1 + \alpha + \beta )}
                { \Gamma(- \lambda - 1 -n + \alpha + \beta )} 
            \;  \frac{1}{q^{2( - \lambda -1 + \alpha + \beta )}} \nn \\
 &=&  \frac{1}{(4 \pi)^{2-\ep} }
   B(\lambda +1 +n - \alpha,  \lambda + 1 - \beta) 
   \frac{ \Gamma( -\lambda - 1 + \alpha + \beta)}{\Gamma(\alpha) \Gamma(\beta)}
   \frac{1}{q^{2( - \lambda -1 + \alpha + \beta )}}   \nn \\
  &=:& \frac{1}{(4 \pi)^{2-\ep}} G^{(n)}(\alpha,\beta) \,.
\ea

\chapter{Das Zwei-Schleifen-Master-Integral}\label{A2loop}
Das Zwei-Schleifen-Master-Integral l"asst sich mit der GPXT-Technik (Anhang \ref{AGPXT}) 
berechnen. 
Die Rechnung ist f"ur $\alpha = \beta = \gamma = \delta = \eta = 1$ skiziert.
\be 
F = \int \frac{d^d\!p \, d^d\!k}{(2 \pi)^{2d}} 
            \frac{1}{ p^2 (q-p)^2 k^2 (q-k)^2 (p-k)^2}
\ee
Die Propagatoren werden mit Gl.~(\ref{fourtrans}) Fourier-transformiert.
\be
F \sim \int \left(\prod_{i=0}^5 d^d\!x_i \right) 
            e^{2 i q \cdot (x_2 + x_4) }  \int d^d\!p\, d^d\!k\, 
            \frac{ e^{2 i p \cdot (x_1 - x_2 + x_5 )} 
               e^{2 i k \cdot (x_3 - x_4  - x_5)} }
           {x_1^{2 \lambda} x_2^{2 \lambda} x_3^{2 \lambda} x_4^{2 \lambda} 
               x_5^{2 \lambda}} \;. 
\ee
Die Integrationen "uber $p$ und $k$ lassen sich mit Gl.~(\ref{delt}) 
ausf"uhren. 
Mit den entstehenden $\delta$-Distributionen 
lassen sich die Integrationen "uber $x_4$ und $x_5$ ausf"uhren.
Durch Ausnutzung der Translationsinvarianz 
$x_3 \rightarrow x_3 - x_1$  ergibt sich
\be
F \sim \int 
   \frac{d^d \!x_1 \, d^d\!x_2 \;d^d\!x_3 \,e^{2 i q \cdot x_3} }
           {x_1^{2\lambda} x_2^{2\lambda} (x_1 - x_2 )^{2\lambda}
    (x_1 - x_3 )^{2\lambda}    (x_2 - x_3 )^{2\lambda}} \: .
\ee
Dies l"asst sich mit Gl.~(\ref{entw1}, \ref{entw2}) 
in Gegenbauerpolynomen entwickeln 
\ba
F &\sim&  \sum_{r=0}^\infty \sum_{s=0}^\infty \sum_{t=0}^\infty \sum_{n=0}^\infty  p^r\int 
     dr_1\,dr_2\,dr_3\, 
               r_1^{2 \lambda+1} \,  r_2^{2 \lambda+1} \,  r_3^{2 \lambda+1} \times  \nn \\
  && \times \frac{r_3^r \left\langle \frac{r_1}{r_2} 
            \right\rangle^s \left\langle \frac{r_1}{r_3} \right\rangle^n
            \left\langle \frac{r_2}{r_3} \right\rangle^t}
{r_1^{2\lambda} r_2^{2 \lambda} 
     ( \max(r_1,r_2))^{2 \lambda}  ( \max(r_1,r_3))^{2 \lambda} 
           ( \max(r_2,r_3))^{2 \lambda}}  
       \;  j_{\lambda + r } (p^2 r_3^2) \times  \nn \\
&& \times \int d \hat{x}_1\, d \hat{x}_2\, d \hat{x}_3 \,
     C_s^\lambda(\hat{x}_1 \cdot \hat{x}_2) C_n^\lambda(\hat{x}_1 \cdot \hat{x}_3) C_t^\lambda(\hat{x}_2 \cdot \hat{x}_3)
      C_r^\lambda(\hat{x}_3 \cdot \hat{p}) \;.
\ea
In dem in Gegenbauerpolynome entwickelten Integral lassen sich 
die Radial- und die Winkelintegration 
trennen. 
Die Winkelintegration kann mit der Orthogonalit"atsrelation 
der Gegenbauerpolynome 
Gl.~(\ref{orthogrel}) ausgef"uhrt werden.
F"ur das Winkelintegral ergibt sich so
\ba
&& \int d \hat{x}_1 \,d \hat{x}_2 \,d\hat{x}_3\, 
     C_s^\lambda(\hat{x}_1 \cdot \hat{x}_2) 
  C_n^\lambda(\hat{x}_1 \cdot \hat{x}_3) C_t^\lambda(\hat{x}_2 \cdot \hat{x}_3)
      C_r^\lambda(\hat{x}_3 \cdot \hat{p})  \nn \\
&=& \frac{\lambda}{n+\lambda} \delta_{sn} 
    \int d \hat{x}_2 \,d\hat{x}_3\, 
     C_n^\lambda(\hat{x}_2 \cdot \hat{x}_3) 
              C_t^\lambda(\hat{x}_2 \cdot \hat{x}_3)
      C_r^\lambda(\hat{x}_3 \cdot \hat{p})  \nn \\ 
 &=& \left(\frac{\lambda}{n+\lambda}\right)^2 \delta_{sn} \delta_{tn}  
    \int  d \hat{x}_3 
     C_n^\lambda( 1 )
      C_r^\lambda(\hat{x}_3 \cdot \hat{p})  \nn \\ 
 &=& \left(\frac{\lambda}{n+\lambda}\right)^2 
   \delta_{sn} \delta_{tn} \frac{\Gamma(n+2\lambda)}{n!\, \Gamma(2 \lambda)} 
    \int  d \hat{x}_3 
      C_r^\lambda(\hat{x}_3 \cdot \hat{p})  
         C_0^\lambda(\hat{x}_3 \cdot \hat{p}) \nn \\
 &=&    \left(\frac{\lambda}{n+\lambda}\right)^2 
            \delta_{sn} \delta_{tn}  \delta_{r0} 
                     \frac{\Gamma(n+2\lambda)}{n!\,\Gamma(2 \lambda)} \;. 
\ea
Im 3. Schritt wurde Gl.~(\ref{spezgeg}, \ref{BX}) verwendet.
Um die Radialintegration auszuf"uhren, muss der 
Integrationsbereich in 6 Regionen aufgeteilt werden. 
F"ur 
$r_1 < r_3 < r_2$ beispielsweise 
muss das folgende Radialintegral gel"ost werden:
\be \label{ri}
\int_0^\infty dr_3 r_3 j_\lambda(r_3^2) 
       \int _0^{r_3} dr_1 r_1^{2n+1} 
           \int_{r_3}^\infty d r_2 r_2^{-2(n+ 2 \lambda)+1 }
  \quad \quad r_1 < r_3 < r_2 \;,
\ee
wobei $p=1$ gesetzt wurde. 
Die $p$-Abh"angigkeit von $F$ l"asst sich aus der Dimension rekonstruieren.
Gl.~(\ref{ri}) l"asst sich mit Hilfe von Gl.~(\ref{mainradint}) berechnen,
indem in jedem Integrationsgebiet die Dimension $d$ so gew"ahlt wird, 
dass die Integrale
konvergieren und das Resultat dann analytisch fortgesetzt wird.
F l"asst sich berechnen, indem die Radialintegrale f"ur alle 6 Integrationsgebiete bestimmt werden.
Man erh"alt schlie\ss lich
\ba
F &=& (4 \pi)^{-2-2\lambda } 
\; \frac{1}{q^{2(1+2\ep)}} \;
\frac{2 \Gamma^3(\lambda) \Gamma(3-2\lambda)}
              {\Gamma(2 \lambda ) \Gamma(3 \lambda -2)}
        \sum_{n=0}^\infty \frac{\Gamma(n+2\lambda)}{n! (n + \lambda)^2}
       \times         \\
 && \times  \Bigg(\frac{1}{(n+1)(n+2-\lambda)} + 
                 \frac{1}{(n+1)( n + 2 \lambda -1)} + \frac{1}{( n + 2 \lambda -1)( n + 3 \lambda -2)}  \Bigg) \;.\nn
\ea
Nach einer Partialbruchzerlegung l"asst sich die Summe
mit der 
Gau"s'schen Summationsformel
\be 
\sum_{n=0}^\infty \frac{\Gamma(n+a) }{n! \, (n+b)} 
     = \frac{\Gamma(a) \Gamma(b) \Gamma(1-a)}{\Gamma(b+1-a)} 
\ee
aufsummieren. F"ur F ergibt sich so die in Gl.~(\ref{falbt}) angegebene Formel
f"ur $\al =\beta=1$ in "Ubereinstimmung mit \cite{GPXT}.

\chapter{Rekursionsbeziehung f"ur das Zwei-Schleifen-Integral} \label{ARb}
Ausgangspunkt f"ur die Herleitung der in Abb.~(\ref{rekursion})
dargestellten Rekursionsbeziehung ist die 
Identit"at
\be
0 = \int \frac{d^d \!p \, d^d \!k}{(2 \pi)^{2 d}} 
         \frac{\partial}{\partial p^\mu} 
              \left( \frac{(p-k)^\mu}{p^2 (p-q)^2 k^2(k-q)^2(p-k)^2} \right)
     =  \int \frac{d^d \!p\, d^d\!k}{(2 \pi)^{2 d}} I(p,k,q) \; .
\ee
Um die Differentation auszuf"uhren, muss man beachten, dass
\be
\frac{\partial}{\partial p^\mu} p^\mu = d = 4 - 2 \ep \quad {\rm und}\quad
\frac{\partial}{\partial p^\mu} \frac{1}{p^2}  = \frac{- 2 p_\mu}{p^4}  
\ee
ist.
Mit der Abk"urzung 
\be
N = p^2 (p-q)^2 k^2 (k-q)^2 (p-k)^2
\ee
ergibt sich f"ur den Integranden
\ba
I(p,k,q)&=& d \frac{1}{N}  -2 (p-k)^\mu \left(\frac{p_\mu}{p^2 N} 
+ \frac{(p-q)_\mu }{(p-q)^2 N} + \frac{(p-k)_\mu}{(p-k)^2 N} \right) \nn \\
&=& d \frac{1}{N} -2 \left( \frac{2}{N} - \frac{p \cdot k}{p^2 N} + \frac{(p-k) \cdot (p-q)}{(p-q)^2 N}\right)\;.  
\ea   
Die Z"ahler lassen sich quadratisch erg"anzen
\ba
p \cdot k &=& -\frac{1}{2} \left( (p-k)^2 -p^2 -k^2  \right) \nn \\ 
(p-k) \cdot (p-q) &=& \frac{1}{2} \left( (p-k)^2 + (p-q)^2 -(k-q)^2 \right)\;.
\ea
K"urzen ergibt
\ba
I(p,k,q) &=&- 2 \ep \frac{1}{N} + \frac{1}{p^4 (p-q)^2(k-q)^2(p-k)^2} - \frac{1}{p^4 (p-q)^2 k^2 (k-q)^2}  \nn \\
&& -  \frac{1}{p^2 (p-q)^4 k^2 (k-q)^2}  + \frac{1}{p^2(p-q)^4 k^2 (p-k)^2} \;.
\ea
Die Translationsinvarianz der dimensionalen Regularisierung l"asst sich ausnutzen, um $I(p,k,q)$ weiter zu 
vereinfachen:
F"ur $p \rightarrow p +q $,  $q \rightarrow -q$ und $k \rightarrow k-q$ im zweiten Summanden 
\be
 \frac{1}{p^4 (p-q)^2(k-q)^2(p-k)^2} \rightarrow \frac{1}{p^2(p-q)^4 k^2 (p-k)^2}
\ee
und im dritten Summanden 
\be 
 \frac{1}{p^4 (p-q)^2 k^2 (k-q)^2} \rightarrow  \frac{1}{p^2 (p-q)^4 k^2 (k-q)^2}
\ee
ergibt sich
\ba
I_{\rm{trans}}(p,k,q) &=& 2 \, \Bigg(-\ep \frac{1}{ p^2 (p-q)^2 k^2 
                              (k-q)^2 (p-k)^2} \nn \\ 
&&- \frac{1}{p^2 (p-q)^4 k^2 (k-q)^2}
          + \frac{1}{p^2(p-q)^4 k^2 (p-k)^2} \Bigg) \;.
\ea 
Die Transformation $q \rightarrow -q$ ist m"oglich, 
da das Integral nur von $q^2$ abh"angen darf.
Integration von $I_{\rm{trans}}(p,k,q)$ ergibt 
die in  Abb.~\ref{rekursion} dargestellte Identit"at in "Ubereinstimmung mit \cite{partint}.
\ba
&& \ep  \int \frac{d^d \!p \,d^d \!k}{(2 \pi)^{2 d}}  \frac{1}{ p^2 (p-q)^2 k^2 (k-q)^2 (p-k)^2} \nn \\ 
&=& \int \frac{d^d \!p \, d^d\!k}{(2 \pi)^{2 d}} \frac{1}{p^2(p-q)^4 k^2 (p-k)^2} - \int \frac{d^d \!p \,d^d\!k}{(2 \pi)^{2 d}} 
        \frac{1}{p^2 (p-q)^4 k^2 (k-q)^2}\;.
\ea
\end{appendix} 

\backmatter

\listoffigures
\chapter{Danksagung}
Ich m"ochte allen danken, die mich 
beim Schreiben meiner Diplomarbeit 
unterst"utzt haben, 
besonders
\begin{itemize}
\item Prof. Dr. J.G.~K"orner
\item Prof. Dr. A.A.~Pivovarov
\item Dr. S.~Groote
\item I.~Bierenbaum, K.~Farouqi, A.~Holfter und J.~Maul
\item meinen Eltern
\end{itemize} 
                 

\begin{thebibliography}{99}
\bibitem{PDG}Particle Data Group, Review of Particle Properties,\\
Eur. \ Phys. \ J. \ C3 (1998) 1.
\bibitem{exp1al}ALEPH  collaboration, Z.\  Phys.\ C76 (1997) 15;
Eur. \ Phys. \ J. \ C4 (1998) 409. 
\bibitem{exp1ms}ALEPH  collaboration,  
CERN-EP/99-026.
\bibitem{exp2}OPAL collaboration, Eur.\ Phys.\ J. \  C7 (1999) 571.
\bibitem{cont}C.~Bernard, A.~Duncan, J.~LoSecco and S.~Weinberg,
Phys.\ Rev. \ D12 (1975) 792;\\ 
E.~Poggio, H.~Quinn and S.~Weinberg,
Phys. \ Rev. \ D13 (1976) 1958.
\bibitem{cont1}R.~Shankar, Phys. \ Rev. \ D15 (1977) 755.
\bibitem{cont2}K.G.~Chetyrkin, N.V.~Krasnikov and A.N.~Tavkhelidze,
Phys.\ Lett. \ 76B (1978) 83;
N.V.~Krasnikov and A.A.~Pivovarov,
Phys.\ Lett.\ B112 (1982) 397;
N.V.~Krasnikov, A.A.~Pivovarov and N.N.~Tavkhelidze,
JETP Lett.\ 36 (1982) 333;
Z.\ Phys.\ C19 (1983) 301;
A.~A.~Penin and A.A.~Pivovarov, Phys.\ Lett.\  B357 (1995) 427.

\bibitem{SchTra84} 
K.~Schilcher and M.D.~Tran, Phys.\ Rev.\ D29 (1984) 570.
\bibitem{Bra88}
E.~Braaten, Phys.\ Rev.\ Lett.\ 53 (1988) 1606.
\bibitem{Bra89} 
E.~Braaten, Phys.\ Rev.\ D39 (1989) 1458.
\bibitem{NarPic88}
S.~Narison and A.~Pich,  Phys. \ Lett. \ B211 (1988) 183.
\bibitem{BraNarPic92} 
E.~Braaten, S.~Narison and A.~Pich, Nucl.\ Phys.\ B373 (1992) 581.
\bibitem{one}J.G.~K\"orner, F.~Krajewski and A.A.~Pivovarov,
Eur. \ Phys. \ J. \ C12 (2000) 461.
\bibitem{two}J.G.~K\"orner, F.~Krajewski and A.A.~Pivovarov,
Eur. \ Phys. \ J. \ C14 (2000) 123.  
\bibitem{three}J.G.~K\"orner, F.~Krajewski and A.A.~Pivovarov,
Phys. Rev. D63 (2001) 036001,\\ hep-ph/0002166.
\bibitem{four}J.G.~K\"orner, F.~Krajewski and A.A.~Pivovarov,
Eur.\ Phys.\ J.\ C20 (2001) 259,\\ hep-ph/0003165.
\bibitem{renRS}N.V.~Krasnikov and A.A.~Pivovarov, 
Mod. \ Phys. \ Lett. A11 (1996) 835; 
Phys.\ Atom.\ Nucl. 64 (2001) 1500;
S.~Groote, J.~G.~K\"orner and A.~A.~Pivovarov,
Phys. Rev. D65 (2002) 036001;
A.~A.~Pivovarov, [arXiv:hep-ph/0104213].

\bibitem{pivrho}A.A.~Pivovarov,
Nucl.\ Phys.\ Proc.\ Suppl.\   64 (1998) 339 [hep-ph/9708461],
Phys. Atom. Nucl. 62 (1999) 1924,
Sov.\ J.\ Nucl.\ Phys.\ 52 (1990) 372;
A.A.~Pivovarov and A.S.~Zubov,
Phys.\ Atom.\ Nucl.\ 63 (2000) 1650.
\bibitem{Pivtau}A.A.~Pivovarov, Sov.\ J. \ Nucl.Phys. \ 54 (1991) 676;
Z. \ Phys. C53 (1992) 461; \\ 
Nuovo \ Cim. 105A (1992) 813;
N.V.~Krasnikov and A.A.~Pivovarov,
Phys.\ Lett.\  B116 (1982) 168.
\bibitem{DP} F.~Le Diberder and A.~Pich,
Phys. \ Lett. \ B286 (1992) 147.
\bibitem{marc} W.M.~ Marciano and H.~Pagels, Phys. \ Rep. \ 36C (1978) 137.
\bibitem{pessch} M.E.~Peskin and D.V.~Schroeder, 
An introduction to quantum field 
theory, Addison-Wesley (1995).
\bibitem{coll} J.~Collins, Renormalisation, Cambridge University Press (1984).
\bibitem{beta4}T. \ van \ Ritbergen, J.A.M.~Vermaseren and S.A.~Larin,  
Phys. \ Lett. \ B400 (1997) 379.
\bibitem{gamma3c}K.G.~Chetyrkin,  Phys. \ Lett. \ B404 (1997) 161. 
\bibitem{gamma3v}T.~van Ritbergen, J.A.M.~Vermaseren and S.A.~Larin,  
Phys. \ Lett. \ B405 (1997) 327.
\bibitem{watermellon}S.~Groote, J.~G.~Korner and A.~A.~Pivovarov,
Eur.\ Phys.\ J.\   C11 (1999) 279,
Nucl.\ Phys.\  B542  (1999) 515,
Phys.\ Lett.\  B443 (1998) 269;
S.~Groote and A.~A.~Pivovarov,
Nucl. Phys. B580 (2000) 459;
S.~Narison and A.A.~Pivovarov,
Phys.\ Lett.\  B327  (1994) 341 [hep-ph/9403225].
\bibitem{GPXT} K.G.~Chetyrkin, 
A.L.~Kataev and F.V.~Tkachov, Nucl. \ Phys. B174 (1980) 345.
\bibitem{mut} T.~Muta, Foundations of Quantum Chromodynamics, 
World Scientific Lecture Notes in Physics, Vol. 5 
\bibitem{partint} K.G.~Chetyrkin and F.V.~Tkachov, 
Nucl. \ Phys. B192 (1981) 159.
\bibitem{RHarl} R.~Harlander, Acta \ Phys. \ Polon. \ B30 (1999) 3443,
hep-ph/9910496.   
\bibitem{eek20}S.G.~Gorishny, A.L.~Kataev and S.A.~Larin,
Phys. \ Lett. \ B259 (1991) 144.
\bibitem{eek21}L.R.~Surguladze and M.A.~Samuel, 
Phys. \ Rev. \ Lett. \ 66 (1991) 560, 2416(E),\\
Phys.\ Rev. \ D44 (1991) 1602.
\bibitem{eek2c}K.G.~Chetyrkin,  Phys. \ Lett. \ B391 (1997) 402.
\bibitem{eek2d}K.G.~Chetyrkin,  Phys. \ Lett. B390 (1997) 309.
\bibitem{eek2e}K.G.~Chetyrkin and J.H.~K\"uhn,  Phys.Lett. B406 (1997) 102. 
\bibitem{effsch}
G.~Grunberg, Phys. \ Lett. \ B95 (1980) 70, Erratum-ibid. B110 (1982) 501. 
\bibitem{HipPaket} A.~Hsieh and E.~Yehudai, Comput. \ Phys.\ 6 (1992) 253
\bibitem{bodrell} J.D.~Bjorken and S.D.~Drell, 
Relativistische Quantenfeldtheorie,\\ B.I.-Hochschultaschenbuch Band 101.
\bibitem{ksch}
N.V.~Krasnikov, Nucl. \ Phys. \ B192 (1981) 497.
\bibitem{kksch}A.L.~Kataev, N.V.~Krasnikov and A.A.~Pivovarov,
Phys. \ Lett. \ B107 (1981) 115;\\ 
Nucl. \ Phys. \ B198 (1982) 508; 
A.A.~Pivovarov,
Phys.\ Atom.\ Nucl.\ 63 (2000) 1646

\bibitem{effDh}A.~Dhar and V.~Gupta, Phys. \ Rev. \ D29 (1984) 2822. 
\bibitem{brodsky}
S.J.~Brodsky, J.R.~Pelaez and N.~Toumbas,
Phys. \ Rev. \  D60 (1999) 037501.
\bibitem{prl}S.~Groote, J.G.~K\"orner, A.A.~Pivovarov and 
K.~Schilcher,
Phys. \ Rev. \ Lett. \ 79 (1997) 2763.
\bibitem{matching} K.G.~Chetyrkin, B.A.~Kniehl, M.~Steinhauser, 
Phys.\ Rev.\ Lett.\ 79 (1997) 2184.  
\bibitem{bbmass}A.A.~Penin, A.A.~Pivovarov, 
Nucl.\ Phys.\  B549 (1999) 217,
Phys.\ Lett.\  B443 (1998) 264,
Phys.\ Lett.\  B435  (1998) 413;
J.H.~K\"uhn, A.A.~Penin and A.A.~Pivovarov,
Nucl.\ Phys.\  B534 (1998) 356;
A.A.~Pivovarov, Phys. Lett. B475 (2000) 135.
\bibitem{groote}S.~Groote, J.G.~K\"orner and A.A.~Pivovarov,
Phys. \ Lett. \ B407 (1997) 66;
Mod. \ Phys. \ Lett. \ A13 (1998) 637.
\bibitem{pichprades}A.~Pich and J.~Prades,  
JHEP 9806:013, (1998); ibid 9910:004, (1999). 
\bibitem{wilczek}F.~Wilczek, 
IASSNS-HEP-99-64, Jun 1999, 
hep-ph/9907340. 
\bibitem{Zakh}V.I.~Zakharov, Prog. \ Theor. \ Phys. \ Suppl. \ 131 (1998) 107.
\bibitem{grootecomp}S.~Groote, ``Compendium on Adler's function",
nicht ver\"offentlichte  pers\"onliche Aufzeichnungen.
\bibitem{stevenson}P.M.~Stevenson, Phys. \ Rev. D23 (1981) 2916;
A.~A.~Penin and A.~A.~Pivovarov,
Phys.\ Lett.\  B367 (1996) 342.

\bibitem{stevenson1}J.~Kubo, S.~Sakakibara and P.M.~Stevenson,
Phys. \ Rev.\ D29 (1984) 1682. 
\bibitem{ewcorr1}W.J.~Marciano and A.~Sirlin, 
Phys. Rev. Lett. 61 (1988) 1815.
\bibitem{ewcorr2}E.~Braaten and C.S.~Li, Phys. \ Rev. \ D42 (1990) 3888. 

\bibitem{HighPres}  S.~Groote, J.G.~K\"orner and A.A.~Pivovarov
Eur. Phys. J. C24 (2002) 393,
A.A.~Pivovarov,
hep-ph/0110249, 
Phys. Atom. Nucl. 65 (2002) 1319
 J.G.~K\"orner, A.A.~Pivovarov and K.~Schilcher 
Eur. Phys. J. C9 (1999) 551.
\bibitem{narga}
C.~Becchi, S.~Narison, E.~de Rafael and F.J.~Yndurain,
Z. \ Phys. \ C8 (1981) 335. 
\bibitem{gammass}
A.A.~Ovchinnikov and A.A.~Pivovarov, Phys. \ Lett. \ B163 (1985) 231;\\
N.V.~Krasnikov and A.A.~Pivovarov,  Nuovo \ Cim. \ 81A (1984) 680.
\bibitem{newgam}Y.~Chung et al. Z.\ Phys.  C25 (1984) 151;
H.G.~Dosch, M.~Jamin and S.~Narison, Phys. Lett. B220 (1989) 251.
\bibitem{msNPB}K.G.~Chetyrkin, J.H.~K\"uhn and A.A.~Pivovarov,
Nucl. \ Phys. \ B533 (1998) 473.
\bibitem{narma}
S.~Narison and E.~de Rafael, Phys. \ Lett. \ B103 (1981) 57.
\bibitem{gassLeut}
J.~Gasser and H.~Leutwyler, Phys. \ Rept. \ 87 (1982) 77. 
\bibitem{oldmass}A.L.~Kataev, N.V.~Krasnikov and A.A.~Pivovarov,
Phys. \ Lett. \ B123 (1983) 93; 
Nuovo \ Cim.\ 76A (1983) 723.
\bibitem{oldmass1}S.G.~Gorishnii, A.L.~Kataev and S.A.~Larin, 
Phys. \ Lett. \ B135 (1984) 457.
\bibitem{recent1}M.~Jamin and M.~M\"unz, Z.\ Phys. \ C66 (1995) 633.
\bibitem{recent2}K.G.~Chetyrkin, D.~Pirjol and K.~Schilcher,
Phys. \ Lett. \ B404 (1997) 337. 
\bibitem{lattice1}
ALPHA and UKQCD Collaboration (Joyce Garden et al.), hep-lat/9906013.
\bibitem{lattice2}
ALPHA Collaboration (Marco Guagnelli et al.), 
Nucl. \ Phys. \ B560 (1999) 465.
\end{thebibliography}
\end{document}